\documentclass[twocolumn]{aastex63}
\usepackage{placeins}
\usepackage{amsmath}
\newcommand{\kms}{\mbox{km\,s$^{-1}$}}
\newcommand{\kkms}{\mbox{K\,km\,s$^{-1}$}}

\newcommand{\mjypb}{\mbox{mJy\,beam$^{-1}$}}
\newcommand{\mlr}{\mbox{M$_\sun$\,yr$^{-1}$}}
\newcommand{\fco}{\mbox{M$_\sun$\,km\,s$^{-1}$\,yr$^{-1}$}}

\received{...., 2023}
\revised{..., 2023}
\submitjournal{ApJS}

\shorttitle{Episodic Molecular Jets}
\shortauthors{Dutta et al.}

\begin{document}


\title{Episodic Accretion in Protostars - An ALMA Survey of Molecular Jets in the Orion Molecular Cloud}

\correspondingauthor{Somnath Dutta}
\email{sdutta@asiaa.sinica.edu.tw}

\author[0000-0002-2338-4583]{Somnath Dutta}
\affiliation{Institute of Astronomy and Astrophysics, Academia Sinica, Roosevelt Rd, Taipei 10617, Taiwan, R.O.C.}

\author[0000-0002-3024-5864]{Chin-Fei Lee}
\affiliation{Institute of Astronomy and Astrophysics, Academia Sinica, Roosevelt Rd, Taipei 10617, Taiwan, R.O.C.}

\author[0000-0002-6773-459X]{Doug Johnstone}
\affiliation{National Research Council of Canada, Herzberg, Astronomy and Astrophysics Research Centre, 5071 West Saanich Road, V9E 2E7 Victoria (BC), Canada}
\affiliation{Department of Physics and Astronomy, University of Victoria, Victoria, BC V8P 5C2, Canada}

\author[0000-0003-3119-2087]{Jeong-Eun Lee}
\affiliation{Department of Physics and Astronomy, Seoul National University, 1 Gwanak-ro, Gwanak-gu, Seoul 08826, Korea}

\author[0000-0001-9304-7884]{Naomi Hirano}
\affiliation{Institute of Astronomy and Astrophysics, Academia Sinica, Roosevelt Rd, Taipei 10617, Taiwan, R.O.C.}

\author{James Di Francesco}
\affiliation{National Research Council of Canada, Herzberg, Astronomy and Astrophysics Research Centre, 5071 West Saanich Road, V9E 2E7 Victoria (BC), Canada}
\affiliation{Department of Physics and Astronomy, University of Victoria, Victoria, BC V8P 5C2, Canada}

\author{Anthony Moraghan}
\affiliation{Institute of Astronomy and Astrophysics, Academia Sinica, Roosevelt Rd, Taipei 10617, Taiwan, R.O.C.}

\author[0000-0002-5286-2564]{Tie Liu}
\affiliation{Shanghai Astronomical Observatory, Chinese Academy of Sciences, 80 Nandan Road, Shanghai 200030, China}

\author[0000-0002-4393-3463]{Dipen Sahu}
\affiliation{Physical Research laboratory, Navrangpura, Ahmedabad, Gujarat 380009, India}
\affiliation{Academia Sinica Institute of Astronomy and Astrophysics, 11F of AS/NTU Astronomy-Mathematics Building, No.1, Sec. 4, Roosevelt Rd, Taipei 10617, Taiwan, R.O.C.}

\author[0000-0003-4603-7119]{Sheng-Yuan Liu}
\affiliation{Institute of Astronomy and Astrophysics, Academia Sinica, Roosevelt Rd, Taipei 10617, Taiwan, R.O.C.}

\author[0000-0002-8149-8546]{Ken'ichi Tatematsu}
\affil{Nobeyama Radio Observatory, National Astronomical Observatory of Japan, 
National Institutes of Natural Sciences, 
462-2 Nobeyama, Minamimaki, Minamisaku, Nagano 384-1305, Japan}
\affiliation{Department of Astronomical Science,
SOKENDAI (The Graduate University for Advanced Studies),
2-21-1 Osawa, Mitaka, Tokyo 181-8588, Japan}

\author{Chang Won Lee}
\affil{Korea Astronomy and Space Science Institute (KASI), 776 Daedeokdae-ro, Yuseong-gu, Daejeon 34055, Republic of Korea}
\affil{University of Science and Technology, Korea (UST), 217 Gajeong-ro, Yuseong-gu, Daejeon 34113, Republic of Korea}

\author[0000-0003-1275-5251]{Shanghuo Li}
\affiliation{Korea Astronomy and Space Science Institute (KASI), 776 Daedeokdae-ro, Yuseong-gu, Daejeon 34055, Republic of Korea}

\author{David Eden}
\affiliation{Astrophysics Research Institute, Liverpool John Moores University, IC2, Liverpool Science Park, 146 Brownlow Hill, Liverpool, L3 5RF, UK}

\author[0000-0002-5809-4834]{Mika Juvela}
\affiliation{Department of Physics, P.O.Box 64, FI-00014, University of Helsinki, Finland}

\author[0000-0002-9574-8454]{Leonardo Bronfman}
\affil{Departamento de Astronomía, Universidad de Chile, Casilla 36-D, Santiago, Chile}

\author{Shih-Ying Hsu}
\affiliation{Institute of Astronomy and Astrophysics, Academia Sinica, Roosevelt Rd, Taipei 10617, Taiwan, R.O.C.}

\author[0000-0003-2412-7092]{Kee-Tae Kim}
\affil{Korea Astronomy and Space Science Institute (KASI), 776 Daedeokdae-ro, Yuseong-gu, Daejeon 34055, Republic of Korea}
\affil{University of Science and Technology, Korea (UST), 217 Gajeong-ro, Yuseong-gu, Daejeon 34113, Republic of Korea}

\author{Woojin Kwon}
\affil{Department of Earth Science Education, Seoul National University, 1 Gwanak-ro, Gwanak-gu, Seoul 08826, Republic of Korea}
\affil{Korea Astronomy and Space Science Institute, 776 Daedeokdae-ro, Yuseong-gu, Daejeon 34055, Republic of Korea}

\author[0000-0002-7125-7685]{Patricio Sanhueza} 
\affiliation{National Astronomical Observatory of Japan, National Institutes of Natural Sciences, 2-21-1 Osawa, Mitaka, Tokyo 181-8588, Japan}
\affiliation{Department of Astronomical Science,
SOKENDAI (The Graduate University for Advanced Studies),
2-21-1 Osawa, Mitaka, Tokyo 181-8588, Japan}

\author[0000-0002-5845-8722]{Jes{\'u}s Alejandro L{\'o}pez-V{\'a}zquez}
\affiliation{Institute of Astronomy and Astrophysics, Academia Sinica, Roosevelt Rd, Taipei 10617, Taiwan, R.O.C.}

\author{Qiuyi Luo}
\affiliation{Shanghai Astronomical Observatory, Chinese Academy of Sciences, 80 Nandan Road, Shanghai 200030, China}

\author[0000-0003-0537-5461]{Hee-Weon Yi}
\affiliation{Korea Astronomy and Space Science Institute (KASI), 776 Daedeokdae-ro, Yuseong-gu, Daejeon 34055, Republic of Korea}



\begin{abstract}
Protostellar outflows and jets are almost ubiquitous characteristics during the mass accretion phase, and encode the history of stellar accretion, complex-organic molecule (COM) formation, and planet formation.  Episodic jets are likely connected to episodic accretion through the disk. Despite the importance, there is a lack of studies of a statistically significant sample of protostars via high-sensitivity and high-resolution observations. To explore episodic accretion mechanisms and the chronologies of episodic events, we investigated 42 fields containing protostars with ALMA observations of CO, SiO, and 1.3\,mm continuum emission. We detected SiO emission in 21 fields, where 19 sources are driving confirmed molecular jets with high abundances of SiO.  Jet velocities,  mass-loss rates, mass-accretion rates, and periods of accretion events are found to be dependent on the driving forces of the jet (e.g., bolometric luminosity, envelope mass). Next, velocities and mass-loss rates are positively correlated with the surrounding envelope mass, suggesting that the presence of high mass around protostars increases the ejection-accretion activity. We determine mean periods of ejection events of 20$-$175 years for our sample, which could be associated with perturbation zones of $\sim$ 2$-$25\,au extent around the protostars. Also, mean ejection periods are anti-correlated with the envelope mass, where high-accretion rates may trigger more frequent ejection events.  The observed periods of outburst/ejection are much shorter than the freeze-out time scale of the simplest COMs like CH$_3$OH, suggesting that episodic events largely maintain the ice-gas balance inside and around the snowline.



\end{abstract}

\keywords{Star formation(1569), Protostars(1302), Stellar jets(1607), Stellar winds(1636), Stellar accretion(1578), Submillimeter astronomy(1647)}

\section{Introduction}\label{sec:intro}
Together, jets and outflows are very intriguing phenomena that appear during the star formation process. They encode the history of accretion onto the protostar through the disk \citep[e.g.,][]{2007prpl.conf..245A,2014prpl.conf..387A,2014prpl.conf..451F,2020A&ARv..28....1L,2022arXiv220311257F} and the formation of complex organic molecules (COMs), in terms of sublimating icy grain mantles with increasing central luminosity  
\citep[e.g.,][]{2015A&A...579A..23J,2016ApJ...821...46T,2022Natur.606..272J}.
Low-density mass ejections are usually referred to as ``outflows", typically exhibiting low velocities ($\sim$ 1 to a few tens $\kms$ ) and wide angles, and are commonly traced by the $^{12}$CO (2$-$1) emission line. In contrast, high-density ejections along the flow axis are referred to as ``jets". Jets typically exhibit high-velocities ($\sim$ 30  to a few hundred $\kms$) and narrow-angle collimation, mostly concentrated around the flow axis, and are commonly traced by high-density tracers like higher transitions of SiO (e.g., J = 5$-$4, 8$-$7), $^{12}$CO (e.g., J = 2$-$1, 3$-$2) or SO (e.g., J = 8$-$7). For context, the critical density for CO (J = 2$-$1) is n$_{\rm cr, CO}$ $\sim$ (1.0 $-$ 1.2 ) $\times$ 10$^{4}$ cm$^{-3}$ for temperature T = 150 K (our assumed temperature of jet gas) to 50\,K (our assumed temperature of outflow gas), respectively (see Table \ref{tab:citical_density}). The critical density of SiO (J = 5$-$4) at a temperature of 150\,K in the jet is n$_{\rm cr, SiO}$ $\sim$ 1.7 $\times$ 10$^{6}$ cm$^{-3}$, a factor of 100 times higher than CO.

Protostellar jets driven by accreting protostars are often observed as episodic \citep[see reviews by][]{2007prpl.conf..245A,2014prpl.conf..387A,2014prpl.conf..451F,2022arXiv220311257F}. Since mass ejection originates from the disk surface and/or inner envelope during accretion, any changes in the accretion will arguably lead to corresponding variability in the ejection phenomena. Indeed,  dense knots or extended bow shocks in the observed jets are often discrete and clumpy, likely due to a discontinuous accretion process. In this picture, the typical duration between consecutive knots is interpreted as being due to periods between short episodes of the ejection, possibly driven by episodic accretion bursts. 
Observed periods of ejection events range from one year to a few hundred years \citep[e.g.,][]{2015Plunkett,2022ApJ...931L...5J}. Sometimes one object exhibits more than one episode, potentially originating  from distinct perturbation distances in the disk \citep[e.g.,][and references threin]{2020A&ARv..28....1L}. A recent study by \citet{2020A&A...636A..38N} with ALMA at 2600 au resolution found mean periods between ejection episodes of $\sim$500$^{+300}_{-100}$ yr in the W43-MM1 protocluster, corrected for a uniform inclination angle of 57.6$\degr$. Investigating ejection events can therefore help probe the corresponding perturbation or accretion shock zones in the disk and the associated increase in the central luminosity, and enable explorations of the changes in ice lines and molecular synthesis from icy grain mantles. 

The JCMT Transient Survey \citep{herczeg17} has been monitoring Class\,0 and I protostars in submillimeter continuum for over four years, with monthly cadence \citep[]{johnstone18,2021ApJ...920..119L}. Eighteen out of 83 protostars, 20\%, are variable, all of which show long-term, many years, secular trends. Among these variables, the Class\,I source EC\,53 in Serpens Main (also known as V371 Serp) reveals a clear 18 month periodicity  \citep{yoo17,lee20}, also seen in the near-IR \citep{hodapp12}. The periodicity of EC\,53 allows for physically modeling the location, $< 0.4\,R_{\odot}$, and strength of the inner accretion disk instability \citep{lee20}. 
The Class I source V1647 Ori, an FUOr-like system \citep{connelley18} with repetitive outbursts \citep{ninan13}, is also seen to vary in the submillimeter \citep[]{2021ApJ...920..119L}, along with four Class\,0 sources HOPS383 \citep[see also][]{safron15}, HOPS373 \citep{2022yoon}, HOPS356, and NGC\,1333 West40.

In the mid-infrared, 6.5 years of six-month cadence NEOWISE all-sky monitoring reveals that 55\% of the 735 Class 0/I protostars within the Gould Belt are variable \citep[][]{2021ApJ...920..132P}. Submillimeter variability primarily correlates with  envelope temperature changes produced by varying accretion luminosity whereas both luminosity and line of sight extinction variations lead to mid-IR variability \citep{johnstone13, macfarlane19a, macfarlane19b, baek20}. Irregular variability therefore dominates the mid-IR light curve classification; however, over similar observing windows the fraction of secular, long-term trending, light curves in the mid-IR, 20\% \citep[][]{2021ApJ...920..132P}, agrees with the submillimeter result.

\begin{deluxetable}{llllll}
\tablenum{1}
\tablecaption{Line Properties targeted in this study\label{tab:citical_density}}
\tablewidth{6pt}
\addtolength{\tabcolsep}{-3pt}
\tablehead{
\colhead{Transition} & \colhead{Frequency\tablenotemark{(a)}} &  \colhead{E$_{\rm up}$\tablenotemark{(a)} } &  \colhead{$\log_{10}(A_{ij})$\tablenotemark{(a)}} & \colhead{Temp}\tablenotemark{(b)}  &  \colhead{n$_{cr}$}  \\
\colhead{} & \colhead{(GHz)} & \colhead{(K)} & \colhead{(s$^{-1}$)} & \colhead{(K)} & \colhead{(cm$^{-3}$)} 
}
\startdata
  SiO (5$-$4) & 217.104919 & 31.26 & $-$3.28 & 150 & 1.7E+6  \\
  CO (2$-$1) & 230.538000 & 16.60 & $-$6.16 & 150  & 1.1E+4  \\
  CO (2$-$1) & 230.538000 & 16.60 & $-$6.16 & 50  & 1.2E+4  \\
  C$^{18}$O (2$-$1) & 219.560354  & 15.81 & $-$6.22 &20  & 1.9E+4  \\
 \enddata
\tablenotetext{}{Notes: (a) Molecular transition parameters are adopted from CDMS database (M{\"u}ller et al. 2001).
(b) The critical densities are calculated for these temperatures using the collisional rate coefficients from \citet[][ SiO]{2018MNRAS.479.2692B} and \citet[][ CO, C$^{18}$O]{2010ApJ...718.1062Y}. In the case of CO, a change in temperature from 50 K to 150 K reduces n$_{cr}$ by only $\sim$ 8.5\%.}
\end{deluxetable}

Episodic accretion could significantly impact the star formation process by modulating radiative feedback. In the case of a continuous accretion process, for example, radiative feedback could suppress fragmentation by maintaining a warm environment in the surrounding disk or cloud core \citep[][]{2009ApJ...703..131O,2014prpl.conf..243K,2015A&A...576A.109Y}. On the other hand, episodic accretion can weaken this effect by allowing this disk to sufficiently cool during the quiescent phase and engendering its fragmentation \citep[][]{2012MNRAS.427.1182S}. Thus, episodic accretion may affect the formation of binary/multiple systems, planets, and initial mass function of protostars  \citep[][]{2011ApJ...730...32S,2017MNRAS.465....2M,2018A&A...614A..53R}. Furthermore, variations in stellar luminosity and the corresponding changes in the surrounding temperature of the disk and envelope can affect the chemical composition and gas-ice balance \citep[][]{2016Natur.535..258C,2019MNRAS.485.1843W,2019ApJ...884..149H,2022Natur.606..272J}. \citet[][]{2016ApJ...821...46T} suggested that during the accretion phase, higher order COMs (e.g., CH$_3$OCH$_3$, CH$_3$OCHO, C$_2$H$_5$OCH$_3$, C$_2$H$_5$OC$_2$H$_5$, C$_2$H$_5$OCHO) could be formed via gas phase reactions around snowlines of the simpler COMs (e.g., CH$_3$OH), i.e., in the hot regions where T $\gtrsim$ 100 K. While cooling down during the more quiescent phase,  CH$_3$OH or water may freeze-out into ice at $\sim$ 100 - 120\,K more quickly than the higher-order COM species (T $\sim$ 70\,K), due to relatively low binding energies of higher-order COMs than CH$_3$OH, water. Hence, the abundance of higher-order species could be higher than those of simpler COMs around the snowlines. Therefore, accretion intervals may reveal the effects of radiative feedback on the COMs and planet formation process. 


After the first discovery of the protostellar outflow by \citet{1980ApJ...239L..17S}, several observational studies have been performed to characterize outflows \citep{1992A&A...261..274C,1996A&A...311..858B,2013ApJ...777...50L,2014ApJ...783...29D,2015A&A...576A.109Y,2019ApJS..240...18K}. Due to limitation of the observations, most of the previous studies were with single dish telescopes with low-resolution, large scale, and poor sensitivity. Recently, \citet[][]{2021A&A...648A..45P} have investigated a sample of 30 sources (classified as 25 Class\,0, 4 Class\,I, one continuum) with IRAM-PdBI as a part of CALYPSO survey. Their study reveals that most of the jets are possibly dust-free winds launched from inside the dust-sublimation radius and the ejection rate decreases as the source evolves and accretion fades. It is time to enlarge the statistics with higher sensitivity and unbiased samples with similar environmental effects (e.g., all sources in a single molecular cloud). 

In this paper, we describe ALMA observations of a statically significant sample of protostars in the Orion molecular cloud. Their location in common puts them at a similar distance ($\sim$ 400 pc) and in a similar environment, minimizing the effects of environmental differences on the observed parameters. First, we probe the morphology of the jet using the most efficient tracers of molecular jets to investigate driving mechanisms (e.g., whether the outflowing material is jet-driven or wind-driven), shock structure and chemistry. Second, we seek to understand the origin of gas phase SiO in the jet,  following the increasing evidence of SiO in the jet. Third, we probe accretion mechanisms based on the distribution of ejected material, especially the episodic knots. Finally, these data allow us to address important questions like: Is there any direct or indirect correlation between episodic jets and COMs formation in the disk? What would be the possible impact of episodic accretion on the emerging planets? 
In a subsequent investigation these same outflow data will be analysed using channel maps and position-velocity diagrams and compared with the theoretical unified jet-wind model developed by \citet{2023Shang}.

We describe the sample selection in Section \ref{sec:sample} and the ensuing observations in Section \ref{sec:observations}. Section \ref{sec:results} describes the results of continuum and line data analyses. In section \ref{sec:discussion},  we discuss the jet morphology and its connection to the accretion process. Section \ref{sec:summary} summarizes the work. In the Appendices, we discuss the properties of individual objects including maps and spectra of each.

\begin{figure}
\fig{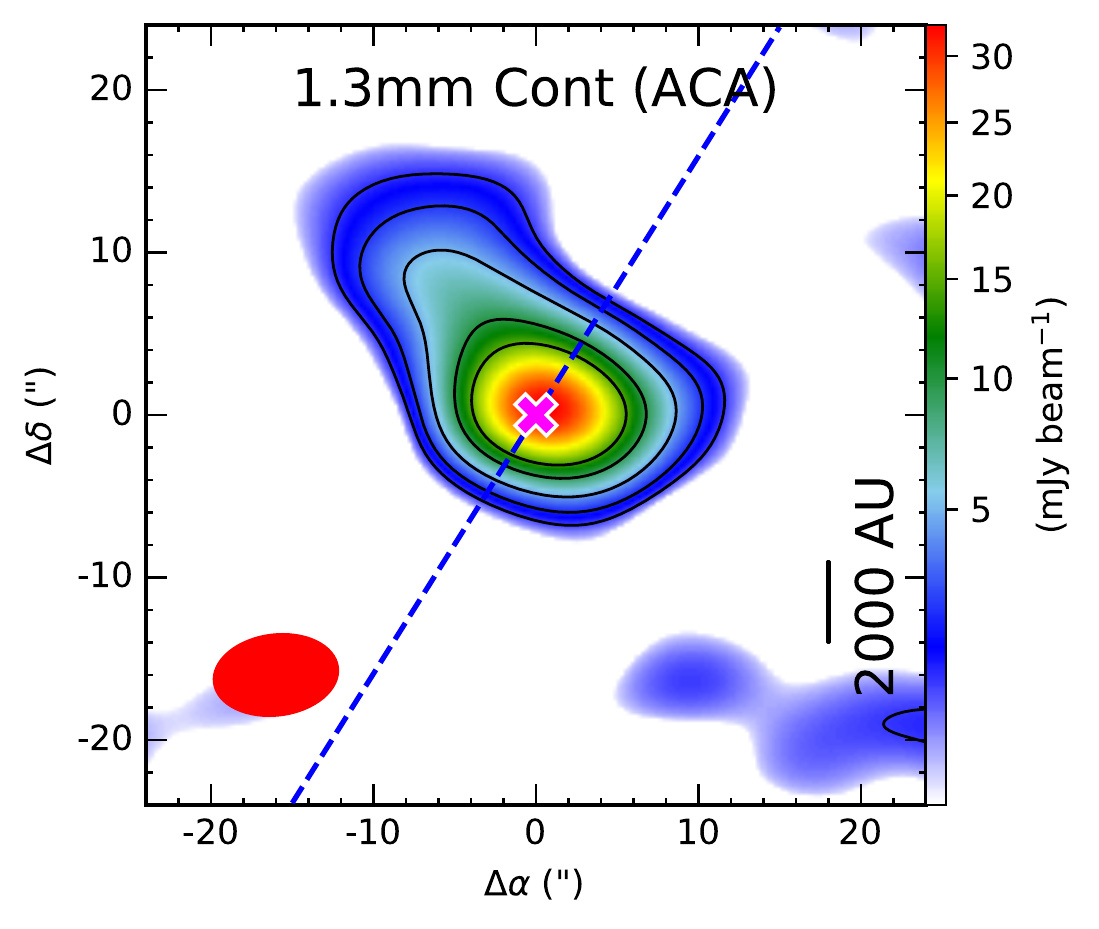}{0.45\textwidth}{}
\caption{
Example image of ALMA 1.3\,mm continuum emission detected toward protostar G203.21-11.20W2. The typical beam size of $\sim$ 7$\farcs$52 $\times$ 5$\farcs$52 (80$\fdg$7) is shown in red in the lower left. The contours are at 3 $\times$(1, 2, 4, 8, 12) $\sigma$, where the sensitivity $\sigma$ = 0.4 $\mjypb$. The cross mark indicates the continuum centre at high-resolution, adopted from \citet{2020ApJS..251...20D}. The blue dashed line indicates the outflow/jet axis, determined from the SiO emission. A scale bar is also shown for context. 
}
\label{fig:G20321_1120W2_ACA_envelope}
\end{figure}

\section{Sample Selection and Observations}\label{sec:sample_observations}

\subsection{Sample}\label{sec:sample}

During ALMA cycle\,6, we initiated a survey-type project (ALMASOP: ALMA Survey of Orion PGCCs) to investigate systematically the fragmentation of dense cores into protostars \citep[][]{2020ApJS..251...20D}. We selected 72 extremely cold young dense cores from the Planck Galactic Cold Cores (PGCC) catalogue, including both candidate starless and protostellar core candidates, therefore, unbiased to any spectral ``Class" of the protostar. The PGCCs are, however, biased to sources with cold dense material around them - so potentialy significantly young! From these observations, we detected 70 substructures within 48 detected dense cores, which include five candidate prestellar cores with centrally dense structures of roughly 2000 au scale where the core shows substructure within the dense portion  \citep[][]{2021ApJ...907L..15S}, and several hot corinos \citep[][]{2020ApJ...898..107H,2022ApJ...927..218H}. Furthermore, some sources were revealed to be multiple systems \citep[][]{2022ApJ...931..158L}. The observations also revealed one of the oldest \citep[G205.46-14.56S3;][]{2022ApJ...925...11D} and one of the youngest protostars \citep[G208.89-20.04Walma;][]{2022ApJ...931..130D} with a bipolar molecular jet in SiO emission.  Interestingly, a few protostars were also found to exhibit monopolar SiO jets \citep[][]{2022ApJ...931L...5J}. 

After removing the candidate prestellar cores from the sample, we selected from \citet[][]{2020ApJS..251...20D} 66 substructures in 42 observing fields containing one or more protostars to investigate outflow and jet characteristics. We identified 42 
outflows detected in CO emission. The rest were likely not detected due to insufficient sensitivity or their outflows could not be separated  from the ambient cloud emission.  Out of 42 outflows, SiO emission was detected in 21. Such SiO emission could trace various shocked regions, e.g., knots or bow shocks, collision zones of adjacent outflows, or  collision zones of outflows on surrounding molecular clouds.  We investigate 39 outflows with CO emission in this paper, of which 18 fields are also associated with SiO emission.  
The results for the remaining three ALMASOP protostellar objects are adopted from previously published literature i.e., \citet[][G205.46-14.56S3 and G206.93-16.61W2]{2022ApJ...925...11D} and \citet[][G208.89-20.04Walma]{2022ApJ...931..130D}. The 39 outflow objects have  bolometric luminosities L$_{bol}$ $\sim$ 0.4 to 180 L$_\sun$ and bolometric temperatures of  T$_{bol}$ $\sim$ 15 to 180 K, though some have undetermined L$_{bol}$ and T$_{bol}$ due to lack of infrared data. Based on the object selection criteria, this sample is one of the youngest, coldest, unbiased protostellar samples within a the molecular environment of the Orion cloud, all located at a similar distance. These features make this sample one of the most suitable samples for investigating variations of the outflow/jet morphology in a consistent manner.

\subsection{Observations}\label{sec:observations}
In this paper we make use of the ALMASOP data observed in Band\,6. Details of the observations and data reduction were presented by \citet{2020ApJS..251...20D}. In brief, we generated 1.3\,mm continuum images using low-resolution ACA observations of typical beam size 7$\farcs$6 $\times$ 4$\farcs$1 to probe large-scale envelope emission. The datacubes of CO(2$-$1), SiO(5$-$4), and C$^{18}$O(2$-$1) line emission are generated using high-resolution observations of typical  beam size $\sim$ 0$\farcs$38 $\times$ 0$\farcs$35 to explore the jet/outflow structures. Here, we combined data from three configurations (TM1, TM2, and ACA) to improve the sensitivity of the line cubes.  The final maps have a primary beam (PB) size $\sim$ 40$\arcsec$, smaller than the ACA PB size $\sim$ 60$\arcsec$. Some of the outflows/jets extend beyond the TM1 PB area, whereas the ACA data are more suitable  for sampling more extended outflow/jet emission. We present, however, the high-sensitivity, combined TM1$+$TM2$+$ACA cubes to characterize outflow cavity walls and shock/knot structure within the jet more precisely. We also compare our results with those from ACA-only observations.  We binned the velocity channels to a resolution of 2 $\kms$ to improve the sensitivity further.

\section{Results}\label{sec:results}


\begin{figure}
\fig{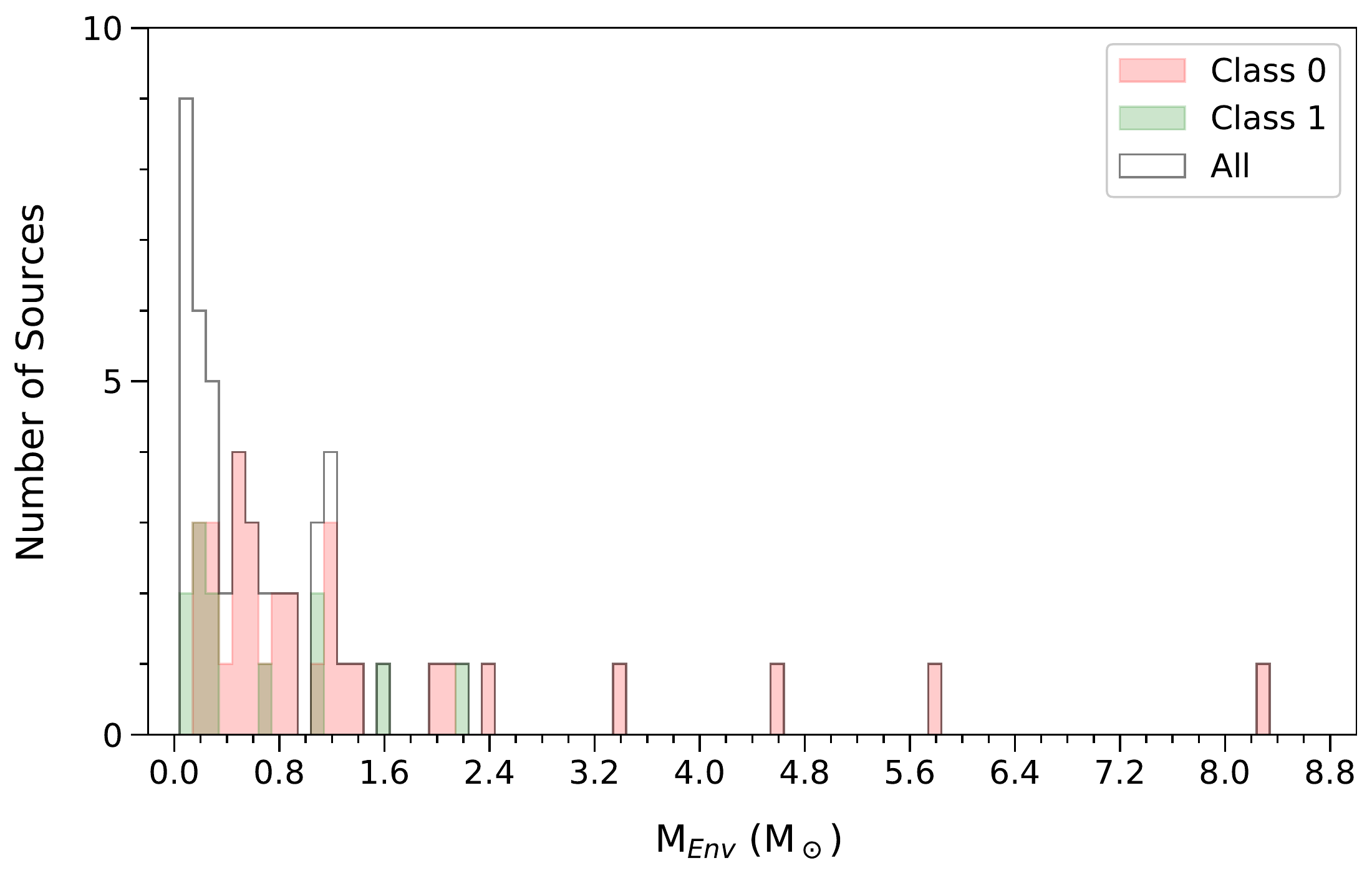}{0.47\textwidth}{}
\caption{
Histogram of envelope masses estimated using 1.3\,mm emission at a resolution of $\sim$ 7$\farcs$6 $\times$ 4$\farcs$1. 
}
\label{fig:hist_aca_mass}
\end{figure}

\subsection{1.3\,mm Continuum}
The low-resolution ACA 1.3\,mm continuum image of the source G203.21-11.20W2 is shown in Figure \ref{fig:G20321_1120W2_ACA_envelope}. The remaining images are presented in the Appendices \ref{sec:Appendix_individual_confirmed_molecular_jet} (objects with confirmed molecular jet), \ref{sec:Appendix_individual_complex_SiO_emission} (objects with complex SiO emission morphology and not considered for jet parameter estimation in this text), \ref{sec:Appendix_individual_NoSiO_CO_definedOutflow} (objects with no SiO emission but well defined CO outflow) and \ref{sec:Appendix_individual_NOSiO_NotCO} (objects with no SiO emission and no well-defined CO outflow).  To determine the properties of the continuum sources, we fit the emission above the 3$\sigma$ contour using CASA {\it imfit} task with one or more 2-dimension Gaussians for one or more, respectively visually identified peaks. Integrated fluxes were obtained from the `flux' component in the output file of {\it imfit} task. All deconvolved Gaussian parameters for each source  are listed in Table \ref{tab:targetobservedContunuum}.

For an extremely young protostar, the 1.3\,mm emission could be optically thick, especially in the central core region.  Hence, this emission may represent a lower limit to the total dust column density from the inner envelope around the protostars. For simplicity, however, we assume the emission is optically thin to infer the  lower limit of inner envelope mass (M$_{\rm Env}$) following the equation: 
\begin{equation}
M_{Env} \sim \frac{D^2 F_\nu(\nu)}{B_\nu (T_{dust}) \kappa_\nu}
\end{equation}
The assumed dust mass opacity is $\kappa_\nu$ $\sim$ 0.00899($\nu$/231 GHz)$^\beta$  cm$^2$ g$^{-1}$, assuming a gas-to-dust mass ratio of 100 \citep[e.g.,][]{2018ApJ...863...94L}. We also assume a dust opacity spectral index $\beta$ $\sim$ 1.5. It is worth noting that $\beta$ could be higher than this value at early protostellar stages due to smaller dust grains, and could be $<$ 1.0 for more evolved Class\,0 or in Class\,I objects where dust grains have grown \citep[][]{2006ApJ...636.1114D}.  Following \citet{2020ApJS..251...20D}, we adopt slightly different distances $D$ for the sources in Orion\,A ($\sim$ 389 $\pm$ 3 pc), Orion\,B (404 $\pm$ 5 pc), and $\lambda$-Ori (404 $\pm$ 4 pc).  $F_\nu$($\nu$) is the measured flux density. $B_\nu$($T_{\rm dust}$) represents the Planck blackbody function at a dust temperature of $T_{\rm dust}$.

We assume an optimal $T_{\rm dust}$ $\sim$ 15\,K for all the inner envelopes \citep[e.g.,][]{2022ApJ...931..130D}.  The resulting masses are also listed in Table \ref{tab:targetobservedContunuum}. Errors are estimated based only on uncertainties in the flux.  The value of $T_{\rm dust}$ could vary from as small as 10 K in the younger protostars to 30 K in the evolved protostars. For an increase of $+$15 K in temperature, the derived mass will decrease by  60\%, and a decrease of $-$5 K will increase the derived mass by 85\%. The histogram of all derived masses is shown in Figure \ref{fig:hist_aca_mass}, including Class\,0, Class\,I and some unclassified cores detected with ACA emission. The masses peak at $\sim$ 0.1 M$_\sun$, where most protostars in this sample have a 1.3\,mm masses $<$ 1.2 M$_\sun$. The mean masses of Class\,0 and Class\,I sources are 1.4  M$_\sun$ and 0.65 M$_\sun$, respectively. The masses $>$ 1 M$_\sun$ are mostly associated with multiple systems (see Appendix for comments on the individual objects).

\subsection{Systemic Velocity from C$^{18}$O}
\begin{figure}
\fig{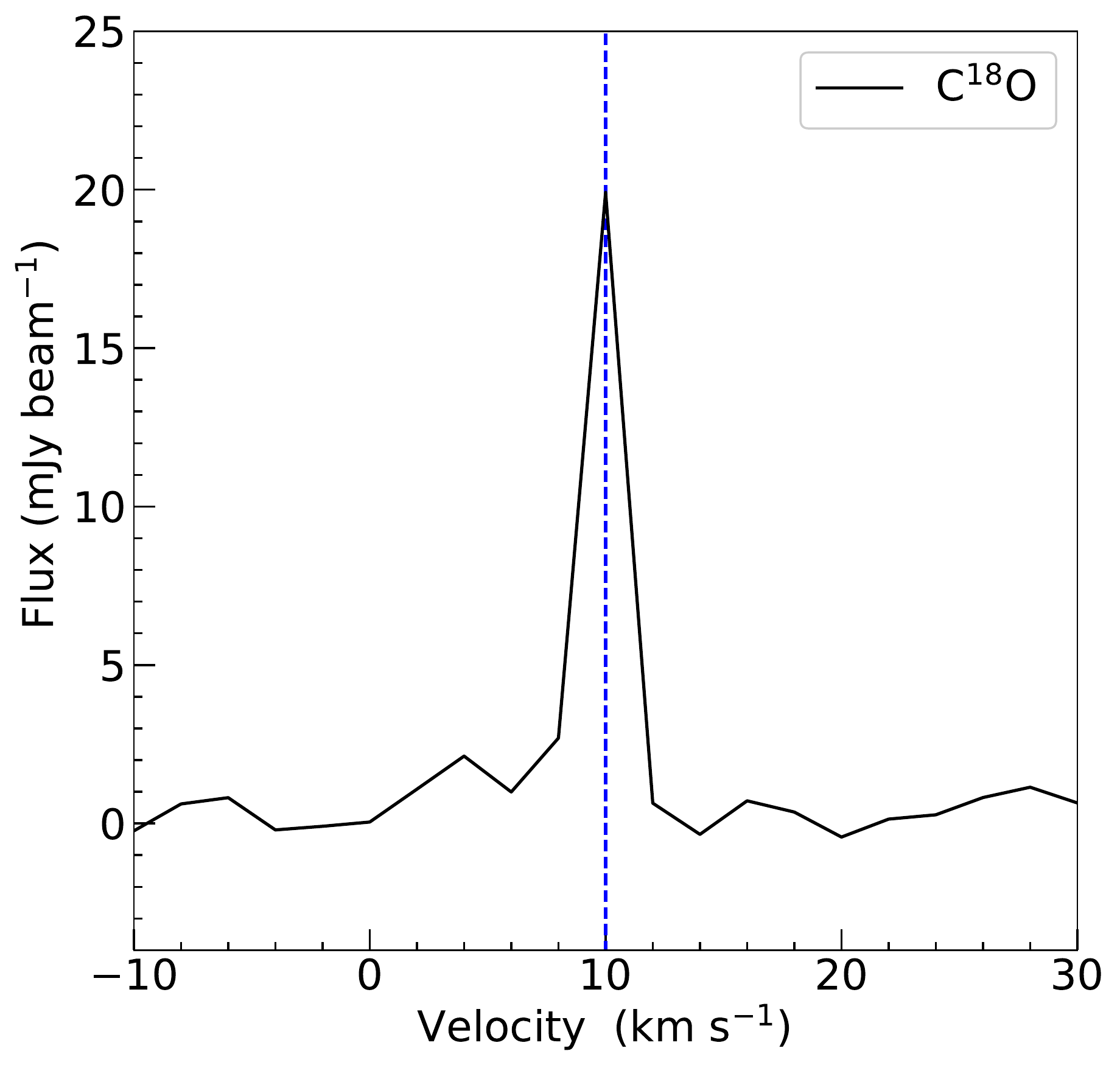}{0.45\textwidth}{}
\caption{
The  C$^{18}$O spectrum of G203.21-11.20W2 extracted from high-resolution maps. The vertical line indicates the systemic velocity of the object determined from the spectrum at $V_{sys}$ = 10 $\kms$. 
}
\label{fig:G20321_1120W2_C18ospectra}
\end{figure}

The systemic velocities ($V_{sys}$) of the sources were estimated from the C$^{18}$O emission observed at the continuum peak in  high-resolution maps, where  C$^{18}$O is most likely tracing the disk/envelope emission around the protostars. It could, however, also trace envelope material entrained by outflow along the boundary wall.  Figure \ref{fig:G20321_1120W2_C18ospectra} shows an example of such C$^{18}$O spectra around the central protostar. The C$^{18}$O spectra of other objects are shown in panels `$b$' of Figures in Appendix \ref{sec:Appendix_individual_confirmed_molecular_jet}, \ref{sec:Appendix_individual_complex_SiO_emission}, \ref{sec:Appendix_individual_NoSiO_CO_definedOutflow} and \ref{sec:Appendix_individual_NOSiO_NotCO}. In most cases, C$^{18}$O spread only over 2-3 channels (one channel corresponding to velocity resolution $\sim$ 2.0 km\,s$^{-1}$), making it challenging to get a reasonable Gaussian fit. Therefore, we chose the velocity of the peak emission as the $V_{sys}$ (Figure  \ref{fig:G20321_1120W2_C18ospectra}) or we average the velocities of the 1st and 2nd brightest velocity-channels from the spectra in cases where the peak emission spanned over more than one channel (as in case G208.89-20.04E in Figure \ref{fig:appendix_G208.89-20.04E}). All these determined $V_{sys}$ values were checked by eye against the spectra.  The respective $V_{sys}$ values are listed in Table \ref{tab:targetobservedContunuum}.



\begin{figure*}
\fig{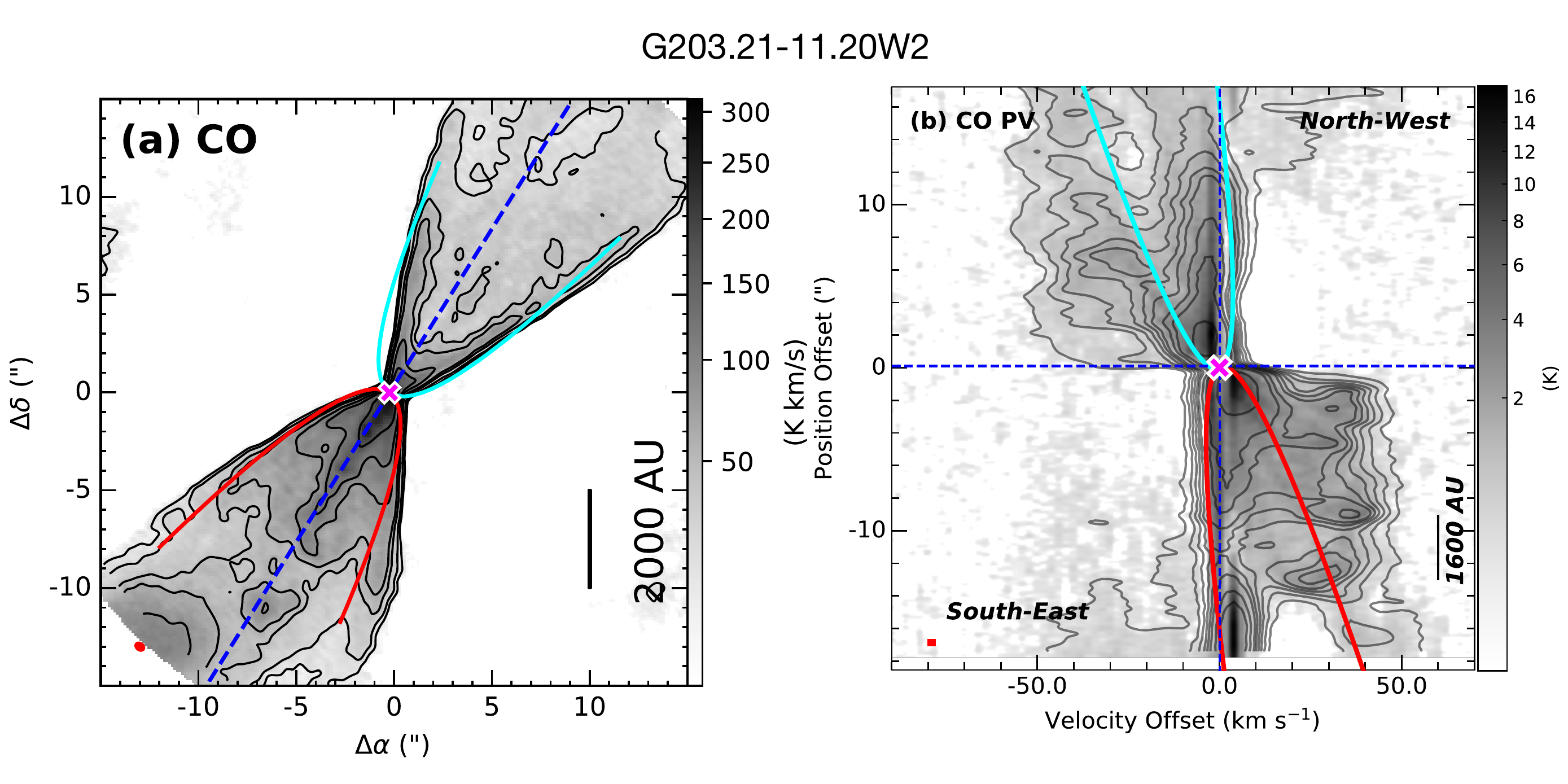}{0.90\textwidth}{}
\caption{(a) High-resolution ALMA  map of G203.21-11.20W2 in CO  emission integrated from $-$50 km\,s$^{-1}$ to $+$50 km\,s$^{-1}$. The contours are over plotted at 3$\times$(1, 2, 3, 4, 6, 8, 12)$\sigma$, where sensitivity $\sigma$ = 4.8 $\kkms$.  The synthesized beam size $\sim$ 0$\farcs$41 $\times$ 0$\farcs$34 is shown in the lower left. CO emission is extended in the North-West to South-East direction with a position angle of $\sim$ 122$\degr$. The cyan (blueshifted) and red (redshifted) parabolas are best fits assuming a proportionality constant c = 0.5 $\rm arcsec^{-1}$.
(b) Position Velocity (PV) diagram along the jet axis of $^{12}$CO(2$-$1) emission. The black contours are at 3$\times$(1, 2, 3, 4, 5, 6)$\sigma$, where sensitivity $\sigma$ = 0.11 K.  The beam sizes are shown in the lower left (in red) emission. The red and cyan parabolas here are for inclination angle, $i$ = 20$\degr$. Cyan and red represent the North-West and South-East direction, respectively, in panel (a). The magenta cross mark in each panel is the location of the continuum peak.
}
\label{fig:G203.21-11.20W2outflow_parabola}
\end{figure*}

\subsection{Physical Structure of outflow}\label{sec:CO_Outflow}

To assess outflow morphologies, we analyzed the CO (2$-$1) emission. We first investigated each velocity channel and spectra along the putative flow axis to identify bipolar emission. We found three kinds of such emission, likely due to outflow inclination angle: (i) the blueshifted and redshifted emissions are well separated on the sky on both sides of the systemic velocity; (ii) there are instances where blueshifted and redshifted emission partially overlap; and (iii) blueshifted and redshifted emission almost entirely overlap.  To explore such different kinds of outflow features, we estimated inclination angles $i_{\rm a}$ using the low-velocity CO emission above 3 $\sigma$. 

\begin{figure*}
\fig{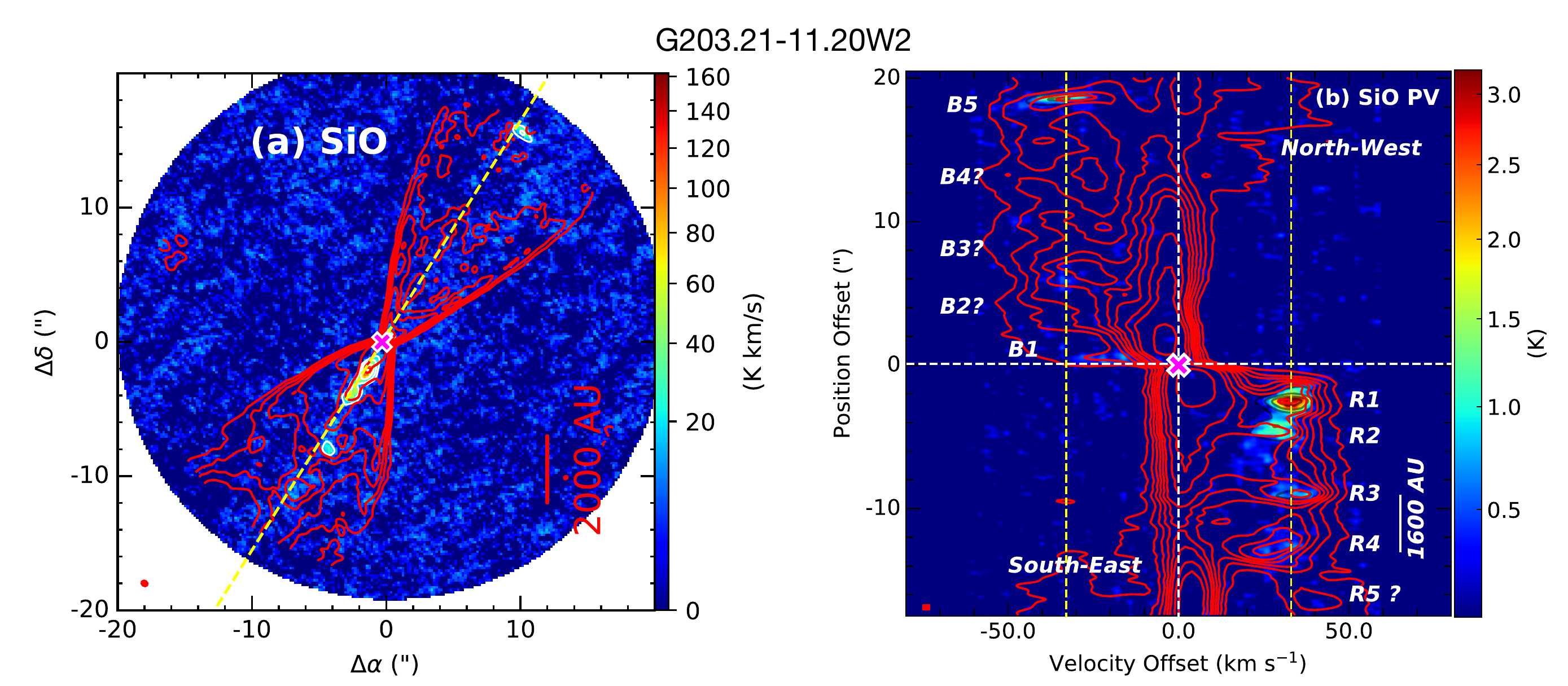}{0.90\textwidth}{}
\caption{(a) High-resolution ALMA  map of G203.21-11.20W2 in SiO  emission integrated from $-$50 km\,s$^{-1}$ to $+$50 km\,s$^{-1}$ with sensitivity  $\sigma$ = 4.6 K\,km\,s$^{-1}$. The CO contours are over-plotted from Figure \ref{fig:G203.21-11.20W2outflow_parabola}a.  The synthesized beam size $\sim$ 0$\farcs$41 $\times$ 0$\farcs$34 is shown in the lower left. The jet axis is shown in a yellow dashed line.
(b) Position-Velocity (PV) diagram along the jet axis of SiO emission with sensitivity $\sigma$ = 0.15 K\,km\,s$^{-1}$. The CO contours are the same as in Figure \ref{fig:G203.21-11.20W2outflow_parabola}b.   The beam sizes are shown in the lower left (in red). The dashed lines at velocity = 33 $\kms$ are the mean jet velocity, V$_j$ (uncorrected for inclination). The knot location is marked in B1, B2, ..... and R1, R2, .... in blue and redshifted lobes, respectively. The magenta cross mark in each panel denotes the location of the continuum peak.
}
\label{fig:G203.21-11.20W2_SiO_integrated-PV}
\end{figure*}

Figure \ref{fig:G203.21-11.20W2outflow_parabola}a  presents the integrated CO (2$-$1) line emission map of the source G203.21-11.20W2. A schematic diagram describing the possible structure of the outflow is shown in Appendix \ref{appendix:schematic_inclination} (Figure \ref{fig:appendix_Schematic_outflow_jet}a). The outermost shell of the outflow is observed to have a parabolic structure.  We assume that the molecular CO outflow shell is a radially expanding parabolic shell of the underlying wide-angle wind, and then accordingly fit a parabolic shape to the moment zero map of CO emission using Equation \ref{equ:inclination_ZCR2} to obtain the curvature ($c$) at the launching point. Using the same $c$ values, we fit another parabola in the PV diagram. Since CO emission may trace the outflow shell along with jet and ambient cloud emission, the fitting depends on the correct assumption of the outflow shell from other components. The likely ambient material near the systemic velocity is distinguished after visually inspecting individual velocity channels in the data cube. Similarly, high-density axial emission and high-velocity emission are considered as likely jet emission. If the objects exhibit high-velocity jets blended with the low-velocity outflow emission, we distinguish the outflow shell by comparing the CO PV diagram with that of SiO emission (Figure \ref{fig:G203.21-11.20W2_SiO_integrated-PV}), assuming SiO traces mostly jet components. In  Figure \ref{fig:G203.21-11.20W2outflow_parabola}a, fitting the parabola to the outermost contour of the outflow yields c = 0.5 $\rm arcsec^{-1}$ for both the Southern and Northern lobes. 
Using these $c$ values, we fitted parabolas to the ``outflow" emission in PV space as shown in Figure \ref{fig:G203.21-11.20W2outflow_parabola}b with Equation \ref{equ:inclination_finalEquation} to create a complete 3D paraboloid structure, where the fitting equations include the inclination of the structure in the plane-of-sky (see Appendix \ref{appendix:schematic_inclination}, for details). For example, the inclination angle for G203.21-11.20W2 is estimated to be $i$ $\sim$ 20$^{+5}_{-5}$ degree. In the $c$ and $d$ panels of Figures in Appendices B, C, D and E, we present all the fitted parabolas in integrated emission as well as in PV diagrams. A total of 20 outflows are fit and the inclinations have been estimated from this model. Some of the outflows are not fit due to indistinguishable outflow shells.  The inclination angles are listed in Table \ref{tab:jet_outflow_Properties} for all the objects.  Here, we note that this model is suitable only for those objects that have well-defined outflow shells, i.e., outflow walls that are well distinguished from the ambient material or outflows of multiple sources.

 



\subsection{Outflow Force}\label{sec:results_massforcemomentum}
Outflow force  $F_{\rm CO}$ is the mean rate of momentum $P$ of outflowing material, derived by estimating the outflow momentum over the dynamical time $\mathcal{T}_{dyn,out}$ of the outflow where $\mathcal{T}_{dyn,out}$ is the length of the jet divided by the velocity of the outflow. Outflow force $F_{\rm CO}$ was calculated from the beam-averaged CO emission across the whole velocity range above 3$\sigma$ on each lobe (blue and red) separately. To avoid contamination from ambient material, we removed a few channels near the systemic velocity after inspecting channel maps. First, outflow emission above 3 $\sigma$ in the whole data cube is converted into mass. For example, the k-th channel has an outflow mass of $M_{k}$:
\begin{equation}\label{equ:outflowmass_equation}
\indent
M_{k} = \mu_{H_2} m_H A \frac{\sum_{l}{N_{CO,l}}}{X_{CO}}\\\\
\end{equation}
where the beam-averaged CO column density $N_{CO}$ is summed over outflow pixels $l$ on the k-th channel and each pixel has an area of $A$. We assumed a CO-to-H$_2$ abundance ratio of X$_{CO}$ $\sim$ 10$^{-4}$ in the outflow for a excitation temperature of $T_{ex}$ $\sim$ 50 K. In the next step, the estimated masses are converted into momentum $P$ $\sim$ $M_k V_k$,  for a mass of $M_k$ on k-th channel with the central velocity of $V_k$ = $|$\,$V_{\rm obs}$ $-$ $V_{\rm sys}$\,$|$.  Finally for a total outflow extension of  $R_{CO}$ with maximum outflow velocity $V_{CO,max}$, $F_{CO}$ is expressed as:
\begin{equation}\label{equ:force_equation}
F_{CO} = \frac{P}{\mathcal{T}_{dyn,out}} = f_{ia}\frac{V_{CO,max}\sum_k{M_k V_k}}{R_{CO}}
%
\end{equation}
where the factor $f_{ia}$ is the correction factor for the inclination of the system in the plane of the sky. The inclination corrected $F_{\rm CO}$ values are listed in Table \ref{tab:jet_outflow_Properties}. 

%

Here we note that we do not cover the full outflow extents within the TM1/TM2 primary beam ($\sim$ 40$\arcsec$). If, however, we increase the FOV (as for ACA maps), the outflow mass and dynamical time, may both increase, and so $F_{\rm CO}$ measurements are not expected to change drastically. We also compared the present $F_{\rm CO}$ values with those derived from ACA maps (FOV $\sim$ 60$\arcsec$), and we see only changes by factors of 1 to 1.5 from the smaller primary beam values. In addition, we could not recover the whole outflow emission from the ACA maps due to their relatively poor sensitivity compared with the combined high-resolution maps. Therefore, we present only the $F_{\rm CO}$ measurements derived from the high-sensitivity combined maps.

Errors of $F_{\rm CO}$ estimation depend on the amount of  missing flux in interferometric observations, the lack of coverage of the full length of the outflow lobe, calibration, the optical depth of CO, the temperature variation of the outflow, and errors in the inclination angle measurements. Since the above information could not be recovered simultaneously from the present observations, we assume a conservative error of 40\% in all plots in the subsequent sections.


\begin{figure}
\fig{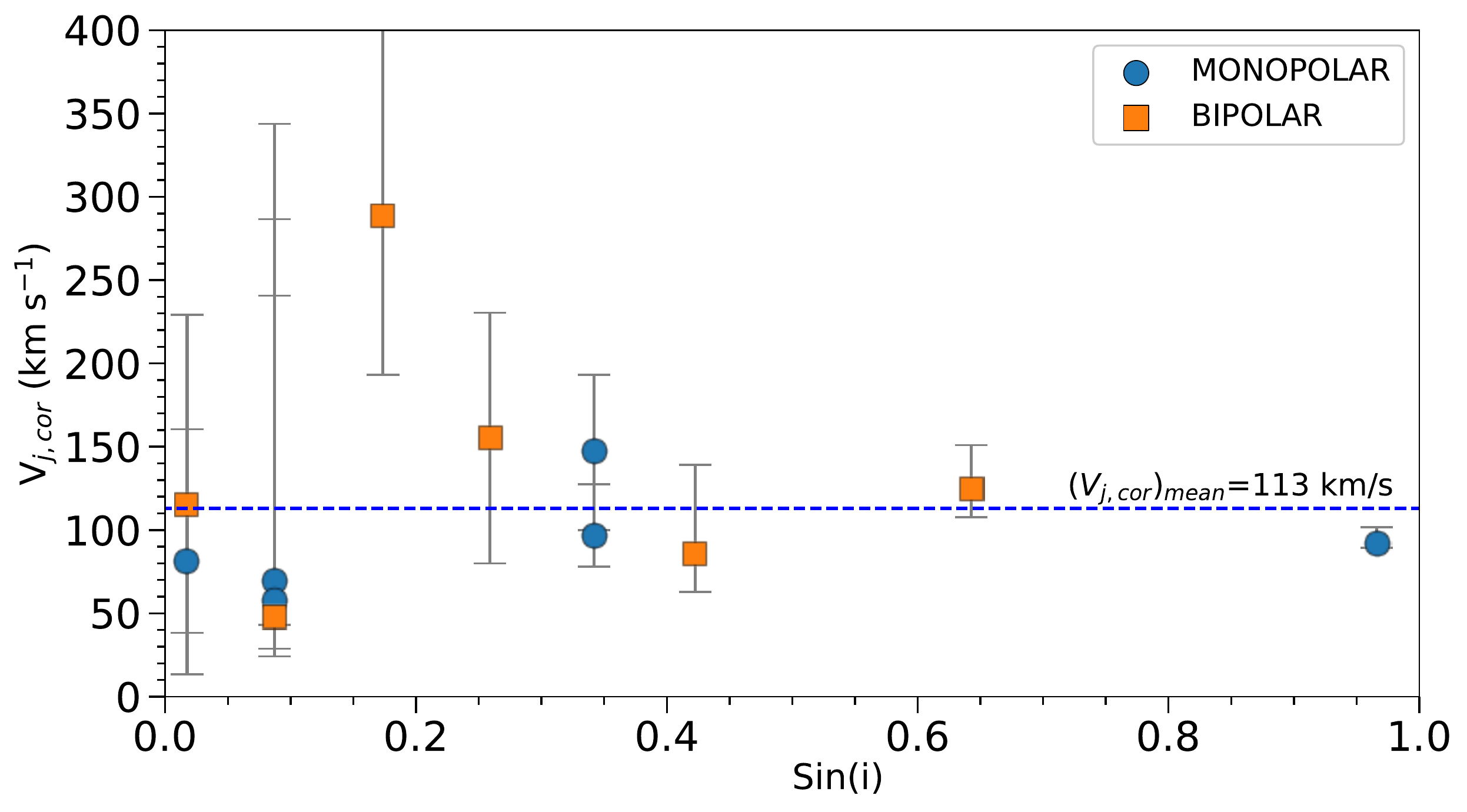}{0.45\textwidth}{}
\caption{
Inclination corrected mean iet velocities $V_{j,cor}$  = $\frac{V_j}{\sin{i}}$ are shown as a function of $\sin{i}$, where $i$ is the measured inclination angle. The mean value of $V_{j,cor} = 113 \kms$ is also shown. Monopolar and bipolar jets are shown as circles and squares. 
}
\label{fig:Mean_jet_velocity}
\end{figure}


\subsection{Molecular jet detection and episodes of accretion}\label{sec:results_episodocknots}
To assess the high-density jets associated with protostars within the outflow cavities, we searched the CO and SiO data cubes at velocities both blue- and redshifted from the $V_{sys}$. Given its higher critical density and an expected increase in SiO from dust grains shattered in shocks, the SiO emission is more suitable for detecting the high-density material and shocks in the jets, where as the CO traces both high-density jet as well as low-density outflow structures.   As shown in Figure \ref{fig:G203.21-11.20W2_SiO_integrated-PV} for G203.21-11.20W2, SiO is clearly observed along the flow axis, where CO is seen more widely i.e., within and surrounding the SiO emission along the flow axis, and in the CO outflow shell. In the red lobe, four knots (R1, R2, R3, and R4) are consistent in both SiO and CO emission, another knot (R5?) is possibly detected in CO emission but not detected  in SiO emission. In the blue lobe, two knots (B1 and B5) are detected in both SiO and CO emission, whereas the CO peaks at B2?, B3?, and B4? are not detected with SiO emission.  For the other outflows, the CO and SiO integrated maps and their respective PV diagrams are are presented in Appendices \ref{sec:Appendix_individual_confirmed_molecular_jet} (objects with confirmed molecular jet) and \ref{sec:Appendix_individual_complex_SiO_emission} (objects with complex SiO emission morphology and not considered for jet parameter estimation in this text).

We observed 18 fields with SiO emission, out of which 14 have confirmed protostellar jets. Three (G200.34-10.97N and G205.46-14.56M2, G205.46-14.56N2) exhibit extended SiO emission along the jet axis, which could be the bow-shocks of jets or collision zones between outflow and ambient material. SiO in one field within G209.55-19.68N1 is clearly a collision zone between two outflows. In the case of G205.46-14.56M2, the driving source of SiO is not clearly identified, and it could be emission from other molecules in this line-rich source (e.g., a hot corino) that mimics SiO emission. It is difficult to conclude the existence of a jet in this particular source from the present observations. All the SiO emission integrated maps and PV diagrams are shown in Figures B1-11 and C1-6.  Six of the sources exhibit monopolar or nearly-monopolar SiO jets (G203.21-11.20W2, G191.90-11.20S, G205.46-14.56M1, G205.46-14.56S1, G209.55-19.68S2, G211.47-19.27S$\_$B). In some cases, CO traces both sides of the jet but SiO is monopolar, such as the blue-shifted component of G203.21-11.20W2. In some monopolar cases, like the source G209.55-19.65S2, there is no jet emission detected in the blue-shifted lobe with SiO as well as in the CO.

The mean of the velocity of the brightest peak knot emission is considered to be the observed jet-velocity $V_j$, and these are listed in Table \ref{tab:jet_outflow_Properties}. A histogram of measured mean jet velocity corrected for inclination angle, i.e., V$_{j, cor}$ (=$\frac{V_j}{\sin{i}}$) is shown in Figure \ref{fig:Mean_jet_velocity}. Jet velocities are distributed over a large range from $\sim$ 50 $\kms$ to 290 $\kms$, where most ($\sim$ 60\%) are less than 100 $\kms$. Proper motion studies of jet knots of the object HH\,212 study also reveal the jet velocity could be as small as $\sim$ 50 $\kms$ \citep[]{2022ApJ...927L..27L}. The mean of all corrected jet velocities are estimated to be ${V_{j, cor}}_{mean}$ = 113 $\kms$. We note that a few of the outflows have very small inclination angles (close to edge-on), but some inclinations could not be derived accurately from present observations of outflow shells. In those cases, the jet velocities could change significantly with only a few degrees of change in inclination angle given the error bars in the measured inclination angles, as shown in Figure \ref{fig:Mean_jet_velocity} (see also Table \ref{tab:jet_outflow_Properties}).  For sources G205.46-14.56S1$\_$A (V$_{j, cor}$ = 80.22 $\kms$) and G208.89-20.04E (V$_{j, cor}$ = 114.60 $\kms$) in particular, the inclination angles could be $<$ 1$\degr$, and so a significant change in velocity could be expected given improved inclination measurements.

The mean timescales between knots are measured as $\mathcal{T}_{knot}$ = $\frac{\delta R}{V_{\rm j}}\tan(i)$, where $\delta$R is the mean distance between two consecutive knots.  The values are listed in Table \ref{tab:jet_outflow_Properties}. The derived 
 $\mathcal{T}_{knot}$ values range from $\sim$ 20 years to 175 years for different jet knot pairs. 

\subsection{Jet Mass-Loss Rate and Kinetic Luminosity}\label{sec:results_JetMassLoss_KineticLuminosity}

The jet mass-loss rates $\dot M_j$ were derived from the CO integrated emission. First, we disentangled the most likely jet emission from the outflow shell by considering SiO emission as being representative for the jet. Figure \ref{fig:G203.21-11.20W2_SiO_integrated-PV}b shows the SiO emission of G203.21-11.20W2. Such emission is also found at high-velocity ($>$ 15 $\kms$). Therefore, we consider the jet velocity range to extend from minimum SiO velocity to the maximum available CO (or SiO) velocity. This particular object is moderately inclined ($i$ $\sim$ 20$\degr$). For a highly inclined object like G191.90-11.20S (close to edge-on; Figure \ref{fig:appendix_G191.90_11.20S} in Appendix \ref{sec:Appendix_G191.90-11.20S}), however, the SiO emission does not extend to the high-velocity range, so here we consider the whole velocity range of SiO emission as jet, independent of higher-velocity consideration. 

Next, nearest to the continuum we choose the peak emission from the knots in CO maps integrated over likely jet velocities. These knots are expected to be less distorted by flow into the ambient medium than farther away knots. The peak emission is converted into beam-averaged CO column density $N_{CO}$ assuming a specific excitation temperature of $T_{ex}$ $\sim$ 150 K within the jet. $N_{CO}$ is utilized to obtain the H$_2$ column density $N_{H_2}$ for a CO abundance ratio, X$_{CO}$  = $N_{CO}$/$N_{H_2}$  \citep[$\sim$ 4 $\times$ 10$^{-4}$,][]{1991ApJ...373..254G}. We consider the molecular jet to be flowing through a uniform cylinder at a constant density at constant speed along the transverse beam direction. Since the jet widths are 
 not spatially resolved even at the high spatial resolution achieved, the beam size $b_m$ is taken to be the jet width.  Thus, $\dot M_j$ can be expressed as:
\begin{equation}\label{equ:mass_equation}
\indent
\dot{M_j} = \frac{1}{3}\mu_{H_2} m_H \frac{N_{CO}}{X_{CO}}\, V_{j,cor}\, b_m,\\\\
\end{equation}
where $\mu_{H_2}$ = 2.8 is the mean molecular weight and $m_H$ is the mass of a  hydrogen atom. $V_{j,cor}$ is the mean deprojected jet velocity (i.e., $V_j$/$\sin(i)$). The knots observed in this spatial resolution likely delineate the internal bow shocks or gas that has been highly compressed by shocks. Since the knots are not resolved,  we assume a compression factor of $1/3$ following \citet[][]{2007ApJ...659..499L,2007ApJ...670.1188L}. The jet kinetic luminosity is defined as $L_{j,kin} = (1/2) \times \dot{M_j} V_{j,cor}^2$.  The inclination corrected  $\dot{M_j}$ and $L_{j,kin}$ values are listed in Table \ref{tab:jet_outflow_Properties}. 

Given that CO abundances could be as small as X$_{CO}$  $\sim$ 10$^{-4}$ instead of the present assumption, the tabulated $\dot{M_j}$ and $L_{j,kin}$ values could be higher. Therefore, we consider our measured values to be lower limits to the mass loss rates.  The error bars of these measurements also depend on various factors in flux measurements, as discussed above in section \ref{sec:results_massforcemomentum}. With these complications, we assumed a conservative error of 40\% for the measured $\dot{M_j}$ values.

\begin{figure}
\fig{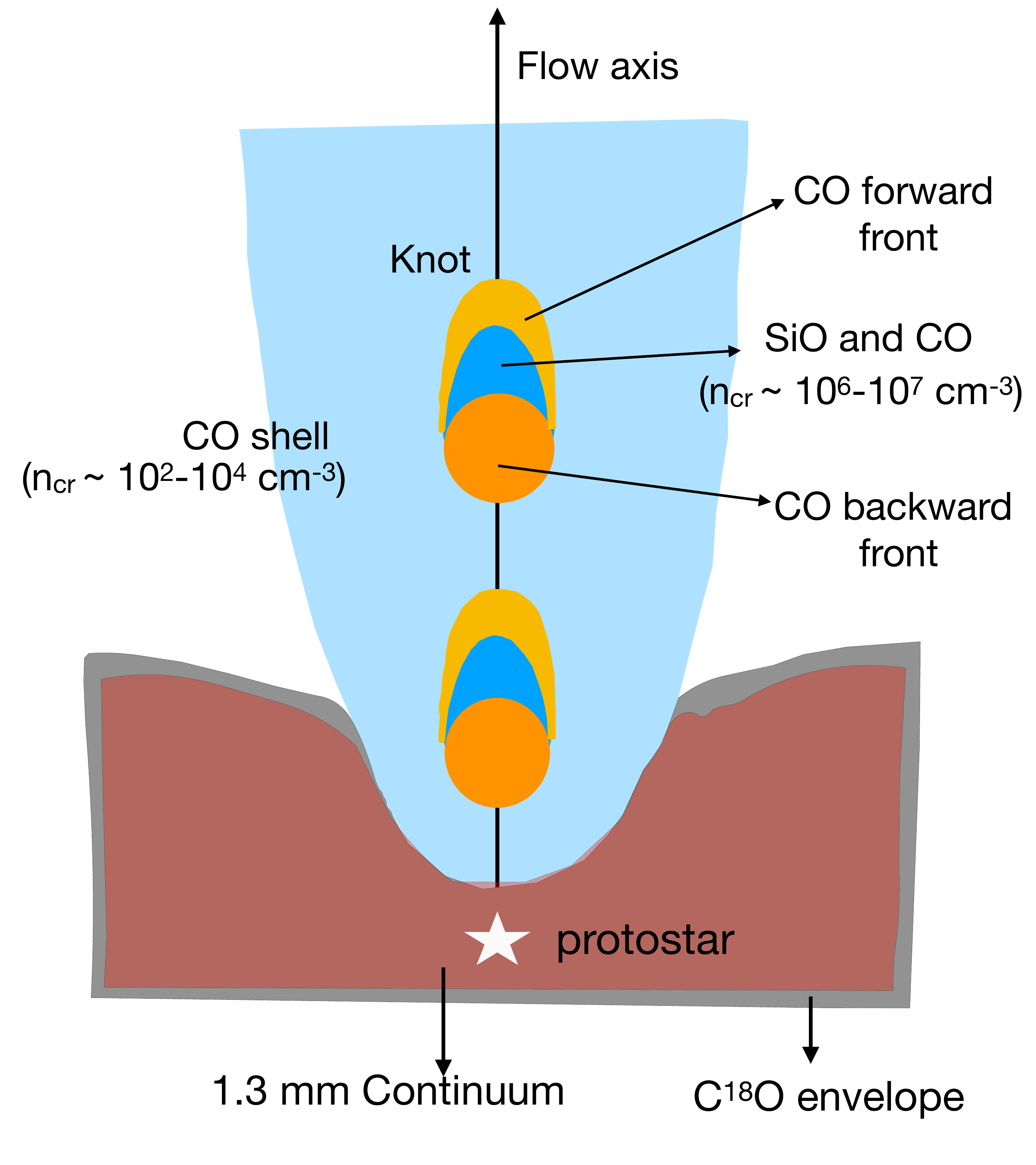}{0.45\textwidth}{}
\caption{A schematic diagram based on observations that illustrates general morphologies of the 1.3\,mm continuum, and the C$^{18}$O, CO and SiO line emission. The 1.3\,mm continuum and C$^{18}$O emission trace the disk and envelope.  CO emission delineates the outflow shell and whole knot structure including a CO emission forward front, a SiO+CO emission middle shock front, and a CO emission backward front. 
}
\label{fig:CO_SIO_shell}
\end{figure}

\begin{figure*}
\fig{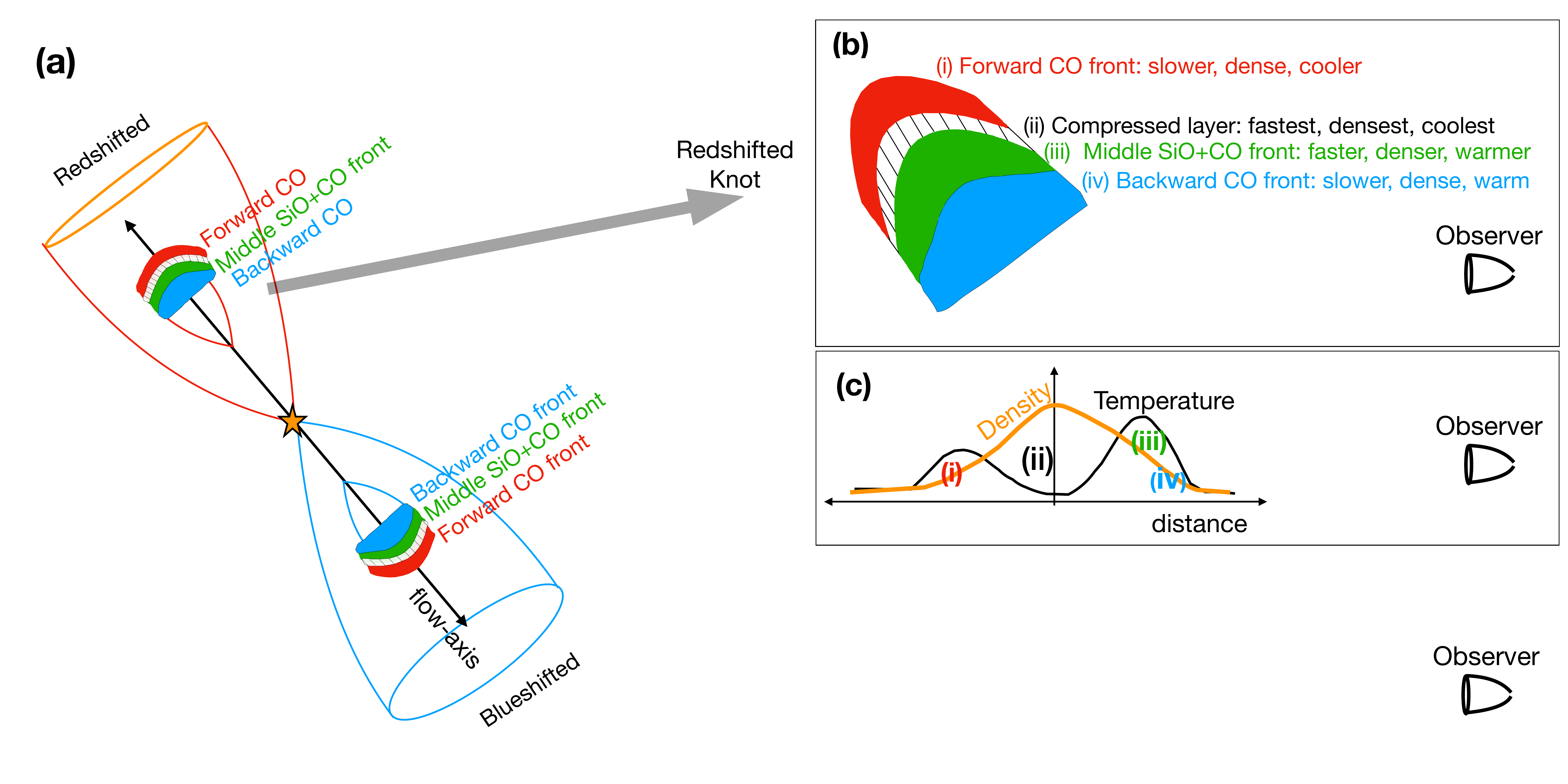}{0.95\textwidth}{}
\caption{A schematic diagram illustrating a monopolar jet. (a) Each knot consists of three observed layers: a forward CO emission front (in red), a middle SiO$+$CO front (in green), and a backward CO emission front (in blue). (b) close up view of a knot in the redshifted lobe. The colours are the same as in panel-$a$. An additional compressed layer is highlighted between the forward CO front and middle SiO$+$CO front. (c) The possible temperature distribution of the knot of the panel-$b$ is plotted, following the simulation results by \citet[][]{2001ApJ...557..429L}. Layers-$iii$ and $iv$ have the highest temperature and layer-$i$ is cooler than $iii$ and $iv$. Layer-$ii$, however, exhibits a temperature discontinuity or is much cooler than other layers. The density of layer-$ii$ is maximum than other layers. For the redshifted lobe, an observer can see the middle SiO$+$CO (layer-$iii$) and backward CO layers (layer-$iv$).  For the blue-shifted lobe, the scenario is opposite, where an observer can see only a forward CO emission front (layer {\it i}). Emission from layers-$iii$ and $iv$ might be shielded by the compressed and coolest layer-$ii$. In that case, only CO jet can be observed in blueshifted emission. If layer-$i$ is not dense enough, then even CO emission can not detect the jet in the blueshifted lobe. 
}
\label{fig:Schematic_monopolarJet}
\end{figure*}

\section{Discussion}\label{sec:discussion}
\subsection{Which emission  traces Which component?}\label{sec:discussion_emission}
The schematic diagram in Figure \ref{fig:CO_SIO_shell}, drawn based on the observations, describes various components of protostars traced by the 1.3\,mm continuum and the C$^{18}$O (2-1), CO (2-1), SiO (5-4) lines. The continuum maps at 140 au and 2200 au resolution delineate the thermal dust emission from the disk and inner envelope, and inner to moderate-scale envelope emission, respectively. At 140 au resolution,  C$^{18}$O emission likely traces the disk and inner envelope.  In a few cases, we may have resolved the larger disks but the velocity resolution of our observations is not sufficient to probe their kinematics and geometry.

As shown in Table \ref{tab:citical_density}, the SiO emission has a higher critical density ($n_{cr}$ $\sim$ 1.66 $\times$10$^{6}$ cm$^{-3}$) than that of CO ($n_{cr}$ $\sim$ 1.06  $\times$10$^{4}$ cm$^{-3}$).  We estimated the optical depth for SiO ($\tau_{SiO}$) and CO ($\tau_{CO}$) emission from the peak emission within the knots, assuming a temperature $T_{ex}$ $\sim$ 150 K, and find that the SiO emission is optically thicker than that of CO ($\tau_{SiO}$ $>$ $\tau_{CO}$) within the knots. So, CO emission arises from deeper layers within the knots than does the SiO emission.  In the ALMASOP sample, we observe that CO emission is tracing the low-density outflow shell. The shock regions or knots of higher density along the jet axis are detected in both SiO and CO emission, and we expect the latter to trace more interior parts of the knots 
due to its lower optical depths. We also see CO emission forward and in front of the SiO$+$CO emission boundary in some knots, which could consist of relatively lower-density and cooler material than the SiO$+$CO shock regions. Another layer of CO emission in a back was also seen in a few knots, which could be the slower-moving material of the shock (see G191.90-11.20S in Figure \ref{fig:appendix_G191.90_11.20S}d and f of the Appendix). In a bow-shock type of knot, we indeed usually see an angular offset in CO and SiO emission (e.g., G191.90-11.20S in Figure \ref{fig:appendix_G191.90_11.20S}f, G208.89-20.04E in  Figure \ref{fig:appendix_G208.89-20.04E}f, G209.55-19.68S1  in  Figure \ref{fig:appendix_G20955_1968S1}f). Such bow-shocks exhibit successively a CO forward front, SiO peak emission and then CO peak emission (see also Figure \ref{fig:Schematic_monopolarJet}). The backward shock detected in CO could be due to either optical depth effects or slower-moving material (discussed later).


\subsubsection{Monopolar Molecular jet}
Some  SiO jets are observed to be only monopolar, or have bright red-shifted knots but faint blueshifted knots. Examples of such asymmetric jets include G203.21-11.20W2, G191.90-11.20S, G205.46-14.56M1, G205.46-14.56S1, G209.55-19.68S2, and G211.47-19.27S$\_$B. In these cases, the redshifted SiO emission can be observed prominently and the blueshifted emission is either  fainter, is missing, or has fewer knots.  As an example, in G203.21-11.20W2 (Figure \ref{fig:G203.21-11.20W2_SiO_integrated-PV}) the blueshifted CO jet is visible and SiO is not detected in most of the CO knot locations (only fainter B1 and B5). In the case of G209.55-19.68S2, however, the blueshifted jet is completely missing in both CO and SiO emission (Figure \ref{fig:appendix_G209.55_19.68S2} in the Appendix). 

A possible scenario for the monopolar jet is illustrated in Figure \ref{fig:Schematic_monopolarJet}. 
In most knots, we observe an SiO forward peak and a CO backward peak, and  additionally the whole SiO+CO peaks are surrounded by a less bright CO front in both space and velocity.  Therefore, we have divided the observed emission in the knots into three sub-layers:  layer-(i), a forward CO front (in red), which is possibly slow-moving, dense, cooler material; layer-(iii), a middle SiO+CO front (in green), which could be the faster moving, denser, and warmer part of the knots; and layer-(iv) a  backward CO front (in blue) which is relatively slow, less dense, and cooler than middle-front. Now, following simulation results by \citet[][]{2001ApJ...557..429L}, we introduce another sub-layer layer-(ii):, a compressed layer (black and white lines) between layers-(i) and (iii), which could be the fastest moving, densest, and coolest layer but one which is not detected in our observations. As predicted by \citet[][]{2001ApJ...557..429L}, the density and temperature profiles of different layers are shown in panel (c). In the general scenario of shock processing, a slow-moving layer-(i) collides with a faster-moving layer-, e.g., (ii)+(iii), and as a result, a very high-density compressed layer-(ii) is originated. Since radiative cooling is inversely proportional to density \citep[e.g.,][]{1990ApJ...360..370B}, layer-(ii) should cool faster than the other layers and produce a temperature discontinuity, as shown in panel (c). Layer-(iii) is the transition region between the compressed layer and the slow-moving CO layer-(iv). It could be detected if it reaches the critical density of the SiO. Layer-(iv) is the slow-moving trail of layers-(ii)+(iii). 

In the redshifted lobe, an observer can see layers-($iii$)+($iv$) in SiO and CO emission, and the forward CO front at layer-($i$) could be shielded by the layer-(ii) of high density and low temperature. In the blue-shifted lobe, the scenario is opposite to that of a redshifted lobe, where an observer can only see the forward CO material front from layer-(i), as in the case of G203.21-11.20W2. Emission from layers-(iii) and (iv) might be shielded in the compressed layer-(ii) of the highest-density and coolest material, therefore, only the CO jet can be observed in the blueshifted lobe. If the forward front (layer-i) is not dense enough and the compressed layer (layer-ii) is highly dense and cool, then the jet can remain undetected even with CO emission, as in the case of the blueshifted lobe of G209.55-19.68S2 (Figure \ref{fig:appendix_G209.55_19.68S2}) where the jet component is completely missing in both the SiO emission as well as in the CO emission. 


The above scenario can explain most of the redshifted SiO monopolar jets in the ALMASOP sample. Another source  NGC 1333-IRAS2A from CALYPSO IRAM$-$PdBI survey exhibits blueshifted jet only \citet[][]{2014A&A...563L...3C}. G205.46-14.56M1 (Figure \ref{fig:appendix_G205.46_14.56M1}) also exhibits one bright blueshifted knot but fainter redshifted SiO emission in the knot location. Indeed, such behaviour scenario is difficult to explain with the above schematic diagram. Therefore, we also suggest that there could be further unknown intrinsic properties driving monopolar jets.


\subsubsection{Molecular jets and protostellar evolutionary phase}\label{sec:discussion_jetevolution}
Class\,0 protostars are observationally defined as being in the youngest phase of star formation, with bolometric temperature $T_{bol}$ $<$ 70 K and infrared spectral index of $\alpha_{IR}$ $>$ 0.3 \citep[][]{2016ApJS..224....5F}. Class\,0 sources exhibit typically very high accretion and mass-loss rates ($\sim$ $10^{-6}$ - $10^{-7}$ M$_\sun$ yr$^{-1}$) onto the protostar, which could produce high-density jets \citep[5$-$10 $\times$ 10$^{6}$ cm$^{-3}$;][]{2013A&A...551A...5E,2020A&ARv..28....1L}. The higher transitions of SiO are the most commonly observed tracer of such high-density material \citep[][]{2014ApJ...797L...9L,2015A&A...581A..85P,2017NatAs...1E.152L,2021A&A...648A..45P}. While the protostar evolves from Class\,0 through Class\,I and on to the Class\,II phase, both the accretion and mass-loss rates typically decrease. Therefore, SiO should be a better tracer of jets in the earlier phases of protostars than the later phases and the detection of SiO emission in jets more likely indicates protostars in a younger phase and at a higher mass-loss rate. 

Molecular jets with SiO emission have been detected in many of the  Class\,0 protostars, e.g., B\,335 \citep[][]{2019ApJ...873L..21I,2019A&A...631A..64B}, 
HH\,212 \citep[][]{2017NatAs...1E.152L,2017SciA....3E2935L}, 
L1157 \citep[][]{2015A&A...573L...2T,2016A&A...593L...4P}, 
HH\,211 \citep[][]{2016ApJ...816...32J,2018ApJ...863...94L}, 
and IRAS\,04166+2706 \citep[][]{2009A&A...495..169S,2017A&A...597A.119T}. 
One  Class\,I multiple system, SVS13A, composed of VLA4A and VLA4B, was also found with SiO knots \citep[][]{2000A&A...362L..33B,2017A&A...604L...1L}, where VLA4B is identified as the base of the jet \citep[][]{2017A&A...604L...1L}. In this case, however, the spectral classification of the components based on infrared observations could be largely affected by the multiplicity. For example, no SiO jet has been previously detected from an isolated Class\,I object.  On the other hand, SiO jets also have not been detected in extremely young objects, such as the candidate first hydrostatic cores \citep[e.g., Per-Bolo 45, Barnard1b-S, Cha-MM1, CB-17 MMS; see ][for more details]{2022ApJ...931..130D}. Recently, the outflow of a very young object, L1451-mm object, which was previously believed to be a candidate FHSC, has been detected in SiO (3-2) emission from a region very close to the source ($<$ 1000 AU) by \citet{2022A&A...666A.191W}.  Successive knot-like structures in the jet axis are yet to be confirmed with detections of emission from higher-critical density tracers like SiO (5-4) and SiO (8-7). 

In the ALMASOP sample, most SiO jet sources are in their Class\,0 phase. The source G208.89-20.04Walma is one of the youngest known objects with a SiO jet \citep[][]{2022ApJ...931..130D}. Nevertheless, G205.46-14.56S3 \citep[see][for more details]{2022ApJ...925...11D} and G208.89-20.04E are two Class\,I objects from the ALMASOP sample with SiO molecular jets observed. The objects with SiO jets exhibit small to high bipolar mass-loss rates ranging from (0.08 - 5.5) $\times$ 10$^{-6}$ M$_\sun$ yr$^{-1}$ (including both blue and redshifted lobes). 

These results challenge previous thinking about the jet launching timescale and sustainability of jets in the molecular phase. Two-dimensional ideal magnetohydrodynamic (MHD) simulations have showed that the outflow is driven by the first core or isothermal core after the first collapse \citep[][]{1969MNRAS.145..271L}. After a few hundred years, the central temperature reaches $\sim$ 2000 K, and a second collapse occurs and a rotationally supported Keplerian disk forms around the protostars \citep[][]{2013MNRAS.431.1719M,2019ApJ...876..149M}. The high-velocity jets are believed to be launched from the deep gravitational potential near this second core (or more plainly the actual protostellar core) with high velocity corresponding to the escape velocity of the protostars. These jets are detectable with molecular transitions. Such molecular jets detected with SiO emission continue up to the end of their Class\,0 life cycle or the early stage of Class\,I phase. Then the increased central luminosity in evolved protostars may photodissociate the molecular jets, but the density of jets is also decreased due to reduced mass loss rates. As a result, jets in more evolved protostars are mostly detected with ionized emission. In the ALMASOP sample, a large fraction of Class\,0 protostars do not exhibit molecular jets (Appendix \ref{sec:Appendix_individual_NoSiO_CO_definedOutflow} and \ref{sec:Appendix_individual_NOSiO_NotCO}), these could be relatively evolved protostars with reduced accretion/ejection activity. Nevertheless, they might have high-density jet components and be optically thick to SiO (5-4) and CO (2-1) emission, higher-transition SiO (e.g, J = 8-7) and CO (e.g., J = 3-2) observations could confirm the absence of a jet. 


\begin{figure}
\fig{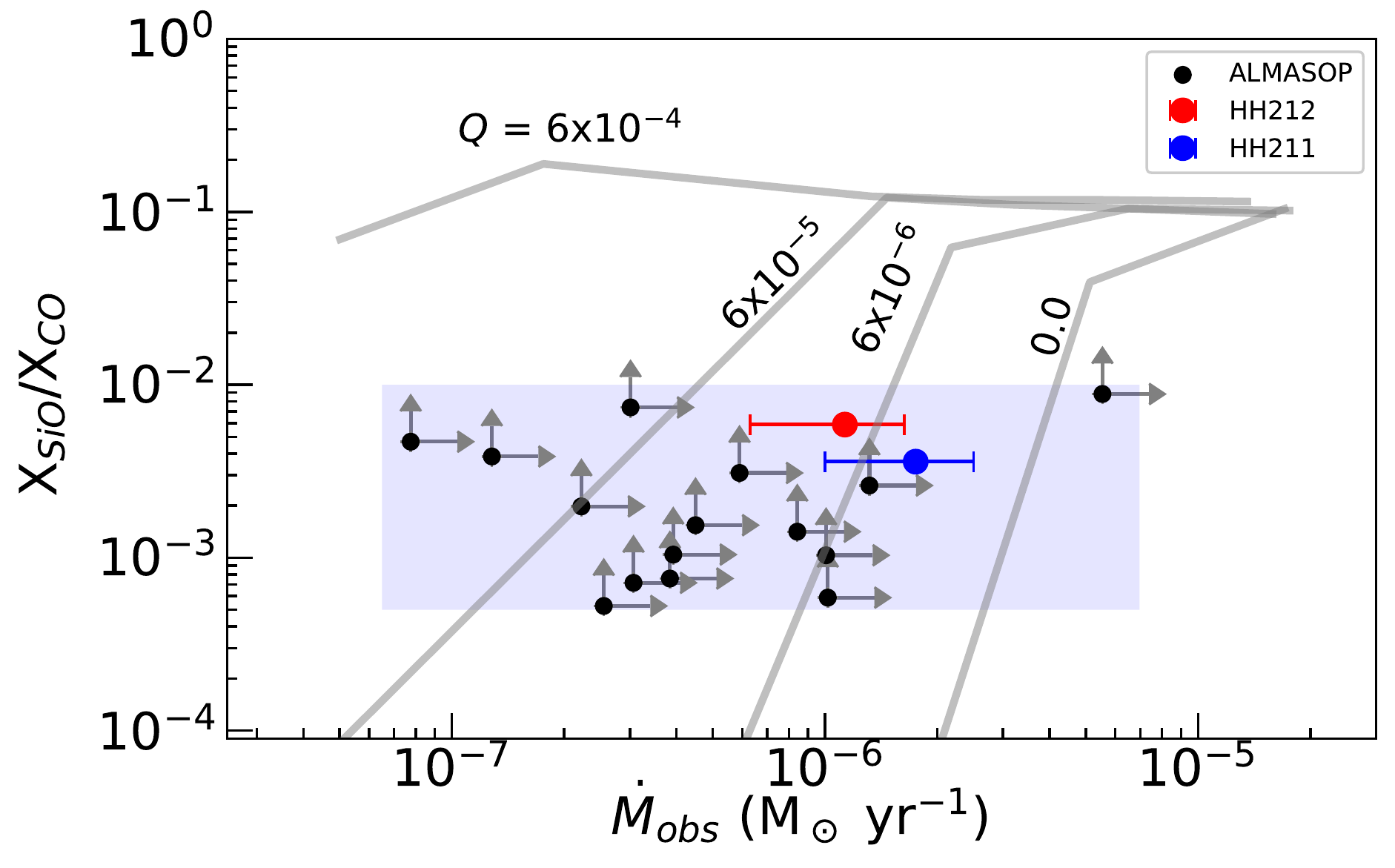}{0.48\textwidth}{}
\caption{SiO-to-CO abundance ratios, X$_{SiO}$/X$_{CO}$ = N$_{SiO}$/N$_{CO}$, are plotted as a function of observed jet mass loss rate ($\dot{M}_j$). The error bars in the abundance ratios represent the lower limit since SiO is optically thicker than CO. The abundance ratios of HH\,212 (in red) and HH\,211 (in blue) are calculated from the column densities in the published literature: \citep[HH\,212,][]{2007ApJ...659..499L} and \citep[HH\,211,][]{2021ApJ...909...11J}.
Asymptotic abundance ratios (grey lines) are adopted from \citep[][]{2020A&A...636A..60T} estimated using the Paris-Durham shock code designed to model irradiated environments and a laminar 1D disk wind model. The model considered a streamline launched at R$_0$ = 0.15 AU with Temperature = 1000 K in the disk for dust-free models (dust-to-gas mass ratio, Q = 0) and dust-poor models (Q = 6 $\times$ 10$^{-6}$, 6 $\times$ 10$^{-5}$ and 6 $\times$ 10$^{-4}$) as well as for different values of wind mass loss rates ($\dot{M}_w$)  referring to theoretical models that account for the origin of jets. The ISM Q ($\sim$ 10$^{-2}$) is indicated to be much higher than the present plotting range (green). The shaded region indicates the location of most of the observed data. 
}
\label{fig:SioCO_abundance_massLoss}
\end{figure}

\subsection{Jet-driven or wind-driven outflow}\label{sec:discussion_jetlaunchingscenario}
%
%

Whether outflowing material is jet or wind-driven is a long-standing issue with no clear conclusion yet due to a lack of a statistically significant observational sample.  We take the opportunity of the large ALMASOP dataset to explore this question while studying this unique sample of protostars with high-resolution and high-sensitivity observations with simultaneous observations of jet and winds (or outflow). In the ALMASOP sample, we observe four different types of jet and outflow ejection:  (i) a narrow CO wind shell but no SiO jet,  (ii) a narrow CO wind shell and a SiO jet, and (iii) wide-angle CO wind shell but no SiO jet (iv) wide-angle CO wind shell and a SiO jet.  We note that SiO (5-4) and CO (2-1), both of which could be optically thick towards the innermost region of the knots, could be detected with higher density tracers like SiO (8-7) and CO (3-2). 


If we accept that outflow is jet-driven, then the SiO jets with CO outflow can be straightforwardly explained (cases {\it ii} and {\it iv}). For objects with no SiO jet and narrow CO wind shell (case $i$), the outflow could be wind-driven. On the other hand, those objects with a wide angle CO shell and no SiO-jet (case $iii$) could be evolved protostars and their jets could be detected with ionized jet tracers. ALMASOP sample, however, consists of mostly very young Class\,0 systems that should drive high-density jets which may not have been detected with the lower transition of SiO and CO (as discussed in section \ref{sec:discussion_jetevolution}). It is also possible these outflows, indeed, do not associate with any jet or no strong jet that could entrain outflow material, which could possibly indicate that outflows are wind-driven.

\begin{figure*}
\fig{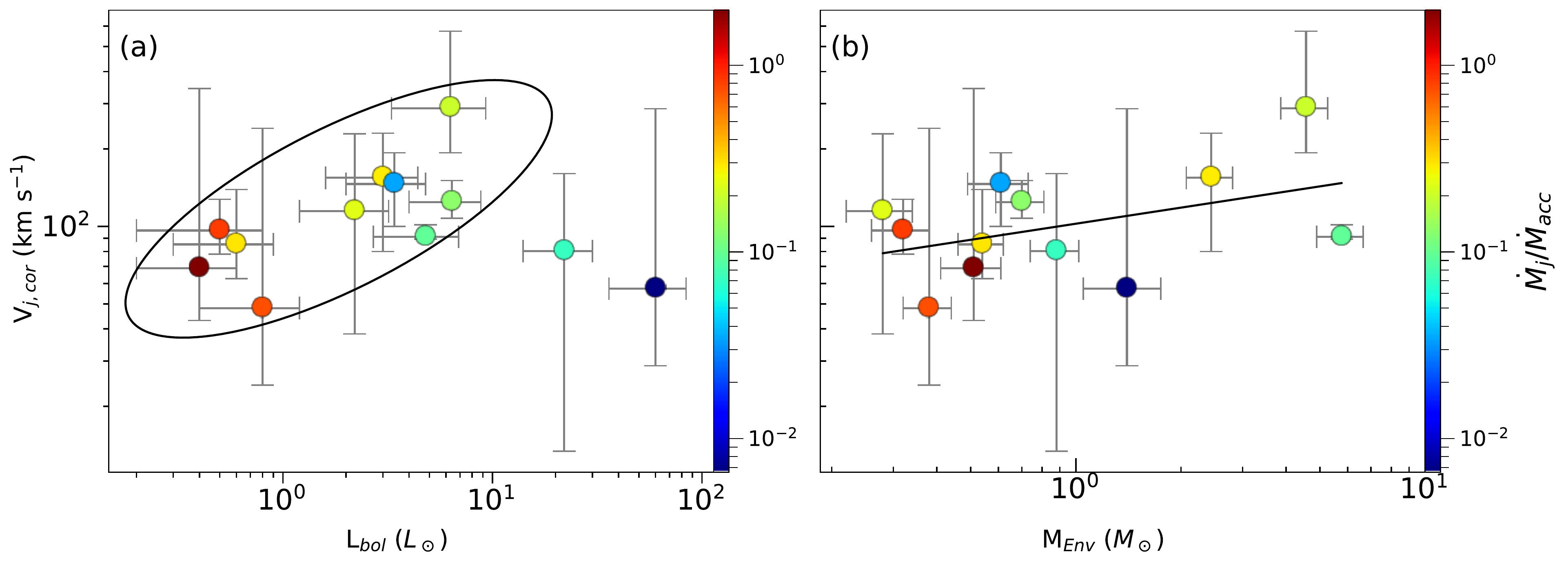}{0.98\textwidth}{}
\caption{(a) Jet velocities (V$_j$) are plotted as a function of bolometric luminositiy (L$_{bol}$) in panel (a) and envelope mass (M$_{Env}$) estimated from ACA continuum in panel (b). The color bar indicates the ratio between mass ejection rate to accretion rate ($\dot{M}_{j}$/$\dot{M}_{acc}$), where the mass accretion rates $\dot{M}_{acc}$ were estimated for a protostellar mass of 0.1 M$_\sun$ and total luminosity using Equation \ref{equ:mass_accretion}. Objects with increasing trends are marked with ellipse to guide the eye. 
}
\label{fig:Vj_LbolEnvelopeMass}
\end{figure*}

\subsection{Origin of gas-phase SiO}
SiO emission has been frequently reported in previous studies as originating in the the shock regions along protostellar jets in their earliest phases. As discussed earlier, we observe 21 fields with SiO emission in the ALMASOP sample (18 in this study, three previously published in \citet[][]{2022ApJ...925...11D,2022ApJ...931..130D}). Of these, 16 protostars are confirmed to have SiO emission along the jet axis. Three other objects display bow-shock-like extended SiO emission along their jet axes, which could be associated with jets but their small velocity ranges or insufficient velocity information make it difficult to be sure. One source exhibits SiO in the collision zone between two outflows and another the driving source of SiO emission is unclear. So far, the precise origin of SiO in the jet is still not constrained due to the lack of significant observations at high resolution and sensitivity. 

There are two competing scenarios in the literature describing the mechanism of  SiO formation. In the ``grain-sputtering" scenario, SiO could originate at the shock region within the jet itself through grain sputtering due to ion-neutral decoupling  \citep[][]{1997A&A...321..293S,2012A&A...538A...2P}.  In the ``dust-sublimation zone" scenario, if the jet is launched from the base or within the dust sublimation radius \citep[R$_{\rm sub}$ $\sim$ 0.15\,au $\times$ $\sqrt{L_{bol}/L_\sun}$; ][]{2016A&A...585A..74Y}, the stellar far-ultraviolet (FUV) emission there can sublimate the silicate dust grains and release Si$^+$ into the gas phase, which later recombines with oxygen to produce SiO. On the other hand, the presence of FUV excess emission can also lead to the photodissociation of molecules and cause a dramatic decrease in SiO and CO abundances \citep[][]{1997A&A...321..293S}. \citet[][]{2020A&A...636A..60T}, however,  suggest in their laminar dust-free disk model that if the jet is launched from the dust sublimation zone at high temperature ($>$ 800 K), SiO and CO could be abundant within the launching material.

Assuming optically thin CO and SiO emission, we measure the X$_{SiO}$/X$_{CO}$ for all the SiO emitting jet sources from the innermost knots (B1 and R1 knots) and plot thse values as a function of jet mass-loss rate in Figure \ref{fig:SioCO_abundance_massLoss}. We note that SiO (5-4) could be optically thicker towards the dense shock than CO (2-1) in the same region. Therefore the SiO emission is likely the lower limit and the X$_{SiO}$/X$_{CO}$ is the lower limit in this case. From our observations, we see that SiO jets tend to have X$_{SiO}$/X$_{CO}$ in the range $\sim$ 5 $\times$ 10$^{-4}$ to 10$^{-2}$. We also plotted two well known Class\,0 objects with SiO jets, HH\,211 from \citet[][]{2021ApJ...909...11J} and HH212 from \citet[][]{2007ApJ...659..499L}, for comparison with the ALMASOP sample.  We then compare this sample to the tracks for different dust-to-gas mass ratios (Q) of \citet[][]{2020A&A...636A..60T} (solid curves). 
We see that all the SiO jets are possibly dust-poor and hence SiO could have originated from within the dust sublimation zone of its associated protostars.

\begin{figure*}
\fig{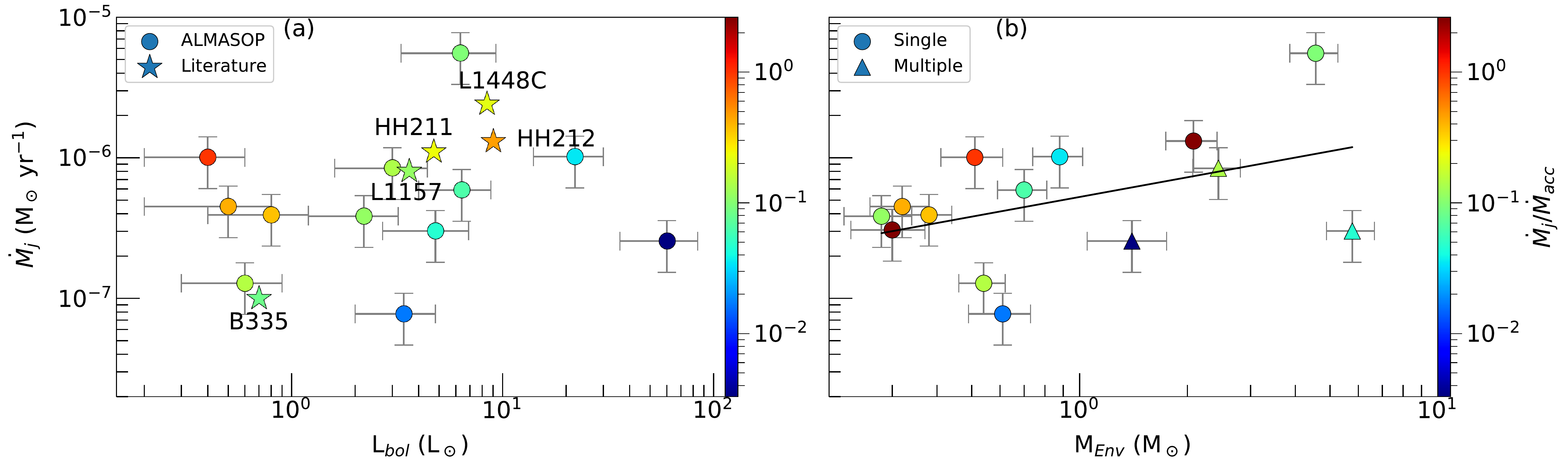}{0.98\textwidth}{}
\caption{(a) Jet mass-loss rates ($\dot{M_j}$) are plotted as a function of bolometric luminosity (L$_{bol}$) in panel (a) and envelope mass M$_{Env}$ in panel (b). The color bar indicates the ratio between mass-loss rate to accretion rate ($\dot{M}_{j}$/$\dot{M}_{acc}$). Stars in panel (a) are data points for different  well-known SiO-jet objects adopted from the literature. The triangles in panel (b) are multiple sources within a common envelope.  
}
\label{fig:Mj_Lbol_Menv}
\end{figure*}


Detectable SiO (5-4) emission within the jet needs very high-density material, given that transition's high-critical density (Table \ref{tab:citical_density}). The majority of the ALMASOP sources have mass-loss rates $\dot{M}_j$ $<$ 10$^{-6}$ $\mlr$ (73\% sources within 0.1 - 0.4 $\times$ 10$^{-6}$ $\mlr$). Such small $\dot{M}_j$ may not create enough dense material to reach up to the critical density of SiO. \citet[][]{1991ApJ...373..254G} estimated X$_{SiO}$/X$_{CO}$ $<$ 5 $\times$ 10$^{-9}$ for a collimated and accelerated wind and a typical mass-loss rate of \.{M}$_j$ $\sim$ 0.3 $\times$ 10$^{-6}$ M$_\sun$ year$^{-1}$ (see their Figure 8). From the ALMASOP sample, we observe SiO-to-CO abundances X$_{SiO}$/X$_{CO}$ $\sim$ 5 $\times$ 10$^{-4}$ to 10$^{-2}$, which could be much higher if we consider that SiO is optically thicker than CO 
(indicated as the lower limit of error bars in Figure \ref{fig:SioCO_abundance_massLoss}). For a typical mass-loss rate of \.{M}$_j$ $\sim$ 3 $\times$ 10$^{-6}$ M$_\sun$ year$^{-1}$, X$_{SiO}$/X$_{CO}$ could be enhanced to $\sim$ 5 $\times$  10$^{-2}$ \citep[see Figure 6 in][]{1991ApJ...373..254G}, then our observational findings could be compared with that of the model prediction. One possible explanation for such high observed SiO abundances than predicted by models is that \citet[][]{1991ApJ...373..254G}  considered mass-loss to be  spatially extended as a wide `wind'. On the other hand, we estimate the jet mass-loss rate from the innermost part of the outflowing cavity, i.e., along the jet axis. Hence, the mass-loss rates along the inner jet-axis in this study could have underestimated the actual mass-loss rates \citep[see Figure 3 of ][]{1995ApJ...455L.155S}. Therefore, our $\dot M_j$ values estimated along the jet axis could be biased too low to compare accurately with \citet[][]{1991ApJ...373..254G}. 


Another possibility to explain the high SiO-to-CO ratios is that the efficiency of CO production is less than SiO if the SiO and CO within the jet originate by the same process. Following \citet[][]{1979ApJ...231...77D,2006ApJ...648..435N}; and \citet[][]{2019MNRAS.487.3252H}, the dust sputtering time-scale can be defined as
\begin{equation}\label{equ:sputtering_rate}
\noindent
\begin{aligned}
& t_{sput} \equiv \frac{a}{3 n_H Y_{tot}}, \\
& \approx 0.33 ~Myr ~\Big(\frac{a}{\mu m}\Big) \Big(\frac{n_H}{cm^{-3}}\Big)^{-1} \Big(\frac{Y_{tot}}{10^{-6} \mu m ~ yr^{-1}~cm^3}\Big)^{-1}, \\\\
\end{aligned}
\end{equation}
where grain size is denoted by $a$ and hydrogen number density of the gas is $n_H$.  The erosion rate $Y_{tot}$ $\equiv$ $(da/dt)/n_H$ is adopted from \citet[][]{2006ApJ...648..435N}. For thermal sputtering, Y$_{tot}$ depends on the temperature of the region and grain sputtering occurs above $\sim$ 10$^{5}$ K. For non-thermal sputtering, the jet velocity should be above $\sim$ 30 $\kms$. In the jets observed here, the temperatures are much smaller than the thermal sputtering limit. Non-thermal sputtering, however, could occur at a high jet velocity.  We estimate the typical critical density $n_{cr,sio}$ $\sim$ 10$^{6}$ cm$^{-3}$ is needed to excite SiO in the shocks (section \ref{sec:intro}), which could be also considered as a typical post-shock density in the jet. Therefore, the pre-shock density is expected to be smaller than n$_{cr,sio}$ along the jet direction, i.e., less than $<$ 10$^{6}$ cm$^{-3}$. 
For a mean jet velocity of $\sim$ 50 $\kms$, we estimate sputtering time scales of 2.75 - 275 years for silicate dust and 5.4 -  540 years in the case of carbon dust, for pre-shock densities $\sim$ 10$^{6}$ to 10$^{4}$ cm$^{-3}$, respectively. So for high pre-shock density and high-velocity jets, sputtering is a plausible mechanism to form SiO and CO, even in the nearest knots with 
 dynamical time scales less than 100 years. The erosion rate of carbonaceous dust grains (producing CO) is smaller than silicate dust grains (producing SiO) for the same jet velocity \citep[][]{2006ApJ...648..435N}.  Therefore, equation \ref{equ:sputtering_rate} suggests that the SiO formation time-scale is half that of CO formation for the same pre-shock density, which could be the reason for higher SiO to CO ratios in the knots. CO, however, can easily sublimate from grain mantles at temperatures of $\sim$ 20$-$25 K, where sputtering is not necessary. Therefore, it is unlikely that the majority of the CO is formed later than SiO.

The dust-sublimation zone scenario is well applicable to the objects with SiO emission in the knots close to the source (e.g., G205.46-14.56M1, G205.46-14.56S1, G205.46-14.56S3, G206.12-15.76, G206.93-16.61W2, G208.68-19.20N3, G208.89-20.04E, G208.89-20.04Walma, G209.55-19.68S1, G209.55-19.68S2, G210.37-19.53S, G211-19.27S$\_$B G215.87-17.62M). Although the dynamical time scales of the nearest knots could not be measured due to the poor spatial resolution of our observations, in most cases it is likely less than 100 years. Such short time scales are long enough to produce the high X$_{SiO}$/X$_{CO}$ ratios observed in this sample, except in case where the preshock density is very high ($\sim$ 10$^{6}$ cm$^{-3}$). As an example, for G191.90-11.20S (Figure \ref{fig:appendix_G191.90_11.20S}f), 
 the jet emitting zone is uncertain, since the nearest knot (R1) is very faint and brightest knots (e.g., R2) are far away from the source, and the grain-sputtering scenario could be a possible mechanism to form SiO within the jet even for a low preshock density  ($\sim$ 10$^{-4}$ cm$^{-3}$).  A few other objects also exhibit SiO emission in the outflow direction such as G192.32-11.88N, G200.34-10.97N, G205.46-14.56M2, G209.55-19.68N1. Meanwhile, SiO emission in the field of G209.55-19.68N1 is clearly in the collision zone between two outflows. The SiO emission in the remaining three cases could be processed in the location through grain-sputtering or collision between the outflow material with ambient clouds. 

To summarize, we suggest that most of the SiO is originating within the dust sublimation zone near the base of the source. We can not, however, exclude the grain-sputtering mechanism from the present state observations. Further high-sensitivity and high-velocity resolution observations could provide more constrain on the SiO formation scenario. Such observations could provide a more stringent view of the nearby knots, and allow measurement of jet rotation that can in turn constrain the jet-launching radius \citep[][]{2017NatAs...1E.152L}.

\subsection{Accretion and Ejection Process}\label{sec:discussion_accretion_inporotostars}
The ejected material along the jet can carry vital information about the present evolutionary status, history of the accretion process, and final mass budget of the protostar. The properties of the jet ($V_{j,cor}$, $\dot{M}_j$, $L_{j,kin}$, $\mathcal{T}_{\rm knot}$) are possibly associated with the driving internal properties of the protostars and physical structure of the surrounding materials.  In this section, we explore the inter-correlation of the jet ejection with the intrinsic properties of the protostars and their disks and envelopes.

We estimated mass acceleration rates $\dot M_{\rm acc}$ based on the total jet mass-loss rates $\dot{M_j}$, bolometric luminosity $L_{bol}$, the total kinetic luminosity of both the blueshifted and redshifted lobes $L_{j,kin}$, as presented in the Table \ref{tab:jet_outflow_Properties}. The accretion rate is defined as,
\begin{equation}\label{equ:mass_accretion}
\indent
\dot{M}_{acc} \sim \frac{(L_{bol} + L_{j,kin}) R_*}{GM_*} ~~ M_\sun ~yr^{-1}\\\\
\end{equation}
 assuming that both the bolometric and kinetic luminosity originate entirely from accretion. For these calculations, we assume a stellar radius of R$_*$ $\sim$ 2R$_\sun$ \citep[][]{1988ApJ...332..804S}. The jet-ejecting protostars consists of mostly Class\,0 and two early Class\,I objects. Therefore, we assume the stellar cores have not evolved much so that their central objects are in the low mass range of $M_*$ = 0.05$-$0.30 M$_\sun$ \citep[][]{2000ApJ...545.1034S,2017ApJ...834..178Y}. Specifically, we consider an average mass of  M$_*$ = 0.1 M$_\sun$ for all the objects to estimate $\dot{M}_{acc}$. The ratio of $\dot M_{\rm acc}$/$\dot M_{\rm j}$ ranges from 0.003 to 2.1. In the subsequent sections, we discuss how these $\dot{M}_{acc}$ values are correlated with the central driving force of the jet and explore possible implications on the chemical composition of the disk and planet formation.

\subsubsection{Jet velocity: dependence on luminosity and envelope mass}\label{sec:discussion_accretion_inporotostars_Vj}
Figure \ref{fig:Vj_LbolEnvelopeMass} shows the jet velocity of the ALMASOP objects as a function of luminosity in panel (a) and envelope mass in panel (b).  In panel (a), most of the objects (marked with ellipse) exhibit a correlation between $V_{j,cor}$ and $L_{bol}$. Large error bars of two objects beyond the ellipse possibly restrict them to locate within the correlation zone. $\dot{M}_{j}$/$\dot{M}_{acc}$ ratio also likely decreases with bolometric luminosity from $L_{bol}$ $\sim$ 0.2 to 10 L$_\sun$. 
With increasing luminosity, the protostellar core is expected to be more evolved. Ejection-to-accretion activity will also decrease. 
Similar correlations of ejection-to-accretion activity were also reported in \citet[][]{2013A&A...551A...5E}; and \citet[][]{2020A&ARv..28....1L} (see their $\dot{M}_{j}$ versus $\dot{M}_{acc}$ plots). 

In panel (b), most of the $V_{j,cor}$ values are correlate with the mass of the envelope $M_{Env}$. A linear fit is consistent with the equation: 
\begin{equation}
   \log(V_{\rm j,cor}) = (0.21 \pm 0.14) \log(M_{\rm Env}) + (2.0 \pm 0.06).
\end{equation}  
We do not see any clear dependence on $\dot{M}_{j}$/$\dot{M}_{acc}$ for such a correlation. Higher envelope masses may increase the accretion rates given the larger local mass reservoir. In addition, larger protostellar masses have higher gravitational potentials from which high-velocity jets are ultimately powered. At evolved protostellar phases, however, the jets become ionized and accretion rates also decrease, but the jet will be still in high velocity due to the higher mass of the central protostars. A consistent measurement of the protostellar masses from kinematics will be more helpful to illustrate this scenario.

\subsubsection{Jet mass-loss: dependence on luminosity and envelope mass}\label{sec:discussion_accretion_inporotostars_Mj}
Figure \ref{fig:Mj_Lbol_Menv}a shows the observed mass ejection rates $\dot{M}_{j}$ as a function of the bolometric luminosity of the objects. We compare ALMASOP objects with some well-known SiO jet objects from the literature. The $\dot{M}_{j}$ and $\dot{M}_{acc}$ of B\,335 is comparable to low-luminosity sources. The objects L\,1157, HH\,212, HH\,211, and L\,1448C share  similar properties with ALMASOP objects of $L_{bol}$ $\sim$ 3 - 10 L$_\sun$. Objects with low $\dot{M}_{j}$/$\dot{M}_{acc}$  (bluish colours) exhibit lower $\dot{M}_{j}$ rates with increases in central luminosity.  The objects with high  $\dot{M}_{j}$/$\dot{M}_{acc}$ (green, yellow, orange, and red colours), however, do not show any obvious correlation. 

Figure \ref{fig:Mj_Lbol_Menv}b shows the $\dot{M_j}$ values as a function of envelope mass, as estimated from the ACA continuum.  Although large scatter in the data points, we see a correlation of jet mass loss rate with envelope mass. Further, we do not see any obvious effect of multiplicity on the mass accretion and ejection phenomena. 
A linear fit reveals a correlation as,
\begin{equation}
   \log(\dot{M}_{\rm j}) = (0.47 \pm 0.27) \log(M_{\rm Env}) - (6.27 \pm 0.11)
\end{equation}  
There are five sources (blue data points in Figure \ref{fig:Mj_Lbol_Menv}b) below the fitting-line with lower $\dot{M}_{j}$/$\dot{M}_{acc}$, those could be more evolved sources. The sources above the fitting-line mostly exhibit higher $\dot{M}_{j}$/$\dot{M}_{acc}$, which could be due to then being much younger sources with high ejection-to-accretion activity.



\begin{figure*}
\fig{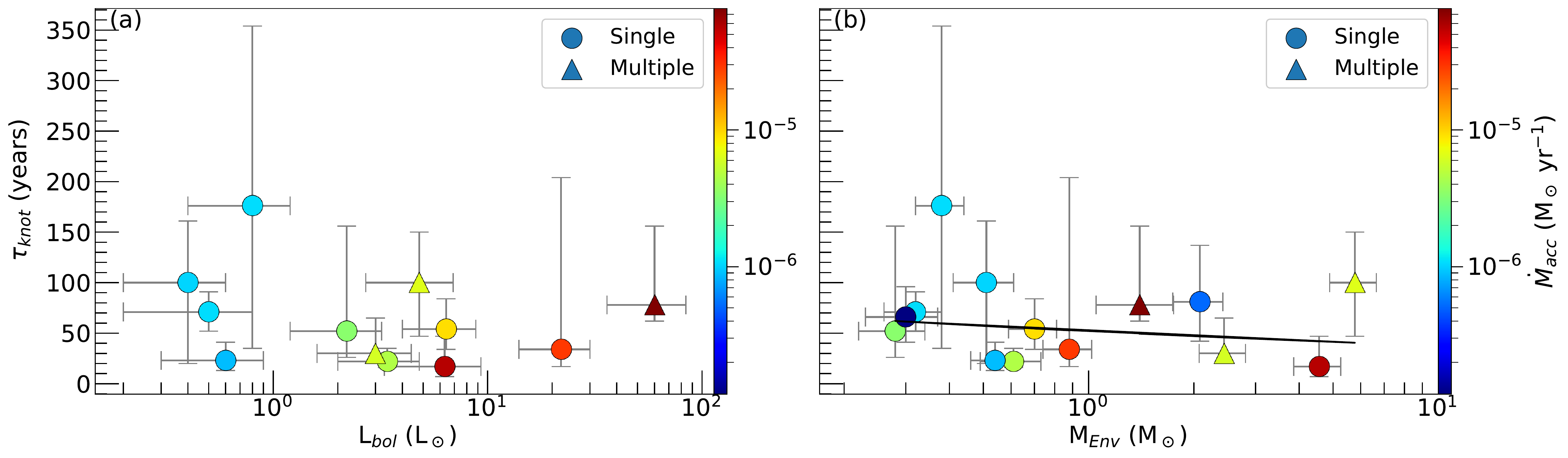}{0.98\textwidth}{}
\caption{Mean ejection periods ($\mathcal{T}_{\rm knot}$) are plotted as a function of bolometric luminosity (L$_{bol}$ in panel (a) and envelope masses (M$_{env}$) in panel (b). Single and multiple sources are shown in circles and triangles, respectively. The colour scale indicates the mass-accretion rate ($\dot{M}_{acc}$). Straight line panel (b) is the linear fit to the data points. 
}
\label{fig:period}
\end{figure*}

\subsubsection{Episodic ejection: dependence on luminosity and envelope mass}\label{sec:discussion_accretion_inporotostars_periods}
As discussed earlier, all of the SiO jets are discrete clumpy knot-like or bow-shock-like structures.  Some are roughly equally spaced and may trace a quasi-periodic ejection. We estimated the mean periods between knots in the range of $\sim$  20$-$175 years, with the error bars based on the uncertainties in the inclination angle, jet velocity, and knot spacing.  These discontinuous structures may be produced due to the temporal variations in the jet velocity and density. Although the mechanisms that drive the variations in jet density and velocity are still not well understood, the general consensus is that they are driven by the quasi-periodic perturbations of the underlying accretion in the disks. A few potential mechanisms have been proposed which could produce variations of accretion. e.g., \citep[][]{2014prpl.conf..387A,2020A&ARv..28....1L,2022arXiv220311257F}: (i) accretion driven by binary interaction where the variations are modulated by the orbital timescale; (ii) gravitational instabilities governed by envelope accretion; (iii) planetesimal accretion onto the central protostar; and (iv) gravitational instabilities produced at dust sublimation zones. 

In a few cases, episodes have been reported in the jets of objects such as HH\,34 \citep[270 yrs and 1400 yrs;][]{2002A&A...395..647R}, HH\,111 \citep[60 yrs and 950 yrs;][]{2002A&A...395..647R}, and HH\,212 \citep[1 yr, 60 yrs and 605 yrs;][]{1998Natur.394..862Z}. Different periods in a single system could be linked to different periodic perturbations of the underlying accretion in the disk.  Further evidence for short timescales comes from monitoring of protostars at  mid-infrared through submm wavelengths. The JCMT Transient Survey \citep{herczeg17} has uncovered submm brightness variability with what appear to be decades long timescales from many deeply embedded protostars \citep{2021ApJ...920..119L} and a curious 18-month periodicity for one particular source \citep{lee20}. Similarly an analysis of mid-infrared monitoring of over 5000 YSOs by NEOWISE recovered a similar large percentage (20\%) of long-term secular variables among the most deeply embedded members of the sample \citep{2021ApJ...920..132P}.

\begin{figure}
\fig{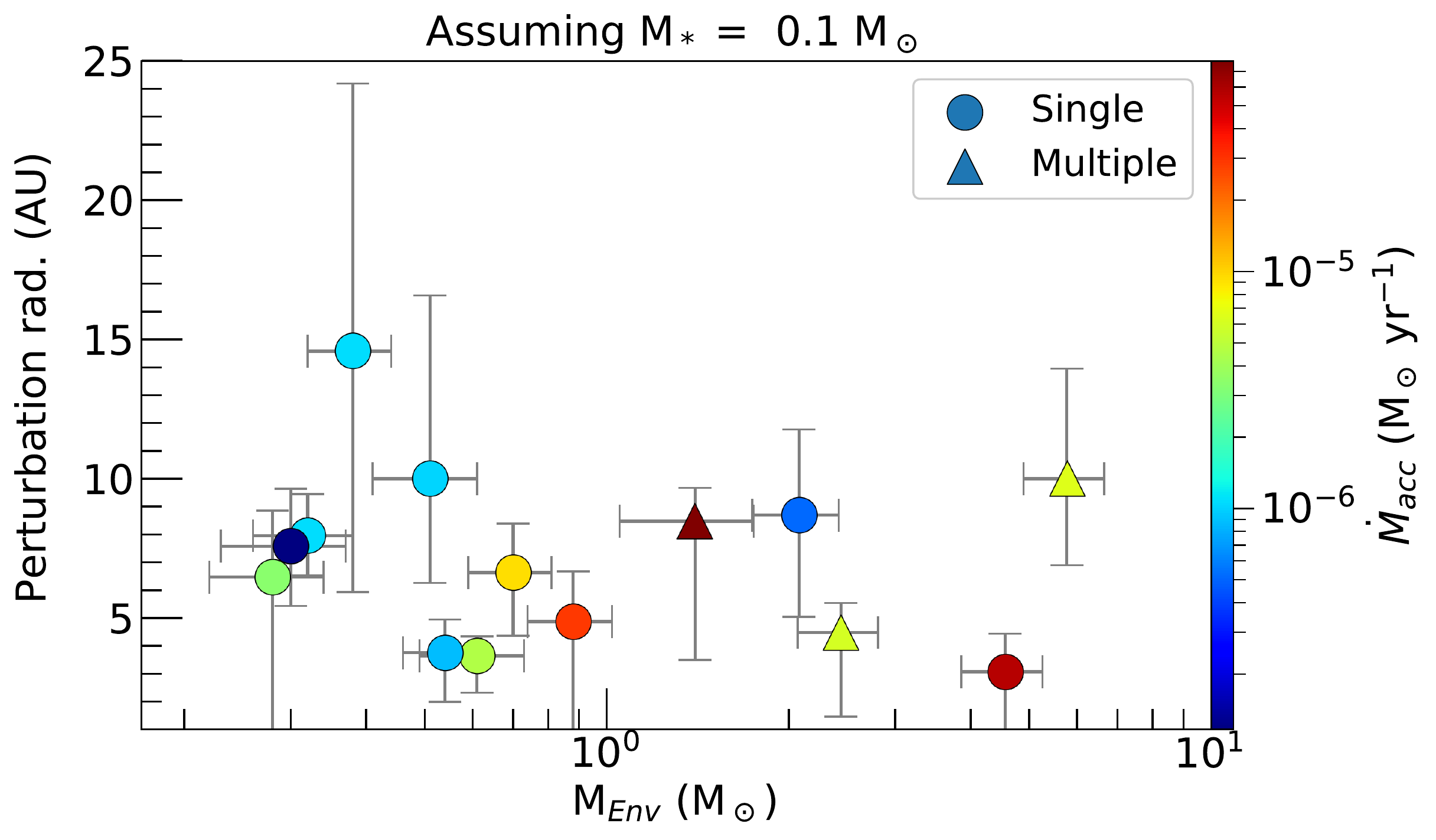}{0.47\textwidth}{}
\caption{Keplerian radius corresponding the observed mean periods of episodic events are plotted as a function of envelope masses. A stellar mass M$_*$ = 0.1 M$_\sun$ is assumed for all protostars. All the notations are same as Figure \ref{fig:period}b. 
}
\label{fig:Perturbation_Menv}
\end{figure}

\begin{figure}
\fig{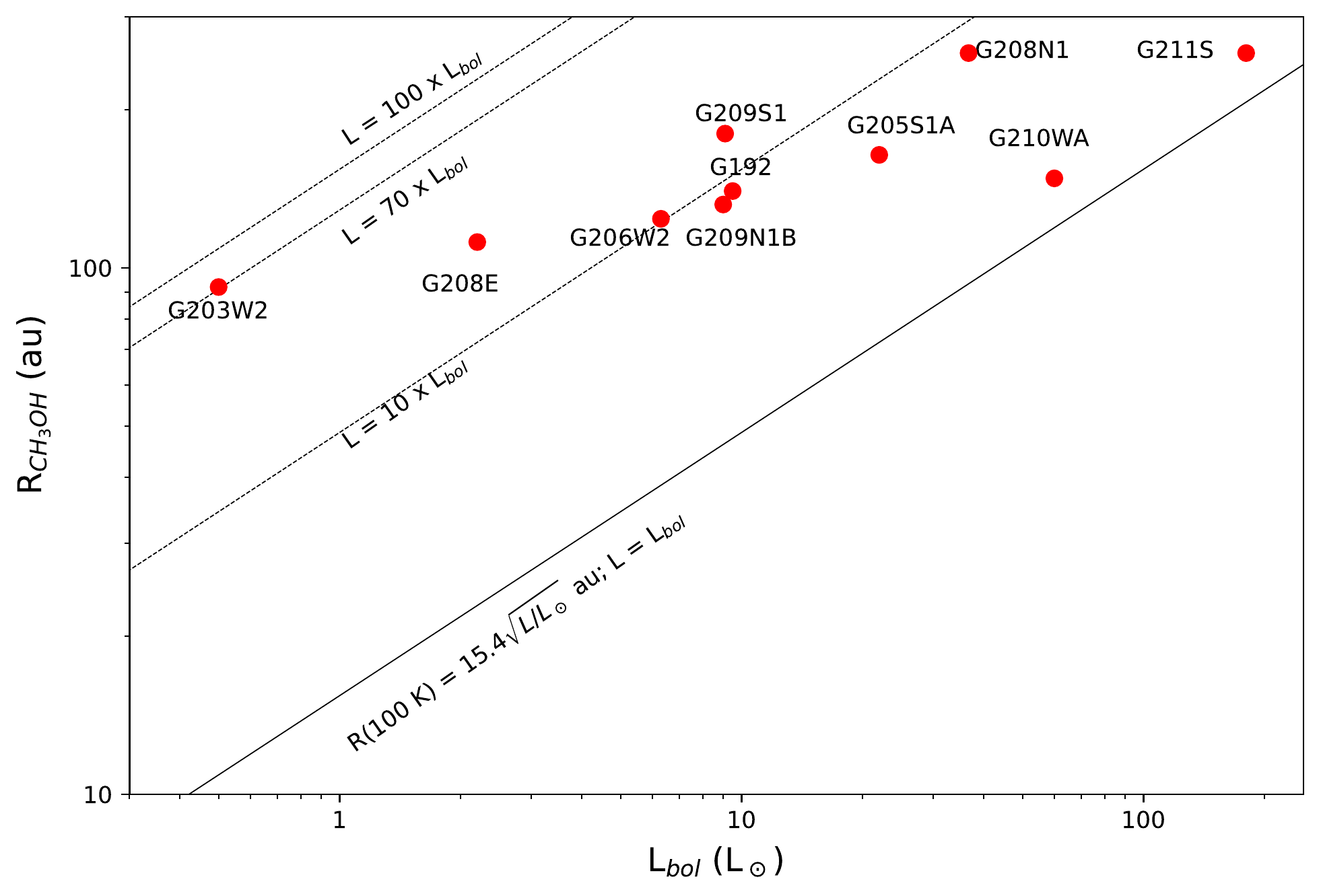}{0.47\textwidth}{}
\caption{CH3OH icelines (R$_{CH_3OH}$) of the known Hot Corinos are plotted as function of bolometric luminosity (L$_{bol}$. Observed CH$_3$OH radius are adopted from \citep[]{2022ApJ...927..218H}. The slanting lines indicate iceline radius at 100 K for luminosity L$_{bol}$ = 1, 10, 70 and 100 L$_\sun$, respectively. 
}
\label{fig:CH3OH_hot_corno}
\end{figure}


In our observations with relatively small fields-of-view, we observe mostly single periods or quasi-periodic knots/bow-shocks. As a result, we estimate only mean periods of events. The corresponding orbital radii for those periods can be estimated  from Kepler's third law of orbital motion following equation:
\begin{equation}
\noindent
\begin{aligned}
a = \Big(\frac{GM_*T^2}{4\pi^2}\Big)^{1/3} = \Big(\frac{M_*}{M_\sun}\Big)^{1/3} \Big(\frac{T}{yr}\Big)^{2/3} AU,
\end{aligned}
\end{equation}
where $T = \mathcal{T}_{\rm knot}$ is the mean episodes of ejection and $M_*$ is the mass of the central protostars, which is assumed to be 0.05$-$0.30 M$_\sun$. The Keplerian radius corresponding to the episodes are estimated to be 2.0$-$25 au (including errorbars) for different objects (Table \ref{tab:keplerian_radius}). We estimated two sets of radii for two masses. In the absence of studies on large-scale episodes beyond our fields-of-view and without confirmations of binary components within those perturbation zones, it is difficult to link accretion perturbation with any particular mechanism. Since ALMASOP sample contains mostly Class\,0 and early Class\,I,  most sources are expected to have small or intermediate disks, and  so the perturbation may have originated within the disk or in the outskirts of the disk. Further studies of disk size and large scale periodicity could confirm the origins of the accretion variability.



We plotted $\mathcal{T}_{\rm knot}$ as a function of bolometric luminosity or envelope mass in Figure \ref{fig:period}a and b, respectively. Although the data points in Figure \ref{fig:period}a exhibit considerable scatter, objects with lower luminosity and lower mass-loss rates (bluish points) have mostly longer periods. The lack of a clear trend with luminosity may be due to  variable accretion. The observed luminosity could represent the true evolutionary sequence of the protostars. In Figure \ref{fig:period}b there is a mild anti-correlation between the $\mathcal{T}_{\rm knot}$ and M$_{Env}$. A linear least square fit provides:
\begin{equation}
   \log({\mathcal{T}}_{\rm knot}) = (-0.14 \pm 0.19) \log(M_{\rm Env}) - (1.71 \pm 0.08).
\end{equation}  
%
In Figure \ref{fig:Perturbation_Menv} the Keplerian perturbation radius corresponding to the periods of ejection is plotted as a function of envelope mass, assuming the same protostellar mass M$_*$ = 0.1 M$_\sun$. For protostars with lower envelope masses, the perturbation radii tend to be larger. There are three wide binary systems with separations $>$ 100 au in the plots. None have separations matching the Keplerian perturbation radius. Hence, it is unlikely these respective accretion/ejection events are driven by binary orbital dynamics. 

\begin{figure*}
\fig{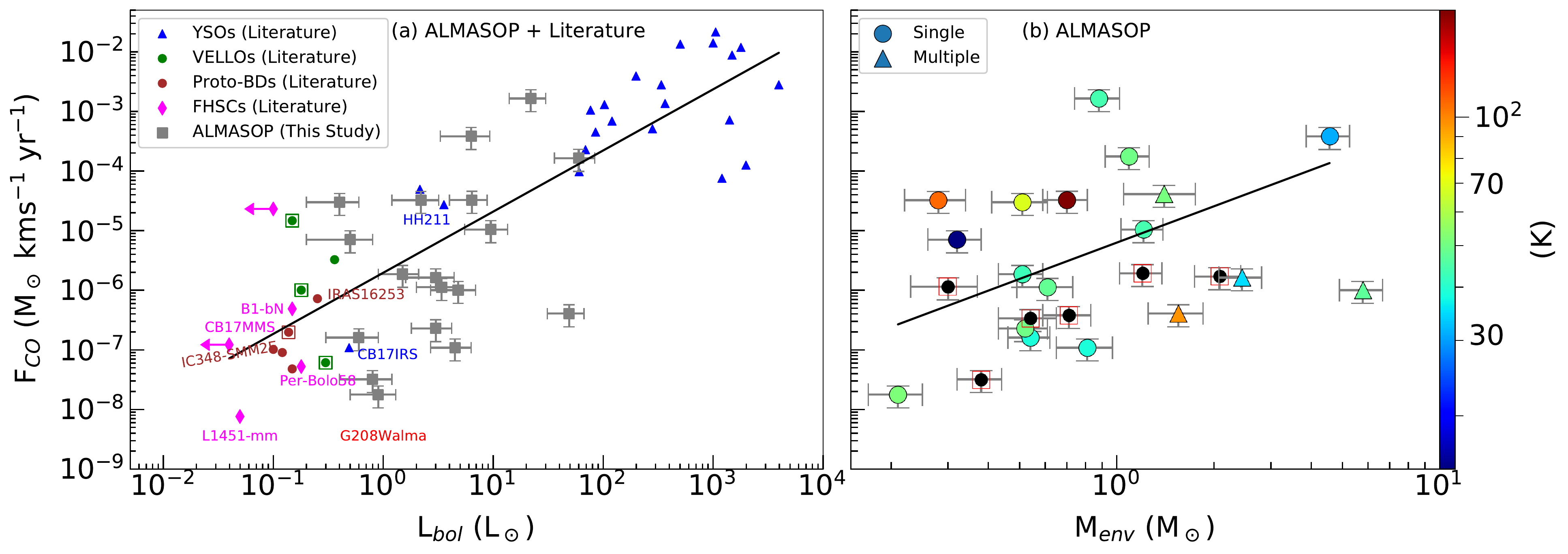}{0.99\textwidth}{}
\caption{
(a) F$_{\rm CO}$ is plotted as a function of L$_{\rm bol}$. ALMASOP observations are shown in grey square symbols. Blue, green, brown, and magenta points represent YSOs, VeLLOs, proto-BDs, and candidate FHSCs, respectively. Proto-BDs are also in the VeLLOs category. These data points have been adopted from  \citet[][]{2014MNRAS.444..833P,2016ApJ...826...68H,2022ApJ...931..130D}. Open squares indicate single-dish observations. A linear fit of the whole sample (ALMASOP and literature) is shown in the black line. 
(b) F$_{CO}$ is shown as a function of M$_{Env}$. The filled circles are single systems. Tringles indicate binary or other multiple systems. Colors are for bolometric temperature (T$_{bol}$), where T$_{bol}$ $>$ 70\,K are condered as Class\,I objects. The black-filled circles with red open squares do not have T$_{bol}$ estimation. 
}
\label{fig:FCO_Lbol_Menv}
\end{figure*}

\subsubsection{Episodic ejection: Implication to Complex Organic Molecule and Planet formation}\label{sec:discussion_COMs_planet}
During each episodic outburst, the central luminosity increases to a maximum $L_{max}$. Subsequently, the surrounding disk temperatures increase and the various snowlines shift outwards. After each outburst, during the quiescent phase, the luminosity decreases exponentially with time to a minimum $L_{min}$, the surrounding disk temperatures decrease and the snowlines shift inward. Therefore episodic outbursts affect the variation of temperature within the disk. When the temperature increases, more molecular ices on grains convert to the gas phase, which increases the possibility of complex molecule formation through gas-phase reactions  \citep[][]{2016ApJ...821...46T,2019NatAs...3..314L,2022Natur.606..272J}. During the post-outburst quiescent phase, gas-phase molecules could  recondense to ice.  Following \citet[][]{2001A&A...378.1024C} and \citet[][]{2003ApJ...585..355R}, the freeze-out rate for a neutral molecule, $X$,  can be estimated as
\begin{equation}
\noindent
\begin{aligned}
\lambda(X) = 2.3 \times 10^{-18} \Big(\frac{T}{m(X)}\Big)^{0.5} n_{H_2} ~ (s^{-1})
\end{aligned}
\end{equation}
To explore variations of chemistry in these systems, we assume a mean grain radius of 0.1 $\mu$m, and a grain abundance of 10$^{-12}$ at the protostellar envelope outside the hot corino.  $T$ is the temperature of the gas in Kelvin, $m(X)$ is the molecular weight of the molecule $X$, and n$_{H_2}$ is the molecular hydrogen density in cm$^{-3}$. CH$_3$OH is one of the simplest COMs, which could produce CH$_3$OCH$_3$, CH$_3$OCHO, C$_2$H$_5$OCH$_3$, C$_2$H$_5$OC$_2$H$_5$, and C$_2$H$_5$OCHO through gas-phase reactions \citep[][]{2016ApJ...821...46T}.  In the case of CH$_3$OH, for an approximate ice-evaporation temperature $T \sim$ 100 K and a density of n$_{H_2}$ $\sim$ 10$^7$ cm$^{-3}$ in the outer disk hot-core boundary region, the freeze-out time scale becomes $\sim$ 780 years (710 years for a maximum temperature of 120\,K or 930 years for 70\,K evaporation temperature of CH$_3$OH). Based on ejection periods, the outburst time scales of 17$-$530 years for the present ALMASOP objects are  less than the freeze-out time scale. Therefore, each episode of outburst will maintain the outer disk temperature and allow the further formation of complex organic molecules through gas-phase reactions.

Some of the studied objects have been classified as ``hot corinos" by \citet[]{2022ApJ...927..218H} due to detection of several COMs around them. We have adopted the CH$_3$OH radius from that work and plotted it in Figure \ref{fig:CH3OH_hot_corno}.  The water sublimation radius R(100) in the envelope at a given luminosity (black line) has been calculated by the equation in Bisschop et al. 2007.  Most of the sources have much larger R$_{CH_3OH}$ than what is expected from their current luminosities. The dashed lines show R(100K) when the luminosity is enhanced by a certain factor. The R$_{CH_3OH}$ of many sources is consistent with the enhancement factor of 10. G210WA and G211S seem to be in their burst phases since R$_{CH_3OH}$ $\sim$ R(100K), and their L$_{bol}$s are much larger than typical protostars. The large discrepancy between the methanol emission size, R$_{CH_3OH}$ and the R(100) estimated from the current luminosity suggests that all sources had prior accretion bursts, and the freeze-out timescale should be longer than the time interval between episodes.

The NEOWISE photometric data in the Orion clouds has been extensively detected the variable stars in the region \citep{2021ApJ...920..132P}. A majority of the jets ejecting objects are observed to be photometric variable.



Since there is growing evidence of grain-growth and planet formation in the early phase \citep[e.g.,][]{2018NatAs...2..646H}, the emerging planet's chemical composition, especially in the outer disk, will be largely impacted by the repetitive outbursts. Evidence of the role of episodic outbursts on the evolution of disks and chemical composition of planets was also reported for a young Class\,0 binary system NGC 1333-IRAS2A \citep[][]{2022Natur.606..272J}.


\subsection{Correlation between outflow force with luminosity and envelope mass} 
Correlations between outflow force $F_{CO}$ and source properties have long been investigated e.g., with bolometric luminosity and envelope mass \citep[][]{1996A&A...311..858B}. We compare the ALMASOP sample with other candidate first hydrostatic cores, very low luminosity Objects (VeLLOs), and known young stars.  
Figure \ref{fig:FCO_Lbol_Menv}a shows $F_{CO}$ as a function of $L_{bol}$.  Despite the limiting values in our estimation, we found that the ALMASOP sample is located between the well-known Class 0/I sample, and the candidate FHSCs and VeLLOs. 
Linear fitting to all the young sources provides a correlation:
\begin{equation}\label{equ:FCO_Lbol_Menv_a}
   \log(F_{\rm CO}) = (1.03 \pm 0.09) \log(L_{\rm bol}) - (5.70 \pm 0.15).
\end{equation}
Figure \ref{fig:FCO_Lbol_Menv}a contains mostly Class\,0 objects with a few Class\,I (three from ALMASOP sample) objects. This slope in Equation {\ref{equ:FCO_Lbol_Menv_a} is similar to those found in other independent studies of Class\,0 samples from  \citet[][]{2022A&A...660A..39S} (slope $\sim$ 1.02 $\pm$ 0.26) and \citet[][]{1996A&A...311..858B} (slope $\sim$ 0.90$\pm$0.15).

Figure \ref{fig:FCO_Lbol_Menv}b shows the outflow force as a function of envelope mass. A correlation of $F_{CO}$ with $M_{env}$ is seen for all sources, given by the linear fit equation:
\begin{equation}\label{equ:FCO_Lbol_Menv_b}
   \log(F_{\rm CO}) = (2.02 \pm 0.89) \log(M_{\rm env}) - (5.20 \pm 0.31).
\end{equation}
This slope is significantly higher than that of \citet[][]{2022A&A...660A..39S} and \citet[][]{1996A&A...311..858B}. 
Among four Class\,I sources ($T_{bol}$ $>$ 70\,K), three show higher values of $F_{CO}$ than for Class\,0 sources of similar $M_{env}$. Due to the absence of a significant Class\,I sample, we can not derive separate correlations for Class\,0 and Class\,I sources. Four multiple sources (three Class\,0 and one Class\,I) exhibit $F_{CO}$ $\sim$ (0.4 - 1.1) $\times$ 10$^{-6}$ $\fco$. We do not observe any special deviation in the $F_{CO}$ of multiple sources from those of single sources.


\section{Summary}\label{sec:summary}
We have analysed 42 outflow fields to investigate protostellar outflows and jets with ALMA observations of SiO (5-4), CO (2-1), C$^{18}$O, and 1.3\,mm continuum. 

(i) Observations of SiO emission mostly trace the dense knots and bow-shock regions within jets. SiO emission also traces the shocks in the collision zone between two outflows or between outflows and the ambient cloud. On the other hand, CO emission  traces both the outflow shell and the jet component. The CO emission is less optically thick than SiO within the dense knots, therefore it is expected to reveal deeper structures within the knots than SiO can. 

(ii) To understand the morphologies of the various outflow or wind components, we fit simple analytical models assuming a parabolic distribution of the outflow shell and determined inclination angles from the best fits. In general, if CO emission is available on both sides of the velocity axis in the PV diagram, the object is closer to being seen where the disk is edge-on.  When the CO emission is well separated on both sides of the velocity axis, i.e., no emission in the positive velocity quadrant in the blue-shifted lobe or vice versa, then the object is closer to being seen where the disk is face-on or alternatively, the outflow is being viewed pole-on. 

(iii) A significant fraction of the jets are observed to be monopolar in SiO emission, with only redshifted components in most cases. We suggest that a specific structure of knots and shocks is responsible for such monopolar jets. 

(iv) We estimated higher SiO-to-CO abundances in the knots than predicted in the model by \citet[][]{1991ApJ...373..254G}. 
The knots in the jets are likely dust poor in general. The knots closest to the YSO are not evolved enough ($<$ 100 years) to produce SiO rich jets. Therefore, the SiO emission could originate in the base of the source or within the dust sublimation zone. 


(v) From our limited sample, we do not observe any clear correlation between the jet velocities and internal luminosity. Objects with lower $\dot{M}_{j}$/$\dot{M}_{acc}$, however, ratio tend to have smaller jet velocities. Furthermore, there is a clear correlation between the jet velocities and surrounding envelope masses, irrespective of $\dot{M}_{j}$/$\dot{M}_{acc}$. 

(v) Mass loss rates $\dot{M}_{j}$ do not show any clear correlation with L$_{bol}$.  In most cases, objects of higher luminosity and smaller $\dot{M}_{j}$/$\dot{M}_{acc}$ tend to exhibit smaller $\dot{M}_{j}$. As with jet velocity, $\dot{M}_{j}$ is clearly correlated with M$_{env}$. 

(vi) The episodes of knots do not show any clear dependence on L$_{bol}$. In most cases, however, low luminosity sources tend to have lower $\dot{M}_{j}$/$\dot{M}_{acc}$ and longer periods. Episodes are explicitly anti-correlated with the M$_{Env}$, which could be due to some underlying mechanism that slows down YSO rotation by the envelope masses. 

(vii) Episodes are usually more frequent than the freeze-out time scales of complex organic molecules. Episodic ejection could significantly impact the gas-ice balance in the disk. Therefore, it will affect the COM formation within the disk and likely the chemical compositions of the forming planets. 

(viii) Outflow forces for different types of protostars and candidate FHSCs are correlated with the bolometric luminosity and envelope masses.


More detailed studies are needed with high-resolution and high-sensitivity multiwavelength observations (e.g., in sub-millimetre with ALMA and infrared with JWST) to characterize the ice features and disk-jet connection. Such studies would allow a more precise investigation of the direct/indirect effect of jets on the COMs and planet formation.


\movetabledown=15mm
\FloatBarrier
\begin{longrotatetable}
\begin{deluxetable*}{lcccccccccc}
\tablenum{2}
\tablecaption{Details of targeted dense cores in the Orion Complex\label{tab:targetobservedContunuum}}
\tablewidth{6pt}
\addtolength{\tabcolsep}{-3pt}
\tablehead{
\colhead{Object} & \colhead{HOPS}  & \colhead{RA}  & \colhead{Dec} & \colhead{Maj} &  \colhead{Min} & \colhead{PA} & \colhead{Peak} & \colhead{Flux} &  \colhead{Mass} & \colhead{V$_{\rm sys}$} \\
\colhead{ } & id & (h:m:s) & (d:m:s) & $^{\prime\prime}$ & $^{\prime\prime}$ & $(\degr$) & \colhead{(mJy\,beam$^{-1}$)} & \colhead{(mJy)} & \colhead{(M$_\sun$)}&  (km\,s$^{-1}$) \\
}
\decimalcolnumbers
\startdata
\multicolumn{11}{c}{$\lambda$-Orionis}\\
\cline{1-11} \\
G191.90-11.21S & $--$  & 05:31:31.58 & +12:56:14.35 & 5.04 $\pm$ 1.7 & 2.71 $\pm$ 1.51 & 36.9 $\pm$ 37.3 & 66.64 $\pm$ 4.83 & 92.58 $\pm$ 10.56 & 0.51 $\pm$ 0.1 & 7 $\pm$ 2 \\
G192.12-11.10 & $--$  & 05:32:19.38 & +12:49:40.98 & 4.17 $\pm$ 0.4 & 2.9 $\pm$ 0.33 & 94.7 $\pm$ 15.0 & 168.12 $\pm$ 2.87 & 217.77 $\pm$ 5.99 & 1.21 $\pm$ 0.18 & 10 $\pm$ 2 \\
G192.32-11.88N & $--$  & 05:29:54.14 & +12:16:53.08 & 1.71 $\pm$ 0.28 & 1.42 $\pm$ 0.23 & 81.3 $\pm$ 60.8 & 202.97 $\pm$ 1.38 & 215.08 $\pm$ 2.52 & 1.2 $\pm$ 0.18 & 12 $\pm$ 2 \\
G192.32-11.88S & $--$  & 05:29:54.41 & +12:16:30.37 & 9.07 $\pm$ 0.41 & 4.98 $\pm$ 0.29 & 58.7 $\pm$ 3.6 & 26.38 $\pm$ 0.58 & 56.5 $\pm$ 1.74 & 0.31 $\pm$ 0.05 & 12 $\pm$ 2 \\
G192.32-11.88S$\_$02  & $--$& 05:29:54.14 & +12:16:53.52 & 1.67 $\pm$ 0.4 & 1.27 $\pm$ 0.44 & 96.5 $\pm$ 83.4 & 197.46 $\pm$ 1.58 & 207.76 $\pm$ 2.88 & 1.15 $\pm$ 0.17 & 12 $\pm$ 2 \\
G196.92-10.37$\_$A  & $--$& 05:44:29.30 & +09:08:51.48 & 7.32 $\pm$ 0.93 & 4.57 $\pm$ 0.89 & 131.6 $\pm$ 15.7 & 68.18 $\pm$ 3.15 & 128.49 $\pm$ 8.6 & 0.71 $\pm$ 0.12 & 11 $\pm$ 2 \\
G196.92-10.37$\_$BC  & $--$& 05:44:29.98 & +09:08:56.36 & 5.55 $\pm$ 1.24 & 2.59 $\pm$ 1.74 & 0.6 $\pm$ 42.8 & 18.93 $\pm$ 1.33 & 29.17 $\pm$ 3.11 & 0.16 $\pm$ 0.03 & 11 $\pm$ 2 \\
G200.34-10.97N  & $--$& 05:49:03.34 & +05:57:57.72 & 5.97 $\pm$ 0.49 & 4.26 $\pm$ 0.32 & 87.0 $\pm$ 12.1 & 55.11 $\pm$ 1.38 & 91.87 $\pm$ 3.45 & 0.51 $\pm$ 0.08 & 14 $\pm$ 2 \\
\cline{1-11} \\
\multicolumn{11}{c}{Orion B}\\
\cline{1-11} \\
G201.52-11.08 & $--$   & 05:50:59.15 & +04:53:49.66 & 2.23 $\pm$ 0.21 & 0.82 $\pm$ 0.34 & 118.2 $\pm$ 7.3 & 23.79 $\pm$ 0.12 & 25.25 $\pm$ 0.23 & 0.14 $\pm$ 0.02 & 9 $\pm$ 2 \\
G203.21-11.20W1 & $--$   & 05:53:42.65 & +03:22:35.04 & 9.46 $\pm$ 1.85 & 5.52 $\pm$ 1.12 & 88.0 $\pm$ 20.8 & 40.62 $\pm$ 3.92 & 97.38 $\pm$ 12.86 & 0.54 $\pm$ 0.11 & 10 $\pm$ 2 \\
G203.21-11.20W2 & $--$   & 05:53:39.48 & +03:22:24.59 & 7.77 $\pm$ 1.31 & 3.16 $\pm$ 2.09 & 42.0 $\pm$ 14.6 & 30.06 $\pm$ 2.19 & 57.94 $\pm$ 6.07 & 0.32 $\pm$ 0.06 & 10 $\pm$ 2 \\
G205.46-14.56M1$\_$A & 317 & $--$ & $--$  & $--$ & $--$ & $--$  & $--$  & $--$  & $--$  & $--$  \\
G205.46-14.56M1$\_$B  & $--$  & 05:46:08.39 & -00:10:43.32 & 2.5 $\pm$ 0.56 & 1.3 $\pm$ 0.57 & 62.6 $\pm$ 26.7 & 940.15 $\pm$ 14.27 & 1039.1 $\pm$ 27.44 & 5.77 $\pm$ 0.88 & 10 $\pm$ 2 \\
G205.46-14.56M2$\_$ABCD & 387,386 & 05:46:08.12 & -00:10:00.61 & 12.55 $\pm$ 1.8 & 2.56 $\pm$ 0.65 & 101.7 $\pm$ 2.4 & 91.07 $\pm$ 6.26 & 186.45 $\pm$ 19.1 & 1.04 $\pm$ 0.19 & 10 $\pm$ 2 \\
G205.46-14.56N1 & 402 & 05:46:10.04 & -00:12:16.79 & 2.04 $\pm$ 0.28 & 1.09 $\pm$ 0.67 & 170.6 $\pm$ 24.5 & 189.98 $\pm$ 1.87 & 207.13 $\pm$ 3.56 & 1.15 $\pm$ 0.17 & 10 $\pm$ 2 \\
G205.46-14.56N2 & 401 &  05:46:07.72 & -00:12:21.14 & 2.33 $\pm$ 0.34 & 2.27 $\pm$ 0.46 & 5.9 $\pm$ 81.6 & 132.48 $\pm$ 1.25 & 152.62 $\pm$ 2.46 & 0.85 $\pm$ 0.13 & 10 $\pm$ 2 \\
G205.46-14.56S1$\_$A & 358 & 05:46:07.24 & -00:13:30.95 & 4.41 $\pm$ 0.62 & 2.72 $\pm$ 1.87 & 36.2 $\pm$ 25.1 & 112.92 $\pm$ 3.59 & 158.63 $\pm$ 7.95 & 0.88 $\pm$ 0.14 & 11 $\pm$ 2 \\
G205.46-14.56S1$\_$B & $--$ & 05:46:07.33 & -00:13:43.47 & 2.35 $\pm$ 0.35 & 0.98 $\pm$ 0.86 & 165.6 $\pm$ 20.8 & 210.13 $\pm$ 2.67 & 232.9 $\pm$ 5.12 & 1.29 $\pm$ 0.2 & 11 $\pm$ 2 \\
G205.46-14.56S2 & $--$  & 05:46:04.77 & -00:14:16.79 & 4.52 $\pm$ 0.21 & 2.68 $\pm$ 0.09 & 106.1 $\pm$ 2.9 & 37.98 $\pm$ 0.31 & 49.49 $\pm$ 0.66 & 0.28 $\pm$ 0.04 & 10 $\pm$ 2 \\
G205.46-14.56S3 & 315 & 05:46:03.63 & -00:14:49.73 & 3.1 $\pm$ 0.46 & 2.53 $\pm$ 0.59 & 141.6 $\pm$ 37.6 & 102.65 $\pm$ 1.53 & 125.49 $\pm$ 3.12 & 0.7 $\pm$ 0.11 & 10 $\pm$ 2 \\
G206.12-15.76 & 400 & 05:42:45.27 & -01:16:13.72 & 0.0 $\pm$ 0.0 & 0.0 $\pm$ 0.0 & 0.0 $\pm$ 0.0 & 480.34 $\pm$ 7.9 & 438.55 $\pm$ 13.69 & 2.44 $\pm$ 0.37 & 8 $\pm$ 2 \\
G206.93-16.61E2$\_$ABCD & 298 & 05:41:37.14 & -02:17:17.16 & 4.71 $\pm$ 0.51 & 2.52 $\pm$ 0.77 & 61.1 $\pm$ 10.8 & 281.57 $\pm$ 5.97 & 386.93 $\pm$ 13.14 & 2.15 $\pm$ 0.33 & 12 $\pm$ 4 \\
G206.93-16.61W2 & 399 & 05:41:24.93 & -02:18:06.78 & 4.1 $\pm$ 0.46 & 2.51 $\pm$ 0.77 & 145.7 $\pm$ 14.7 & 621.15 $\pm$ 11.68 & 820.16 $\pm$ 24.97 & 4.56 $\pm$ 0.7 & 9 $\pm$ 2 \\
 \cline{1-11} \\
\multicolumn{11}{c}{Orion A}\\
\cline{1-11} \\
G207.36-19.82N1$\_$AB & $--$  & 05:30:51.25 & -04:10:35.46 & 4.31 $\pm$ 0.51 & 2.14 $\pm$ 0.41 & 82.5 $\pm$ 9.7 & 55.13 $\pm$ 0.94 & 69.06 $\pm$ 1.97 & 0.38 $\pm$ 0.06 & 11 $\pm$ 2 \\
G208.68-19.20N1 & 87 & 05:35:23.43 & -05:01:30.39 & 0.0 $\pm$ 0.0 & 0.0 $\pm$ 0.0 & 0.0 $\pm$ 0.0 & 1389.02 $\pm$ 30.21 & 1487.11 $\pm$ 56.35 & 8.26 $\pm$ 1.28 & 11 $\pm$ 2 \\
G208.68-19.20N3$\_$A & $--$  & 05:35:18.07 & -05:00:18.44 & 10.06 $\pm$ 0.93 & 4.75 $\pm$ 0.52 & 119.2 $\pm$ 5.6 & 164.65 $\pm$ 6.98 & 374.3 $\pm$ 22.15 & 2.08 $\pm$ 0.34 & 12 $\pm$ 2 \\
G208.68-19.20N3$\_$BC & 92 & 05:35:18.33 & -05:00:34.24 & 4.78 $\pm$ 0.99 & 1.84 $\pm$ 2.02 & 167.2 $\pm$ 27.2 & 134.33 $\pm$ 7.05 & 191.71 $\pm$ 15.76 & 1.07 $\pm$ 0.18 & 12 $\pm$ 2 \\
G208.68-19.20S$\_$AB & 84 & 05:35:26.50 & -05:03:55.61 & 5.63 $\pm$ 1.53 & 3.15 $\pm$ 2.17 & 39.4 $\pm$ 44.0 & 171.99 $\pm$ 13.82 & 279.37 $\pm$ 33.79 & 1.55 $\pm$ 0.3 & 10 $\pm$ 2 \\
G208.89-20.04E &  $--$  & 05:32:48.23 & -05:34:41.66 & 8.89 $\pm$ 2.48 & 3.04 $\pm$ 1.07 & 98.5 $\pm$ 11.3 & 28.76 $\pm$ 2.88 & 49.69 $\pm$ 7.64 & 0.28 $\pm$ 0.06 & 8 $\pm$ 2 \\
G208.89-20.04Walma & $--$  & 05:32:28.25 & -05:34:19.51 & 8.77 $\pm$ 0.77 & 4.45 $\pm$ 0.34 & 94.4 $\pm$ 4.7 & 34.47 $\pm$ 1.23 & 68.92 $\pm$ 3.55 & 0.38 $\pm$ 0.06 & 7 $\pm$ 2 \\
G209.55-19.68N1$\_$ABC & 12 & 05:35:08.71 & -05:55:54.50 & 9.58 $\pm$ 0.43 & 4.37 $\pm$ 0.16 & 97.2 $\pm$ 1.7 & 171.49 $\pm$ 3.38 & 357.13 $\pm$ 10.11 & 1.98 $\pm$ 0.3 & 6 $\pm$ 2 \\
G209.55-19.68S1 & 11 & 05:35:13.40 & -05:57:57.71 & 4.64 $\pm$ 0.65 & 3.11 $\pm$ 0.57 & 126.1 $\pm$ 18.0 & 140.2 $\pm$ 3.4 & 196.98 $\pm$ 7.63 & 1.09 $\pm$ 0.17 & 8 $\pm$ 2 \\
G209.55-19.68S2 & 10 & 05:35:08.96 & -05:58:28.37 & 8.06 $\pm$ 1.85 & 5.61 $\pm$ 2.07 & 140.1 $\pm$ 39.1 & 49.36 $\pm$ 4.04 & 110.21 $\pm$ 12.5 & 0.61 $\pm$ 0.12 & 8 $\pm$ 2 \\
G210.37-19.53S & 164 & 05:37:00.44 & -06:37:10.92 & 3.14 $\pm$ 0.24 & 1.89 $\pm$ 1.21 & 19.1 $\pm$ 20.8 & 78.45 $\pm$ 1.27 & 97.1 $\pm$ 2.61 & 0.54 $\pm$ 0.08 & 6 $\pm$ 2 \\
G210.49-19.79W & 168 & 05:36:18.91 & -06:45:24.47 & 8.86 $\pm$ 2.82 & 3.41 $\pm$ 2.94 & 46.9 $\pm$ 27.4 & 117.21 $\pm$ 16.54 & 252.45 $\pm$ 49.88 & 1.4 $\pm$ 0.35 & 12 $\pm$ 2 \\
G210.82-19.47S & $--$  & 05:38:03.48 & -06:58:16.61 & 6.02 $\pm$ 1.11 & 4.15 $\pm$ 0.56 & 102.8 $\pm$ 32.6 & 7.44 $\pm$ 0.38 & 12.32 $\pm$ 0.95 & 0.07 $\pm$ 0.01 & 6 $\pm$ 2 \\
G210.82-19.47S$\_$02  & $--$& 05:38:03.67 & -06:58:23.80 & 10.02 $\pm$ 1.37 & 6.86 $\pm$ 1.22 & 66.5 $\pm$ 20.1 & 8.15 $\pm$ 0.52 & 23.54 $\pm$ 1.97 & 0.13 $\pm$ 0.02 & 6 $\pm$ 2 \\
G210.97-19.33S2$\_$AB & $--$  & 05:38:45.40 & -07:01:02.38 & 11.75 $\pm$ 1.46 & 2.71 $\pm$ 0.88 & 80.8 $\pm$ 4.0 & 15.45 $\pm$ 0.93 & 32.8 $\pm$ 2.86 & 0.18 $\pm$ 0.03 & 3 $\pm$ 2 \\
G210.97-19.33S2$\_$02 & $--$ & 05:38:46.21 & -07:00:48.22 & 6.75 $\pm$ 1.01 & 2.38 $\pm$ 1.28 & 53.3 $\pm$ 12.1 & 9.74 $\pm$ 0.49 & 16.25 $\pm$ 1.23 & 0.09 $\pm$ 0.02 & 3 $\pm$ 2 \\
G210.97-19.33S2$\_$03 & $--$ & 05:38:43.78 & -07:01:13.24 & 0.0 $\pm$ 0.0 & 0.0 $\pm$ 0.0 & 0.0 $\pm$ 0.0 & 10.1 $\pm$ 0.46 & 7.58 $\pm$ 0.76 & 0.04 $\pm$ 0.01 & 3 $\pm$ 2 \\
G211.01-19.54N & $--$  & 05:37:57.22 & -07:06:57.50 & 12.73 $\pm$ 2.36 & 3.84 $\pm$ 1.05 & 115.4 $\pm$ 6.4 & 60.94 $\pm$ 5.71 & 146.58 $\pm$ 19.22 & 0.81 $\pm$ 0.16 & 6 $\pm$ 2 \\
G211.01-19.54S & $--$  & 05:37:58.77 & -07:07:27.71 & 7.85 $\pm$ 2.19 & 6.43 $\pm$ 3.46 & 36.9 $\pm$ 82.8 & 15.06 $\pm$ 1.63 & 36.91 $\pm$ 5.41 & 0.21 $\pm$ 0.04 & 7 $\pm$ 2 \\
G211.16-19.33N2 & 133 & 05:39:05.83 & -07:10:40.87 & 7.53 $\pm$ 1.45 & 4.32 $\pm$ 2.13 & 34.9 $\pm$ 30.8 & 12.72 $\pm$ 1.0 & 26.71 $\pm$ 2.93 & 0.15 $\pm$ 0.03 & 4 $\pm$ 2 \\
G211.47-19.27N$\_$AB & 290 & 05:39:57.29 & -07:29:33.01 & 5.51 $\pm$ 1.38 & 3.61 $\pm$ 2.41 & 154.6 $\pm$ 35.1 & 53.35 $\pm$ 3.11 & 84.1 $\pm$ 7.53 & 0.47 $\pm$ 0.08 & 4 $\pm$ 2 \\
G211.47-19.27N$\_$02 & $--$ & 05:39:56.99 & -07:29:46.65 & 4.62 $\pm$ 0.45 & 2.66 $\pm$ 1.23 & 165.8 $\pm$ 13.8 & 4.76 $\pm$ 0.1 & 6.76 $\pm$ 0.22 & 0.04 $\pm$ 0.01 & 4 $\pm$ 2 \\
G211.47-19.27S$\_$01 & 288 & 05:39:56.01 & -07:30:27.75 & 5.22 $\pm$ 0.81 & 2.24 $\pm$ 0.79 & 83.6 $\pm$ 11.8 & 464.69 $\pm$ 11.33 & 606.19 $\pm$ 24.8 & 3.37 $\pm$ 0.52 & 4 $\pm$ 2 \\
G212.10-19.15N2$\_$AB & 263,262  & 05:41:23.72 & -07:53:47.02 & 8.56 $\pm$ 1.17 & 4.35 $\pm$ 1.76 & 23.0 $\pm$ 23.7 & 17.77 $\pm$ 1.07 & 40.39 $\pm$ 3.34 & 0.22 $\pm$ 0.04 & 4 $\pm$ 2 \\
G212.10-19.15N2$\_$02 & $--$  & 05:41:24.04 & -07:53:35.62 & 0.0 $\pm$ 0.0 & 0.0 $\pm$ 0.0 & 0.0 $\pm$ 0.0 & 7.87 $\pm$ 0.3 & 18.45 $\pm$ 0.98 & 0.1 $\pm$ 0.02 & 4 $\pm$ 2 \\
G212.10-19.15N2$\_$03 & $--$ & 05:41:24.80 & -07:54:09.09 & 0.0 $\pm$ 0.0 & 0.0 $\pm$ 0.0 & 0.0 $\pm$ 0.0 & 19.55 $\pm$ 0.38 & 19.94 $\pm$ 0.72 & 0.11 $\pm$ 0.02 & 4 $\pm$ 2 \\
G212.10-19.15S & 247 & 05:41:26.21 & -07:56:51.94 & 3.84 $\pm$ 0.95 & 2.81 $\pm$ 2.19 & 152.2 $\pm$ 37.5 & 110.81 $\pm$ 3.48 & 144.9 $\pm$ 7.5 & 0.81 $\pm$ 0.13 & 4 $\pm$ 2 \\
G212.84-19.45N & 224 & 05:41:32.04 & -08:40:08.90 & 0.0 $\pm$ 0.0 & 0.0 $\pm$ 0.0 & 0.0 $\pm$ 0.0 & 159.37 $\pm$ 5.93 & 93.74 $\pm$ 8.1 & 0.52 $\pm$ 0.09 & 4 $\pm$ 2 \\
G215.87-17.62M & $--$  & 05:53:32.53 & -10:25:07.83 & 6.73 $\pm$ 2.37 & 2.58 $\pm$ 3.34 & 46.8 $\pm$ 38.4 & 31.52 $\pm$ 3.69 & 53.54 $\pm$ 9.35 & 0.3 $\pm$ 0.07 & 10 $\pm$ 2 \\
G215.87-17.62N  & $--$ & 05:53:42.33 & -10:23:59.22 & 6.56 $\pm$ 1.65 & 4.66 $\pm$ 3.12 & 55.8 $\pm$ 38.7 & 8.46 $\pm$ 0.63 & 15.62 $\pm$ 1.69 & 0.09 $\pm$ 0.02 & 10 $\pm$ 2 \\
G215.87-17.62N$\_$02 & $--$  & 05:53:41.85 & -10:24:04.77 & 15.1 $\pm$ 1.46 & 8.28 $\pm$ 1.13 & 65.7 $\pm$ 8.7 & 5.16 $\pm$ 0.31 & 21.45 $\pm$ 1.59 & 0.12 $\pm$ 0.02 & 10 $\pm$ 2 \\
\enddata

\end{deluxetable*}
\FloatBarrier
\end{longrotatetable}


\movetabledown=15mm
\FloatBarrier
\begin{longrotatetable}
\begin{deluxetable*}{l@{\extracolsep{3pt}}c@{\extracolsep{5pt}}c@{\extracolsep{5pt}}c@{\extracolsep{5pt}}c@{\extracolsep{5pt}}c@{\extracolsep{5pt}}c@{\extracolsep{5pt}}c@{\extracolsep{5pt}}c@{\extracolsep{5pt}}c@{\extracolsep{5pt}}c@{\extracolsep{5pt}}c@{\extracolsep{5pt}}c@{\extracolsep{5pt}}c@{\extracolsep{5pt}}c@{\extracolsep{5pt}}c@{\extracolsep{5pt}}}
\tablenum{3}
\tablecaption{Properties of the sources with jet and outflows\label{tab:jet_outflow_Properties}}
\tablewidth{0pt}
\addtolength{\tabcolsep}{-2pt}
\tabletypesize{\scriptsize}
\tablewidth{6pt}
\addtolength{\tabcolsep}{-4pt}
\tablehead
{
\colhead{ALMA}&
  \multicolumn{1}{c}{HOPS}&  \multicolumn{1}{c}{Class}&   \multicolumn{1}{c}{L$_{\rm bol}$}&  \multicolumn{1}{c}{T$_{\rm bol}$} &  \multicolumn{1}{c}{$i$} &  \multicolumn{1}{c}{lobe} & \multicolumn{1}{c}{F$_{\rm CO}$\tablenotemark{*}} &  \multicolumn{1}{c}{N$_{\rm SiO}$} &  \multicolumn{1}{c}{N$_{\rm CO}$\tablenotemark{}} &  \multicolumn{1}{c}{X$_{SiO}$/X$_{CO}$\tablenotemark{}} & \multicolumn{1}{c}{V$_{\rm j}$\tablenotemark{\dag}}&  \multicolumn{1}{c}{V$_{\rm j,cor}$}& \multicolumn{1}{c}{M$_{\rm j}$\tablenotemark{*}} &   \multicolumn{1}{c}{L$_{j,kin}$\tablenotemark{*}} & \multicolumn{1}{c}{T$_{perd}$\tablenotemark{*}} \\ 
  \colhead{id}&
   \multicolumn{1}{c}{id}&  \multicolumn{1}{c}{}&   \multicolumn{1}{c}{(L$_\sun$)}&  \multicolumn{1}{c}{(K)} &  \multicolumn{1}{c}{($\degr$)} &  \multicolumn{1}{c}{} & \multicolumn{1}{c}{(10$^{-8}$ $\fco$)} &  \multicolumn{1}{c}{(10$^{-14}$ cm$^{-3}$)} &  \multicolumn{1}{c}{(10$^{-16}$ cm$^{-3}$)\tablenotemark{}} &  \multicolumn{1}{c}{{(10$^{-4}$)}\tablenotemark{}} & \multicolumn{1}{c}{($\kms$)}& \multicolumn{1}{c}{($\kms$)}& \multicolumn{1}{c}{(10$^{-7}$ $\mlr$)} &   \multicolumn{1}{c}{(10$^{-2}$ L$_\sun$)} & \multicolumn{1}{c}{(years)} \\ 
}
\decimalcolnumbers
\startdata
%
\cline{1-16} \\
\multicolumn{16}{c}{SiO Monopolar or significantly fainter than other pole}\\
\cline{1-16} \\
G191.90-11.21S&       &0&0.4$\pm$0.2&69$\pm$17&5$^{+3}_{-4}$&Blue&922.52&0.85&45.96&10.322&6$\pm$3&69$^{+275}_{-26}$&4.88&18.99&100$^{+61}_{-80}$\\  
       &       &       &      &      &                      &Red&2070.4&8.94&48.89&       &      &                      &5.19&20.2&                      \\  \hline
G203.21-11.20W2&       &0&0.5$\pm$0.3&15$\pm$5&20$^{+5}_{-5}$&Blue&348.68&1.65&8.65&15.408&33$\pm$5&96$^{+32}_{-18}$&1.29&9.83&71$^{+20}_{-19}$\\  
       &       &       &      &      &                      &Red&358.98&3.01&21.59&       &      &                      &3.21&24.56&                      \\  \hline
G205.46-14.56M1&317&0&4.8$\pm$2.1&47$\pm$12&75$^{+5}_{-15}$&Blue&90.76&7.93&14.27&73.938&88$\pm$10&91$^{+11}_{-2}$&2.0&13.67&100$^{+50}_{-53}$\\ 
       &       &       &      &      &                      &Red&9.7&0.0&7.17&       &      &                      &1.01&6.86&                      \\  \hline
G205.46-14.56S1&358&0&22.0$\pm$8.0&44$\pm$19&1$^{+5}_{-0}$&Blue&153837.41&1.42&34.35&5.873&1$\pm$4&57$^{+0}_{-47}$&4.25&22.45&34$^{+170}_{-17}$\\   
       &       &       &      &      &                      &Red&11035.28&3.41&47.91&       &      &                      &5.92&31.31&                      \\  \hline
G209.55-19.68S2&10&0&3.4$\pm$1.4&48$\pm$11&20$^{+10}_{-5}$&Blue&50.89&0.0&0.0&46.845&50$\pm$5&146$^{+47}_{-46}$&0.0&0.0&22$^{+13}_{-6}$\\  
       &       &       &      &      &                      &Red&61.7&1.61&3.44&       &      &                      &0.78&13.61&                      \\  \hline     
G192.32-11.88N&       &0&       &      &9$^{+5}_{-8}$&Blue&184.69&       &       &       &      &                      &       &       &                      \\  
       &       &       &      &      &                      &Red&7.89&       &       &       &      &                      &       &       &                      \\  \hline   
G211.47-19.27S&   288    & 0 &       180$\pm$70 &  49$\pm$21    &&Blue&&       &       &       &      &                      &       &       &                      \\  
       &       &       &      &      &                      &Red&&       &       &       &      &                      &       &       &                      \\  \hline       
\cline{1-16} \\
\multicolumn{16}{c}{SiO Bipolar}\\
\cline{1-16} \\       
G205.46-14.56S3&315&1&6.4$\pm$2.4&178$\pm$33&40$^{+8}_{-8}$&Blue&2150.0&3.7&9.5&30.89&80$\pm$10&124$^{+27}_{-16}$&2.9&36.83&54$^{+30}_{-20}$\\  
       &       &       &      &      &                      &Red&1100.0&2.2&9.6&       &      &                      &3.0&38.1&                      \\  \hline
G206.12-15.76&400&0&3.0$\pm$1.4&35$\pm$9&15$^{+15}_{-5}$&Blue&57.77&2.41&9.86&14.128&40$\pm$15&155$^{+75}_{-75}$&2.35&46.09&30$^{+35}_{-10}$\\  
       &       &       &      &      &                      &Red&105.4&2.58&25.51&       &      &                      &6.08&119.21&                      \\  \hline
G206.93-16.61W2&399&0&6.3$\pm$3.0&31$\pm$10&10$^{+5}_{-5}$&Blue&9500.0&13.0&15.0&88.38&50$\pm$10&288$^{+286}_{-95}$&28.8&1957.97&17$^{+30}_{-10}$\\ 
       &       &       &      &      &                      &Red&28840.0&12.1&13.4&       &      &                      &26.6&1808.4&                      \\  \hline
G208.68-19.20N3&       &0&       &      &35$^{+15}_{-15}$&Blue&96.75&21.25&69.9&26.131&35$\pm$15&61$^{+41}_{-15}$&6.58&20.11&81$^{+56}_{-39}$\\  
       &       &       &      &      &                      &Red&73.51&15.28&69.9&       &      &                      &6.58&20.11&                      \\  \hline
G208.89-20.04E&       &1&2.2$\pm$1.0&108$\pm$25&1$^{+2}_{-0}$&Blue&1562.03&0.92&12.0&7.572&2$\pm$2&115$^{+0}_{-77}$&2.12&22.86&52$^{+104}_{-26}$\\  
       &       &       &      &      &                      &Red&1680.27&0.73&9.73&       &      &                      &1.72&18.55&                      \\  \hline
G208.89-20.04Walma&       &0&0.8$\pm$0.4&31$\pm$10&5$^{+5}_{-4}$&Blue&2.0&0.27&2.3&10.417&4$\pm$2&46$^{+183}_{-23}$&2.02&1.9&176$^{+178}_{-141}$\\  
       &       &       &      &      &                      &Red&1.2&0.18&2.02&       &      &                      &1.9&1.8&                      \\  \hline
G209.55-19.68S1&11&0&9.1$\pm$3.6&50$\pm$15&  &Blue&&3.28&11.74&19.79&27$\pm$4&&&&\\  
       &       &       &      &      &                      &Red&&4.2&26.03&       &      &                      &&&                      \\  \hline
G210.37-19.53S&164&0&0.6$\pm$0.3&39$\pm$10&25$^{+10}_{-10}$&Blue&10.31&2.21&5.01&38.474&36$\pm$5&85$^{+54}_{-22}$&0.66&3.92&23$^{+18}_{-10}$\\  
       &       &       &      &      &                      &Red&5.76&1.54&4.74&       &      &                      &0.62&3.71&                      \\  \hline
G210.49-19.79W&168&0&60.0$\pm$24.0&51$\pm$20&5$^{+5}_{-4}$&Blue&5030.76&0.72&21.31&5.255&5$\pm$4&57$^{+229}_{-28}$&1.89&5.1&78$^{+78}_{-16}$\\ 
       &       &       &      &      &                      &Red&11410.83&0.8&7.56&       &      &                      &0.67&1.81&                      \\  \hline
G215.87-17.62M&       &0&       &      &30$^{+10}_{-10}$&Blue&36.01&1.46&17.22&7.152&30$\pm$10&60$^{+28}_{-13}$&1.59&4.71&66$^{+30}_{-25}$\\  
       &       &       &      &      &                      &Red&79.06&0.92&15.94&       &      &                      &1.47&4.36&                      \\  \hline
\cline{1-16} \\
\multicolumn{16}{c}{SiO Complicated}\\
\cline{1-16} \\  
G200.34-10.97N&       &0&1.5$\pm$0.6&43$\pm$10&12$^{+10}_{-5}$&Blue&102.38&       &       &       &      &                      &       &       &                      \\  
       &       &       &      &      &                      &Red&83.76&       &       &       &      &                      &       &       &                      \\  \hline
G205.46-14.56M2N&       &0& & & &Blue&&       &       &       &      &                      &       &       &                      \\  
       &       &       &      &      &                      &Red&&       &       &       &      &                      &       &       &                      \\  \hline
G205.46-14.56N2N&       &0& & & &Blue&&       &       &       &      &                      &       &       &                      \\  
       &       &       &      &      &                      &Red&&       &       &       &      &                      &       &       &                      \\  \hline
 G208.68-19.20S&84&1&49.0$\pm$18.0&96$\pm$25&10$^{+10}_{-5}$&Blue&37.74&       &       &       &      &                      &       &       &                      \\ 
       &       &       &      &      &                      &Red&2.97&       &       &       &      &                      &       &       &                      \\ \hline
\cline{1-16} \\
\multicolumn{16}{c}{No SiO Emission, Only CO outflows}\\
\cline{1-16} \\ 
G192.12-11.10&       &0&9.5$\pm$4.0&44$\pm$15&15$^{+10}_{-5}$&Blue&435.86&       &       &       &      &                      &       &       &                      \\  
       &       &       &      &      &                      &Red&610.18&       &       &       &      &                      &       &       &                      \\  \hline
G196.92-10.37&       &0&       &      &70$^{+10}_{-10}$&Blue&25.52&       &       &       &      &                      &       &       &                      \\  
       &       &       &      &      &                      &Red&12.52&       &       &       &      &                      &       &       &                      \\  \hline
G203.21-11.20W1&       &0&       &      &20$^{+10}_{-5}$&Blue&22.2&       &       &       &      &                      &       &       &                      \\ 
       &       &       &      &      &                      &Red&11.71&       &       &       &      &                      &       &       &                      \\  \hline
G209.55-19.68N1&       &0&       &      &27$^{+15}_{-10}$&Blue&187.43&       &       &       &      &                      &       &       &                      \\ 
       &       &       &      &      &                      &Red&18.87&       &       &       &      &                      &       &       &                      \\ \hline
G211.01-19.54N&153&0&4.5$\pm$1.8&39$\pm$12&50$^{+10}_{-20}$&Blue&2.14&       &       &       &      &                      &       &       &                      \\ 
       &       &       &      &      &                      &Red&8.72&       &       &       &      &                      &       &       &                      \\ \hline
G211.01-19.54S&152&0&0.9$\pm$0.4&52$\pm$8&&Blue&&       &       &       &      &                      &       &       &                      \\ 
       &       &       &      &      &                      &Red&&       &       &       &      &                      &       &       &                      \\ \hline
G212.84-19.45N&224&0&3.0$\pm$1.2&50$\pm$13&25$^{+15}_{-10}$&Blue&8.03&       &       &       &      &                      &       &       &                      \\ 
       &       &       &      &      &                      &Red&14.88&       &       &       &      &                      &       &       &                      \\
\enddata
\tablenotetext{*}{Corrected for inclination angle $i$}
\tablenotetext{\dag}{NOT corrected for inclination angle $i$}
\end{deluxetable*}
\end{longrotatetable}

\begin{deluxetable}{lll}
\tablenum{4}
\tablecaption{Perturbed (Keplerian) radius in the disk-plane for the knots in the jet\label{tab:keplerian_radius}}
\tablewidth{12pt}
\addtolength{\tabcolsep}{1pt}
\tablehead{
\multicolumn{1}{c}{Source}&  \multicolumn{2}{c}{Keplerian Radius (AU)}\\
\multicolumn{1}{c}{}&  \multicolumn{1}{c}{(0.05 M$_\sun$)} &  \multicolumn{1}{c}{(0.30 M$_\sun$)}
}
\startdata
G191.90-11.21S&7.9$^{+3.0}_{-5.2}$&14.4$^{+5.4}_{-9.5}$\\ 
G203.21-11.20W2&6.3$^{+1.1}_{-1.2}$&11.5$^{+2.1}_{-2.2}$\\ 
G205.46-14.56M1&7.9$^{+2.5}_{-3.1}$&14.4$^{+4.5}_{-5.7}$\\ 
G205.46-14.56S1&3.9$^{+8.9}_{-1.4}$&7.0$^{+16.2}_{-2.6}$\\ 
G205.46-14.56S3&5.3$^{+1.8}_{-1.4}$&9.6$^{+3.3}_{-2.5}$\\ 
G206.12-15.76&3.6$^{+2.4}_{-0.8}$&6.5$^{+4.4}_{-1.5}$\\ 
G206.93-16.61W2&2.4$^{+2.4}_{-1.1}$&4.4$^{+4.3}_{-2.0}$\\ 
G208.68-19.20N3&6.9$^{+2.9}_{-2.4}$&12.5$^{+5.3}_{-4.4}$\\ 
G208.89-20.04E&5.1$^{+5.5}_{-1.9}$&9.3$^{+10.1}_{-3.5}$\\ 
G208.89-20.04Walma&11.6$^{+6.9}_{-7.6}$&21.0$^{+12.5}_{-13.9}$\\
G209.55-19.68S2&2.9$^{+1.0}_{-0.6}$&5.3$^{+1.9}_{-1.0}$\\ 
G210.37-19.53S&3.0$^{+1.4}_{-0.9}$&5.4$^{+2.5}_{-1.7}$\\ 
G210.49-19.79W&6.7$^{+4.0}_{-1.0}$&12.2$^{+7.2}_{-1.7}$\\ 
G215.87-17.62M&6.0$^{+1.7}_{-1.6}$&10.9$^{+3.1}_{-3.0}$\\ 
\enddata
\end{deluxetable}


\facility{ALMA}\\
\software{Astropy \citep[][]{2013A&A...558A..33A}, APLpy \citep[][]{2012ascl.soft08017R}, Matplotlib \citep[][]{2007CSE.....9...90H}, CASA \citep[][]{2007ASPC..376..127M}}

\newpage
\appendix



\setcounter{figure}{0} 
\renewcommand{\thefigure}{A\arabic{figure}} 
\begin{figure}
\fig{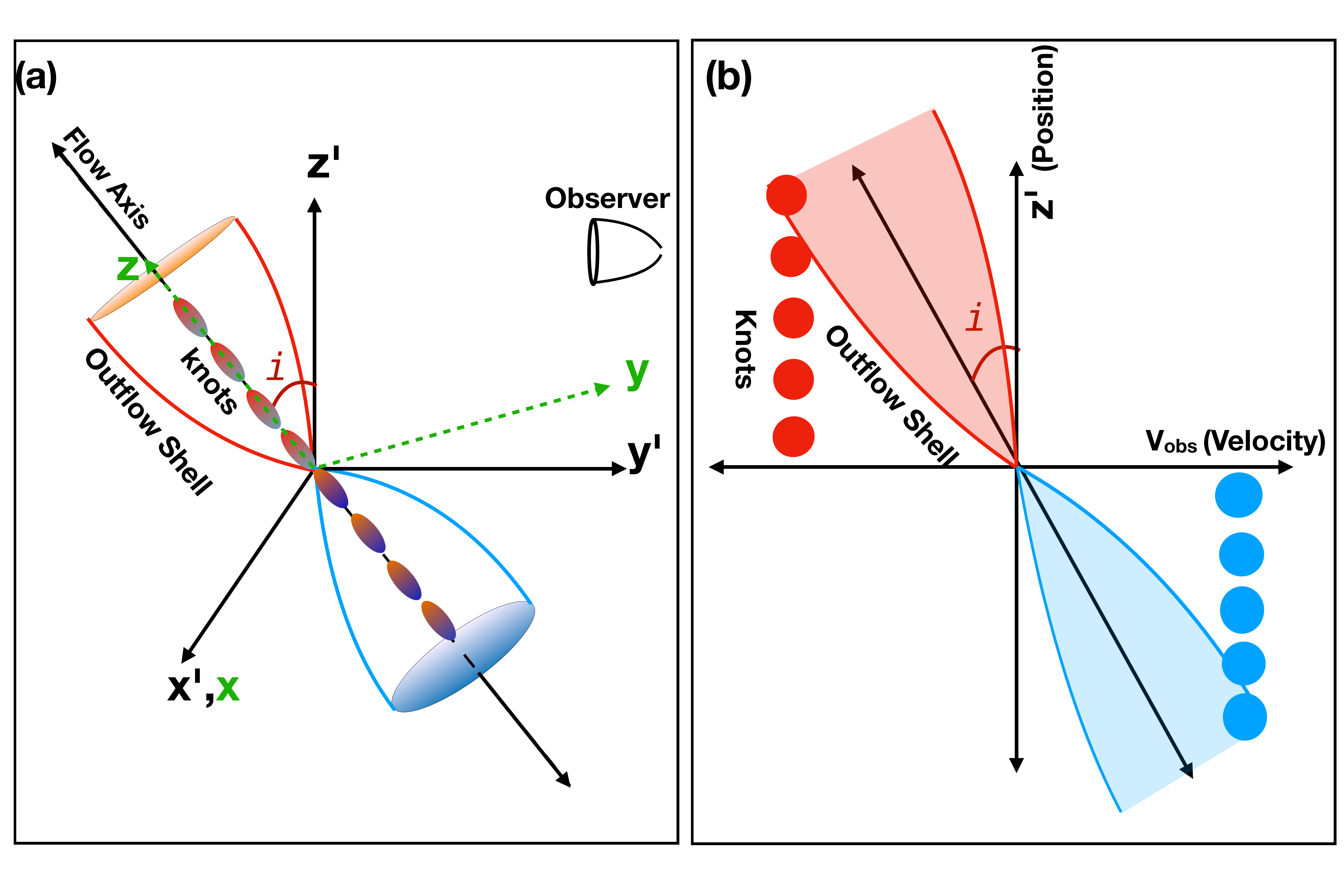}{0.98\textwidth}{}
\caption{
(a) Schematic Diagram of outflow-jet system. The system lies in the $xyz$ plane, which is inclined at an angle $i$ to the plane-of-sky plane, $x^\prime y^\prime z^\prime$. The blue color indicates the blue-shifted lobe and the red color indicates red-shifted lobe.  (b) PV diagram along the flow axis of panel-$a$. The blue-shifted outflow shell and knots are depicted in blue, while the red-shifted outflow shell and knots are depicted in red. The system is inclined at an angle $i$ along the $z^\prime$ position axis. The typical velocities of the knots are represented by red and blue circular patches. These assume that all knots are equidistant and have equal velocities in one case study.
}
\label{fig:appendix_Schematic_outflow_jet}
\end{figure}

\section{Inclination Angle from CO outflow shell}\label{appendix:schematic_inclination}
The molecular CO outflow shell is assumed to be a radially expanding parabolic shell generated by the underlying wide-angle wind. By following the simple analytical model proposed by \citet[][]{2000ApJ...542..925L}, the inclination angle ($i$) of the sources in the plane-of-sky can be derived from the physical structure of the wind/outflow shell as observed in CO emission. Figure \ref{fig:appendix_Schematic_outflow_jet}a provides a schematic diagram that depicts the integrated outflow emission (here, CO zeroth moment map). The outflow system lies in the $xyz$ plane, which is inclined at an angle $i$ to the plane-of-sky plane, $x^\prime y^\prime z^\prime$.  The outflow shells could be expressed in a cylindrical coordinate system:
\begin{equation}\label{equ:inclination_ZCR2}
    z = cR^2
\end{equation}
, where $R$ (= $\sqrt{x^2 + y^2}$) denotes the radial distance of the shell from the outflow axis $z$, and $c$ is the curvature of the parabola near the origin of the outflow or launching zone. The transformations between the two coordinate systems are $x = x^\prime$, $y^\prime = y \cos{i} + z \sin{i}$, $z^\prime = z \cos{i} - y \sin{i}$. For a radially expanding shell with a velocity proportional to the radial distance $v_R = R/t_0$, where $t_0$ is the dynamical time of expansion. In the case of only expansion, the observed velocity ($v_{obs}$) in the sky with a coordinate system ($x^\prime$, $z^\prime$) can be expressed as $y^\prime = -v_{obs} t_0$. By substituting these values into equation \ref{equ:inclination_ZCR2}, we obtain:   
\begin{equation}\label{equ:inclination_final_previous_Equation}
    z^\prime \cos{i} - v_{obs}t_0 \sin{i} = c[x^{\prime 2} + (-v_{obs} t_0 \cos{i} - z^\prime \sin{i})^2]
\end{equation}

 In a narrow position-velocity (PV) cut along the jet-axis, $y^\prime$ (velocity) and $z^\prime$ (position) become the variables and, the emission along $x^\prime$ (or x) is essentially zero. The schematic diagram in Figure \ref{fig:appendix_Schematic_outflow_jet}b displays a PV diagram with an outflow shell and jet (knots). Under these conditions, equation \ref{equ:inclination_final_previous_Equation} simplifies to: 
\begin{equation}\label{equ:inclination_finalEquation}
    z^\prime \cos{i} - v_{obs}t_0 \sin{i} = c[(-v_{obs} t_0 \cos{i} - z^\prime \sin{i})^2]
\end{equation}
The curvature $c$ is determined from fitting to the integrated emission (zeroth moment) maps (see equation \ref{equ:inclination_ZCR2}). By varying $i$ and $t_0$ we can obtain a parabolic fit of the PV diagram.

For the observed set of velocity (v$^\prime$) and distance along the flow axis (l$^\prime$) the inclination correction terms are v = v$^\prime$/$\sin{i}$, l = l$^\prime$/$\cos{i}$. 


\section{Objects with confirmed molecular jet}\label{sec:Appendix_individual_confirmed_molecular_jet}
\subsection{G191.90-11.20S}\label{sec:Appendix_G191.90-11.20S}
Figure \ref{fig:appendix_G191.90_11.20S} presents the low-resolution ACA 1.33\,mm continuum map (panel $a$), high-resolution C$^{18}$O spectra (panel $b$), CO map integrated over the whole velocity dispersion (panel $c$), position-velocity (PV) of CO along flow-axis (panel $d$), SiO map integrated over whole velocity dispersion (panel $e$), and SiO PV diagram cut along flow axis of G191.90$-$11.20S. The observed object is a Class\,0 $-$ I transition protostar with T$_{bol}$ $\sim$ 69 $\pm$ 17\,K and L$_{bol}$ $\sim$ 0.4$\pm0.2 $ L$_\sun$.  It has an envelope mass of the $\sim$ 0.51$\pm$0.1\,M$_\sun$. 

The protostar drives a CO outflow with a well-defined cavity wall and dense jet along the flow axis detected with both SiO and CO.  The source is nearly edge-on (i $\sim$ $7^{+8}_{-3}$ degree). The velocity dispersion of CO emission and SiO emission is nearly identical. A mean jet velocity V$_j$ = 6$\pm$3 $\kms$ was observed. Due to the edge-on orientation, the jet velocity appears small. The nearest knot at the redshifted lobe ($R1$) is faint. The brightest knot at $R2$ resembles a bow-shock kind of structure. The other two redshifted knots ($R3$ and $R4$) do not show a dense structure, but closer inspection of the SiO PV diagram confirms the presence of separate knots. The SiO emission is primarily observed in the redshifted lobe. Only one faint SiO knot (B1) was detected in the blue-shifted lobe positioned 10$-$12 $\arcsec$ ($\sim$ 4000 $-$ 4800 AU) away from the central source. The mean episodic intervals were estimated to be $\mathcal{T}_{knot}$ $\sim$ 140$^{+167}_{-60}$ years, corresponding to a Keplerian radius in the range $\sim$ 7.0 to 30 AU (including estimated error) for a mass range of 0.05 to 0.30 M$_\sun$. Therefore, the instabilities causing episodic accretion and ejection possibly originated in the disk or its outskirts. The objects exhibit a very high $\dot{M}_{j}$/$\dot{M}_{acc}$ ratio of $\sim$ 2.1, indicating it is a highly active ejection/accretion object.


\subsection{G205.46-14.56M1 (HOPS 317)}\label{sec:Appendix_G205.46-14.56M1} 
The observations of the continuum, C$^{18}$O, CO, and SiO emissions for G205.46$-$14.56M1 are shown in Figure \ref{fig:appendix_G205.46_14.56M1}, following the same order as in Figure \ref{fig:appendix_G191.90_11.20S}. The system comprises two sources, A and B, separated by $\sim$ 6.14$\arcsec$ ($\sim$ 2460 AU). Source `A' aligns with the position of HOPS\,317 \citep[][]{2020ApJ...890..130T}, a Class\,0 object with T$_{bol}$ $\sim$ 47$\pm$12\,K, L$_{bol}$ $\sim$ 4.8$\pm$2.1\,L$_\sun$. This binary system possesses a combined envelope mass M$_{env}$ $\sim$ 5.77$\pm$0.88 M$_\sun$. G205.46-14.56M1$\_$B, brighter in 1.3\,mm continuum is located at the envelope's center.

Source G205.46-14.56M1$\_$A exhibits a small north-east outflow lobe, with the south-western lobe potentially blended with the outflow from G205.46-14.56M1$\_$B. However, from our current velocity resolution, we were unable to confirm the presence of an outflow from G205.46-14.56M1$\_$B or discern between the outflow components of the sources. The presence of episodic shells in the blue-shifted lobe similar to the episodic Wide-angle Outflow of HH\,46/47 \citep[][]{2019ApJ...883....1Z}, is a possibility. When calculating F$_{CO}$ from the south-west lobe of G205.46-14.56M1$\_$A, we consider all the observed CO emissions from the lobe. The southern component (blueshifted lobe) of the source G205.46-14.56M1$\_A$ reveals a single SiO knot ($B1$), which is corroborated by CO emission. A CO knot-like structure ($R1?$) is visible in the redshifted lobe, although it is undetected in SiO emission. The mean jet velocity derived from peak SiO emission is V$_j$ $\sim$ 88$\pm$10 $\kms$. With an inclination of i $\sim$ $75^{+5}_{-15}$ degrees, the source is closer to face-on orientation. The mean episodic intervals of the nearest knots are approximately $\mathcal{T}_{knot}$ $\sim$ $100^{+50}_{-53}$ years, yielding a Keplerian radius of perturbation between $\sim$ 4 to 18 AU. The objects exhibit significant accretion/ejection phenomena, as indicated by the high $\dot{M}_{j}$/$\dot{M}_{acc}$ ratio of $\sim$ 0.67. 

\subsection{G205.46-14.56S1 (HOPS 358)}\label{sec:Appendix_G205.46-14.56S1}
Various components of the observations of G205.46-14.56S1 are displayed in Figure \ref{fig:appendix_G20546_1456S1} following the same sequence as in Figure \ref{fig:appendix_G191.90_11.20S}. This system comprises a wide binary with a separation of 13.37$\arcsec$ ($\sim$ 5349 AU). Source G205.46-14.56S1$\_$A drives a nearly north-south outflow and jet, while the outflow of G205.46-14.56S1$\_$B is blended with the redshifted lobe (southern lobe) of G205.46-14.56S1$\_$A. In this paper, we discuss the outflow of G205.46-14.56S1$\_$A, which is identified as HOPS\,358 in \citet[][]{2020ApJ...890..130T}. It is a Class\,0 source with T$_{bol}$ = 44$\pm$19, L$_{bol}$ = 22$\pm$8 L$_\sun$, and the envelope mass of G205.46-14.56S1$\_$A is 0.88$\pm$0.14 M$_\sun$. 

The outflow from G205.46-14.56S1$\_$A demonstrates a narrow inner section and a suddenly widening outflow lobe. The thick envelope likely suppresses the outflow opening near the source, allowing it to abruptly expand beyond the envelope. The CO outflow kinematics suggest that the object is nearly edge-on with an inclination $i$ $\sim$ $5^{+5}_{-4}$ degrees. Notably, the source exhibits strong SiO emission in the southern (redshifted) lobe, whereas the northern (blueshifted) lobe only has faint SiO emission near the source, resembling a monopolar SiO jet. Owing to edge-on orientation the observed mean V$_j$ is quite small $\sim$ 1.4$\pm$4.0. The error bar indicates the jet velocity could become negative, implying the redshifted lobe might transform into a blueshifted one, and due to edge-on view, the redshifted and blueshifted velocities could be reliably defined. Several knots are detected in the southern part (R1, R2, ..... R14). Here, it is important to note that knots are identified based on peak emission, and we were unable to determine whether they form part of the forward and backward components of a single bow shock. The average episodic intervals of the knots knots were estimated to be $\mathcal{T}_{knot}$ $\sim$ $170^{+163}_{-136}$ years, which corresponds to a Keplerian radius ranging from $\sim$ 5.0 to 31 AU (including estimated error, for a mass range of 0.05 to 0.30 M$_\sun$). The objects exhibit intermediate accretion/ejection phenomena ($\dot{M}_{j}$/$\dot{M}_{acc}$ $\sim$ 0.09). 


\subsection{G206.12-15.76 (HOPS 400)}\label{sec:Appendix_G206.12-15.76}
Figure \ref{fig:appendix_G206.12-15.76} shows different components of the G206.12-15.76 system following the same order as Figure \ref{fig:appendix_G191.90_11.20S}. This system, HOPS\,400 \citet[][]{2020ApJ...890..129K,2020ApJ...890..130T}, is a close binary, although our observed resolution and sensitivity were not sufficient to resolve the two components. It is a Class\,0 system with T$_{bol}$ = 35$\pm$9 K and L$_{bol}$ = 3.0$\pm$1.4 L$_\sun$. The combined envelope mass was estimated at M$_{env}$ $\sim$ 2.44$\pm$0.37 M$_\sun$.

The length of the CO outflow is relatively small (L$_{red}$ $\sim$ ${16}\farcs{0}$ $\equiv$ 6400 AU; L$_{blue}$ $\sim$ ${10}\farcs{0}$ $\equiv$ 4000 AU). Both the CO and SiO emission have a similar range of velocity dispersion, making it difficult to distinguish outflow shells from jet components. Assuming low-velocity CO emission is solely a wind/outflow component, we estimated the inclination angle to be $i$ $\sim$ $15^{+20}_{-5}$ degrees. Both SiO and CO exhibit high- and low-velocity components. A few knots are identified along the jet axis. From the high-velocity knots, the mean observed jet velocity is estimated as V$_j$ $\sim$ 40$\pm$15 $\kms$. The mean episodes of the knots are $\mathcal{T}_{knot}$  $\sim$ $30^{+35}_{-10}$ years, corresponding to a Keplerian radius of perturbation zone between 3 $-$ 11 AU (including estimated error, for a mass range of 0.05 to 0.30 M$_\sun$). This object exhibits intermediate ejection phenomena relative to accretion with $\dot{M}_{j}$/$\dot{M}_{acc}$ $\sim$ 0.3. The object could be at a very young phase of protostellar evolution with a dynamical age of $\mathcal{T}_{DynAge,blue}$ $\sim$  $203^{+329}_{-70}$ years for the blueshifted lobe and $\mathcal{T}_{DynAge,red}$ $\sim$  $127^{+205}_{-44}$ years for the redshifted lobe (Both estimated ages corrected for inclination angle). Thus, we suggest this system to be less than 1000 years of old, as a conservative estimate.

\subsection{G208.68-19.20N3}\label{sec:Appendix_G208.68-19.20N3}
The details of different observed components are shown in Figure \ref{fig:appendix_G208.68-19.20N3} in the same order of Figure \ref{fig:appendix_G191.90_11.20S}. There are three sources in the field, suggesting the presence of a multiple system. The source G208.68-19.20N3$\_$A is located at $\sim$ 16$\arcsec$ away from the close binary system (G208.68-19.20N3$\_$B and G208.68-19.20N3$\_$C). In this paper, we focus on the outflow/jet of G208.68-19.20N3$\_$A. Neither T$_{bol}$ nor L$_{bol}$ for this source are available in the literature. The envelope mass of the object is estimated to be M$_{env}$ $\sim$ 2.08$\pm$0.34 M$_\sun$. The location of G208.68-19.20N3$\_$B and G208.68-19.20N3$\_$C align with HOPS\,92. The outflows of this close binary system are blended together and situated on the edge of primary beam, thus, we exclude this source from our study.

The velocity dispersion of SiO and CO are nearly identical. Based on the position-velocity diagrams, it is challenging to differentiate between wind/outflow and jet components, leading to a poor fit of the parabola to the CO shell. We assumed that the majority of low-velocity components originate from the wind, and are likely wind components. The inclination angle is measured to be $i$ $\sim$ $35^{+15}_{-15}$ degrees. The mean jet velocity is V$_j$ $\sim$ 35$\pm$15. There are a few knot-like structures appearing in both the outflow lobes. Although `R1' appears to be double-headed, it does not present itself as a single knot with a bow-shock structure in the PV diagrams. The system exhibits a high total mass loss rate of $\dot{M}_j$ $\sim$ 1.3 $\times$ 10$^{-6}$ $\mlr$. The mean ejection episodes were estimated to be $\mathcal{T}_{knot}$  $\sim$ $137^{+56}_{-39}$ years, corresponding to a perturbation zone of Keplerian radius $\sim$ 4.5 $-$ 18 AU. The dynamical age of the blueshifted lobe is $\mathcal{T}_{DynAge,blue}$ $\sim$  $228^{+160}_{-110}$ years and $\mathcal{T}_{DynAge,red}$ $\sim$  $190^{+133}_{-91}$ years for the redshifted lobe, with both adjusted for the inclination angle. We propose that the system is less than 1000 years old (a conservative estimate). The smaller dynamical age, higher jet mass ejection rate, and presence of an SiO jet indicate that the source is in an extremely young phase of protostellar evolution.

\subsection{G208.89-20.04E}\label{sec:Appendix_G208.89-20.04E}
The details of continuum and emission components of the G208.89-20.04E system are depicted in Figure \ref{fig:appendix_G208.89-20.04E}, and arranged similarly to Figure \ref{fig:appendix_G191.90_11.20S}. It is a Class\,I system having T$_{bol}$ $\sim$ 108$\pm$25 K and L$_{bol}$ $\sim$ 2.2$\pm$1.0 L$_\sun$. The envelope mass of the source is M$_{env}$ $\sim$ 0.28$\pm$0.06 M$_\sun$. The elliptical structure of the inner part of the continuum emission, and a small extension along one side (western part) as compared to the other (north-east and eastern part), suggest that this object could be a highly inclined system.

The structure of the CO outflow is complex and irregularly shaped. The redshifted lobe (south-western) appears wider than the blueshifted (north-east) lobe. The redshifted lobe is possibly attached to an ambient cloud or outflow of an unknown source. The redshifted lobe is smaller than the blueshifted lobe, but scattered emission is noticeable along the redshifted lobe in both CO and SiO emissions. PV diagrams with CO and SiO indicate that the source is likely nearly edge-on. We measured the inclination angle to be $i$ $\sim$ $10^{+10}_{-5}$ degrees. Due to its edge-on orientation, the blueshifted and redshifted wind/outflow components are difficult to distinguish. Based on the peak emission of CO and SiO, we defined the blue and redshifted velocity components. We estimated the mean jet velocity to be V$_j$ $\sim$ 2.0$\pm$2.0 $\kms$ (negative error bar leads to zero velocity, indicating a fully edge-on orientation). Different knots are identified from the SiO emission with mean episodes of $\mathcal{T}_{knot}$  $\sim$ $523^{+557}_{-263}$ years. The Keplerian radius for a potential perturbation zone is $\sim$ 15 - 70 AU, which is a slightly larger range than that of Class\,0 sources.

The ALMASOP sample includes two Class\,I molecular jets with SiO emission along the flow axis. One Class\,I source, G205.46-14.56S3, was reported with very high-velocity SiO emission \citep[][]{2022ApJ...925...11D}. Another molecular jet detected with SiO emission is the G208.89-20.04E source. This is a Class\,I source, or could be in the transition phase from Class\,0 to Class\,I. The jet velocity for this source is very low, potentially due to its edge-on orientation. 


\subsection{G209.55-19.68S1 (HOPS 11)}\label{sec:Appendix_G209.55-19.68S1}
The details of different continuum and emission components of G209.55-19.68S1 are illustrated in Figure \ref{fig:appendix_G20955_1968S1}, following the same order as Figure \ref{fig:appendix_G191.90_11.20S}. This source coincides with the HOPS\,11 source \citep[][]{2020ApJ...890..130T} and has T$_{bol}$ $\sim$ 50$\pm$15 K and L$_{bol}$ $\sim$ 9.1$\pm$3.6 L$_\sun$. The mass of the envelope is estimated to be M$_{env}$ $\sim$ 1.09$\pm$0.17 M$_\sun$. The 1.33\,mm emission is extended along the outflow direction, potentially implying that there is entrained envelope material along the outflow wall driven by the outflow.

The CO outflow's northern part (redshifted lobe) bends towards the north-west from the flow axis, and the southern outflow CO emission appears very wide. The CO and SiO PV diagrams along the flow axis suggest that the source is not an edge-on object. The outflow shells are challenging to distinguish for this source, so the exact fitting is not shown. A tentative fitting to the blueshifted lobe of the CO outflow shells suggests an inclination angle $i$ $\sim$ $45^{+15}_{-15}$ degrees. A few SiO knots were observed along the flow axis, and the mean jet velocities are estimated to be V$_j$ $\sim$ 27$\pm$8 $\kms$. The mean ejection periods are $\mathcal{T}_{knot}$  $\sim$ $300^{+125}_{-80}$ years, corresponding to a perturbation zone at a Keplerian radius of $\sim$ 13 - 38 AU (including error bars). The objects exhibit a smaller ratio of $\dot{M}_{j}$/$\dot{M}_{acc}$ $\sim$ 0.03, indicating that it could be at a relatively evolved phase of protostellar evolution. Due to the unreliable inclination measurement, this source is not included in the statistical analyses in the main text. 


\subsection{G209.55-19.68S2 (HOPS 10)}\label{sec:Appendix_G209.55-19.68S2} 
The continuum and different emission components of G209.55-19.68S2 are displayed in Figure \ref{fig:appendix_G209.55_19.68S2} in the same order as Figure \ref{fig:appendix_G191.90_11.20S}. The object's location coincides with HOPS\,10 \citep[][]{2020ApJ...890..130T}. The source has a T$_{bol}$ $\sim$ 48$\pm$11 K and a L$_{bol}$ $\sim$ 3.4$\pm$1.4 L$_\sun$. The envelope mass is estimated to be M$_{env}$ $\sim$ 0.61$\pm$0.12 M$_\sun$. The peak of the continuum is shifted towards north-east from the envelope's centre. Such continuum emission could be due to an edge-on orientation of the object.

The wind/outflow component in CO emission is well separated from the high-velocity jet component. The object exhibits a redshifted monopolar jet, detectable with both SiO and CO, making it one of the most beautiful high-velocity monopolar jets discovered to date. The inclination angle is measured to be $i$ $\sim$ $15^{+15}_{-5}$ degree. The observed jet velocity is V$_j$ $\sim$ 50$\pm$5 $\kms$. Several knots are detected in the redshifted lobe, with ejection episodes estimated to be $\mathcal{T}_{knot}$  $\sim$ $16^{+19}_{-6}$ years. This corresponds to a perturbation zone at a Keplerian radius $\sim$ 1.7 - 7.2 AU.


\subsection{G210.37-19.53S (HOPS 164)}\label{sec:Appendix_G210.37-19.53S}
Figure \ref{fig:appendix_G210.37-19.53S} shows the continuum, C$^{18}$O spectra, integrated CO and SiO maps, and PV diagrams of G210.37-19.53S in the same order as Figure \ref{fig:appendix_G191.90_11.20S}. This source has T$_{bol}$ $\sim$ 39$\pm$10 K and L$_{bol}$ $\sim$ 0.6$\pm$0.3 L$_\sun$, making it one of the very low luminosity objects. The location of this object matches with HOPS\,164 \citep[][]{2020ApJ...890..130T}. The surrounding envelope mass is estimated to be $\sim$ M$_{env}$ $\sim$ 0.54$\pm$0.08 M$_\sun$.

The SiO knots are considered to delineate the jet axis or flow axis. The CO emission is asymmetrically distributed on both sides of the flow axis. We calculated an inclination angle of $i$ $\sim$ $25^{+15}_{-10}$ degrees. Several knots are detected from integrated maps and PV diagrams of SiO and CO, close to the source on both the blueshifted ($R1$, $R2$, $R3$) and redshifted ($B1$, $B2$, $B3$) lobes. Knots $R1$, $R2$ and $R3$ appear to represent different components of bow shock (forward shock, backward shock) in a single knot from the SiO PV diagram, even though integrated maps do not present a bow shock-like appearance. Given the present sensitivity and velocity resolution of our observations, it is difficult to determine whether it is really a single knot or three different knots. We estimated a mean velocity of V$_j$ $\sim$ 36$\pm$5 $\kms$. The mean episode of the ejection events is $\mathcal{T}_{knot}$  $\sim$ 23$^{+18}_{-10}$ years. The perturbation zone for the episodic accretion could be at a Keplerian radius of $\sim$ 2 - 8 AU. The relatively higher ratio of $\dot{M}_{j}$/$\dot{M}_{acc}$ $\sim$ 0.3 exhibited by this object indicates that it might be at a relatively younger phase of protostellar evolution.

\subsection{G210.49-19.79W (HOPS 168)}\label{sec:Appendix_G210.49-19.79W}
 Figure \ref{fig:appendix_G210.49_19.79W} shows various components of the continuum and emission in the same order as Figure \ref{fig:appendix_G191.90_11.20S}.  High-resolution maps ($\sim$ 140 AU) reveal two continuum peaks, G210.49-19.79W$\_$A and G210.49-19.79W$\_$B \citep[][]{2020ApJS..251...20D}. The fainter continuum peak of G210.49-19.79W$\_$B is yet to be confirmed as a protostar. The location of G210.49-19.79W$\_$A coincides with HOPS\,168 \citep[][]{2020ApJ...890..130T}. It has a T$_{bol}$ $\sim$ 51$\pm$20 K and a L$_{bol}$ $\sim$ 60$\pm$24 L$_\sun$. The envelope mass is calculated to be M$_{env}$ $\sim$ 1.4$\pm$0.35 M$_\sun$. The envelope material is possibly entrained by an outflow and extends towards the southwest along the boundary wall of the redshifted outflow. The inclination angle is measured to be $i$ $\sim$ $45^{+10}_{-10}$ degrees from the outermost CO shell. 

A close look at the CO emission near the source suggests a twisted CO shell at its base. The SiO emission does not align along a particular axis (as marked in the SiO integrated map). One SiO knot is observed in the north-east of the jet axis in the blue-shifted lobe. In the redshifted lobe, one SiO knot close to the source appears to be shifted to the southwest. All SiO knots form an `S'-shaped structure, possibly due to wiggling of the jet. Similar wiggling structures have also been reported in molecular jets from other protostars, such as HH\,211 in \citep[][]{2010ApJ...713..731L,2016MNRAS.460.1829M}. We measure the mean jet velocity $V_j$ $\sim$ 5$\pm$4 $\kms$ (V$_{j,cor}$ $\sim$ $V_j$/$\sin{i}$). This small velocity range is not in agreement with the SiO emission in the jet if we consider a dust-sputtering scenario. SiO emission is plausible if it originates at the dust sublimation zone. Otherwise, there could be an undetected close binary with two jets originating at two different position angles.  

Another possible explanation could be the scenario of episodic CO shells and SiO knots launched at different times, as in the case of the episodic wide-angle outflow in HH\,46/47  \citep[][]{2019ApJ...883....1Z}. The CO PV diagram in Figure \ref{fig:appendix_G210.49_19.79W}d and f also hints at such episodic shell occurrences (marked with thicker blue and red curves in CO PV diagram). Now consider a cloud core contracting along the magnetic field lines toward the midplane (equatorial plane) and forming a flattened envelope or pseudo-disk. In this scenario, the flattened envelope is warped because of the contraction direction of the cloud core and thus the minor axis of the flattened envelope is changed gradually from parallel to the magnetic field axis in the outer region to parallel to the rotation axis in the inner region \citep[][]{2019MNRAS.485.4667H,2019ApJ...879..101L}. Therefore, the position angle of the jet axis is also expected to be rotated according to the rotation direction of the disk. The SiO jet could be launched at different orientations of the jet at different timescales. Then the inclination angle measured above could not be a true value from a single outermost CO shell but rather a combination of multiple episodic shells. The orientation ($i$) of those shells could be much smaller (towards the edge-on) than estimated above. Considering the inclination angle and estimated jet velocities of other jets in our ALMASOP sample, we assume an inclination angle of $i$ $\sim$ $10^{+10}_{-9}$ degrees for this object. We recommend further investigation of this source with higher-angular, higher-velocity resolution, and higher-sensitivity observations to better understand its true nature.


\subsection{G215.87-17.62M}\label{sec:Appendix_G215.87-17.62M}
Figure \ref{fig:appendix_G215.87-17.62M} shows the continuum and different components of emissions for G215.87-17.62M in the same order as Figure \ref{fig:appendix_G191.90_11.20S}. Currently, there are no available measurements of T$_{bol}$ and L$_{bol}$ for this source in the literature. The envelope mass is estimated to be M$_{env}$ $\sim$ 0.3$\pm$0.07 M$_\sun$. The 1.3\,mm continuum is extended along the north-east to south-west direction, possibly representing the entrained material along the outflow cavity wall. However, the continuum emission is extended beyond the length of the outflow/jet, thus it is not clear whether the entirety of the extended continuum emission is due to the entrained material or if there is some other mechanism involved.

The CO and SiO velocity dispersion and spatial extensions are relatively similar. From position-velocity diagrams, it is difficult to separately identify the wind/outflow and jet components. Assuming that most of the low-velocity material is from wind/outflow shell, the inclination angle is estimated to be $i$ $\sim$ $30^{+10}_{-10}$ degree. The SiO emission is extended from low velocity to high velocity, making it difficult to measure the jet velocity. The mean jet velocity is V$_j$ $\sim$ 30$\pm$10 $\kms$. A few knot-like structures appear in both the outflow lobes.  It exhibits a relatively low mass loss rate $\dot{M}_j$ $\sim$ 3.2 $\times$ 10$^{-7}$ $\mlr$. The mean episodes of the ejection are $\mathcal{T}_{knot}$  $\sim$ $66^{+30}_{-25}$ years, corresponding to a perturbation zone of Keplerian radius $\sim$ 4.5 $-$ 14 AU.  The  dynamical age of blueshifted lobe is $\mathcal{T}_{DynAge,blue}$ $\sim$  $219^{+99}_{-81}$ years, and for redshifted lobe it is $\mathcal{T}_{DynAge,red}$ $\sim$  $292^{+133}_{-108}$ years, each corrected for inclination angle. We estimate that this system is less than 1000 years old, a conservative value. A smaller spatial extension, smaller dynamical age, lower jet mass ejection rate, and presence of a brighter SiO jet suggest that this source could be a very low-mass and low-luminosity object in an extremely early stage of protostellar evolution. 

\section{Objects with Complex SiO Emission Morphology and not considered for jet parameter estimation}\label{sec:Appendix_individual_complex_SiO_emission}

\subsection{G192.32-11.88N}\label{sec:Appendix_G192.32-11.88N}
Figure \ref{fig:appendix_G192.32_11.88N} shows different components of the studied continuum and emission maps in the same order as Figure \ref{fig:appendix_G191.90_11.20S}. No T$_{bol}$ and L$_{bol}$ estimations are currently available for this object in the literature. The envelope mass deduced from 1.3\,mm continuum is estimated to be $\sim$ 1.2$\pm$0.18 M$_\sun$.

CO integrated map shows that the outflow shell structure is blended with ambient molecular cloud in the southeast (blueshifted) lobe. Therefore, to probe the outflow shell, we consider high-velocity components, represented by the higher contours in the CO integrated map. The redshifted lobe appears to be much shorter in length than the blueshifted lobe, possibly due to the presence of the ambient cloud in the blueshifted lobe, forming an extended blended structure. We measured the average inclination angle to be $\sim$ $i$ $\sim$ $50^{+10}_{-10}$ degree. In the SiO integrated emission, a bright SiO emission is evident at the tip of the blueshifted lobe, but absent in the redshifted lobe. Scattered SiO emission is also observed along the whole blueshifted lobe. The SiO PV diagram shows SiO emission along the outflow axis, close to the systemic velocity of the source. If this emission originates from the jet, then the jet velocity would be V$_j$ $\sim$ 2$\pm$2 $\kms$, where V$_{j,cor}$ = V$_j$/$\sin{i}$. Such a low velocity is unlikely to synthesize a high-density SiO jet. Therefore, we suggest that the SiO emission at the tip of the blueshifted lobe could be the interaction zone between two outflows and the ambient material, as in the case of G208.89-20.04Walma \citep[][]{2022ApJ...931..130D}. To confirm the origin of the SiO emission - subsection{G192.32-11.88N}\label{sec:Appendix_G192.32-11.88N}
Figure \ref{fig:appendix_G192.32_11.88N} shows different components of our studied continuum and emission maps in the same order as Figure \ref{fig:appendix_G191.90_11.20S}. No T$_{bol}$ and L$_{bol}$ estimations are available for this object in the literature. The envelope mass from 1.3\,mm continuum is estimated to be $\sim$ 1.2$\pm$0.18 M$_\sun$.

CO integrated map shows that the outflow shell structure is blended with ambient molecular cloud in the south-east (blueshifted) lobe. Therefore, we consider high-velocity components to probe the outflow shell, that is the higher contours in the CO integrated map. The redshifted lobe appears to be much shorter in length than the blueshifted lobe, which is possibly due to the presence of ambient cloud in the blueshifted lobe, making it an extended blended structure. We measured the average inclination angle to be $\sim$ $i$ $\sim$ $50^{+10}_{-10}$ degree. In the SiO integrated emission, a bright SiO emission is seen in the tip of the blueshifted lobe, but no emission in the redshifted lobe. Scattered SiO emission is also observed along the whole blueshifted lobe.  The SiO PV diagram shows SiO emission  along the outflow axis close to the systemic velocity of the source. If it is originating due to jet, then the jet would have a  velocity V$_j$ $\sim$ 2$\pm$2 $\kms$, where V$_{j,cor}$ = V$_j$/$\sin{i}$, which is too small to synthesize a high-density SiO jet. Therefore, we suggest that the SiO emission at the tip of the blueshifted lobe could be the interaction zone between two outflows and the ambient material as in the case of G208.89-20.04Walma \citep[][]{2022ApJ...931..130D}. To confirm the origin of the SiO emission - whether it is indeed a jet or a collision zone - further high-sensitivity observations are required.


\subsection{G200.34-10.97N}\label{sec:Appendix_G200.34-10.97N}
Figure \ref{fig:appendix_G20034_1097N} shows the continuum, C$^{18}$O, CO and SiO emission of the object G200.34-10.97N in the same order as Figure \ref{fig:appendix_G191.90_11.20S}. The object is a Class\,0 source with T$_{bol}$ $\sim$ 43$\pm$10\,K and L$_{bol}$ $\sim$ 1.5$\pm$0.6 L$_\sun$. The envelope mass is estimated to be M$_{env}$ $\sim$ 0.51$\pm$0.08 M$_\sun$.

The outflow opening near the source is narrow, however, as we move outward, the outflow appears wider. One possible explanation for this could be that the thick envelope near the source compresses the outflow opening and upon moving away from the envelope, the outflow suddenly opens up and becomes wide-angle. Another possibility is that the outflow is blending with the ambient cloud, so it appears wider as a combination of outflow and ambient cloud. Higher spatial resolution and higher-velocity resolution observations would be beneficial to study the kinematics and disentangle these components. We are not able to find a good fit for the outer shell to determine the inclination angle. A tentative fitting suggests that the inclination could be $i$ $\sim$ $40^{+15}_{-10}$ degrees. 

The SiO emission along the jet axis appears bow-shock like, with no dense knot-like structure appearing in this observation. The SiO emission is not at a very high-velocity range. The mean SiO velocity is V$_j$ $\sim$ 4$\pm$2 $\kms$, where V$_{j,cor}$ =  V$_j$ /$\sin{i}$. This suggests the SiO emission may not be associated with the jet, but could rather be from the collision zone between the outflow and the ambient cloud. Further higher velocity and higher sensitivity could help to disentangle these two scenarios.

\subsection{G205.46-14.56M2 (HOPS 387 and HOPS 386)}\label{sec:Appendix_G205.46-14.56M2} 
Figure \ref{fig:appendix_G205.46_14.56M2} displays the 1.3\,mm continuum (panel $a$), C$^{18}$O spectra (panel $b$), integrated CO map (panel $c$), and integrated SiO map (panel $d$) of the G205.46-14.56M2 system. It contains five sources, of which, G205.46-14.56M2$\_$A, B, C, D are in a common envelope of mass M$_{env}$ $\sim$ 1.04$\pm$0.19 M$_\sun$. Sources A and B form part of the Class\,I system HOPS\,387, while sources C and D are components of another Class\,I system HOPS\,386 \citep[][]{2020ApJ...890..130T}.

From the CO map, the outflow components from different sources are not clearly identified. The outflow direction of `B' can be determined, but the outflow components from different sources are not distinguishable. A knot-like emission in the SiO emission is observed, which could be a jet component of source D or C. Alternatively, it could be possible that sources C and/or D are the line-rich akin to hot-corinos. As such, the detected emission is mimicking like SiO emission, however, it could actually represent other line emissions emanating from the hot-corinos near the source. The possibility of another unidentified origin is also viable. From the present status of the observations, it is not possible to conclusively determine the origin of this emission.

\subsection{G205.46-14.56N2 (HOPS 401)}\label{sec:Appendix_G205.46-14.56N2}
Figure \ref{fig:appendix_G205.46_14.56N2} shows the continuum and emission components of G205.46-14.56N2 in the same order as Figure \ref{fig:appendix_G205.46_14.56M2}. This is a Class\,0 source with T$_{bol}$ $\sim$ 32$\pm$8\,K and L$_{bol}$ $\sim$ 0.8$\pm$0.3 L$_\sun$, has an envelope mass of M$_{env}$ $\sim$ 0.85$\pm$0.13 M$_\sun$. The source also coincides with the location of HOPS\,401 \citep[][]{2020ApJ...890..129K,2020ApJ...890..130T}.

The outflow seen in the CO map is difficult to distinguish from the ambient cloud. We detected SiO emission in the tip of the eorth-eastern lobe of the outflow, which could either be a bow-shock structure of the jet, or a collision zone between the outflow and ambient cloud. Unfortunately, due to the low sensitivity in both the SiO and CO maps, we were unable to determine the outflow and jet parameters for this source.  

\subsection{G209.55-19.68N1 (HOPS 12)}\label{sec:Appendix_G209.55-19.68N1}
Figure \ref{fig:appendix_G209.55_19.68N1} shows different continuum and emission components of G209.55-19.68N1 in the same order as Figure \ref{fig:appendix_G191.90_11.20S}. G209.55-19.68N1 is a multiple system, where source `A' is responsible for driving the largest outflow. Two other sources, `B' and `C' form a close system that drives a smaller-scale outflow. High-resolution maps of G209.55-19.68N1 reveal that `A' is itself a close binary system, and `B-C' is a single source \citep[HOPS\,12][]{2020ApJS..251...20D}. In the ACA resolution, all three sources are within the same common envelope of mass M$_{env}$ $\sim$ 1.98$\pm$0.3 M$_\sun$. In this paper, we consider T$_{bol}$ $\sim$ 47$\pm$13\,K and L$_{bol}$ $\sim$ 9.0$\pm$3.7 L$_\sun$ for the main outflow source `A'.

The northern component of the outflow shell from source `A' appears wider than the southern part. Within the northern outflow shell, there is a secondary outflow from the `B' or/and `C' system. The interaction between that two outflows creates a shock region, which is faintly detected with SiO emission. This SiO emission is not located along the jet axis. Due to the presence of the secondary source and blending with the ambient cloud, the northern outflow appears wider than the southern component. A similar outflow collision zone was reported for another protostellar system BHR\,71 \citep[][]{2018AJ....156..239Z}. The southern component of CO emission (redshifted) is narrower than the northern lobe. Here, we visually checked each channel to identify the likely outflow shell. In the northern lobe, we chose a CO emission width close to the source and, in the southern lobe, we consider up to 2 $\sigma$ to define the outflow shell. We measured the average inclination angle from the CO shell to be $\sim$   $i$ $\sim$ $27^{+15}_{-10}$ degrees. The tiny knot-like structure in SiO emission (PV diagram) could not be confirmed as a real knot from the current observations. 

\subsection{G211.47-19.27S (HOPS 288)}\label{sec:Appendix_G211.47-19.27S}
Figure \ref{fig:appendix_G211.47_19.27S} shows the continuum and emission components of  G211.47-19.27S in the same order as Figure   \ref{fig:appendix_G205.46_14.56M2}. From the CO velocity channel maps, we have detected three distinct outflow sources, all of which are within a common envelope of mass M$_{env}$ $\sim$ 3.37$\pm$0.52 M$_\sun$. The location of the system coincides with HOP\,288. The combined system has  T$_{bol}$ $\sim$ 49$\pm$21\,K and L$_{bol}$ $\sim$ 180$\pm$70 L$_\sun$.  Such a high bolometric luminosity and temperature suggest that the system is in the Class\,I phase, however, disentangling the nature of individual components within this multiple source system is challenging due to the low-resolution infrared observations. 

The properties of the three CO outflows could not be separated based on the current state of our observations.  Bright monopolar SiO emission is observed along the jet axis of source `B'. The SiO emission is in the redshifted outflow lobe of `B' with a mean velocity $V_j$ $\sim$ 5$\pm$2 $\kms$.  The source `A' is possibly a line-rich `hot-corino' object \citep[][]{2022ApJ...927..218H}. Given that the outflow/jet components for the different sources could not be differentiated, the outflow/jet parameters have not been estimated.


\section{Objects with No SiO Emission but well-defined CO outflow}\label{sec:Appendix_individual_NoSiO_CO_definedOutflow}

\subsection{G192.12-11.10}\label{sec:Appendix_G192.12-11.10}
Figure \ref{fig:appendix_G192.12-11.10} shows the 1.3\,mm continuum (panel $a$), C$^{18}$O spectra (panel $b$), integrated CO map (panel $c$), and CO position-velocity diagram (panel $d$) of G192.12$-$11.20. It is a Class\,0 source with T$_{bol}$ $\sim$ 44$\pm$15\,K,  L$_{bol}$ $\sim$ 9.5$\pm$4.0 L$_\sun$, and a envelope mass of M$_{env}$ $\sim$ 1.21$\pm$0.18 M$_\sun$. Although the source is in the Class\,0 phase, we do not observe any jet-like components along the flow-axis in either SiO or CO.  However if the jet is very high-density and optically thicker for SiO (5-4) and CO(2-1), then the jet components could be detected in higher transitions, such as SiO (8-7) and CO(3-2). The object appears to be orientated nearly edge-on with an inclination angle of $i$ $\sim$ $15^{+10}_{-5}$ degrees.

\subsection{G196.92-10.37}\label{sec:Appendix_G196.92-10.37}
G196.92-10.37 (Figure \ref{fig:appendix_G196.92-10.37}; in the same sequence as Figure \ref{fig:appendix_G192.12-11.10}) is a wide multiple system, where the strongest outflow is driven by source `A' along northeast to southwest direction. T$_{bol}$ and L$_{bol}$ of the source `A' are currently unknown. The outflows from sources `B' and `C' are less spatially extended following a north-south direction and they blend with the blueshifted lobe from source `A'. There is no way to disentangle the outflow components from different sources from the present data set. Assuming minimal contamination from the outflows of `B' and `C' to the `A' outflow, we have estimated the outflow force for source `A'. The envelope mass for source `A' is M$_{env}$ $\sim$ 0.71$\pm$0.12 M$_\sun$. It is possibly highly inclined, close to face-on (i $\sim$ $70^{+10}_{-10}$ degree). There is no detection of SiO emission in this target field.

\subsection{G203.21-11.20W1}\label{sec:Appendix_G203.21-11.20W1}
Figure \ref{fig:appendix_G203.21-11.20W1} shows various components of G203.21-11.20W1 in the same sequence as Figure \ref{fig:appendix_G192.12-11.10}. An outflow is emanating from the source, which has an envelope mass of  M$_{env}$ $\sim$ 0.54$\pm$0.11 M$_\sun$. Notably, there is no reported infrared emission for this source in the literature. The outflow is characterized by a narrow angle and has a well-defined cavity wall. No jet-like high-velocity component has been detected in either SiO or CO. We measured the average inclination angle to be $\sim$  $i$ $\sim$ $20^{+10}_{-5}$ degrees.



\subsection{G208.68-19.20S (HOPS 84)}\label{sec:Appendix_G208.68-19.20S}
G208.68-19.20S (HOPS\,84; Figure \ref{fig:appendix_G208.68-19.20S}) is a close binary system with T$_{bol}$ $\sim$ 96$\pm$25\,K,  L$_{bol}$ $\sim$ 49$\pm$18 L$_\sun$, and the envelope mass is M$_{env}$ $\sim$ 1.55$\pm$0.3 M$_\sun$. The outflow is likely being driven by the source `A'. The CO PV diagram suggests the object is nearly edge-on with an inclination angle of $\sim$  $i$ $\sim$ $10^{+10}_{-5}$ degrees. No SiO emission was detected in this target field. 


\subsection{G211.01-19.54N}\label{sec:Appendix_G211.01-19.54N}
Figure \ref{fig:appendix_G211.01-19.54N} shows the observed components of G211.01-19.54N (also known as HOPS\,153) in the same sequence as Figure \ref{fig:appendix_G192.12-11.10}. It is catagorised as a Class\,0 system with T$_{bol}$ $\sim$ 39$\pm$12\,K and L$_{bol}$ $\sim$ 4.5$\pm$1.8 L$_\sun$. The envelope mass is M$_{env}$ $\sim$ 0.81$\pm$0.16 M$_\sun$. Continuum emission extends along the same direction as the blueshifted CO outflow, suggesting that this could represent envelope material being entrained by the outflow.

No SiO emission was detected in this target field.  The blueshifted outflow shell in the PV diagram is difficult to distinguish from the ambient cloud.  We assumed the likely shell structure from near-source emission and fitted it with a parabolic equation, which provides an inclination angle of $\sim$  $i$ $\sim$ $50^{+10}_{-20}$ degrees.

\subsection{G211.01-19.54S}\label{sec:Appendix_G211.01-19.54S}
Figure \ref{fig:appendix_G211.01-19.54S}) depicts G211.01-19.54S (also known as HOPS\,152), a Class\,0 system with T$_{bol}$ $\sim$ 52$\pm$8\,K, and L$_{bol}$ $\sim$ 0.9$\pm$0.4 L$_\sun$. This is one of the very low-luminosity objects. It possesses an envelope with a mass of M$_{env}$ $\sim$ 0.21$\pm$0.04 M$_\sun$. Only a redshifted (southern) outflow shell is detected with CO-integrated emission. This could be a monopolar outflow source unless the northern outflow is blended with the ambient cloud.  In the PV diagram, the emission in the blueshifted (northern) part is likely tracing ambient material. The outflow shell, even in the redshifted section, could not be distinguished in the PV diagram. Consequently, the inclination angle measurement for this source is highly biased due to the unreliable outflow shell consideration. No SiO emission is detected in this observed field.	


\subsection{G212.84-19.45N}\label{sec:Appendix_G212.84-19.54N}
G212.84-19.45N (HOPS\,224; Figure \ref{fig:appendix_G212.84-19.45N}) is a Class\,0 system with  T$_{bol}$ $\sim$ 50$\pm$13\,K,  L$_{bol}$ $\sim$ 3.0$\pm$1.2 L$_\sun$, and envelope of mass M$_{env}$ $\sim$ 0.52$\pm$0.09 M$_\sun$. In the CO-integrated emission of this system, the redshifted outflow appears wider than the blueshifted lobe. We have estimated an average inclination angle $\sim$ $i$ $\sim$ $25^{+15}_{-10}$ degrees. Interstingly, a high-velocity CO component can be observed in the redshifted lobe, which may be indicative of a jet. However, no SiO emission has been detected in this field.

\section{Objects with No SiO Emission and No well-defined CO outflow}\label{sec:Appendix_individual_NOSiO_NotCO}
In some identified sources, CO outflows were detected, however, due to poor signal-to-noise, deriving the PV diagram along the outflow axis was not possible. These sources include G201.52-11.08 (a Class\,I source; Figure \ref{fig:appendix_G201.52-11.08}), G205.46-14.56S2 (a class source HOPS\,385; Figure \ref{fig:appendix_G205.46-14.56S2}), G210.82-19.47S (a class source HOPS\,156, Figure \ref{fig:appendix_G210.82-19.47S}; Outflow driven by G210.82-19.47S, not by secondary source G210.82-19.47S$\_$02), G211.16-19.33N2 (a transition Class\,0 to Class\,I source HOPS\,133; Figure \ref{fig:appendix_G211.16-19.33N2}), G211.47-19.27N (a Class\,0 close binary source HOPS\,290; Figure \ref{fig:appendix_G211.47-19.27N}), and G212.10-19.15S (a class\,0 source; HOPS\,247; Figure \ref{fig:appendix_G212.10-19.15S}). 

CO emission is observed in the field of a few other sources, however, in these cases the flow direction can not be resolved due to poor sensitivity. These sources are G192.32-11.88S (a Class\,I source; Figure \ref{fig:appendix_G192.32-11.88S}; and source $\sharp$02 is the same of G192.32-11.88N), G205.46-14.56N1 (a Class\,0 protostar HOPS\,402; Figure \ref{fig:appendix_G205.46-14.56N1}), G206.93-16.61E2 (a class\,I multiple system HOPS\,298;  Figure \ref{fig:appendix_G206.93-16.61E2}), G207.36-19.82N1 (a binary system, outflow driven bu source `A'; Figure \ref{fig:appendix_G207.36-19.82N1}), G208.68-19.20N1 (a Class\,0 system HOPS\,87; Figure \ref{fig:appendix_G208.68-19.20N1}), G210.97-19.33S2 (a wide binary system source `A' or HOPS\,377 and `B' or HOPS\,144; Figure \ref{fig:appendix_G210.97-19.33S2}, G212.10-19.15N2 (a close binary system source `A' or HOPS\,263 and `B' or HOPS\,262; Figure \ref{fig:appendix_G212.10-19.15N2}), and G215.87-17.62N (a Class\,II system; Figure \ref{fig:appendix_G215.87-17.62N}).

High-sensitivity observations with other molecular transitions and higher transitions of CO and SiO could be beneficial for studying the outflow/jet morphology of these objects. For instance, 208.68-19.20N1 (HOPS\,87) exhibits a compact outflow in HCN (4-3) and CO (3-2), reported in \citep[][]{2012ApJ...745L..10T}.


\setcounter{figure}{0} 
\renewcommand{\thefigure}{B\arabic{figure}}
\begin{figure*}
\fig{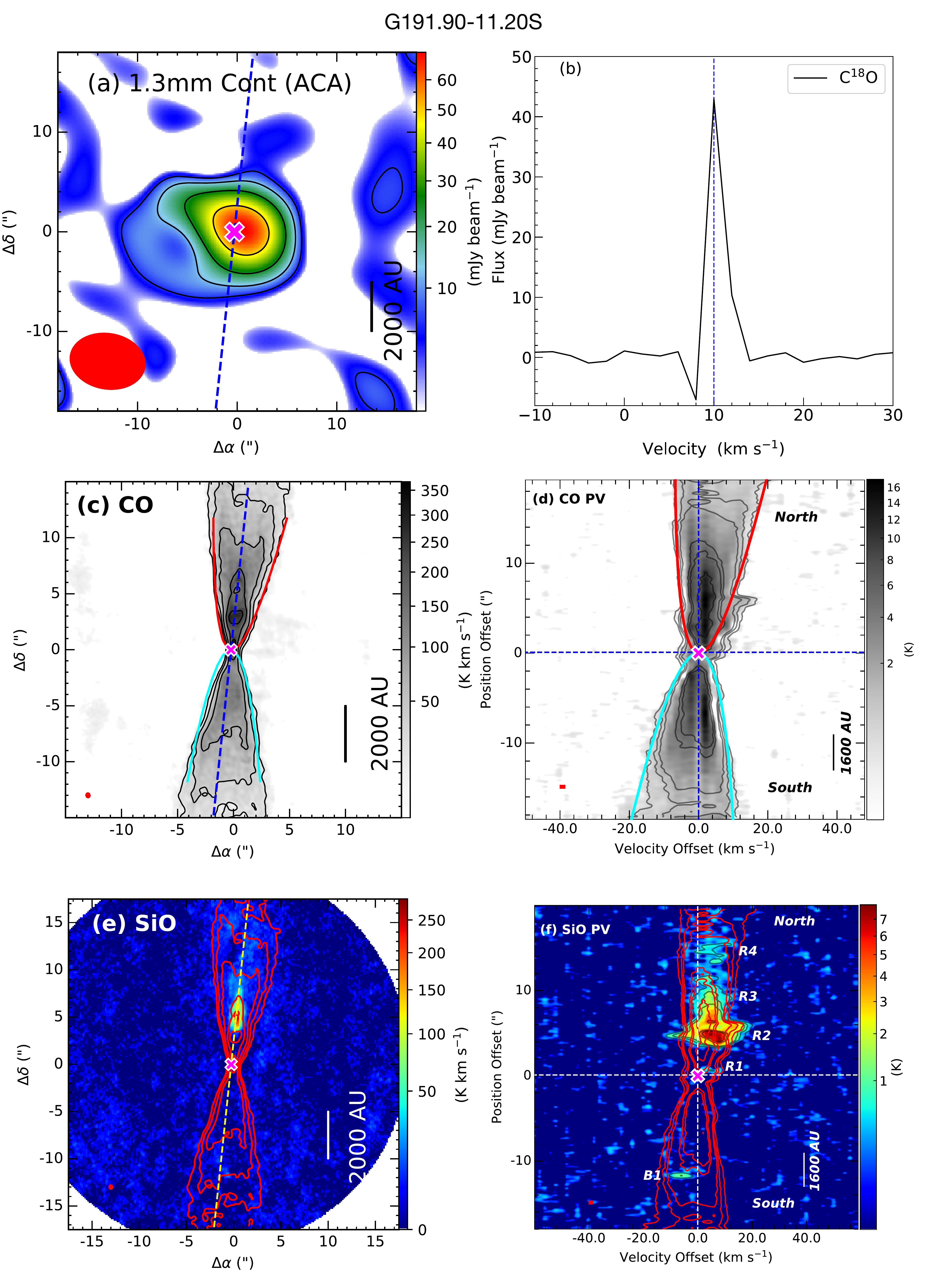}{0.85\textwidth}{}
\caption{G191.90-11.20S: (a) 1.3\,mm continuum map at ACA resolution with sensitivity $\sim$ 2.0 $\mjypb$. The symbols and contour levels are the same as Figure \ref{fig:G20321_1120W2_ACA_envelope}.
(b) The  C$^{18}$O spectra extracted from high-resolution maps, and V$_{sys}$ = 10 $\kms$. All symbols are the same as Figure \ref{fig:G20321_1120W2_C18ospectra}. (c) CO emission integrated over [$-$4 to $+$8] $\kms$ with sensitivity 6.48 $\kkms$ with similar symbols and contour from Figure \ref{fig:G203.21-11.20W2outflow_parabola}a. (d) CO PV diagram with sensitivity 0.16 $K$ with similar symbols and contour as Figure \ref{fig:G203.21-11.20W2outflow_parabola}b. (e) SiO emission integrated over [$-$10 to $+$18] $\kms$ with sensitivity 3.7 $\kkms$ with similar symbols and contours as Figure \ref{fig:G203.21-11.20W2_SiO_integrated-PV}a. (f) SiO PV diagram  with sensitivity 0.16 $K$ with similar symbols and contours as Figure \ref{fig:G203.21-11.20W2_SiO_integrated-PV}b.}
\label{fig:appendix_G191.90_11.20S}
\end{figure*}

\begin{figure*}
\fig{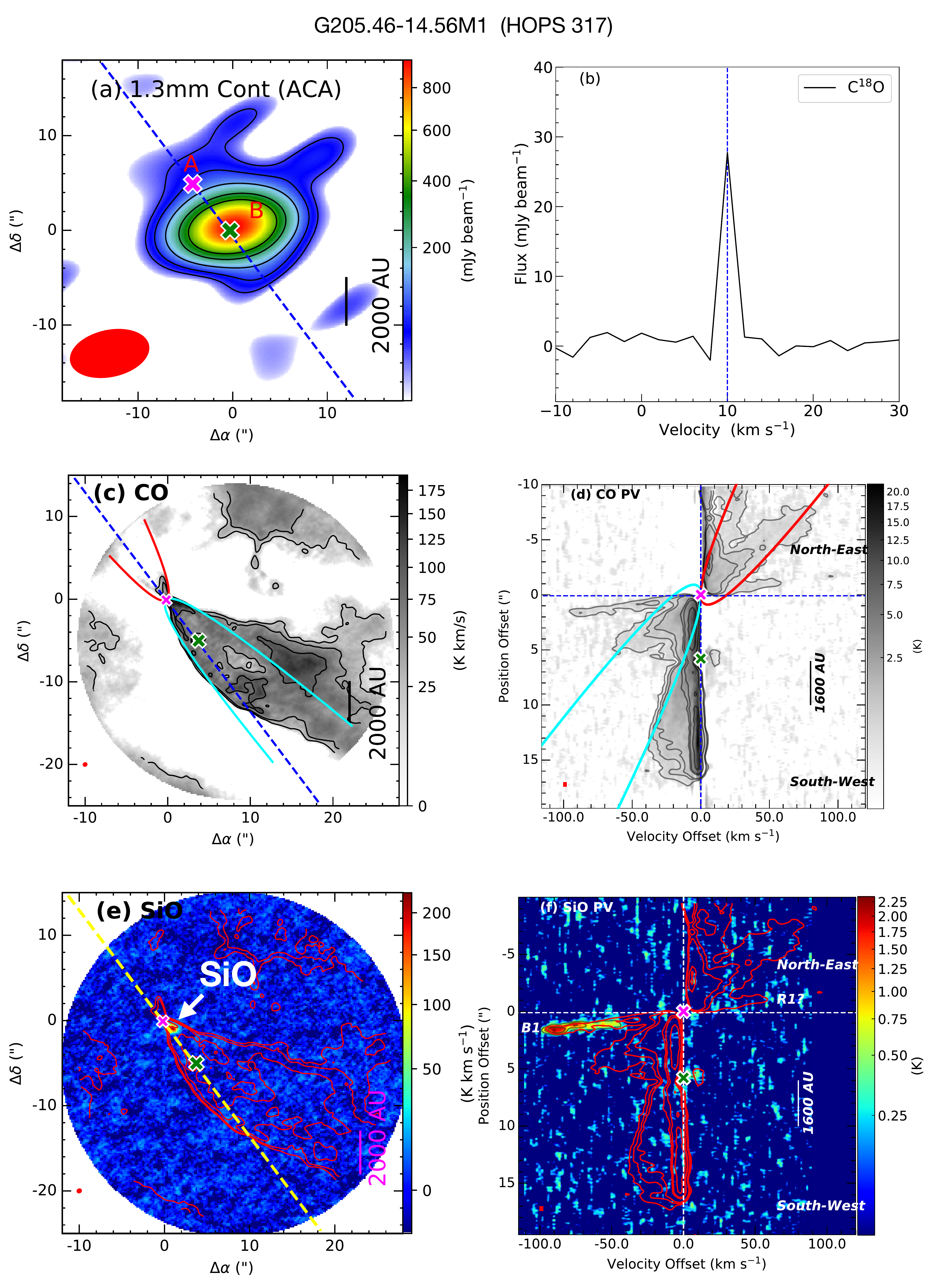}{0.85\textwidth}{}
\caption{G205.46-14.56M1 (HOPS\,317): (a) 1.3mm continuum map at ACA resolution with sensitivity $\sim$ 2.0 $\mjypb$. The symbols and contour levels are the same as Figure \ref{fig:G20321_1120W2_ACA_envelope}.
(b) The  C$^{18}$O spectra extracted from high-resolution maps, and V$_{sys}$ = 10 $\kms$. All symbols are the same as Figure \ref{fig:G20321_1120W2_C18ospectra}. (c)  CO emission maps integrated over [$-$18 to $-$2] and [$+$2 to $+$12] $\kms$ with sensitivity 7.4 $\kkms$ with similar symbols and contours as Figure \ref{fig:G203.21-11.20W2outflow_parabola}a. (d) CO PV diagram with sensitivity 0.11 $K$ with similar symbols and contours as Figure \ref{fig:G203.21-11.20W2outflow_parabola}b. (e) SiO emission integrated over [$-$102 to $-$28] and [$+$4 to $+$12] $\kms$  with sensitivity 3.7 $\kkms$ with similar symbols and contours as Figure \ref{fig:G203.21-11.20W2_SiO_integrated-PV}a. (f) SiO PV diagram with sensitivity 0.16 $K$ with similar symbols and contour of Figure \ref{fig:G203.21-11.20W2_SiO_integrated-PV}b.}
\label{fig:appendix_G205.46_14.56M1}
\end{figure*}

\begin{figure*}
\fig{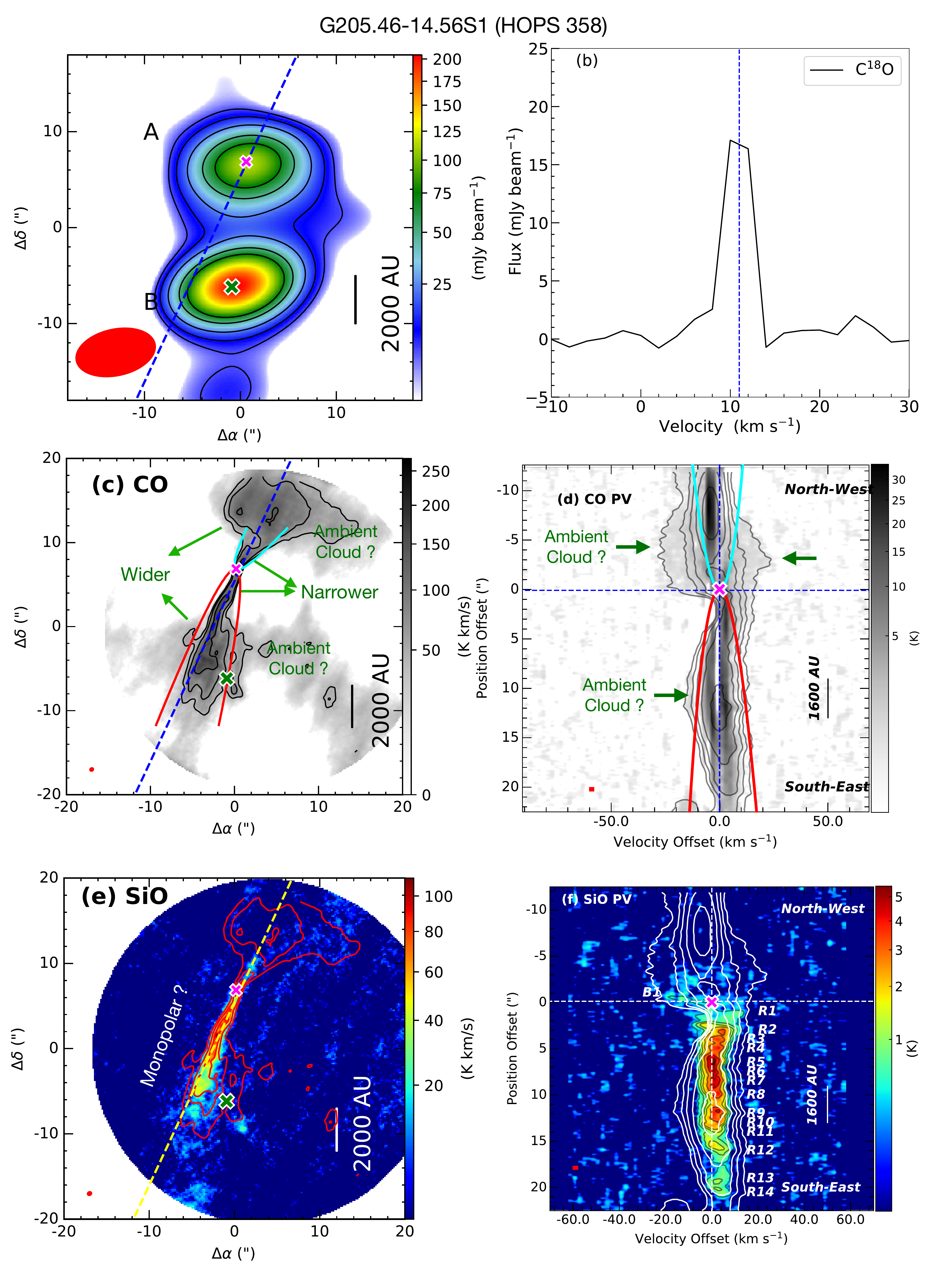}{0.85\textwidth}{}
\caption{G205.46-14.56S1 (HOPS\,358): (a) 1.3\,mm continuum map at ACA resolution with sensitivity $\sim$ 2.0 $\mjypb$. The symbols and contour levels are the same as Figure \ref{fig:G20321_1120W2_ACA_envelope}.
(b) The  C$^{18}$O spectra extracted from high-resolution maps, and V$_{sys}$ = 11 $\kms$. All symbols are the same as Figure \ref{fig:G20321_1120W2_C18ospectra}. (c) CO emission integrated over [$-$20 to $-$4] and [$+$2 to $+$20] $\kms$  with sensitivity $\sigma$ $\sim$ 14.8 $\kkms$ with similar symbols are Figure \ref{fig:G203.21-11.20W2outflow_parabola}a. The contours are at (3, 9, 18, 36, 72) $\times$ $\sigma$ . (d) CO PV diagram with sensitivity 0.11 $K$ with similar symbols and contours as Figure \ref{fig:G203.21-11.20W2outflow_parabola}b. (e) SiO emission integrated over [$-$20 to $+$20] $\kms$ with sensitivity 3.4 $\kkms$ with similar symbols and contours as Figure \ref{fig:G203.21-11.20W2_SiO_integrated-PV}a. (f) SiO PV diagram with sensitivity 0.14 $K$ with similar symbols and contours as Figure \ref{fig:G203.21-11.20W2_SiO_integrated-PV}b.}
\label{fig:appendix_G20546_1456S1}
\end{figure*}

\begin{figure*}
\fig{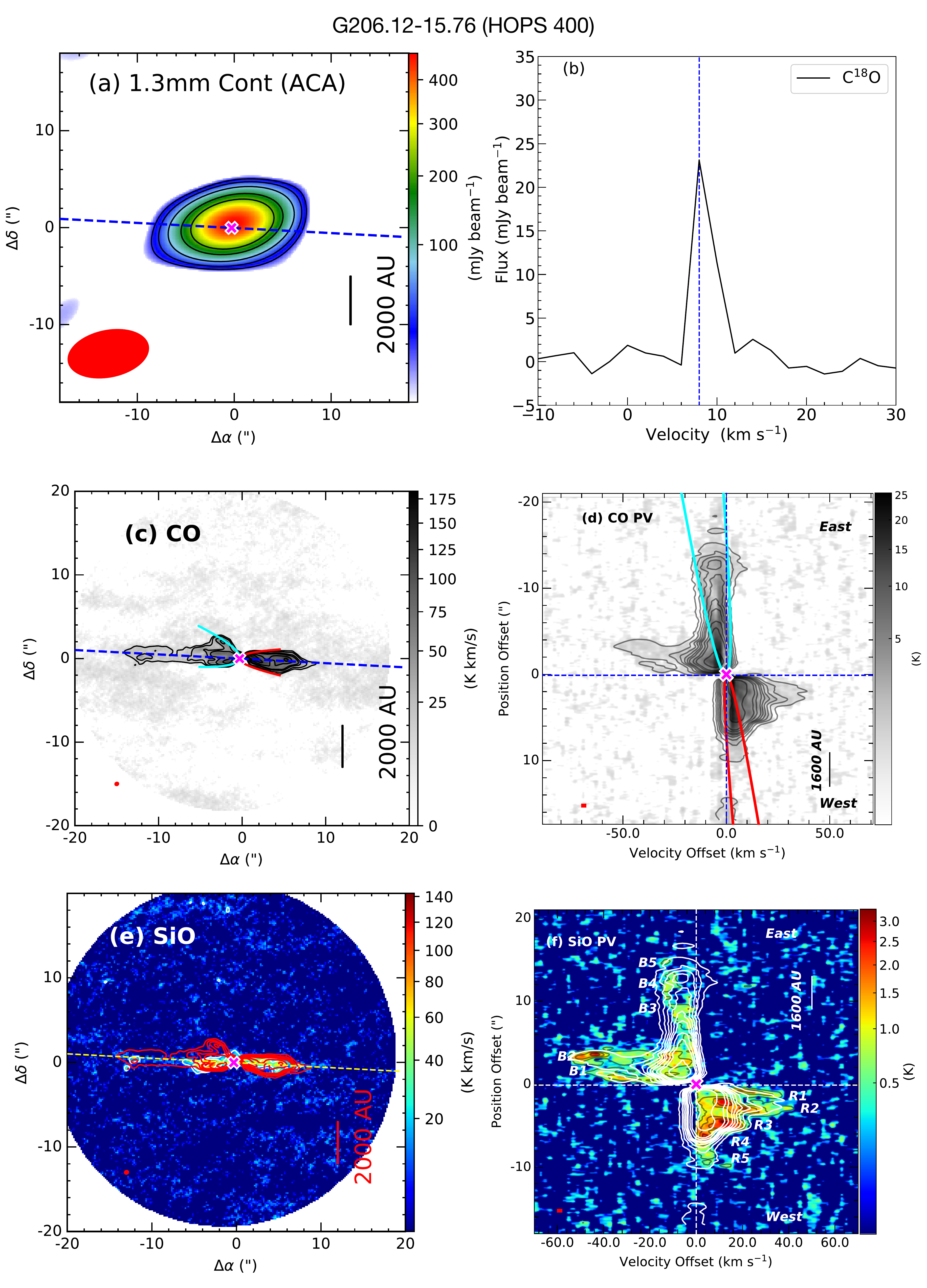}{0.85\textwidth}{}
\caption{G206.12-15.76: (a) 1.3mm continuum map at ACA resolution with sensitivity $\sim$ 6.0 $\mjypb$. The symbols and contour levels are the same as Figure \ref{fig:G20321_1120W2_ACA_envelope}.
(b) The  C$^{18}$O spectra extracted from high-resolution maps, and V$_{sys}$ = 8 $\kms$. All symbols are the same as Figure \ref{fig:G20321_1120W2_C18ospectra}. (c) CO emission integrated over [$-$4 to $+$8] $\kms$ with sensitivity $\sigma$ $\sim$ 4.63 $\kkms$ with similar symbols are Figure \ref{fig:G203.21-11.20W2outflow_parabola}a.  (d) CO PV diagram with sensitivity 0.16 $K$ with similar symbols and contours as Figure \ref{fig:G203.21-11.20W2outflow_parabola}b. (e) SiO emission integrated over [$-$52 to $+$42] $\kms$ with sensitivity 4.6 $\kkms$ with similar symbols and contours as Figure \ref{fig:G203.21-11.20W2_SiO_integrated-PV}a. (f) SiO PV diagram emission with sensitivity 0.14 $K$ with similar symbols and contours as Figure \ref{fig:G203.21-11.20W2_SiO_integrated-PV}b.}
\label{fig:appendix_G206.12-15.76}
\end{figure*}

\begin{figure*}
\fig{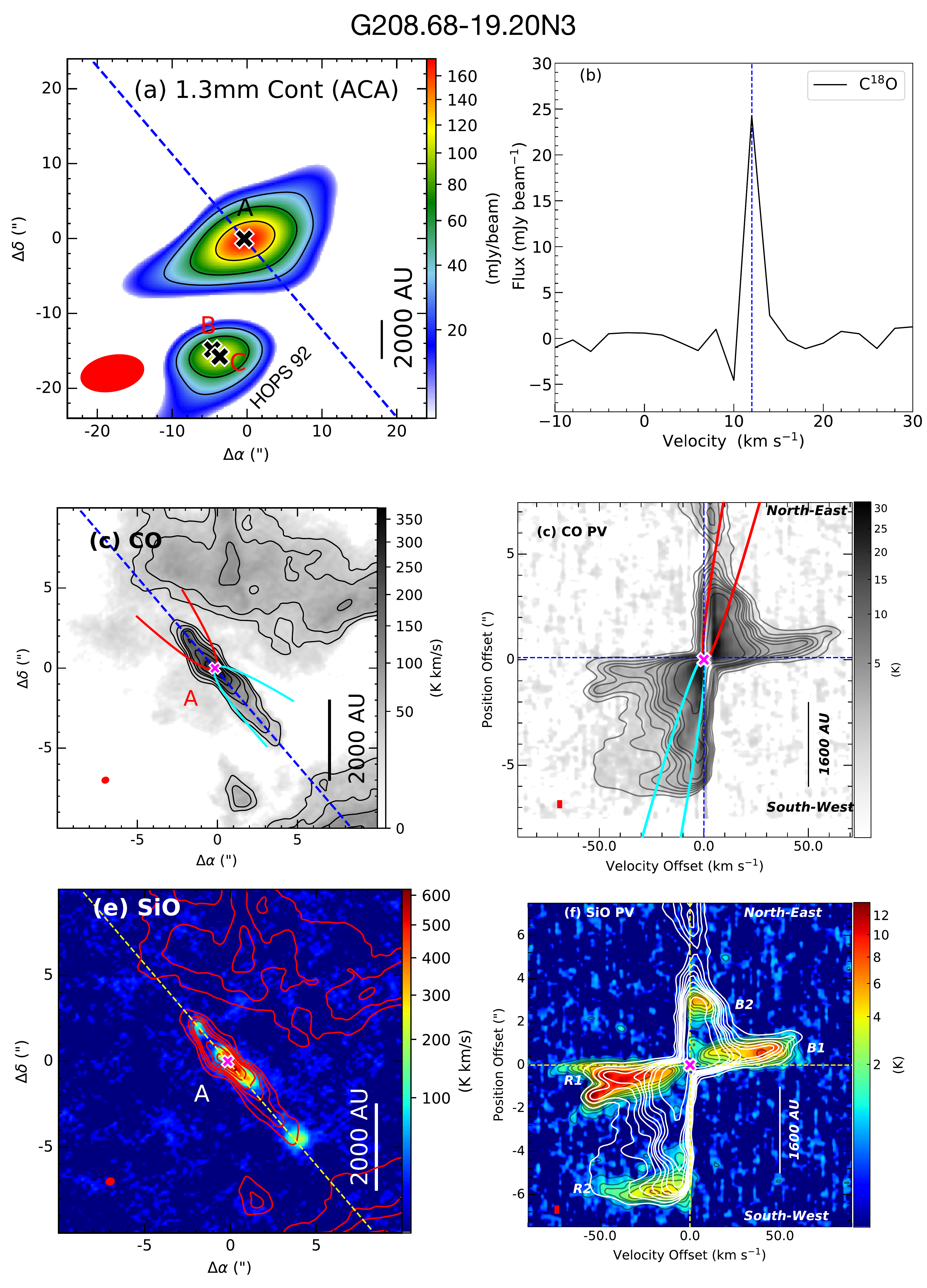}{0.85\textwidth}{}
\caption{G208.68-19.20N3: (a) 1.3mm continuum map at ACA resolution with sensitivity $\sim$ 10.0 $\mjypb$. The symbols and contour levels are the same as Figure \ref{fig:G20321_1120W2_ACA_envelope}.
(b) The  C$^{18}$O spectra extracted from high-resolution maps, and V$_{sys}$ = 12 $\kms$. All symbols are the same as Figure \ref{fig:G20321_1120W2_C18ospectra}. (c) CO emission integrated over [$-$12 to $+$12] $\kms$ with sensitivity $\sigma$ $\sim$ 9.65 $\kkms$ with similar symbols are Figure \ref{fig:G203.21-11.20W2outflow_parabola}a.  (d) CO PV diagram with sensitivity 0.16 $K$ with similar symbols and contours as Figure \ref{fig:G203.21-11.20W2outflow_parabola}b. (e) SiO emission inetgrated over [$-$66 to $+$58] $\kms$ with sensitivity 4.6 $\kkms$ with similar symbols and contours as Figure \ref{fig:G203.21-11.20W2_SiO_integrated-PV} (f) SiO PV diagram with sensitivity 0.15 $K$ with similar symbols and contours as Figure \ref{fig:G203.21-11.20W2_SiO_integrated-PV}b.}
\label{fig:appendix_G208.68-19.20N3}
\end{figure*}

\begin{figure*}
\fig{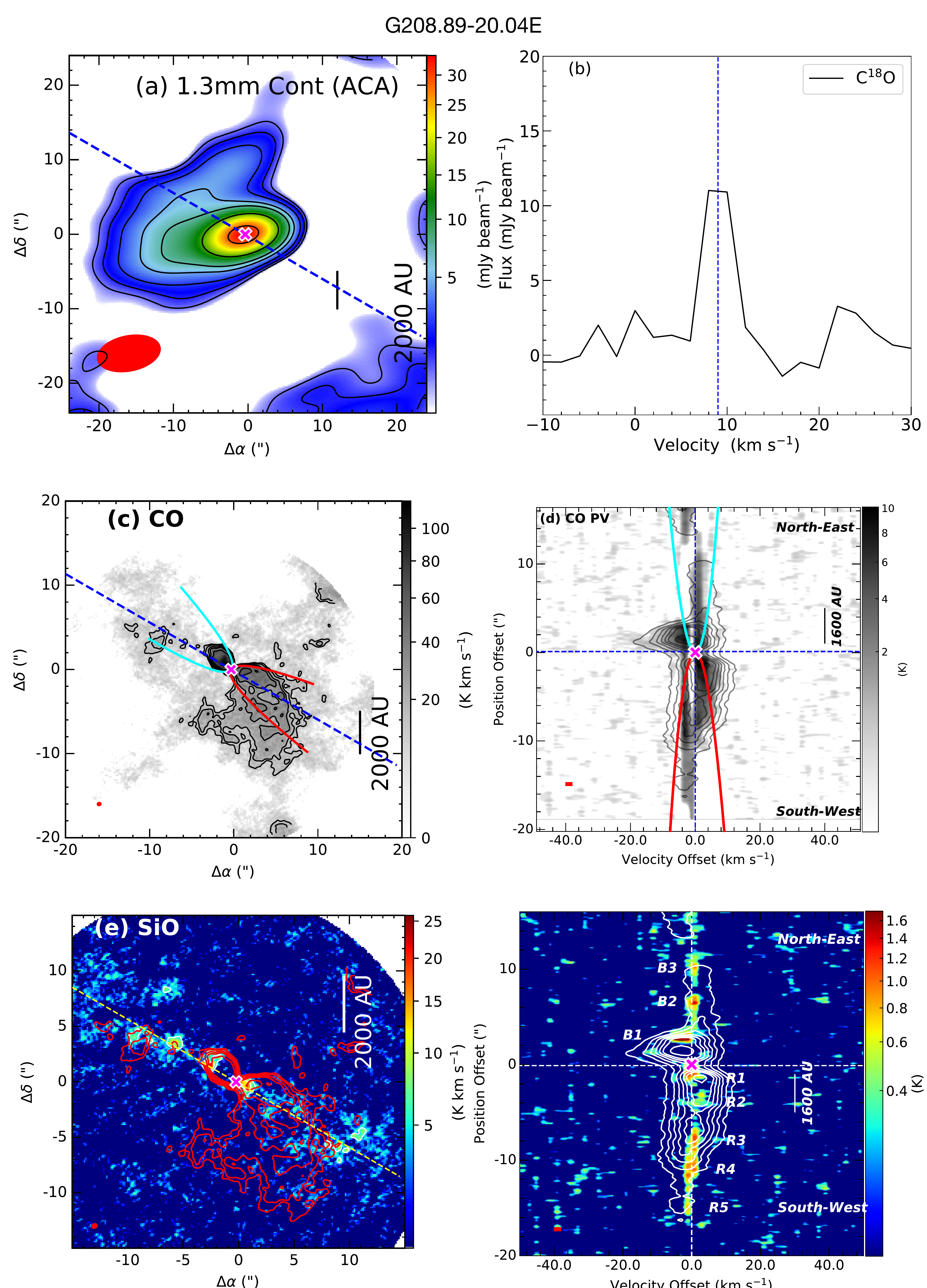}{0.85\textwidth}{}
\caption{G208.89$-$20.04E: (a) 1.3mm continuum map at ACA resolution with sensitivity $\sim$ 0.2 $\mjypb$. The symbols and contour levels are the same as Figure \ref{fig:G20321_1120W2_ACA_envelope}.
(b) The  C$^{18}$O spectra extracted from high-resolution maps, and V$_{sys}$ = 9 $\kms$. All symbols are the same as Figure \ref{fig:G20321_1120W2_C18ospectra}. (c) CO emission integrated over [$-$4, and $+$2 to $+$6] $\kms$ with sensitivity $\sigma$ $\sim$ 1.52 $\kkms$ with similar symbols are Figure \ref{fig:G203.21-11.20W2outflow_parabola}a.  (d) CO PV diagram with sensitivity 0.16 $K$ with similar symbols and contours as Figure \ref{fig:G203.21-11.20W2outflow_parabola}b. (e) SiO emission integrated over [$-$4 to $+$4] $\kms$ with sensitivity 1.7 $\kkms$ with similar symbols and contours as Figure \ref{fig:G203.21-11.20W2_SiO_integrated-PV}a. (f) SiO PV diagram with sensitivity 0.32 $K$ with similar symbols and contours as Figure \ref{fig:G203.21-11.20W2_SiO_integrated-PV}b.}
\label{fig:appendix_G208.89-20.04E}
\end{figure*}

\begin{figure*}
\fig{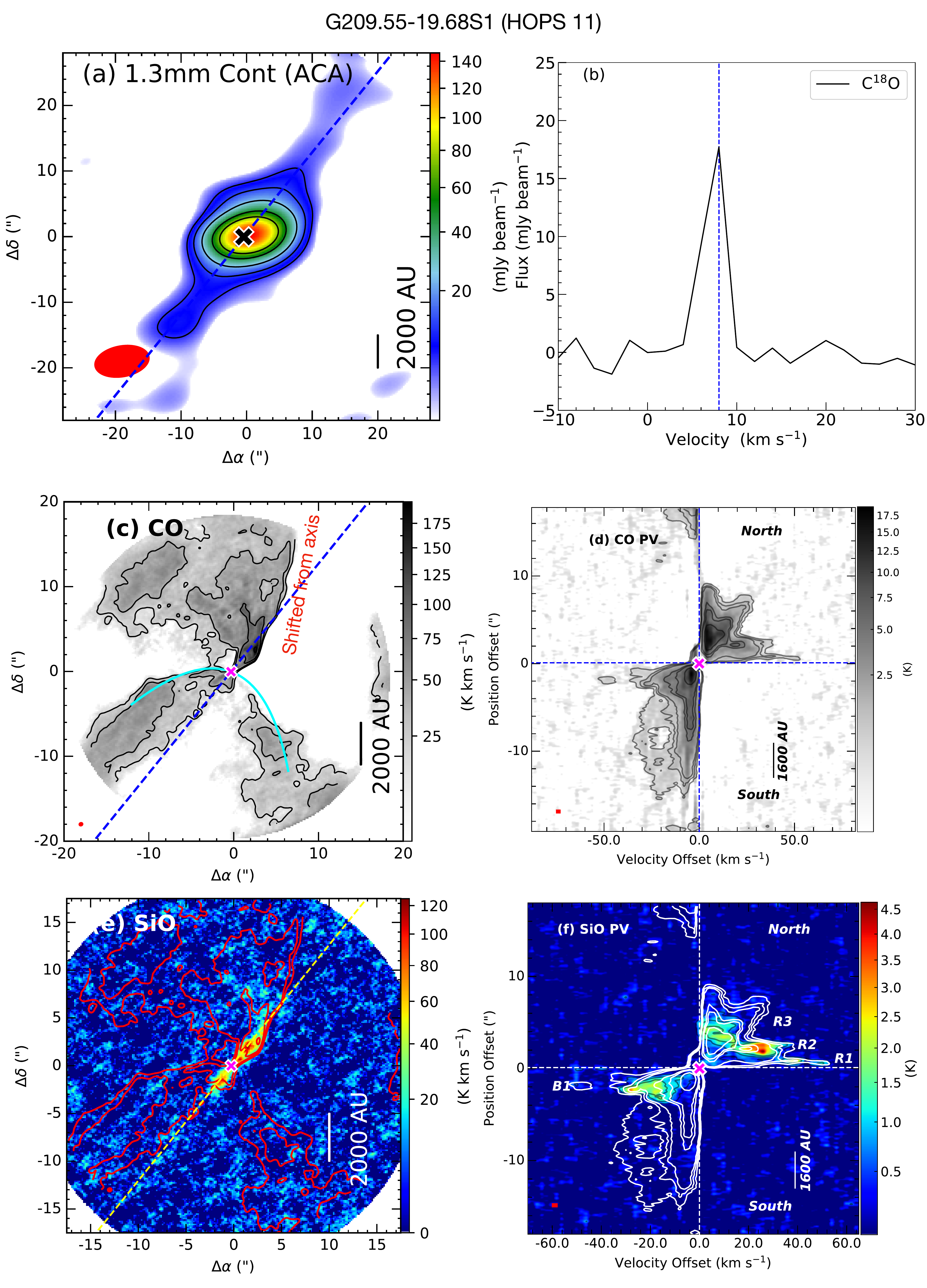}{0.85\textwidth}{}
\caption{G209.55-19.68S1 (HOPS\,11): (a) 1.3mm continuum map at ACA resolution with sensitivity $\sim$ 2.0 $\mjypb$. The symbols and contour levels are the same as Figure \ref{fig:G20321_1120W2_ACA_envelope}.
(b) The  C$^{18}$O spectra extracted from high-resolution maps, and V$_{sys}$ = 8 $\kms$. All symbols are the same as Figure \ref{fig:G20321_1120W2_C18ospectra}. (c) CO emission integrated over [$-$8 to $-$4 and, $+$4 to $+$8] $\kms$ with sensitivity $\sigma$ $\sim$ 6.6 $\kkms$ with similar symbols are Figure \ref{fig:G203.21-11.20W2outflow_parabola}a.  (d) CO PV diagram with sensitivity 0.16 $K$ with similar symbols and contours as Figure \ref{fig:G203.21-11.20W2outflow_parabola}b. (e) SiO emission integrated over [$-$60 to $+$60] $\kms$ with sensitivity 8.0 $\kkms$ with similar symbols and contours as Figure \ref{fig:G203.21-11.20W2_SiO_integrated-PV}a. (f) SiO PV diagram with sensitivity 0.15 $K$ with similar symbols and contours as Figure \ref{fig:G203.21-11.20W2_SiO_integrated-PV}b.}
\label{fig:appendix_G20955_1968S1}
\end{figure*}

\begin{figure*}
\fig{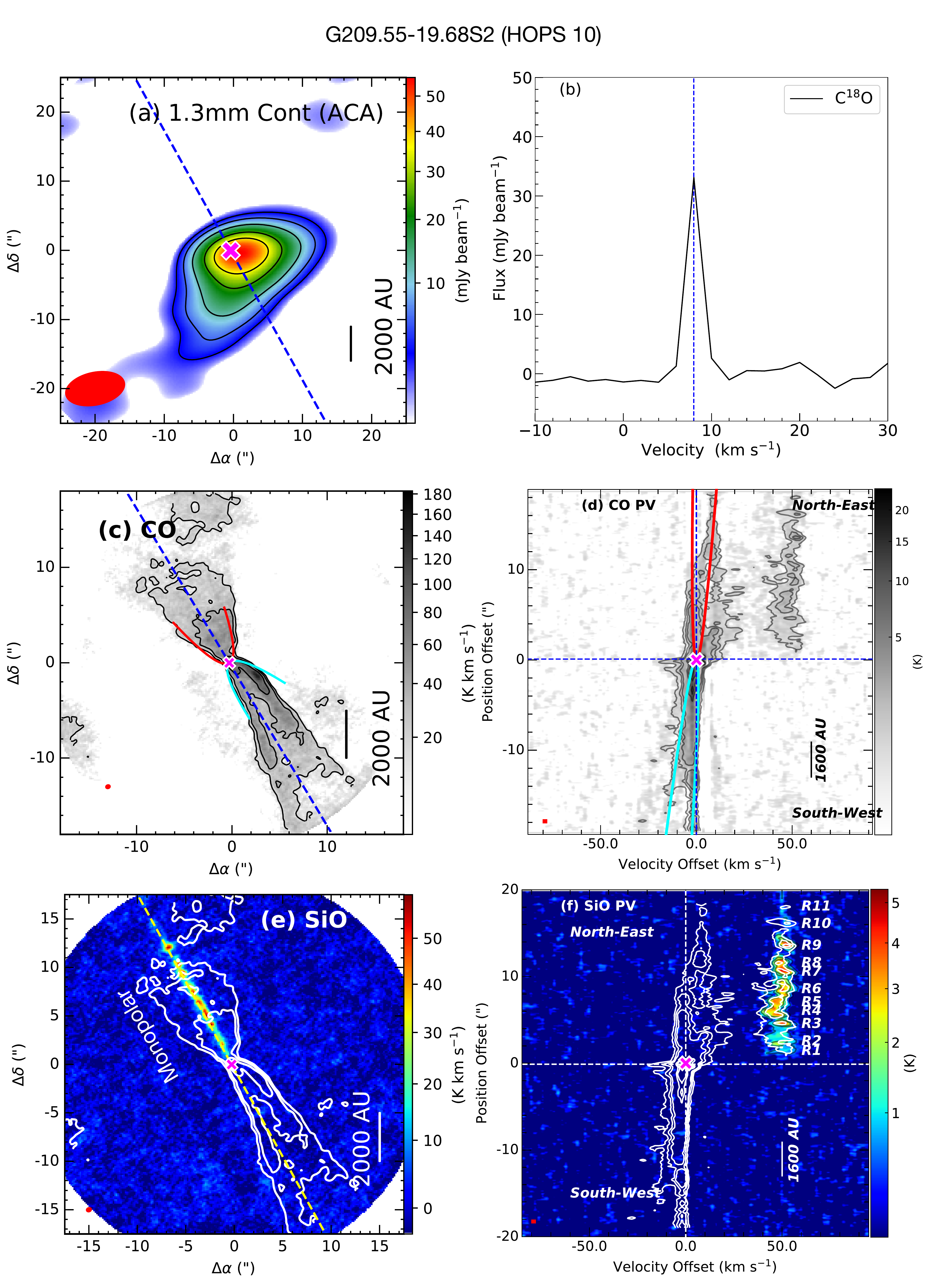}{0.80\textwidth}{}
\caption{G209.55-19.68S2 (HOPS\,10): (a) 1.3mm continuum map at ACA resolution with sensitivity $\sim$ 1.0 $\mjypb$. The symbols and contour levels are the same as Figure \ref{fig:G20321_1120W2_ACA_envelope}.
(b) The  C$^{18}$O spectra extracted from high-resolution maps, and V$_{sys}$ = 8 $\kms$. All symbols are the same as Figure \ref{fig:G20321_1120W2_C18ospectra}. (c) CO emission integrated over [$-$4 to $-$2 and $+$2 to $+$4] $\kms$ with sensitivity $\sigma$ $\sim$ 5.8 $\kkms$ with similar symbols are Figure \ref{fig:G203.21-11.20W2outflow_parabola}a.  (d) CO PV diagram with sensitivity 0.21 $K$ with similar symbols and contours as Figure \ref{fig:G203.21-11.20W2outflow_parabola}b. (e) SiO emission integrated over [$+$40 to $+$58] $\kms$ with sensitivity 2.5 $\kkms$ with similar symbols and contours as Figure \ref{fig:G203.21-11.20W2_SiO_integrated-PV}a. (f) SiO PV diagram with sensitivity 0.21 $K$ with similar symbols and contours as Figure \ref{fig:G203.21-11.20W2_SiO_integrated-PV}b.}
\label{fig:appendix_G209.55_19.68S2}
\end{figure*}

\begin{figure*}
\fig{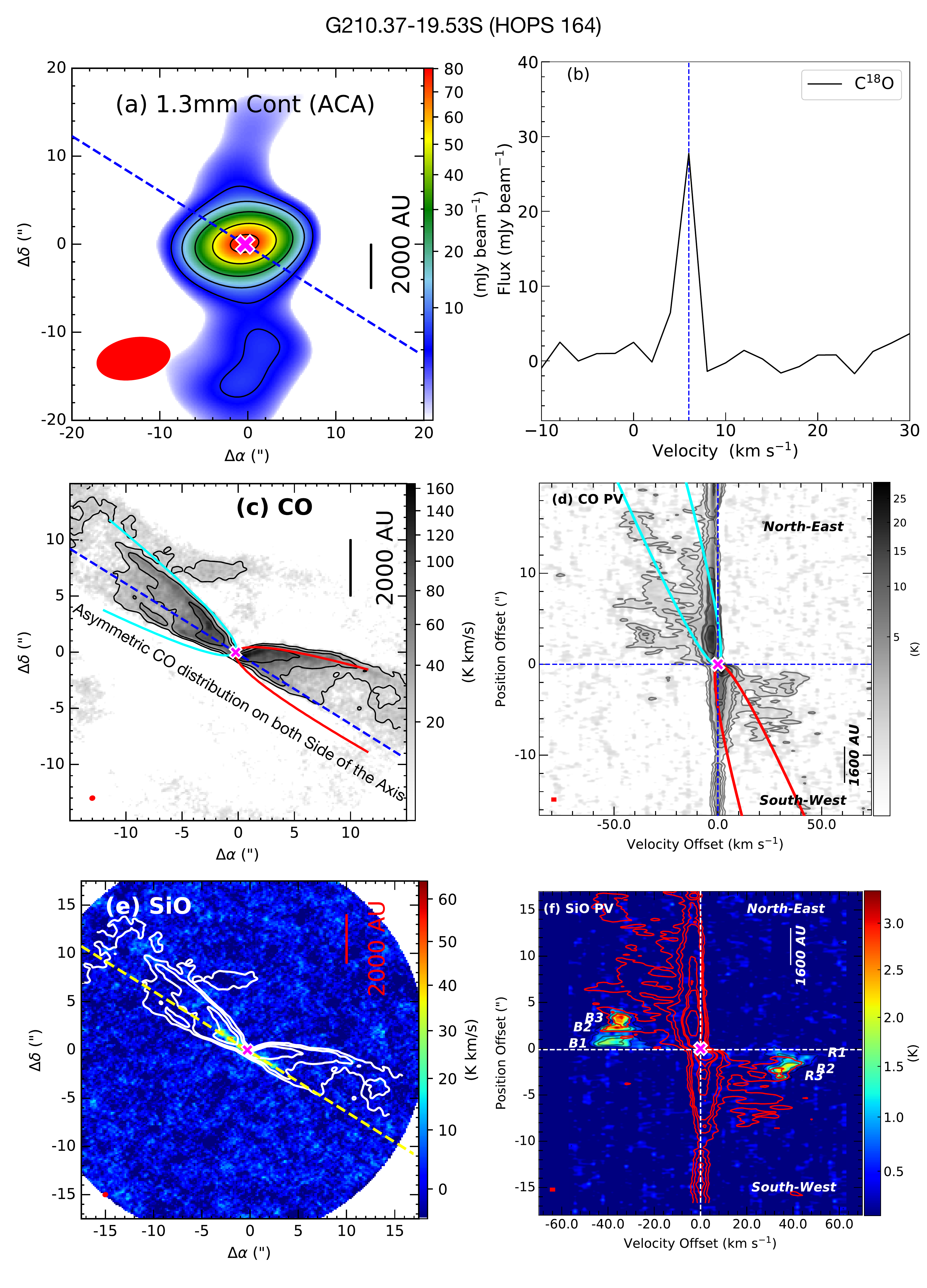}{0.85\textwidth}{}
\caption{G210.37-19.53S (HOPS\,164): (a) 1.3mm continuum map at ACA resolution with sensitivity $\sim$ 1.0 $\mjypb$. The symbols and contour levels are the same as Figure \ref{fig:G20321_1120W2_ACA_envelope}.
(b) The  C$^{18}$O spectra extracted from high-resolution maps, and V$_{sys}$ = 8 $\kms$. All symbols are the same as Figure \ref{fig:G20321_1120W2_C18ospectra}. (c) CO emission integrated over [$-$6 to $-$2 and 0 to $+$4] $\kms$ with sensitivity $\sigma$ $\sim$ 5.8 $\kkms$ with similar symbols are Figure \ref{fig:G203.21-11.20W2outflow_parabola}a.  (d) CO PV diagram with sensitivity 0.21 $K$ with similar symbols and contours as Figure \ref{fig:G203.21-11.20W2outflow_parabola}b. (e) SiO emission integrated over [$-$46 to $-$24 and, $+$30 to $+$48] $\kms$ with sensitivity 2.5 $\kkms$ with similar symbols and contours as Figure \ref{fig:G203.21-11.20W2_SiO_integrated-PV}a. (f) SiO PV diagram with sensitivity 0.21 $K$ with similar symbols and contours as Figure \ref{fig:G203.21-11.20W2_SiO_integrated-PV}b.}
\label{fig:appendix_G210.37-19.53S}
\end{figure*}

\begin{figure*}
\fig{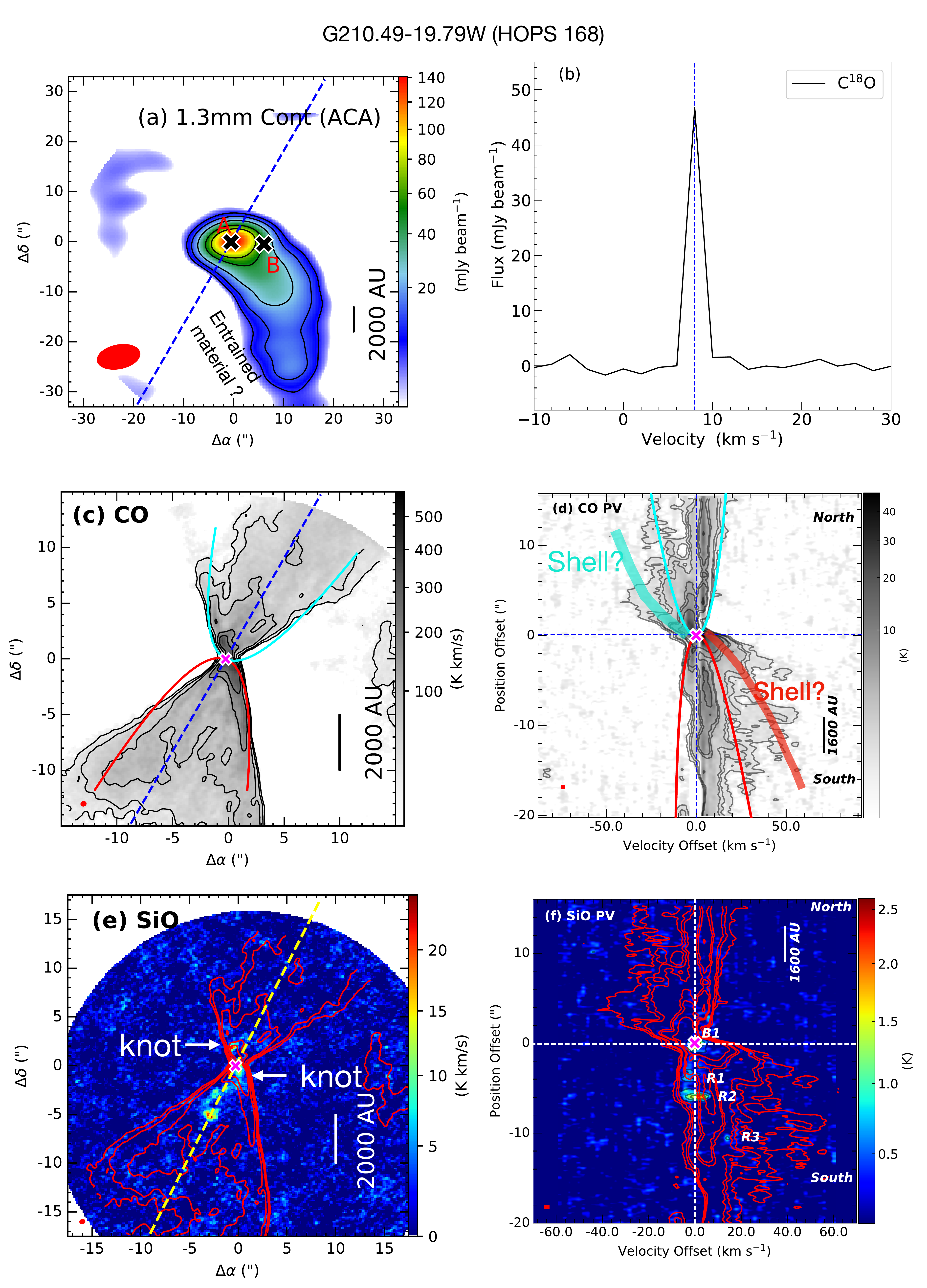}{0.85\textwidth}{}
\caption{G210.49-19.79W (HOPS\,168): (a) 1.3mm continuum map at ACA resolution with sensitivity $\sim$ 2.0 $\mjypb$. The symbols and contour levels are the same as Figure \ref{fig:G20321_1120W2_ACA_envelope}.
(b) The  C$^{18}$O spectra extracted from high-resolution maps, and V$_{sys}$ = 7 $\kms$. All symbols are the same as Figure \ref{fig:G20321_1120W2_C18ospectra}. (c) CO emission map integrated over [-45 to +55] $\kms$ with sensitivity 6.0 $\kkms$ with similar symbols and contours as Figure \ref{fig:G203.21-11.20W2outflow_parabola}a. (d) CO PV diagram with sensitivity 0.15 $K$ with similar symbols and contours as Figure \ref{fig:G203.21-11.20W2outflow_parabola}b. The additional likely episodic shells are marked with thick blue and red lines in the blueshifted and redshifted lones, respectively. (e) SiO map integrated over [$-$20 to +20] $\kms$ with sensitivity 1.9 $\kkms$ with similar symbols and contours as Figure \ref{fig:G203.21-11.20W2_SiO_integrated-PV}a. The knots shifted from jet-axis are marked. (f) SiO PV diagram with sensitivity $\sigma$ = 0.15 $K$ with similar symbol of Figure \ref{fig:G203.21-11.20W2_SiO_integrated-PV}b.}
\label{fig:appendix_G210.49_19.79W}
\end{figure*}

\begin{figure*}
\fig{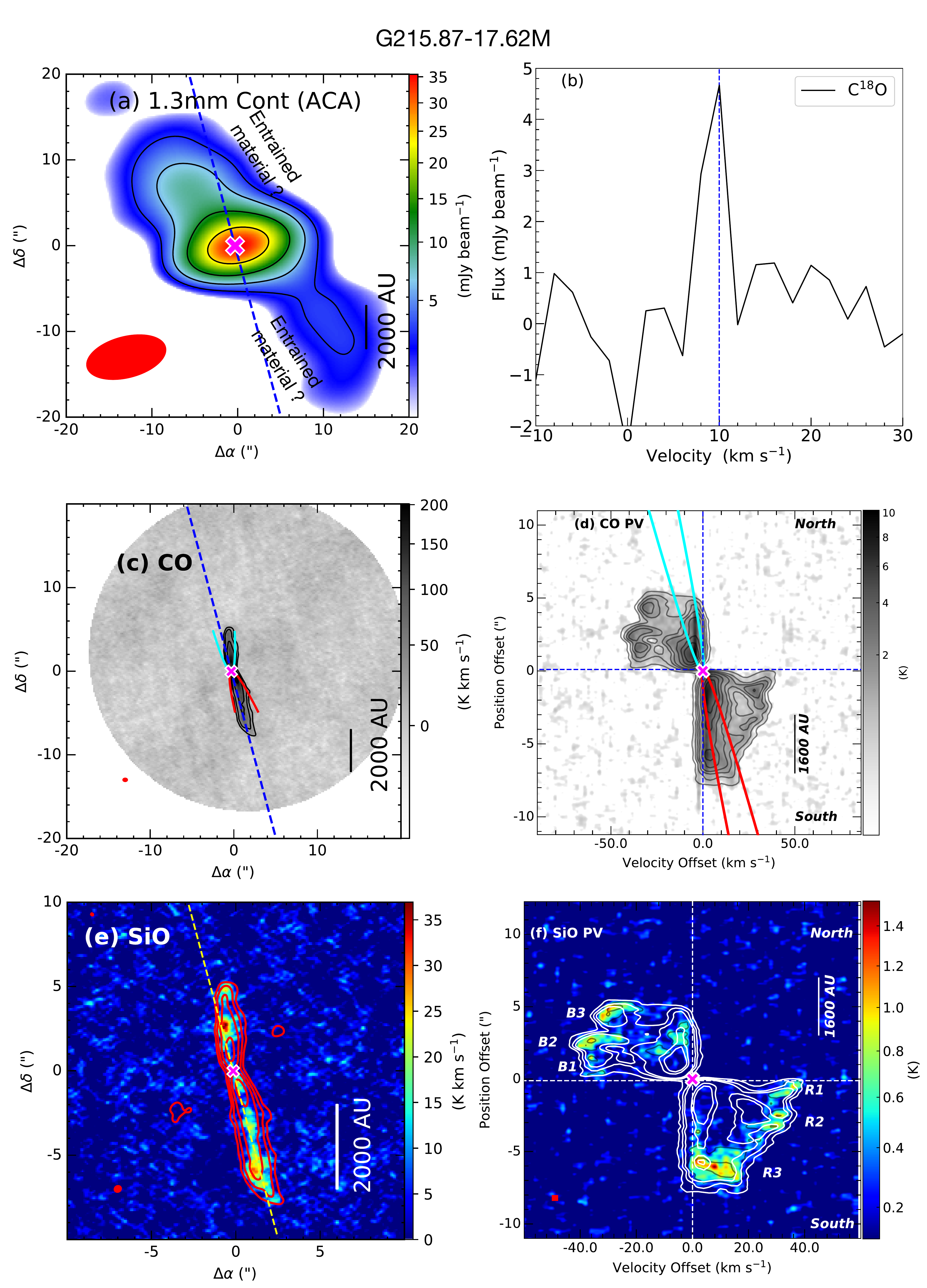}{0.85\textwidth}{}
\caption{G215.87-17.62M: (a) 1.3mm continuum map at ACA resolution with sensitivity $\sim$ 1.0 $\mjypb$. The symbols and contour levels are the same as Figure \ref{fig:G20321_1120W2_ACA_envelope}.
(b) The  C$^{18}$O spectra extracted from high-resolution maps, and V$_{sys}$ = 10 $\kms$. All symbols are the same as Figure \ref{fig:G20321_1120W2_C18ospectra}. (c) CO emission integrated over [$-$10 to $-$4 and, $+$2 to $+$10] $\kms$ with sensitivity $\sigma$ $\sim$ 5.8 $\kkms$ with similar symbols are Figure \ref{fig:G203.21-11.20W2outflow_parabola}a.  (d) CO PV diagram with sensitivity 0.21 $K$ with similar symbols and contours as Figure \ref{fig:G203.21-11.20W2outflow_parabola}b. (e) SiO emission integrated over [$-$40 to $+$38] $\kms$ with sensitivity 2.5 $\kkms$ with similar symbols and contours as Figure \ref{fig:G203.21-11.20W2outflow_parabola}a. (f) SiO emission with sensitivity 0.13 $K$ with similar symbols and contours as Figure \ref{fig:G203.21-11.20W2outflow_parabola}b.}
\label{fig:appendix_G215.87-17.62M}
\end{figure*}


\setcounter{figure}{0}  \renewcommand{\thefigure}{C\arabic{figure}} 
\begin{figure*}
\fig{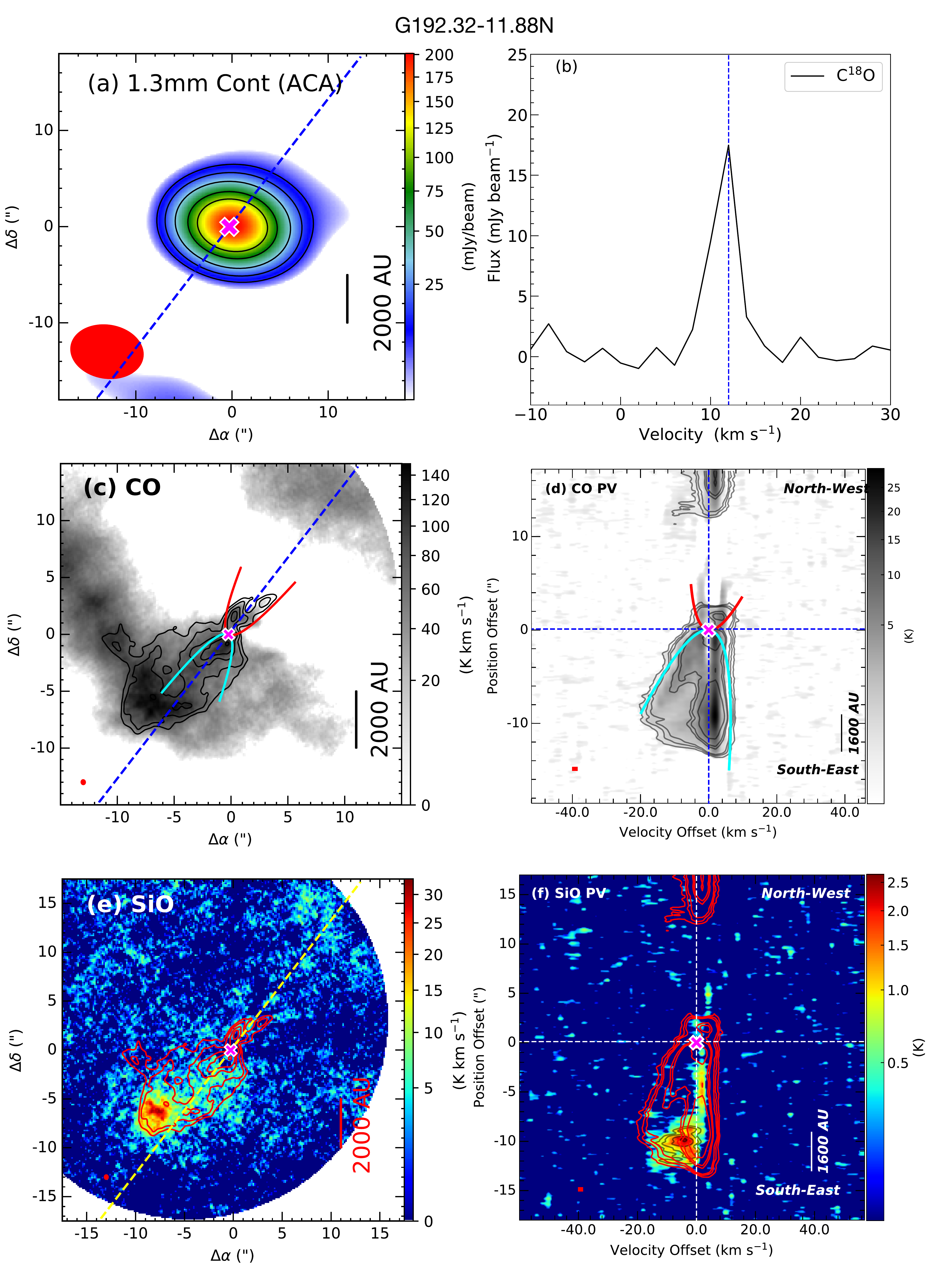}{0.85\textwidth}{}
\caption{G192.32-11.88N: (a) 1.3mm continuum map at ACA resolution with sensitivity $\sim$ 2.9 $\mjypb$. The symbols and contour levels are the same as Figure \ref{fig:G20321_1120W2_ACA_envelope}.
(b) The  C$^{18}$O spectra extracted from high-resolution maps, and V$_{sys}$ = 12 $\kms$. All symbols are the same as Figure \ref{fig:G20321_1120W2_C18ospectra}. (c) Integrated CO emission with sensitivity 6.4 $\kkms$ with similar symbols and contours as Figure \ref{fig:G203.21-11.20W2outflow_parabola}a. (d) CO PV diagram with sensitivity 0.13 $K$ with similar symbols and contours as Figure \ref{fig:G203.21-11.20W2outflow_parabola}b. (e) Integrated SiO emission with sensitivity 2.7 $\kkms$ with similar symbols and contours as Figure \ref{fig:G203.21-11.20W2_SiO_integrated-PV}a. (f) SiO PV diagram with sensitivity $\sigma$ = 0.47 $K$ with similar symbol of Figure \ref{fig:G203.21-11.20W2_SiO_integrated-PV}b. The SiO contour labels are at (1.5, 2, 3, 4)$\times$ $\sigma$. }
\label{fig:appendix_G192.32_11.88N}
\end{figure*}

\begin{figure*}
\fig{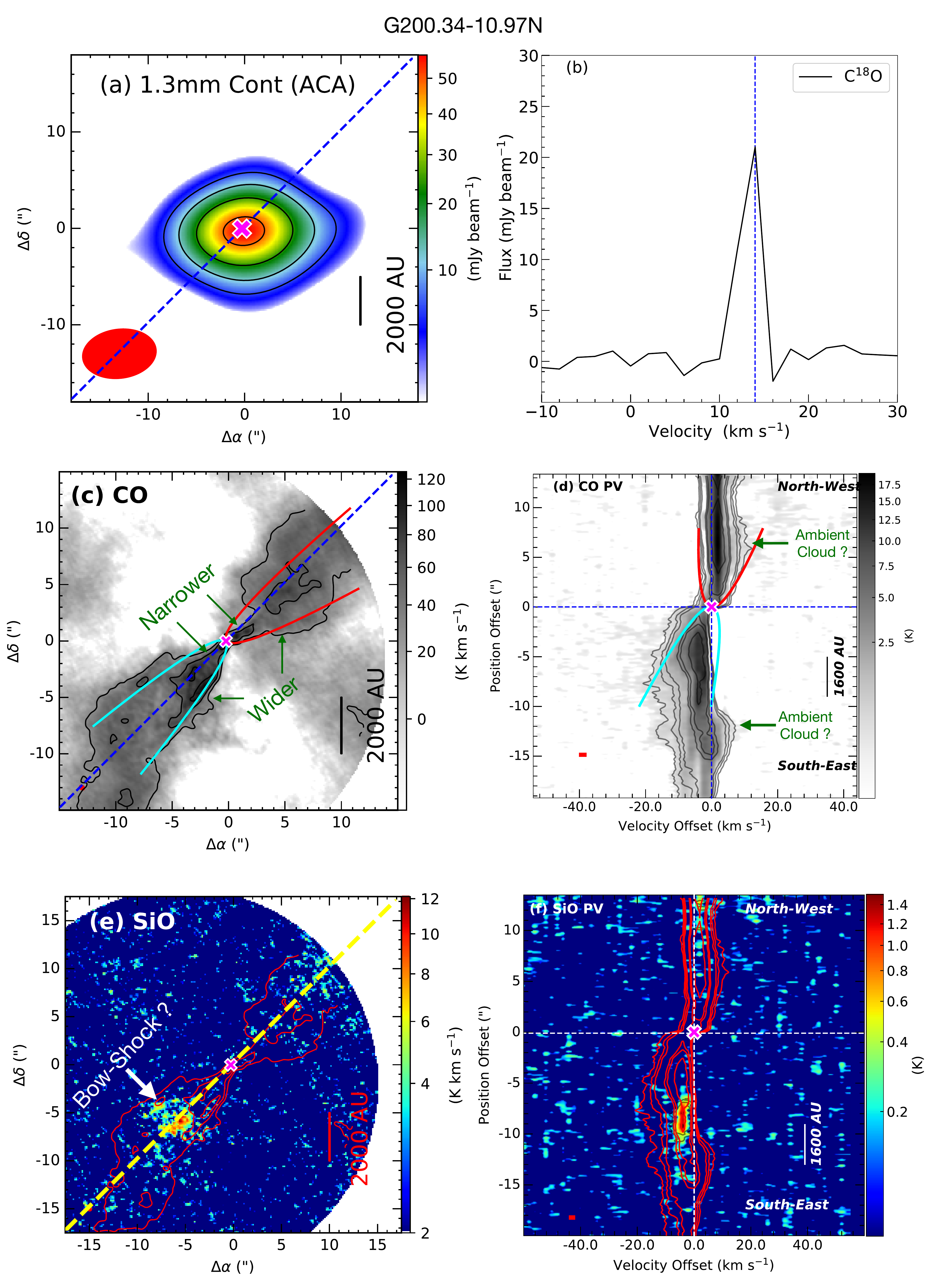}{0.85\textwidth}{}
\caption{G200.34-10.97N: (a) 1.3mm continuum map at ACA resolution with sensitivity $\sim$ 2.0 $\mjypb$. The symbols and contour levels are the same as Figure \ref{fig:G20321_1120W2_ACA_envelope}. (b) The  C$^{18}$O spectra extracted from high-resolution maps, and V$_{sys}$ = 14 $\kms$. All symbols are the same as Figure \ref{fig:G20321_1120W2_C18ospectra}. (c) CO map integrated over [-20 to +12] $\kms$ with sensitivity 9.9 $\kkms$ with similar symbols and contours as Figure \ref{fig:G203.21-11.20W2outflow_parabola}a. (d) CO PV diagram with sensitivity 0.14 $K$ with similar symbols and contours as Figure \ref{fig:G203.21-11.20W2outflow_parabola}b. (e) SiO emission map integrated over [-10 to +10] $\kms$ emission with sensitivity 3.7 $\kkms$ with similar symbols and contours as Figure \ref{fig:G203.21-11.20W2_SiO_integrated-PV}a. (f) SiO PV diagram with sensitivity $\sigma$ = 0.12 $K$ with similar symbol of Figure \ref{fig:G203.21-11.20W2_SiO_integrated-PV}b. The SiO contour labels are at (2, 3, 4)$\times$ $\sigma$.}
\label{fig:appendix_G20034_1097N}
\end{figure*}

\begin{figure*}
\fig{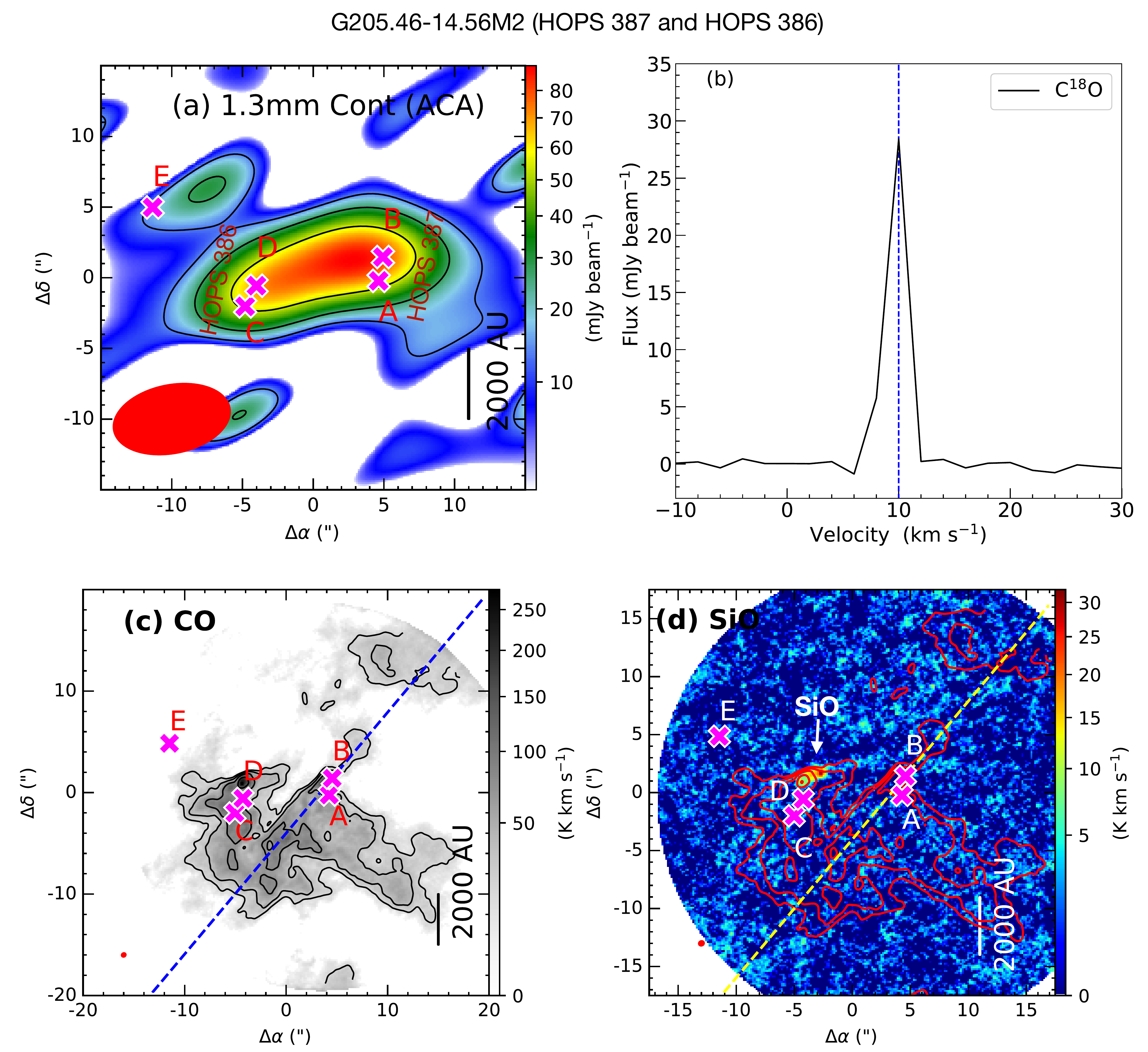}{0.85\textwidth}{}
\caption{G205.46-14.56M2: (a) 1.3mm continuum map at ACA resolution with sensitivity $\sim$ 4.5 $\mjypb$. The symbols and contour levels are the same as Figure \ref{fig:G20321_1120W2_ACA_envelope}.
(b) The  C$^{18}$O spectra extracted from high-resolution maps, and V$_{sys}$ = 14 $\kms$. All symbols are the same as Figure \ref{fig:G20321_1120W2_C18ospectra}. (c) Integrated CO emission with sensitivity 6.15 $\kkms$ with similar symbols and contours as Figure \ref{fig:G203.21-11.20W2outflow_parabola}a.  (d) Integrated SiO emission with sensitivity 2.1 $\kkms$ with similar symbols and contours as Figure \ref{fig:G203.21-11.20W2_SiO_integrated-PV}a. }
\label{fig:appendix_G205.46_14.56M2}
\end{figure*}

\begin{figure*}
\fig{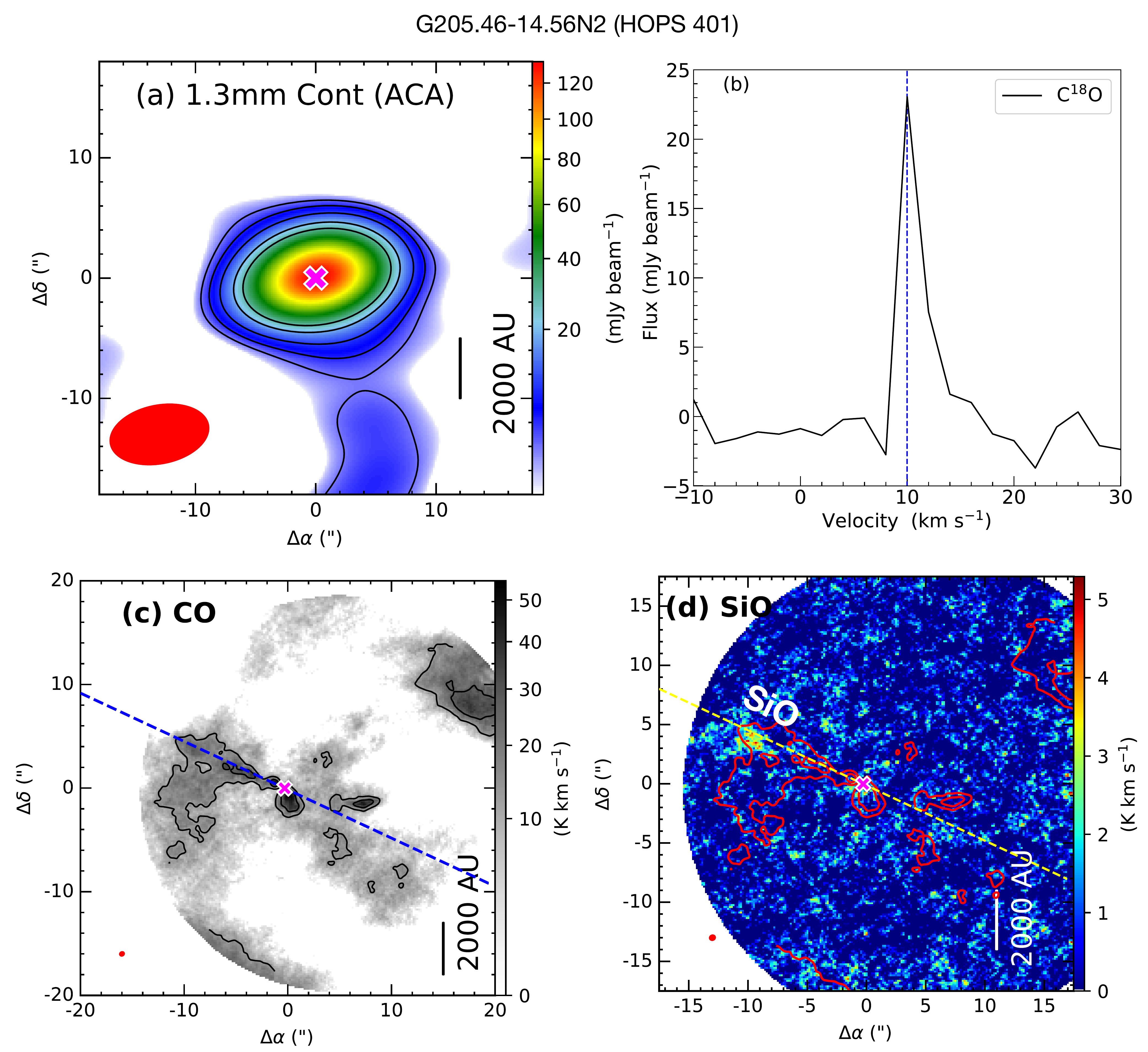}{0.85\textwidth}{}
\caption{G205.46-14.56N2: (a) 1.3mm continuum map at ACA resolution with sensitivity $\sim$ 0.8 $\mjypb$. The symbols and contour levels are the same as Figure \ref{fig:G20321_1120W2_ACA_envelope}.
(b) The  C$^{18}$O spectra extracted from high-resolution maps, and V$_{sys}$ = 10 $\kms$. All symbols are the same as Figure \ref{fig:G20321_1120W2_C18ospectra}. (c) Integrated CO emission with sensitivity 4.5 $\kkms$ with similar symbols and contours as Figure \ref{fig:G203.21-11.20W2outflow_parabola}a.  (d) Integrated SiO emission with sensitivity 1.15 $\kkms$ with similar symbols and contours as Figure \ref{fig:G203.21-11.20W2_SiO_integrated-PV}a. }
\label{fig:appendix_G205.46_14.56N2}
\end{figure*}

\begin{figure*}
\fig{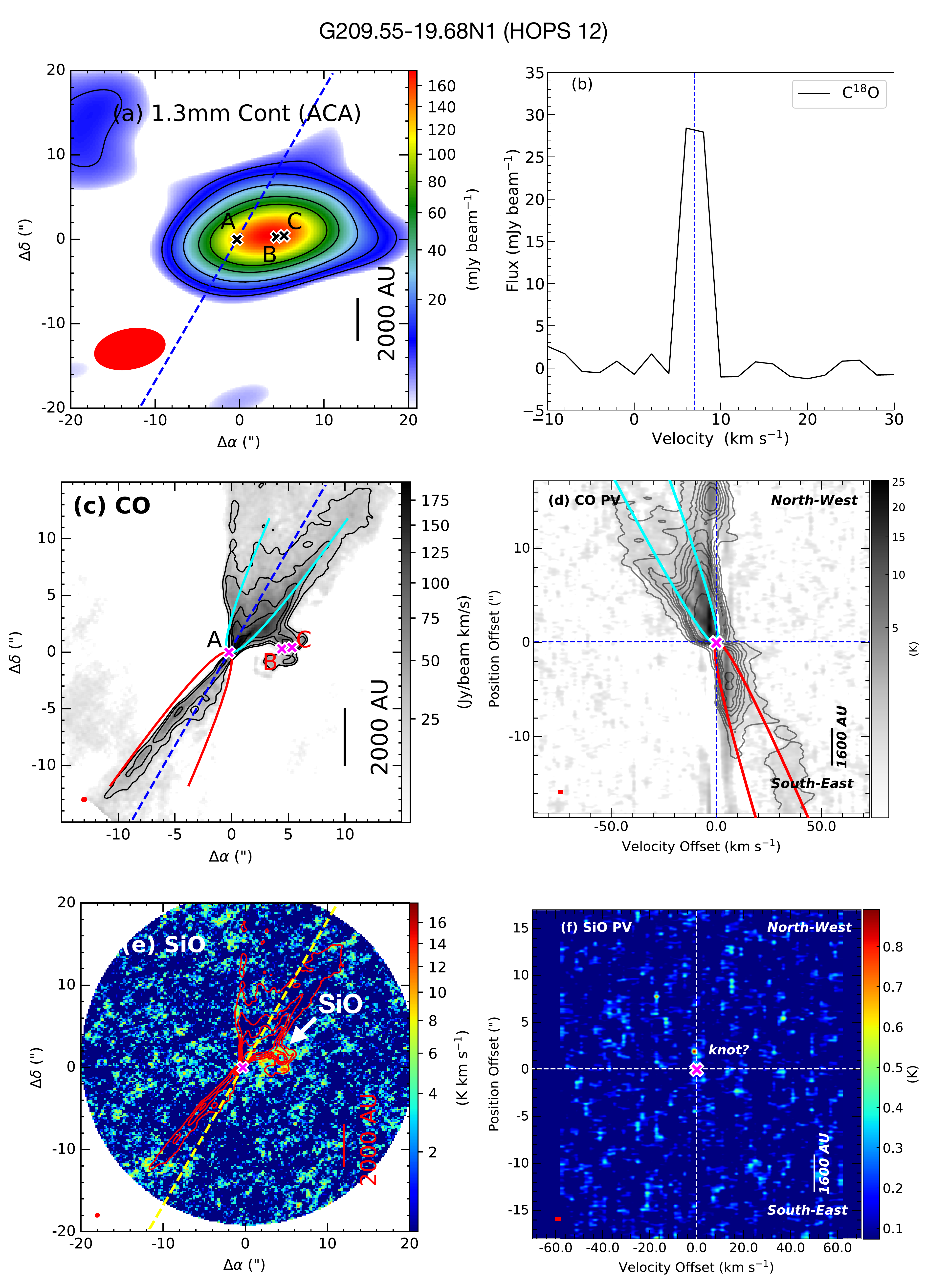}{0.85\textwidth}{}
\caption{G209.55-19.68N1: (a) 1.3mm continuum map at ACA resolution with sensitivity $\sim$ 2.0 $\mjypb$. The symbols and contour levels are the same as Figure \ref{fig:G20321_1120W2_ACA_envelope}.
(b) The  C$^{18}$O spectra extracted from high-resolution maps, and V$_{sys}$ = 7 $\kms$. All symbols are the same as Figure \ref{fig:G20321_1120W2_C18ospectra}. (c) Integrated CO emission with sensitivity 7.8 $\kkms$ with similar symbols and contours as Figure \ref{fig:G203.21-11.20W2outflow_parabola}a. (d) CO PV diagram with sensitivity 0.12 $K$ with similar symbols and contours as Figure \ref{fig:G203.21-11.20W2outflow_parabola}b. (e) Integrated SiO emission with sensitivity 2.6 $\kkms$ with similar symbols and contours as Figure \ref{fig:G203.21-11.20W2_SiO_integrated-PV}a. (f) SiO PV diagram with sensitivity $\sigma$ = 0.15 $K$ with similar symbol of Figure \ref{fig:G203.21-11.20W2_SiO_integrated-PV}b.}
\label{fig:appendix_G209.55_19.68N1}
\end{figure*}

\begin{figure*}
\fig{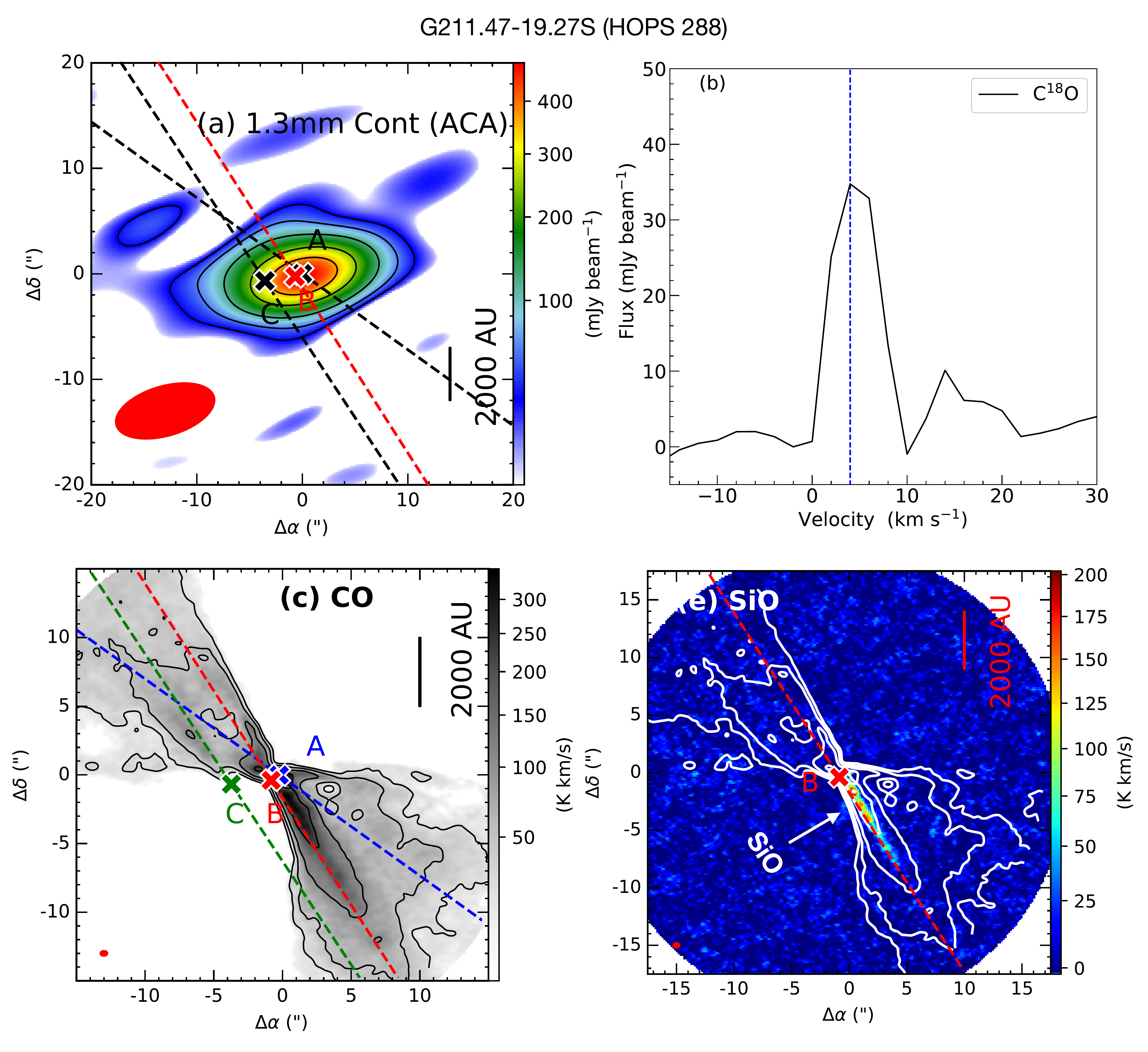}{0.85\textwidth}{}
\caption{G211.47-19.27S: (a) 1.3mm continuum map at ACA resolution with sensitivity $\sim$ 2.0 $\mjypb$. The symbols and contour levels are the same as Figure \ref{fig:G20321_1120W2_ACA_envelope}.
(b) The  C$^{18}$O spectra extracted from high-resolution maps, and V$_{sys}$ = 4 $\kms$. All symbols are the same as Figure \ref{fig:G20321_1120W2_C18ospectra}. (c) Integrated CO emission with sensitivity 8.6 $\kkms$ with similar symbols and contours as Figure \ref{fig:G203.21-11.20W2outflow_parabola}a. (d) CO PV diagram with sensitivity 0.15 $K$ with similar symbols and contours as Figure \ref{fig:G203.21-11.20W2outflow_parabola}b. (e) Integrated SiO emission with sensitivity 1.9 $\kkms$ with similar symbols and contours as Figure \ref{fig:G203.21-11.20W2_SiO_integrated-PV}a.}
\label{fig:appendix_G211.47_19.27S}
\end{figure*}

\setcounter{figure}{0}  \renewcommand{\thefigure}{D\arabic{figure}}
\begin{figure*} 
\fig{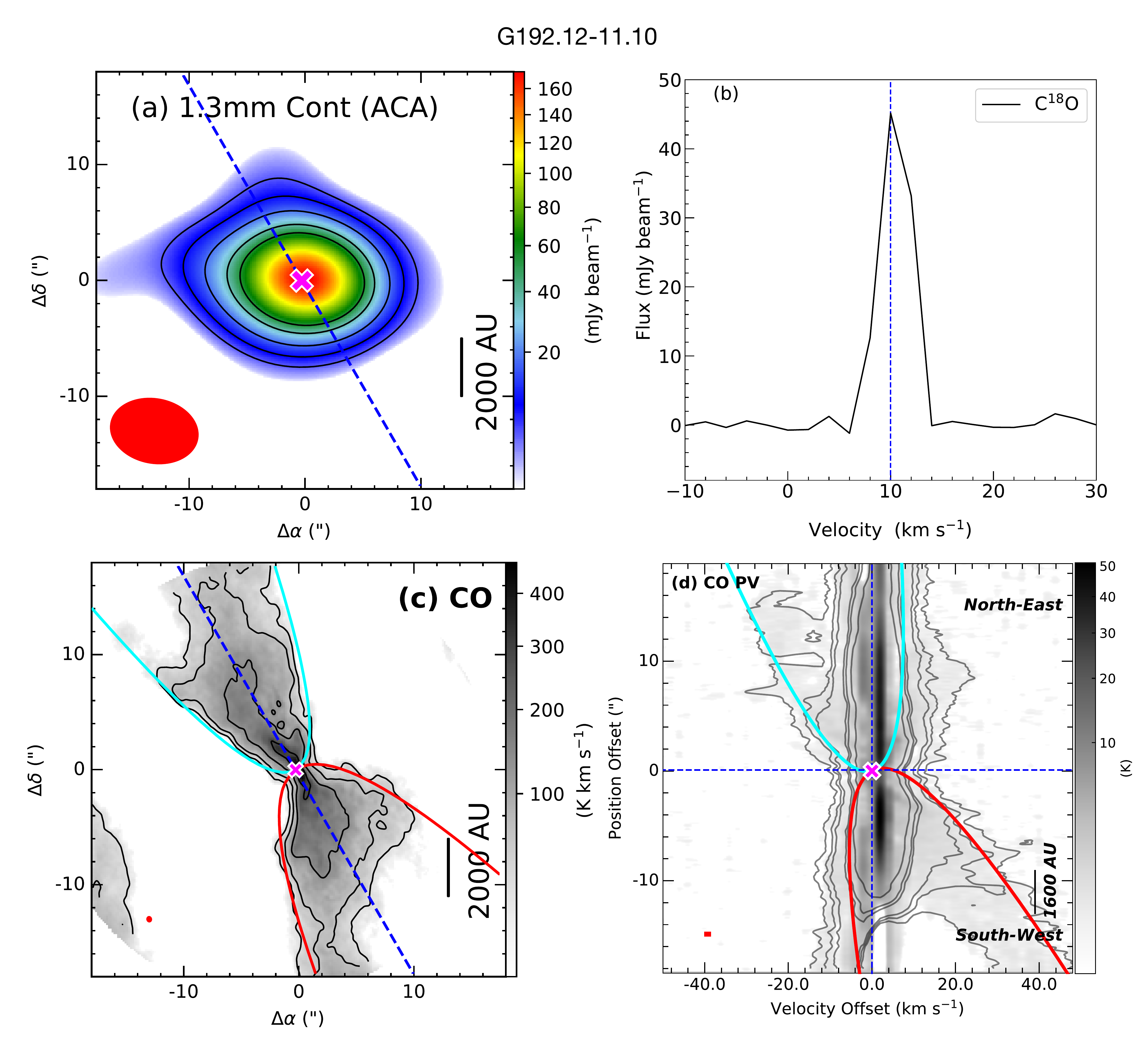}{0.85\textwidth}{}
\caption{G192.12-11.10: (a) 1.3mm continuum map at ACA resolution with sensitivity $\sim$ 1.5 $\mjypb$. The symbols and contour levels are the same as Figure \ref{fig:G20321_1120W2_ACA_envelope}.
(b) The  C$^{18}$O spectra extracted from high-resolution maps, and V$_{sys}$ = 10 $\kms$. All symbols are the same as Figure \ref{fig:G20321_1120W2_C18ospectra}. (c) Integrated CO emission with sensitivity 11.5 $\kkms$ with similar symbols and contours as Figure \ref{fig:G203.21-11.20W2outflow_parabola}a. (d) CO PV diagram with sensitivity 0.16 $K$ with similar symbols and contours as Figure \ref{fig:G203.21-11.20W2outflow_parabola}b.}
\label{fig:appendix_G192.12-11.10}
\end{figure*}

\begin{figure*}
\fig{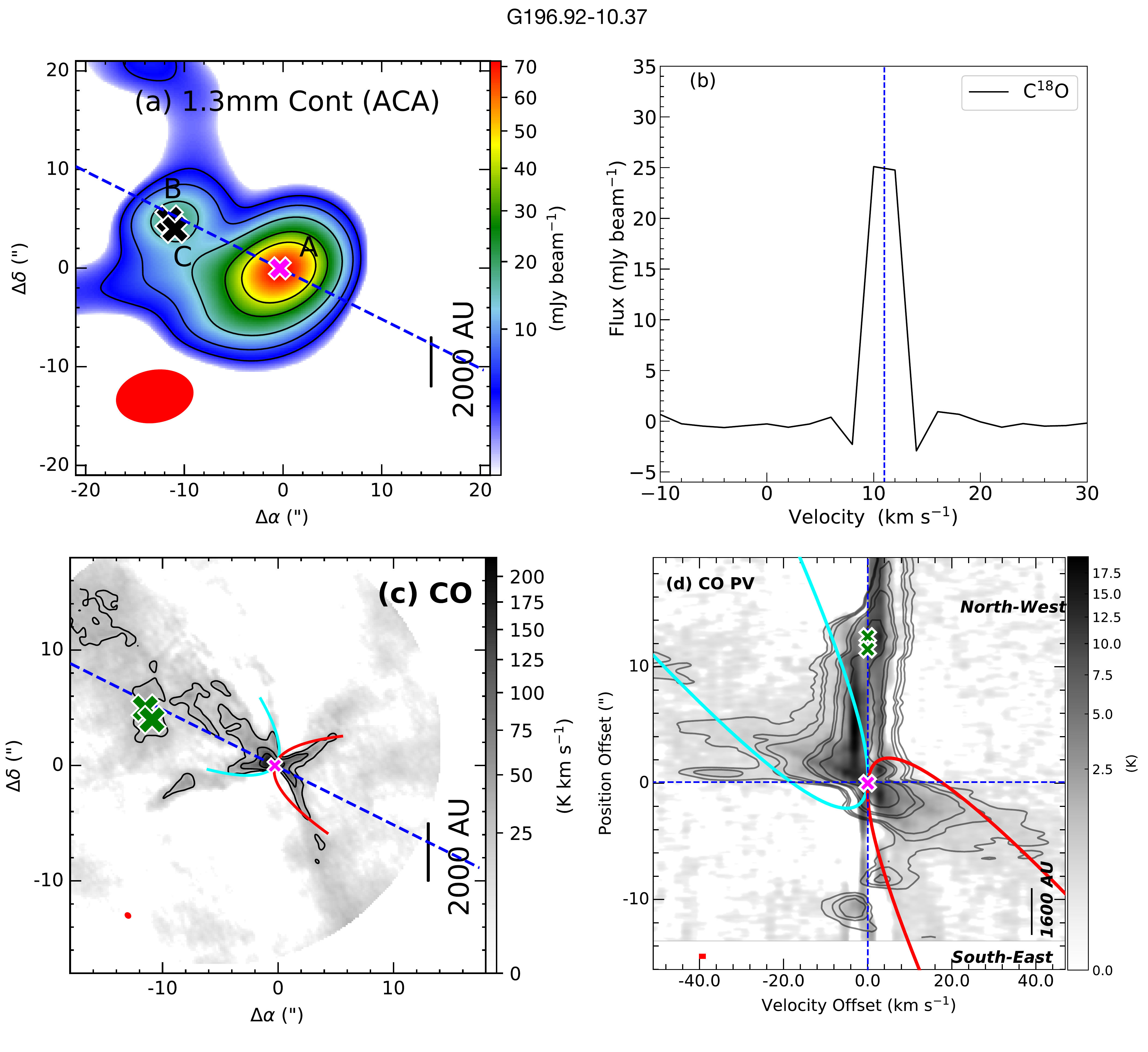}{0.85\textwidth}{}
\caption{G196.92-10.37: (a) 1.3mm continuum map at ACA resolution with sensitivity $\sim$ 1.2 $\mjypb$. The symbols and contour levels are the same as Figure \ref{fig:G20321_1120W2_ACA_envelope}.
(b) The  C$^{18}$O spectra extracted from high-resolution maps, and V$_{sys}$ = 11 $\kms$. All symbols are the same as Figure \ref{fig:G20321_1120W2_C18ospectra}. (c) Integrated CO emission with sensitivity 9.6 $\kkms$ with similar symbols and contours as Figure \ref{fig:G203.21-11.20W2outflow_parabola}a. (d) CO PV diagram with sensitivity 0.1 $K$ with similar symbols and contours as Figure \ref{fig:G203.21-11.20W2outflow_parabola}b.}
\label{fig:appendix_G196.92-10.37}
\end{figure*}

\begin{figure*}
\fig{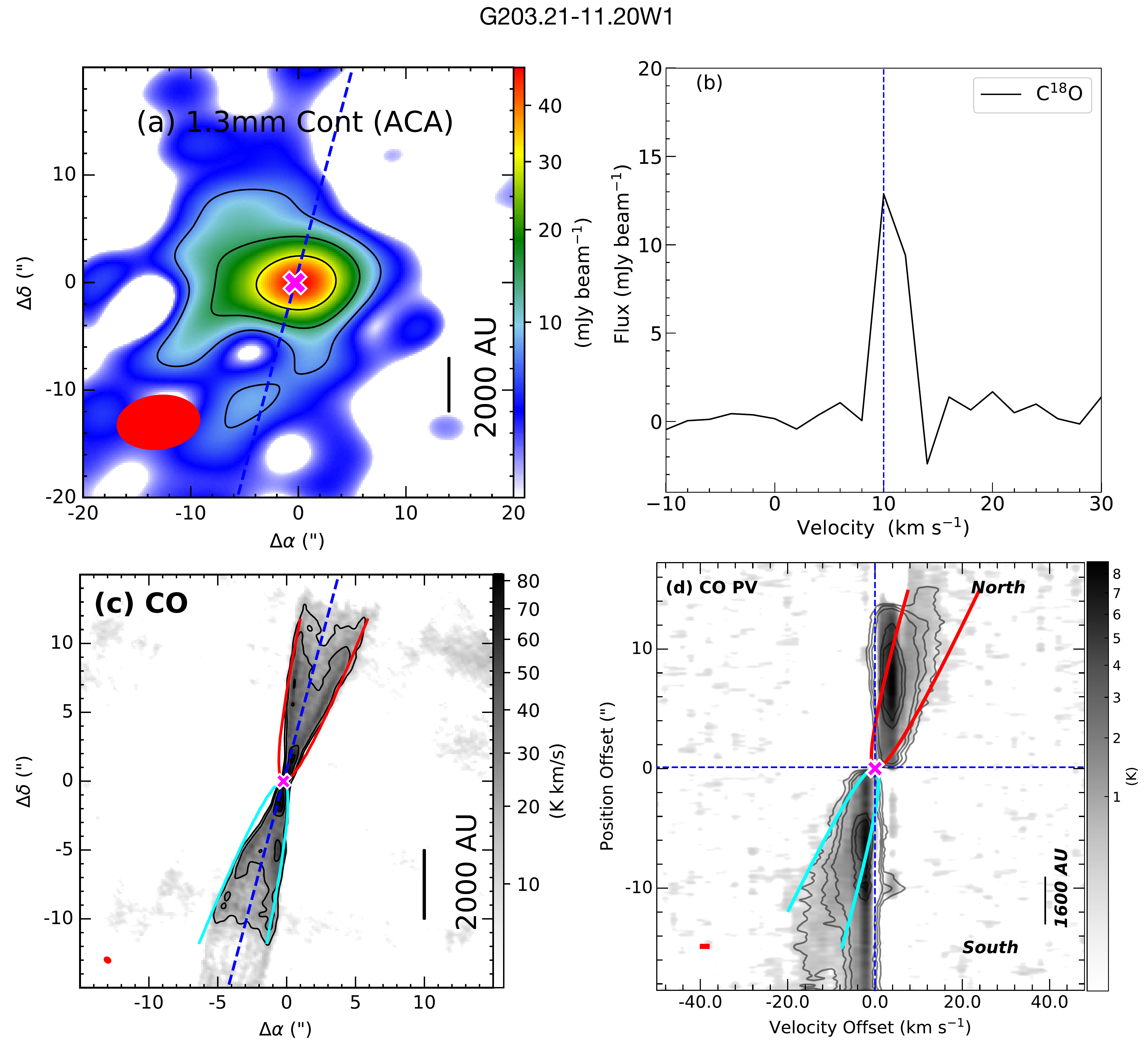}{0.85\textwidth}{}
\caption{G203.21-11.20W1: (a) 1.3mm continuum map at ACA resolution with sensitivity $\sim$ 3.36 $\mjypb$. The symbols and contour levels are the same as Figure \ref{fig:G20321_1120W2_ACA_envelope}.
(b) The  C$^{18}$O spectra extracted from high-resolution maps, and V$_{sys}$ = 10 $\kms$. All symbols are the same as Figure \ref{fig:G20321_1120W2_C18ospectra}. (c) Integrated CO emission with sensitivity 5.14 $\kkms$ with similar symbols and contours as Figure \ref{fig:G203.21-11.20W2outflow_parabola}a. (d) CO PV diagram with sensitivity 0.11 $K$ with similar symbols and contours as Figure \ref{fig:G203.21-11.20W2outflow_parabola}b.}
\label{fig:appendix_G203.21-11.20W1}
\end{figure*}

\begin{figure*}
\fig{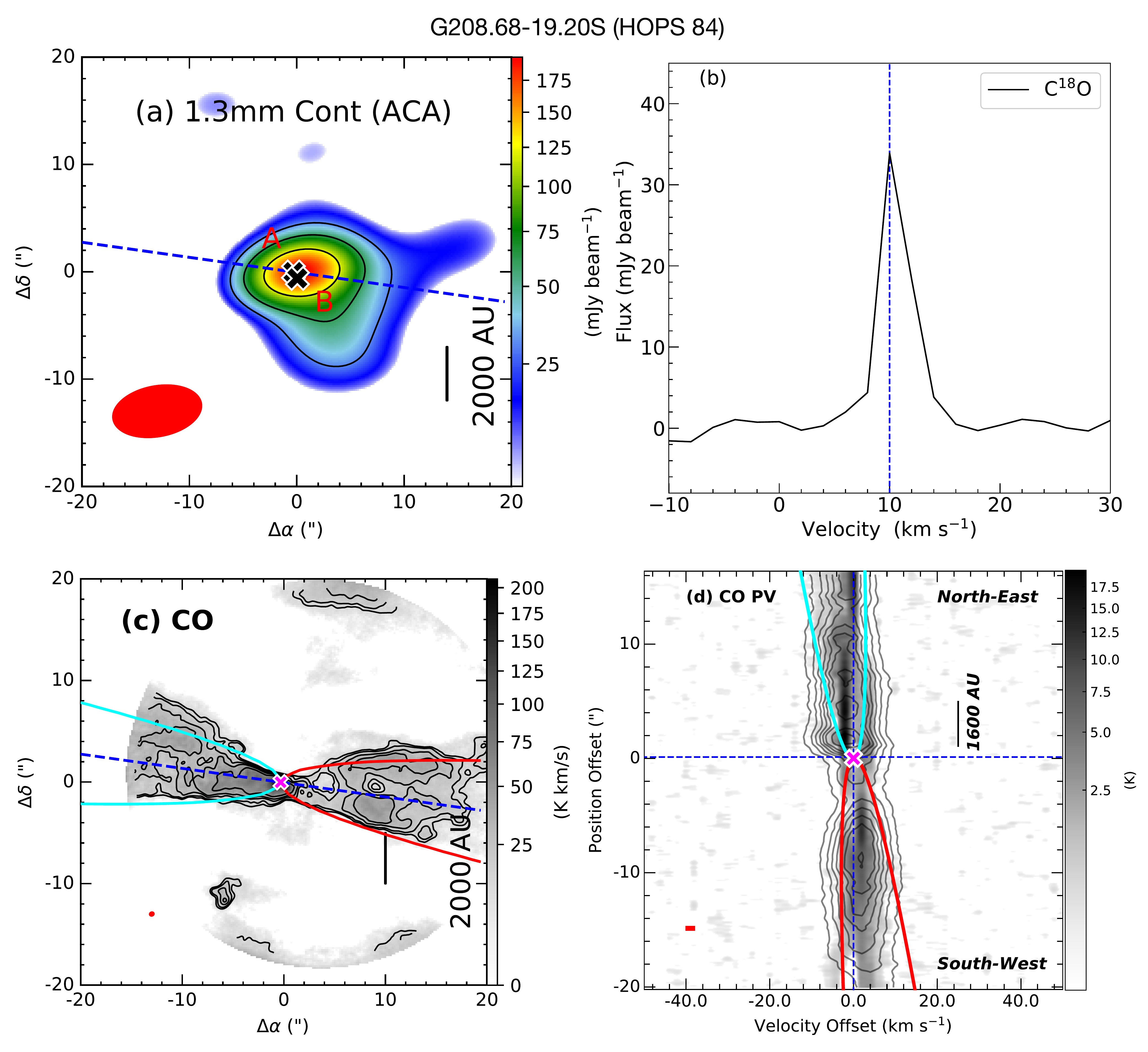}{0.85\textwidth}{}
\caption{G208.68-19.20S: (a) 1.3mm continuum map at ACA resolution with sensitivity $\sim$ 10 $\mjypb$. The symbols and contour levels are the same as Figure \ref{fig:G20321_1120W2_ACA_envelope}.
(b) The  C$^{18}$O spectra extracted from high-resolution maps, and V$_{sys}$ = 10 $\kms$. All symbols are the same as Figure \ref{fig:G20321_1120W2_C18ospectra}. (c) Integrated CO emission with sensitivity 1.5 $\kkms$ with similar symbols and contours as Figure \ref{fig:G203.21-11.20W2outflow_parabola}a. (d) CO PV diagram with sensitivity 0.16 $K$ with similar symbols and contours as Figure \ref{fig:G203.21-11.20W2outflow_parabola}b.}
\label{fig:appendix_G208.68-19.20S}
\end{figure*}

\begin{figure*}
\fig{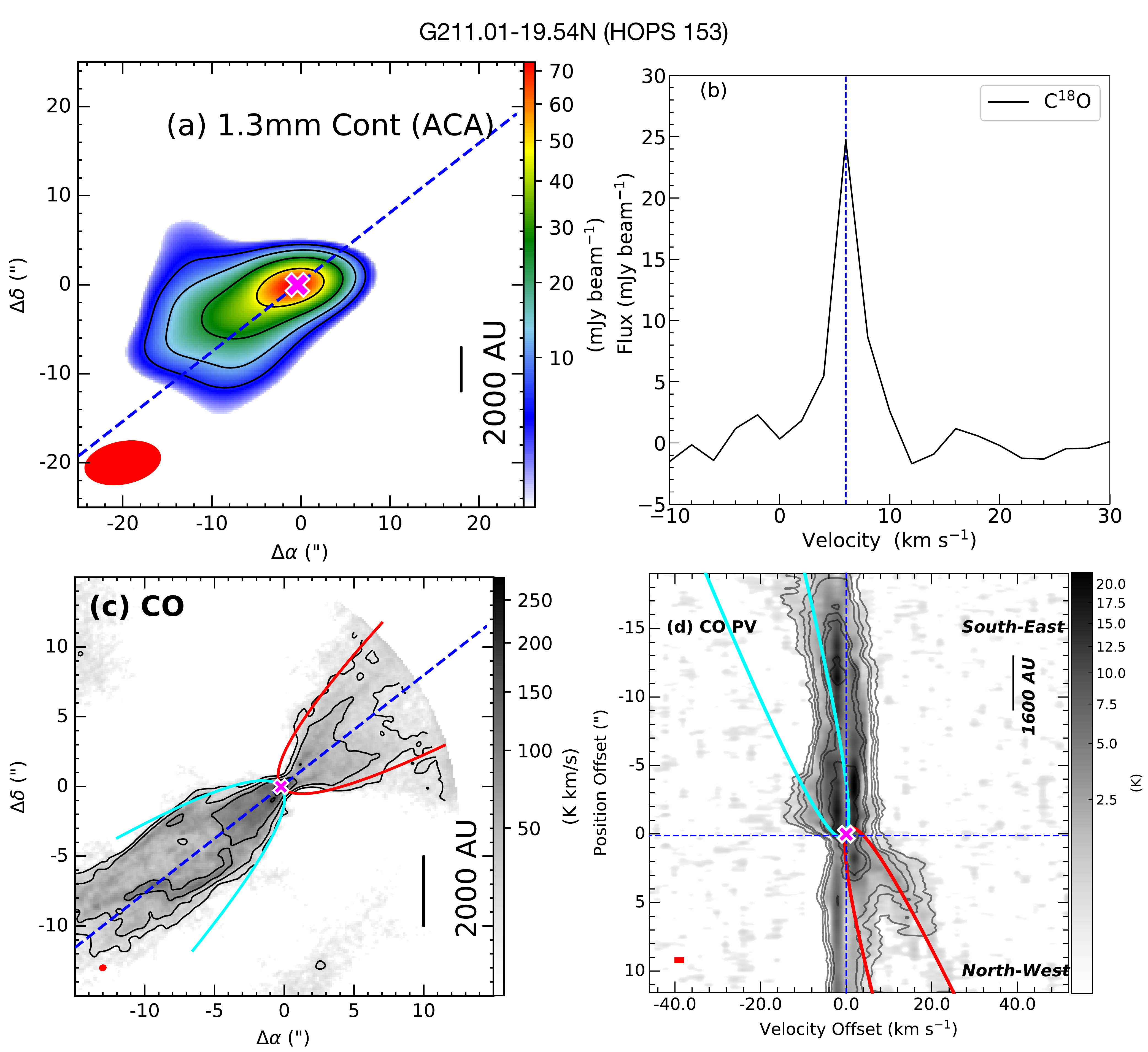}{0.85\textwidth}{}
\caption{G211.01-19.54N: (a) 1.3mm continuum map at ACA resolution with sensitivity $\sim$ 2 $\mjypb$. The symbols and contour levels are the same as Figure \ref{fig:G20321_1120W2_ACA_envelope}.
(b) The  C$^{18}$O spectra extracted from high-resolution maps, and V$_{sys}$ = 6 $\kms$. All symbols are the same as Figure \ref{fig:G20321_1120W2_C18ospectra}. (c) Integrated CO emission with sensitivity 6.3 $\kkms$ with similar symbols and contours as Figure \ref{fig:G203.21-11.20W2outflow_parabola}a. (d) CO PV diagram with sensitivity 0.16 $K$ with similar symbols and contours as Figure \ref{fig:G203.21-11.20W2outflow_parabola}b.}
\label{fig:appendix_G211.01-19.54N}
\end{figure*}

\begin{figure*}
\fig{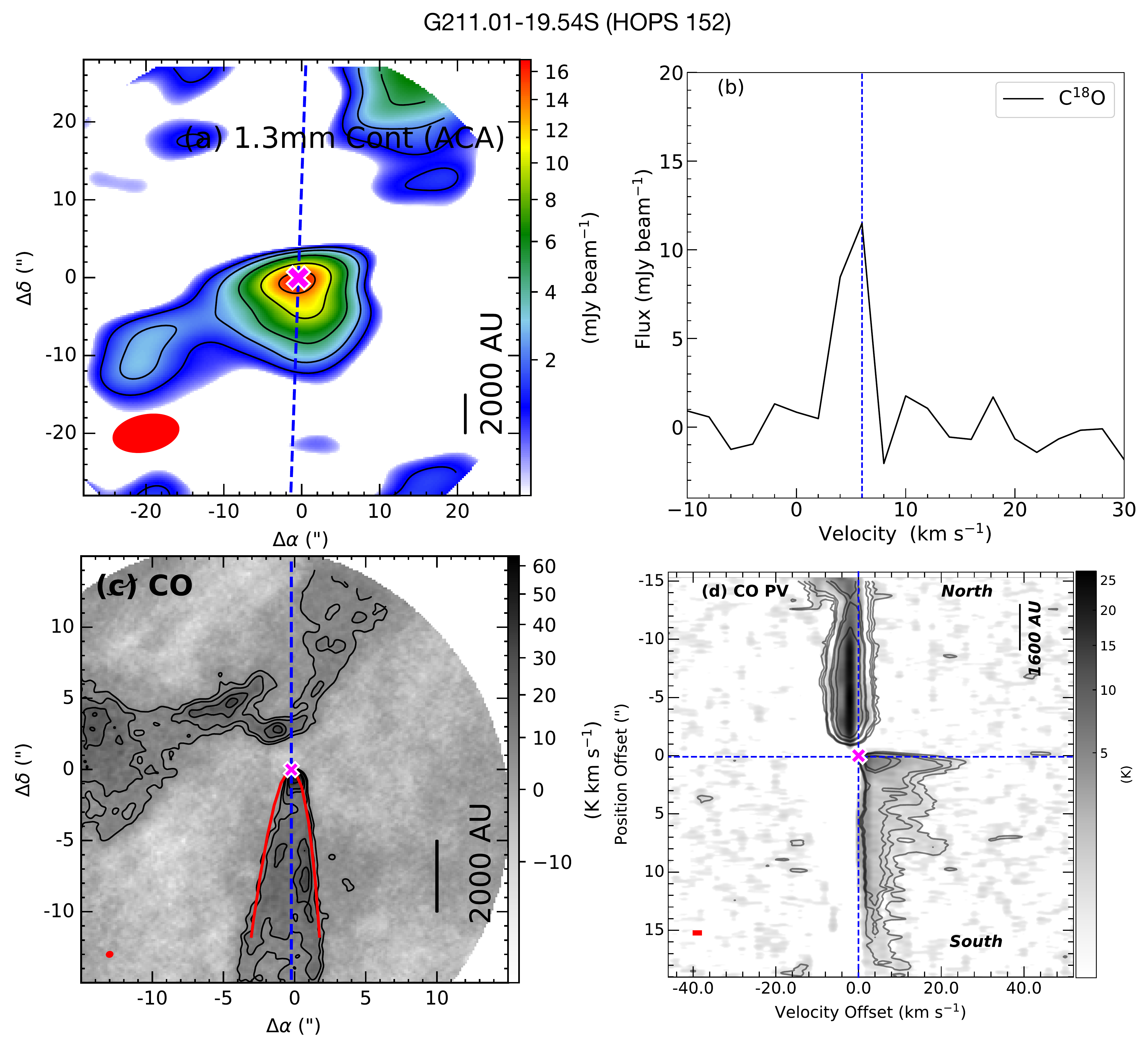}{0.85\textwidth}{}
\caption{G211.01-19.54S: (a) 1.3mm continuum map at ACA resolution with sensitivity $\sim$ 0.4 $\mjypb$. The symbols and contour levels are the same as Figure \ref{fig:G20321_1120W2_ACA_envelope}.
(b) The  C$^{18}$O spectra extracted from high-resolution maps, and V$_{sys}$ = 6 $\kms$. All symbols are the same as Figure \ref{fig:G20321_1120W2_C18ospectra}. (c) Integrated CO emission with sensitivity 2.21 $\kkms$ with similar symbols and contours as Figure \ref{fig:G203.21-11.20W2outflow_parabola}a. (d) CO PV diagram with sensitivity 0.13 $K$ with similar symbols and contours as Figure \ref{fig:G203.21-11.20W2outflow_parabola}b.}
\label{fig:appendix_G211.01-19.54S}
\end{figure*}

\begin{figure*}
\fig{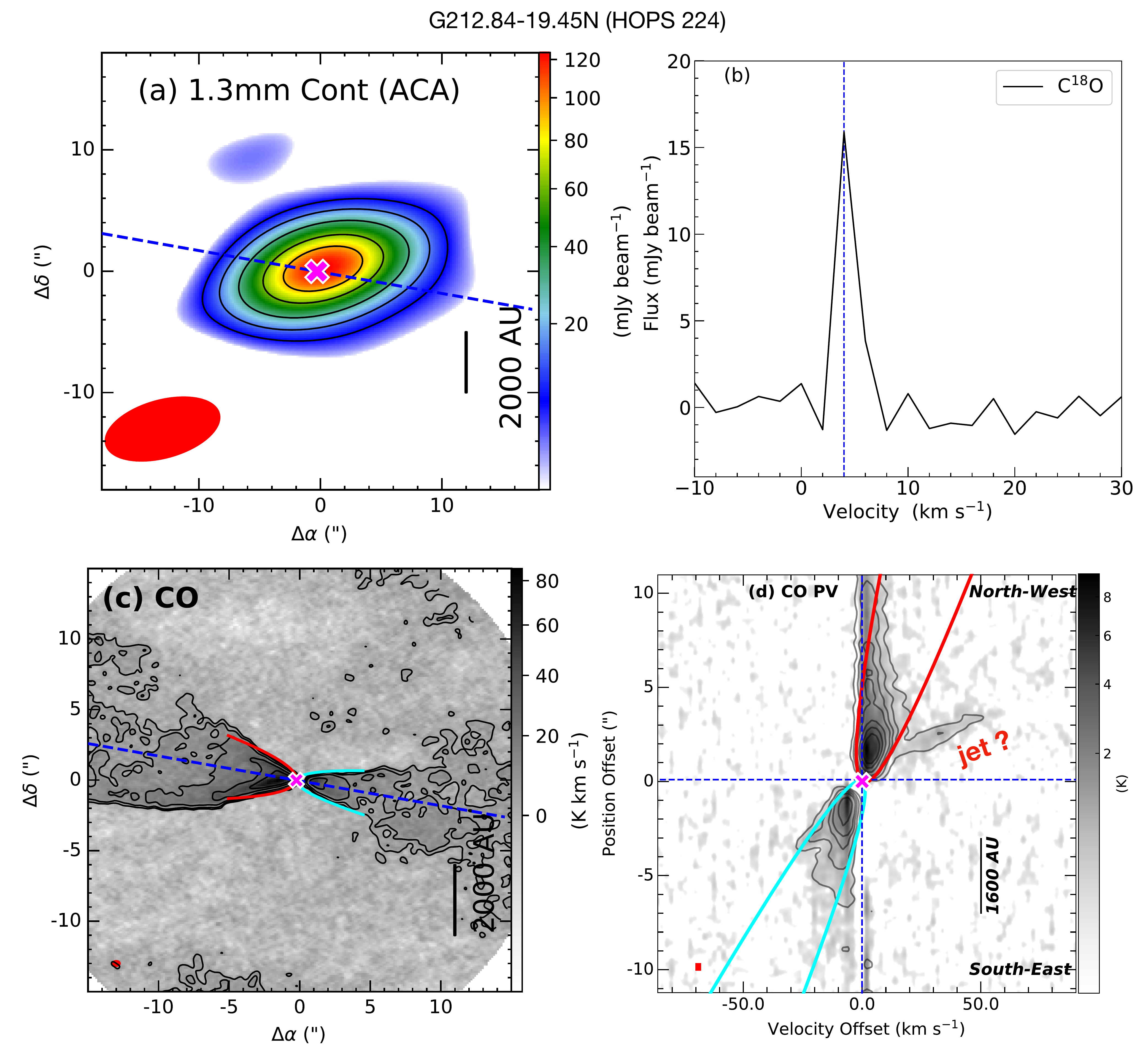}{0.85\textwidth}{}
\caption{G212.84-19.45N: (a) 1.3mm continuum map at ACA resolution with sensitivity $\sim$ 2.5 $\mjypb$. The symbols and contour levels are the same as Figure \ref{fig:G20321_1120W2_ACA_envelope}.
(b) The  C$^{18}$O spectra extracted from high-resolution maps, and V$_{sys}$ = 4 $\kms$. All symbols are the same as Figure \ref{fig:G20321_1120W2_C18ospectra}. (c) Integrated CO emission with sensitivity 1.24 $\kkms$ with similar symbols and contours as Figure \ref{fig:G203.21-11.20W2outflow_parabola}a. (d) CO PV diagram with sensitivity 0.12 $K$ with similar symbols and contours as Figure \ref{fig:G203.21-11.20W2outflow_parabola}b.}
\label{fig:appendix_G212.84-19.45N}
\end{figure*}


\setcounter{figure}{0}  \renewcommand{\thefigure}{E\arabic{figure}}
\begin{figure*}
\fig{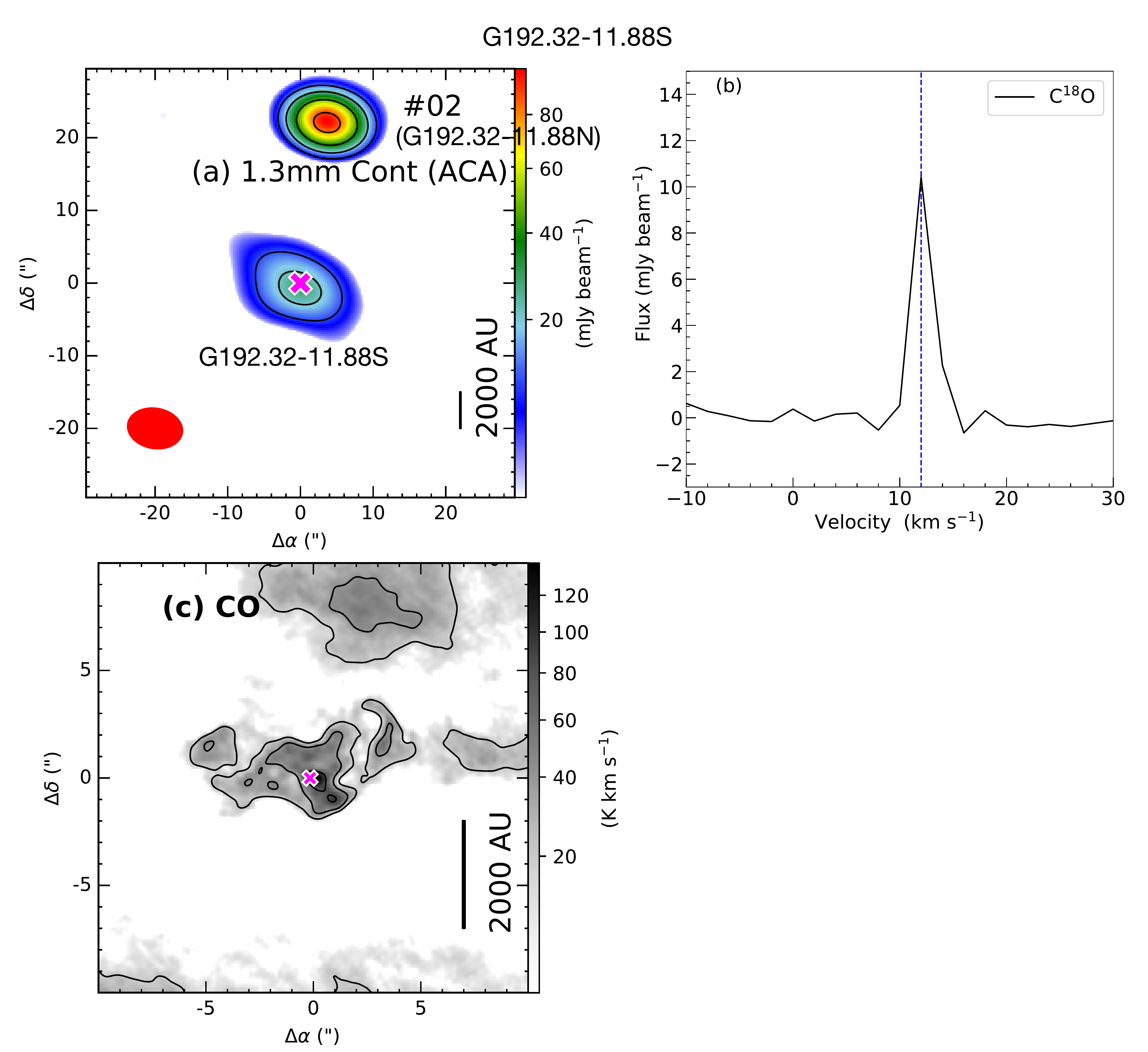}{0.85\textwidth}{}
\caption{G192.32-11.88S: (a) 1.3mm continuum map at ACA resolution with sensitivity $\sim$ 3.5 $\mjypb$. The symbols and contour levels are the same as Figure \ref{fig:G20321_1120W2_ACA_envelope}.
(b) The  C$^{18}$O spectra extracted from high-resolution maps, and V$_{sys}$ = 10 $\kms$. All symbols are the same as Figure \ref{fig:G20321_1120W2_C18ospectra}. (c) Integrated CO emission with sensitivity 6.6 $\kkms$ with similar symbols and contours as Figure \ref{fig:G203.21-11.20W2outflow_parabola}a.}
\label{fig:appendix_G192.32-11.88S}
\end{figure*}

\begin{figure*}
\fig{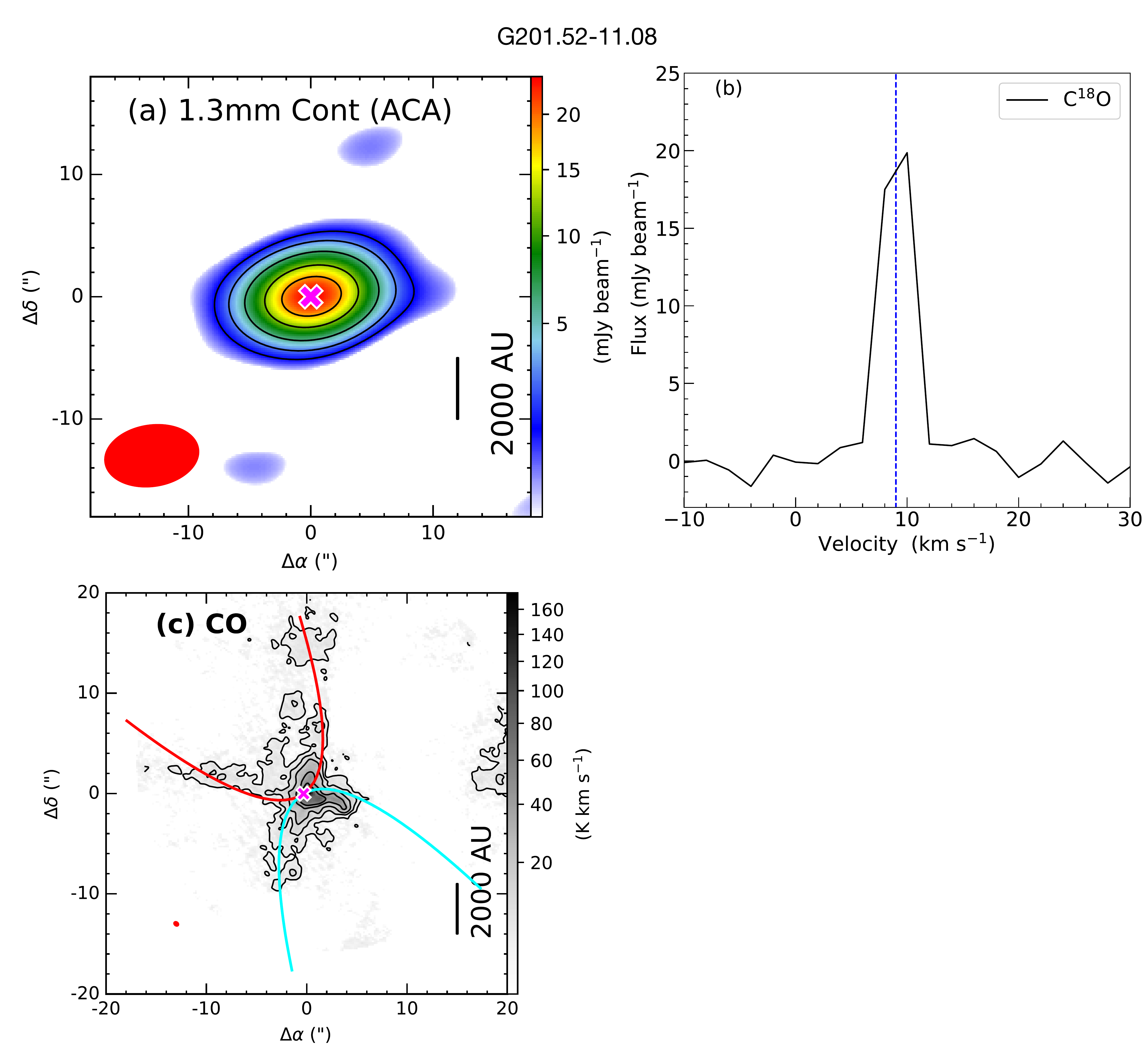}{0.85\textwidth}{}
\caption{G201.52-11.08: (a) 1.3mm continuum map at ACA resolution with sensitivity $\sim$ 0.5 $\mjypb$. The symbols and contour levels are the same as Figure \ref{fig:G20321_1120W2_ACA_envelope}.
(b) The  C$^{18}$O spectra extracted from high-resolution maps, and V$_{sys}$ = 9 $\kms$. All symbols are the same as Figure \ref{fig:G20321_1120W2_C18ospectra}. (c) Integrated CO emission with sensitivity 1.4 $\kkms$ with similar symbols and contours as Figure \ref{fig:G203.21-11.20W2outflow_parabola}a.}
\label{fig:appendix_G201.52-11.08}
\end{figure*}

\begin{figure*}
\fig{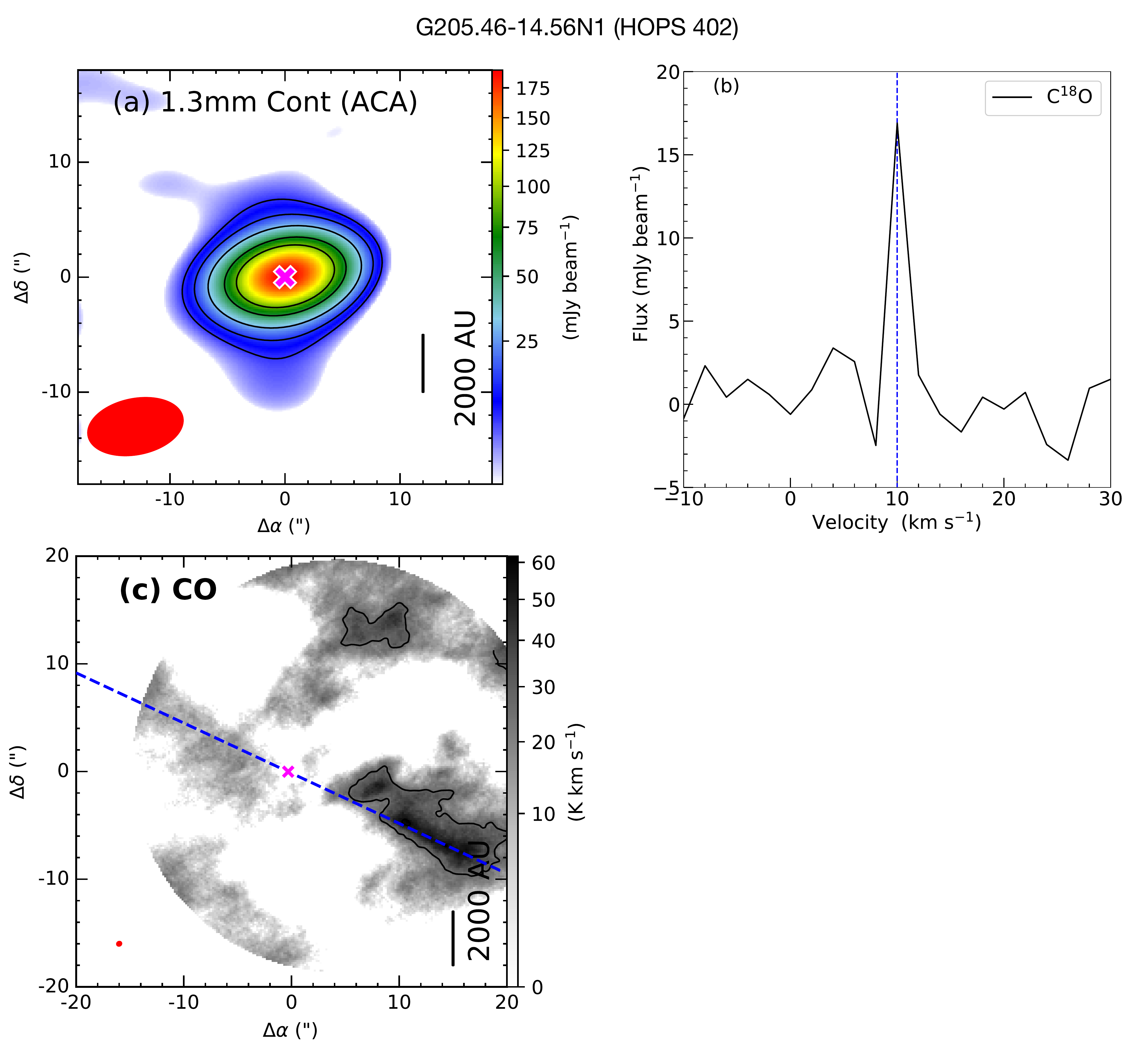}{0.85\textwidth}{}
\caption{G205.46-14.56N1: (a) 1.3mm continuum map at ACA resolution with sensitivity $\sim$ 2.5 $\mjypb$. The symbols and contour levels are the same as Figure \ref{fig:G20321_1120W2_ACA_envelope}.
(b) The  C$^{18}$O spectra extracted from high-resolution maps, and V$_{sys}$ = 10 $\kms$. All symbols are the same as Figure \ref{fig:G20321_1120W2_C18ospectra}. (c) Integrated CO emission with sensitivity 9.2 $\kkms$ with similar symbols and contours as Figure \ref{fig:G203.21-11.20W2outflow_parabola}a.}
\label{fig:appendix_G205.46-14.56N1}
\end{figure*}

\begin{figure*}
\fig{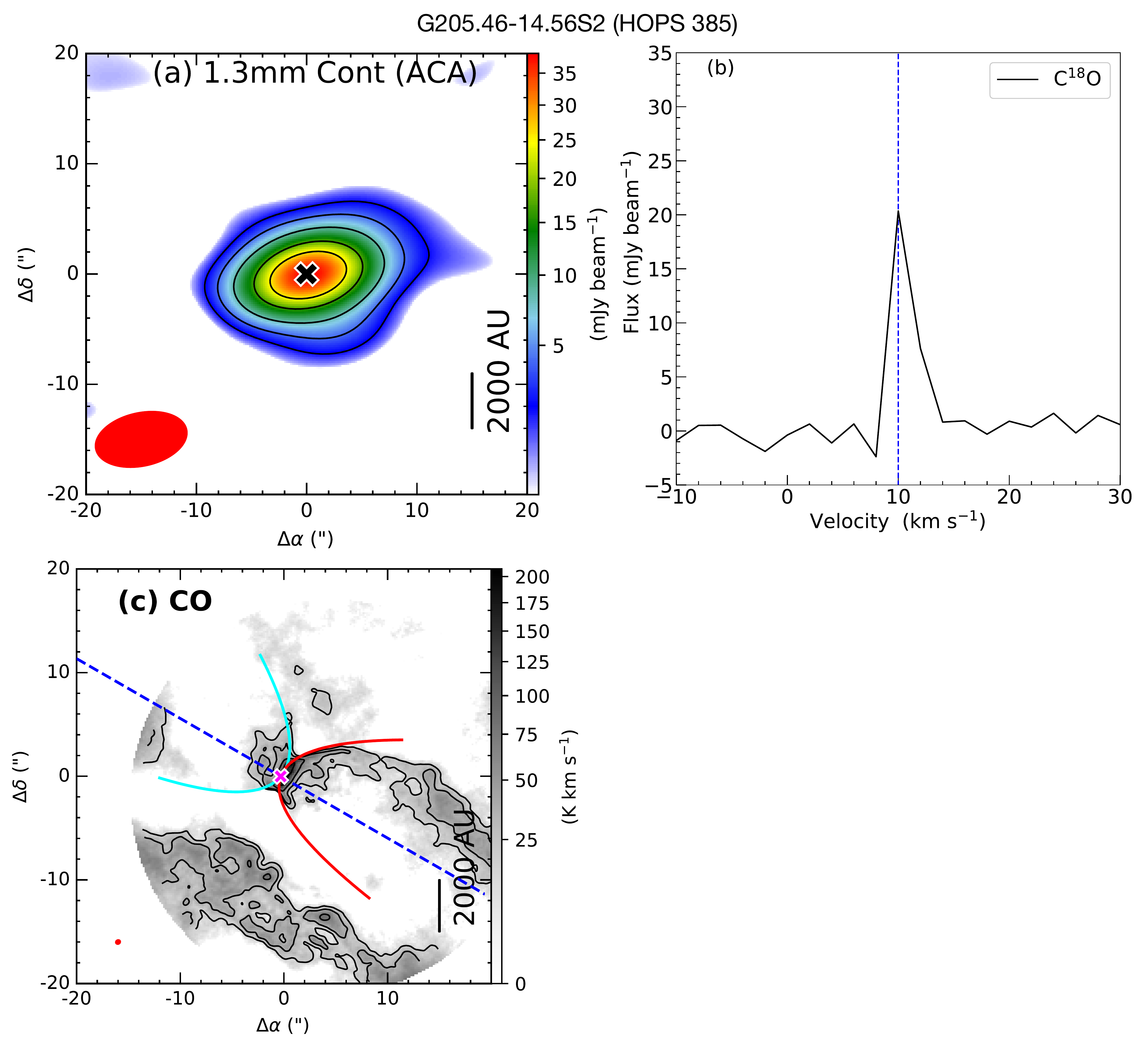}{0.85\textwidth}{}
\caption{G205.46-14.56S2: (a) 1.3mm continuum map at ACA resolution with sensitivity $\sim$ 0.7 $\mjypb$. The symbols and contour levels are the same as Figure \ref{fig:G20321_1120W2_ACA_envelope}.
(b) The  C$^{18}$O spectra extracted from high-resolution maps, and V$_{sys}$ = 10 $\kms$. All symbols are the same as Figure \ref{fig:G20321_1120W2_C18ospectra}. (c) Integrated CO emission with sensitivity 6.0 $\kkms$ with similar symbols and contours as Figure \ref{fig:G203.21-11.20W2outflow_parabola}a.}
\label{fig:appendix_G205.46-14.56S2}
\end{figure*}

\begin{figure*}
\fig{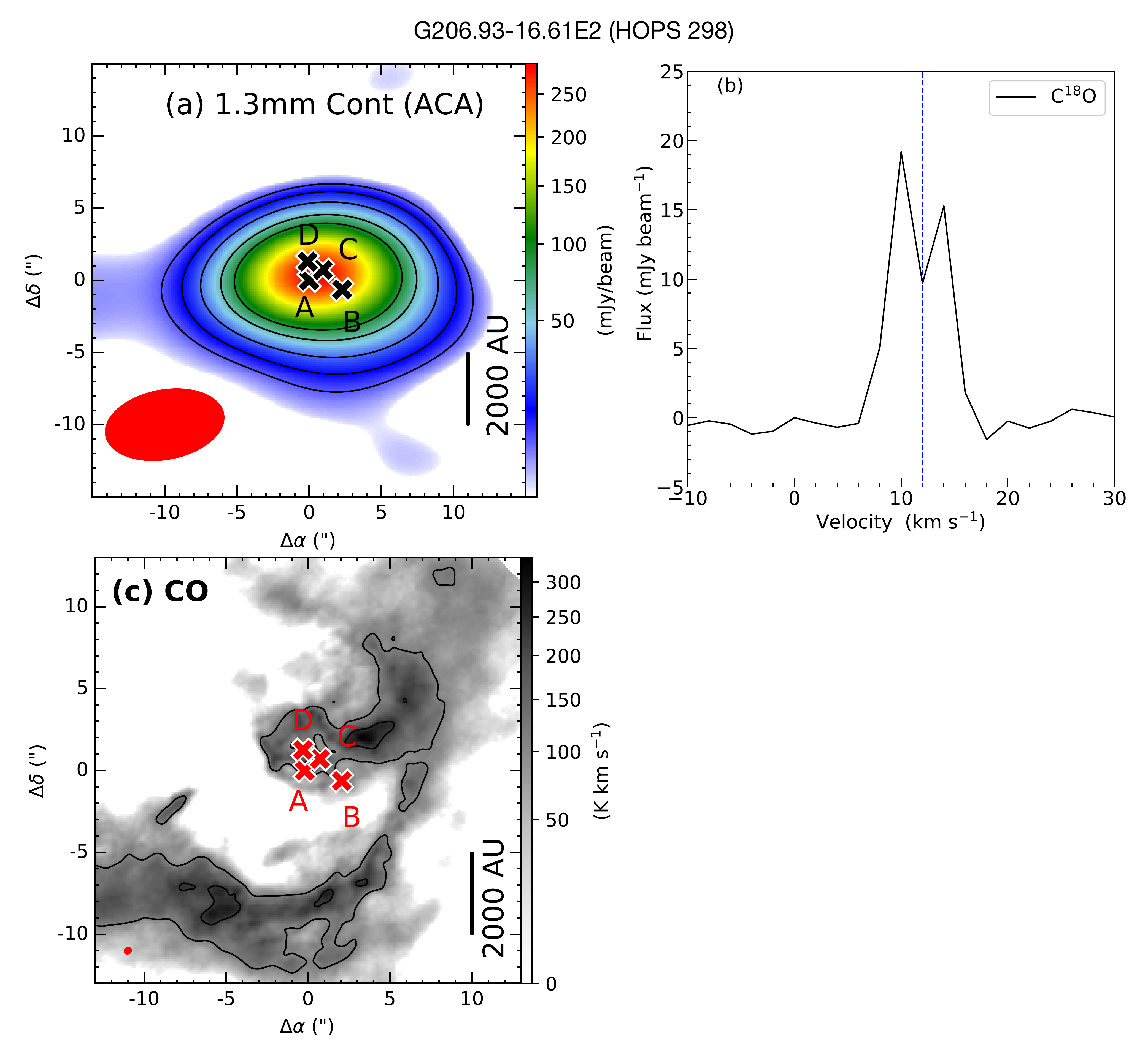}{0.85\textwidth}{}
\caption{G206.93-16.61E2: (a) 1.3mm continuum map at ACA resolution with sensitivity $\sim$ 2.5 $\mjypb$. The symbols and contour levels are the same as Figure \ref{fig:G20321_1120W2_ACA_envelope}.
(b) The  C$^{18}$O spectra extracted from high-resolution maps, and V$_{sys}$ = 12 $\kms$. All symbols are the same as Figure \ref{fig:G20321_1120W2_C18ospectra}. (c) Integrated CO emission with sensitivity 37.5 $\kkms$ with similar symbols and contours as Figure \ref{fig:G203.21-11.20W2outflow_parabola}a.}
\label{fig:appendix_G206.93-16.61E2}
\end{figure*}

\begin{figure*}
\fig{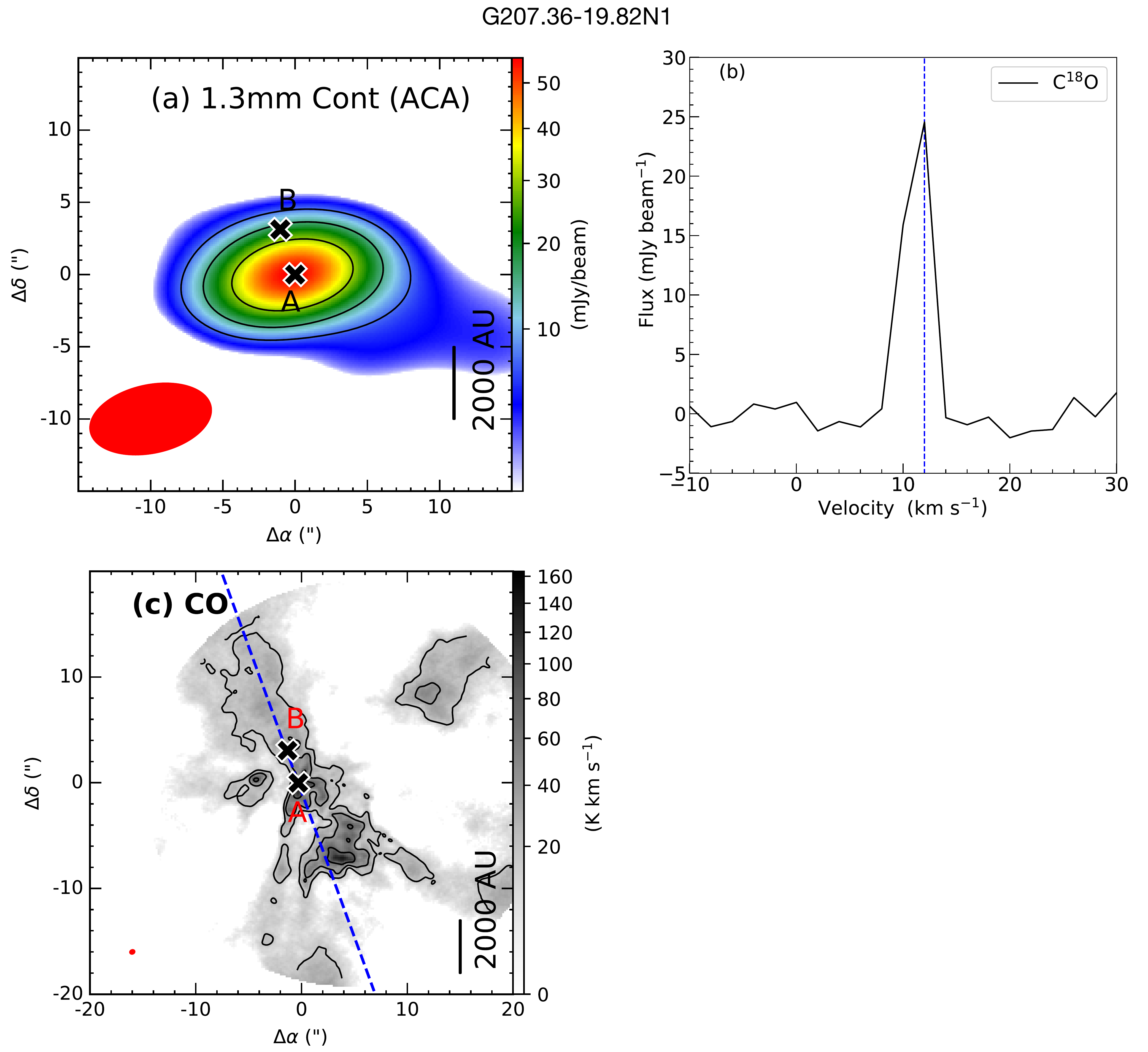}{0.85\textwidth}{}
\caption{G207.36-19.82N1: (a) 1.3mm continuum map at ACA resolution with sensitivity $\sim$ 2.5 $\mjypb$. The symbols and contour levels are the same as Figure \ref{fig:G20321_1120W2_ACA_envelope}.
(b) The  C$^{18}$O spectra extracted from high-resolution maps, and V$_{sys}$ = 12 $\kms$. All symbols are the same as Figure \ref{fig:G20321_1120W2_C18ospectra}. (c) Integrated CO emission with sensitivity 6.13 $\kkms$ with similar symbols and contours as Figure \ref{fig:G203.21-11.20W2outflow_parabola}a.}
\label{fig:appendix_G207.36-19.82N1}
\end{figure*}

\begin{figure*}
\fig{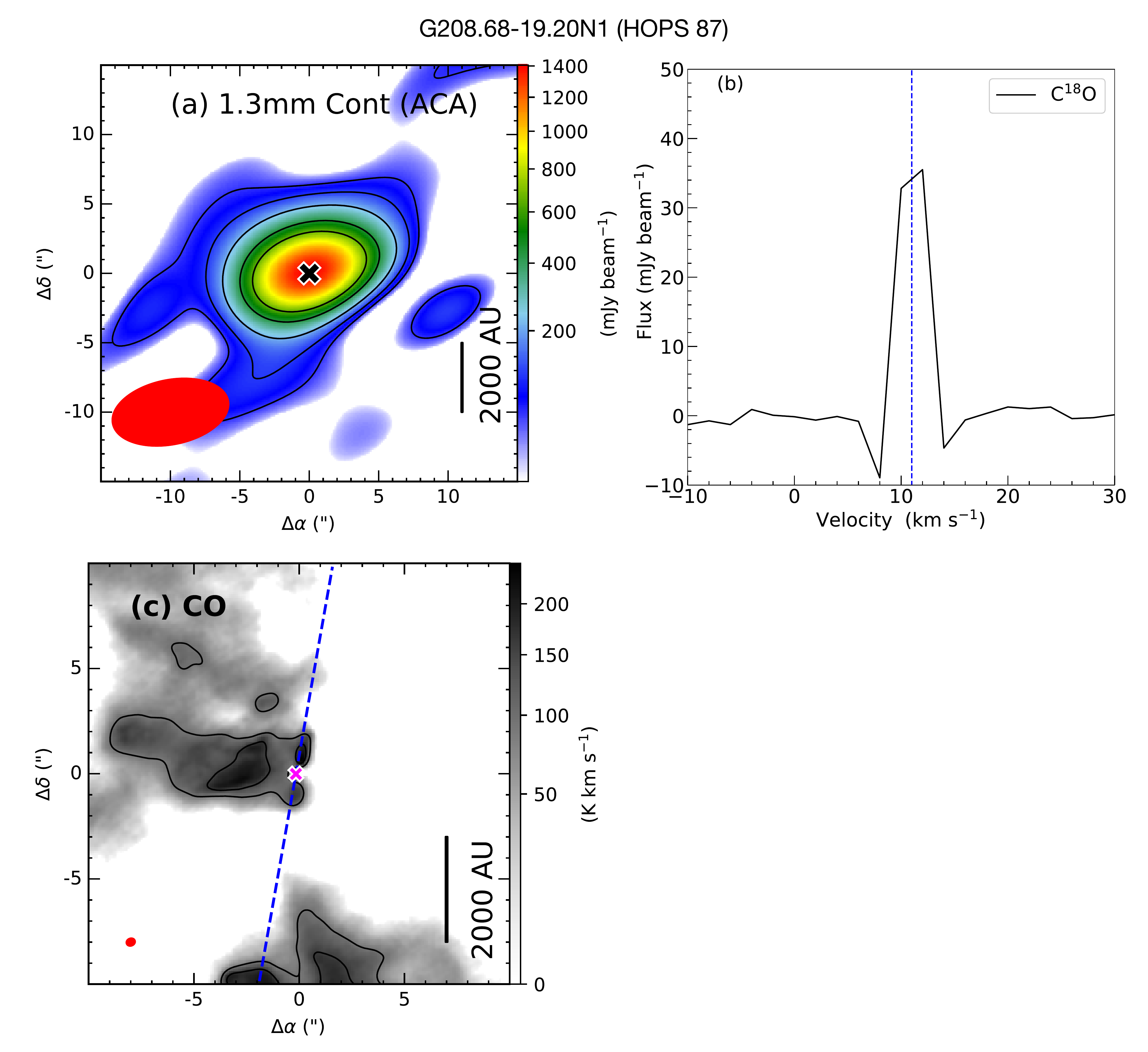}{0.85\textwidth}{}
\caption{G208.68-19.20N1: (a) 1.3mm continuum map at ACA resolution with sensitivity $\sim$ 18 $\mjypb$. The symbols and contour levels are the same as Figure \ref{fig:G20321_1120W2_ACA_envelope}.
(b) The  C$^{18}$O spectra extracted from high-resolution maps, and V$_{sys}$ = 11 $\kms$. All symbols are the same as Figure \ref{fig:G20321_1120W2_C18ospectra}. (c) Integrated CO emission with sensitivity 31.6 $\kkms$ with similar symbols are Figure \ref{fig:G203.21-11.20W2outflow_parabola}a. The contours are at (3, 5)$\times$ 31.6 K }
\label{fig:appendix_G208.68-19.20N1}
\end{figure*}

\begin{figure*}
\fig{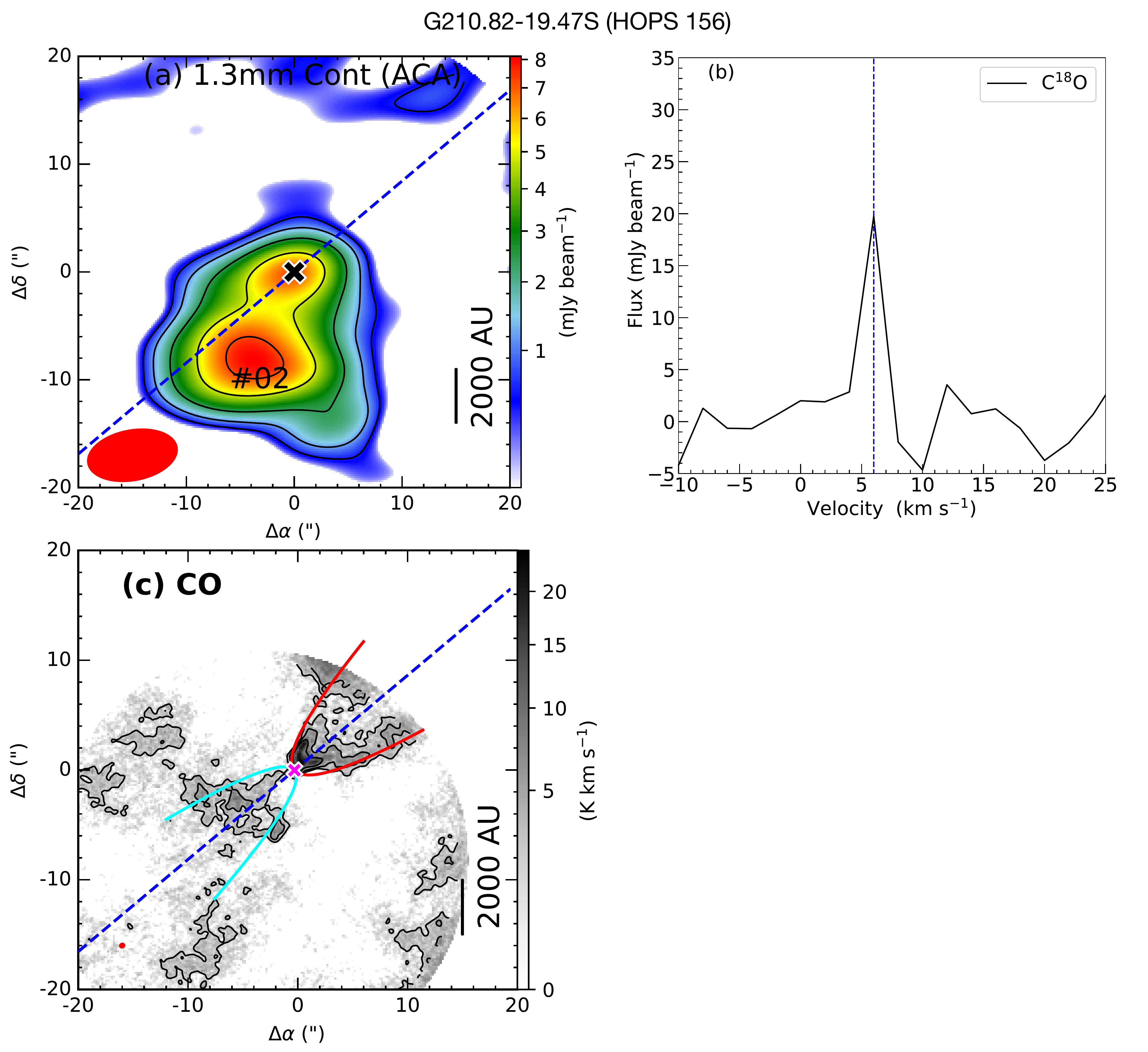}{0.85\textwidth}{}
\caption{G210.82-19.47S: (a) 1.3mm continuum map at ACA resolution with sensitivity $\sim$ 0.2 $\mjypb$. The symbols and contour levels are the same as Figure \ref{fig:G20321_1120W2_ACA_envelope}.
(b) The  C$^{18}$O spectra extracted from high-resolution maps, and V$_{sys}$ = 6 $\kms$. All symbols are the same as Figure \ref{fig:G20321_1120W2_C18ospectra}. (c) Integrated CO emission with sensitivity 0.8 $\kkms$ with similar symbols are Figure \ref{fig:G203.21-11.20W2outflow_parabola}a.}
\label{fig:appendix_G210.82-19.47S}
\end{figure*}

\begin{figure*}
\fig{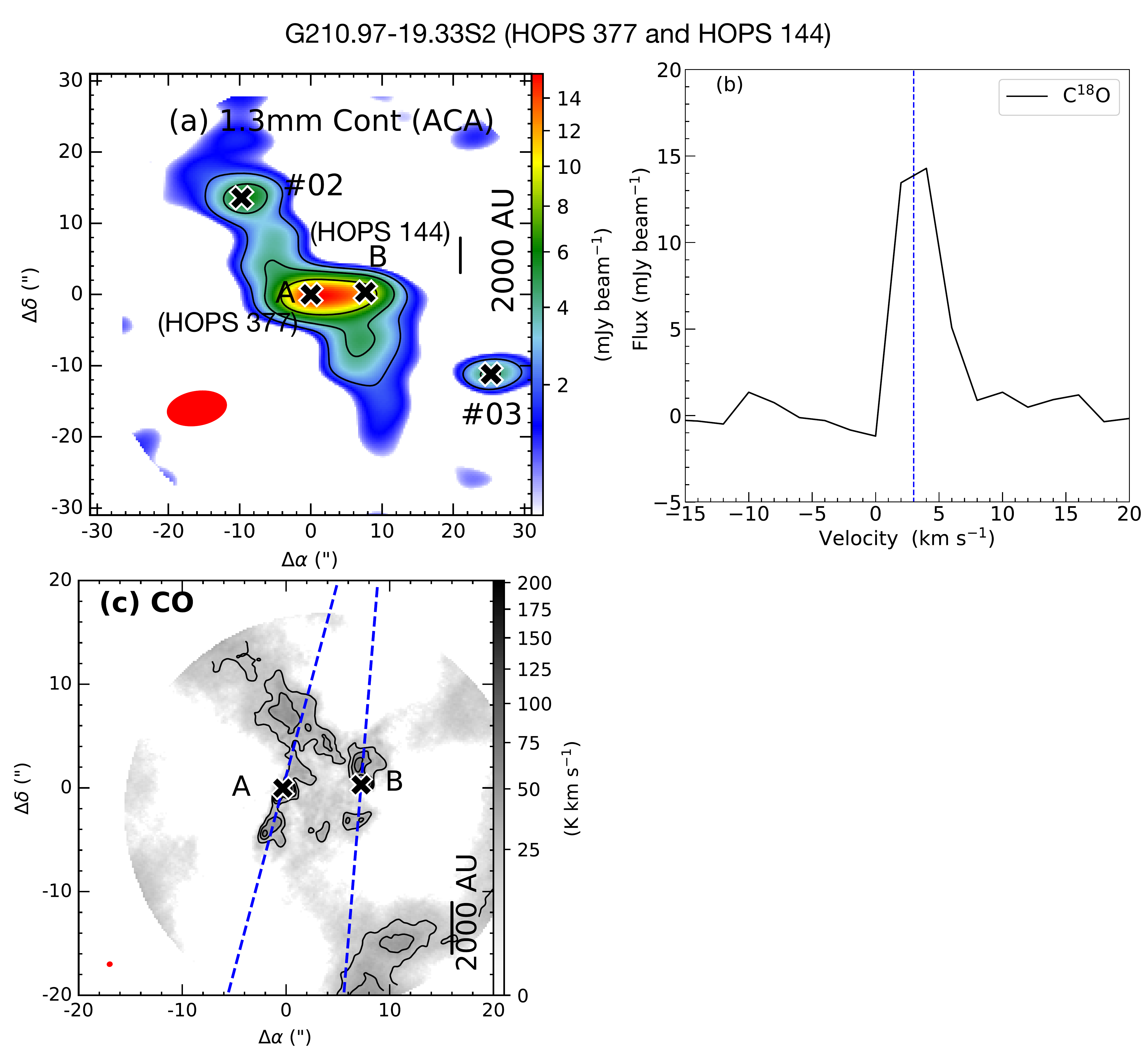}{0.85\textwidth}{}
\caption{G210.97-19.33S2: (a) 1.3mm continuum map at ACA resolution with sensitivity $\sim$ 0.7 $\mjypb$. The symbols and contour levels are the same as Figure \ref{fig:G20321_1120W2_ACA_envelope}.
(b) The  C$^{18}$O spectra extracted from high-resolution maps, and V$_{sys}$ = 3.0 $\kms$. All symbols are the same as Figure \ref{fig:G20321_1120W2_C18ospectra}. (c) Integrated CO emission with sensitivity 8.0 $\kkms$ with similar symbols are Figure \ref{fig:G203.21-11.20W2outflow_parabola}a.}
\label{fig:appendix_G210.97-19.33S2}
\end{figure*}

\begin{figure*}
\fig{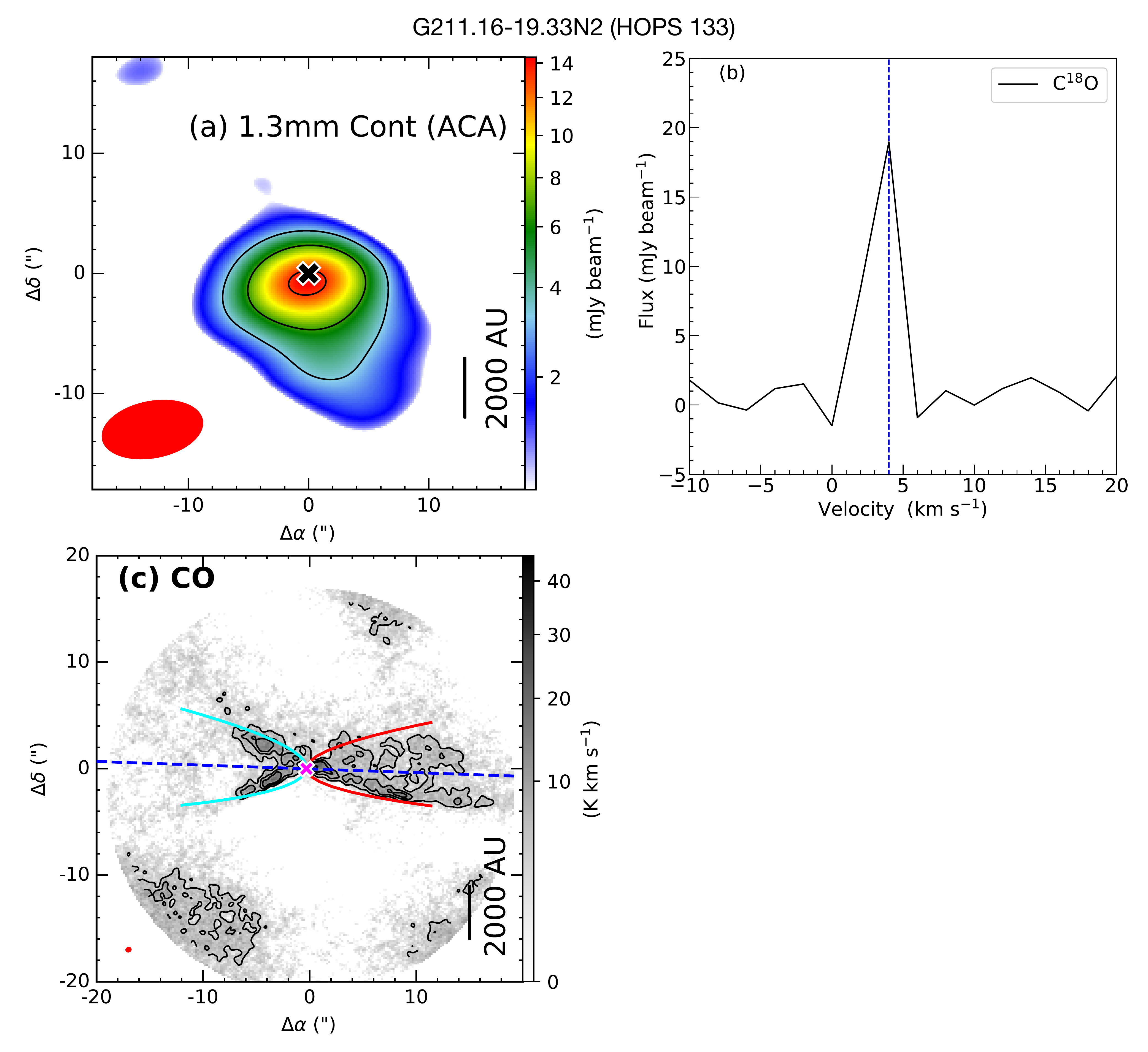}{0.85\textwidth}{}
\caption{G211.16-19.33N2: (a) 1.3mm continuum map at ACA resolution with sensitivity $\sim$ 1.1 $\mjypb$. The symbols and contour levels are the same as Figure \ref{fig:G20321_1120W2_ACA_envelope}.
(b) The  C$^{18}$O spectra extracted from high-resolution maps, and V$_{sys}$ = 4.0 $\kms$. All symbols are the same as Figure \ref{fig:G20321_1120W2_C18ospectra}. (c) Integrated CO emission with sensitivity 1.6 $\kkms$ with similar symbols are Figure \ref{fig:G203.21-11.20W2outflow_parabola}a.}
\label{fig:appendix_G211.16-19.33N2}
\end{figure*}

\begin{figure*}
\fig{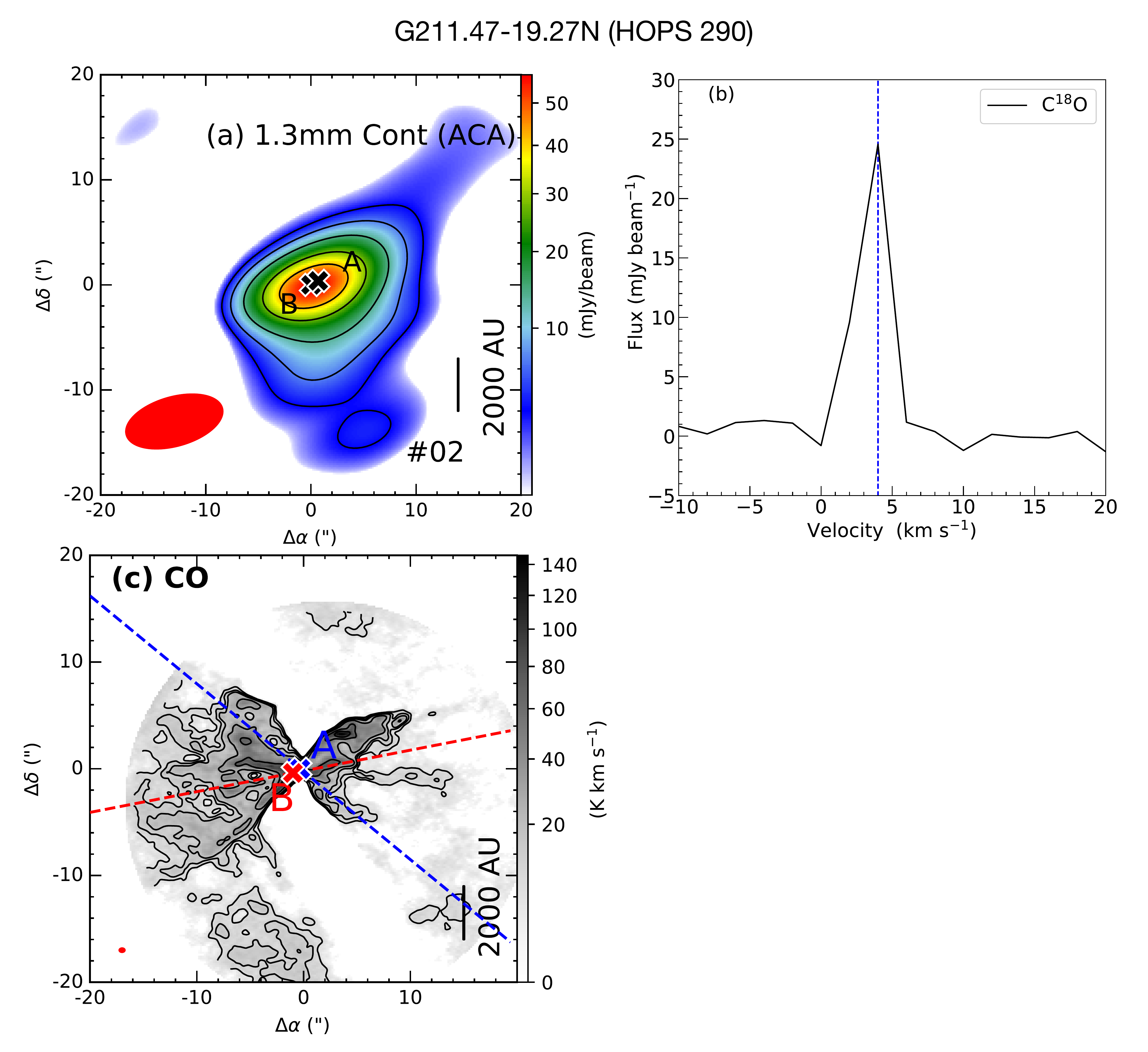}{0.85\textwidth}{}
\caption{G211.47-19.27N: (a) 1.3mm continuum map at ACA resolution with sensitivity $\sim$ 1.15 $\mjypb$. The symbols and contour levels are the same as Figure \ref{fig:G20321_1120W2_ACA_envelope}.
(b) The  C$^{18}$O spectra extracted from high-resolution maps, and V$_{sys}$ = 4.0 $\kms$. All symbols are the same as Figure \ref{fig:G20321_1120W2_C18ospectra}. (c) Integrated CO emission with sensitivity 2.5 $\kkms$ with similar symbols are Figure \ref{fig:G203.21-11.20W2outflow_parabola}a.}
\label{fig:appendix_G211.47-19.27N}
\end{figure*}

\begin{figure*}
\fig{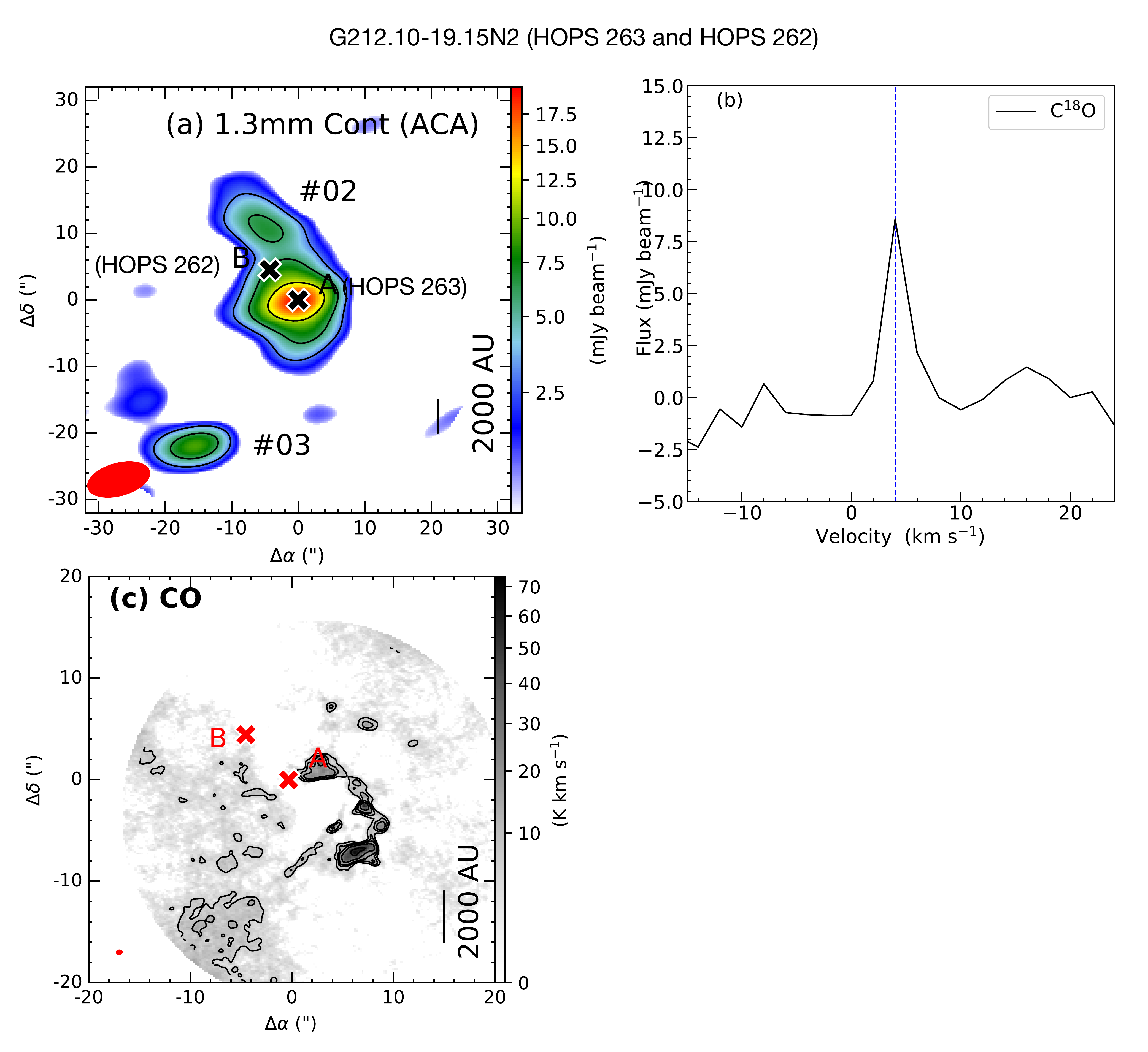}{0.85\textwidth}{}
\caption{G212.10-19.15N2: (a) 1.3mm continuum map at ACA resolution with sensitivity $\sim$ 1.0 $\mjypb$. The symbols and contour levels are the same as Figure \ref{fig:G20321_1120W2_ACA_envelope}.
(b) The  C$^{18}$O spectra extracted from high-resolution maps, and V$_{sys}$ = 4.0 $\kms$. All symbols are the same as Figure \ref{fig:G20321_1120W2_C18ospectra}. (c) Integrated CO emission with sensitivity 2.14 $\kkms$ with similar symbols are Figure \ref{fig:G203.21-11.20W2outflow_parabola}a.}
\label{fig:appendix_G212.10-19.15N2}
\end{figure*}

\begin{figure*}
\fig{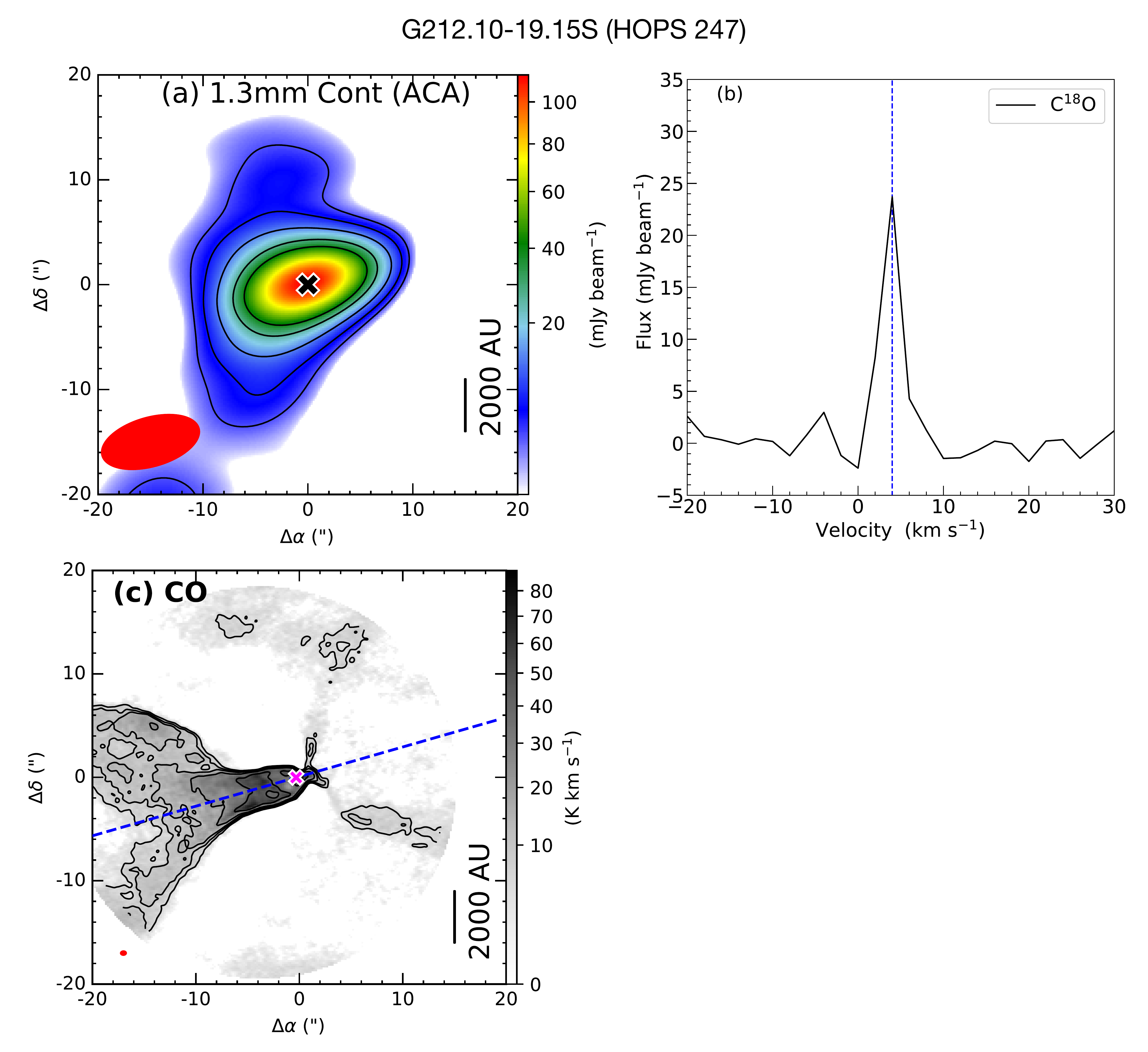}{0.85\textwidth}{}
\caption{G212.10-19.15S: (a) 1.3mm continuum map at ACA resolution with sensitivity $\sim$ 1.1 $\mjypb$. The symbols and contour levels are the same as Figure \ref{fig:G20321_1120W2_ACA_envelope}.
(b) The  C$^{18}$O spectra extracted from high-resolution maps, and V$_{sys}$ = 4.0 $\kms$. All symbols are the same as Figure \ref{fig:G20321_1120W2_C18ospectra}. (c) Integrated CO emission with sensitivity 1.85 $\kkms$ with similar symbols are Figure \ref{fig:G203.21-11.20W2outflow_parabola}a.}
\label{fig:appendix_G212.10-19.15S}
\end{figure*}

\begin{figure*}
\fig{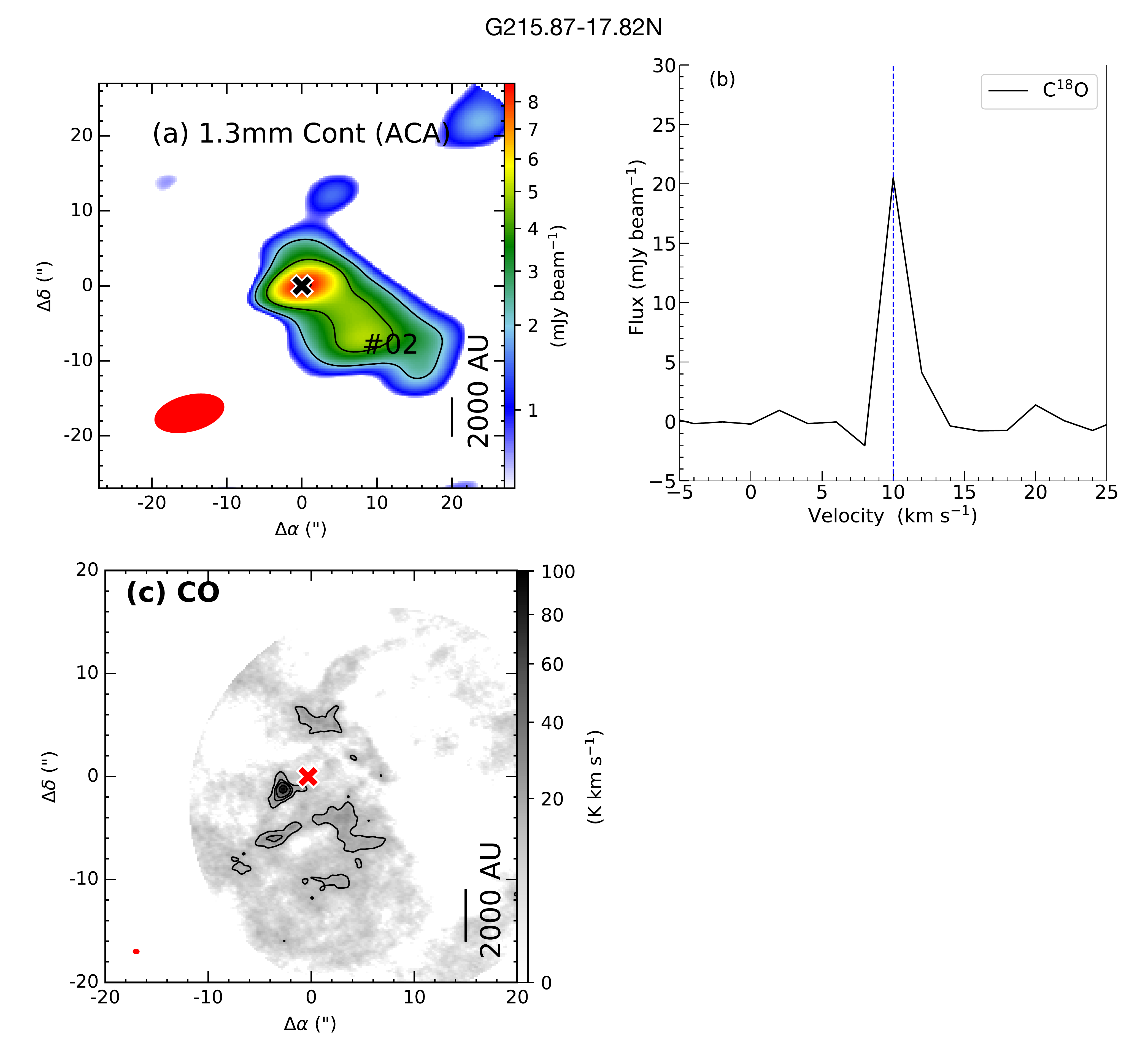}{0.85\textwidth}{}
\caption{G215.87-17.62N: (a) 1.3mm continuum map at ACA resolution with sensitivity $\sim$ 0.7 $\mjypb$. The symbols and contour levels are the same as Figure \ref{fig:G20321_1120W2_ACA_envelope}.
(b) The  C$^{18}$O spectra extracted from high-resolution maps, and V$_{sys}$ = 10.0 $\kms$. All symbols are the same as Figure \ref{fig:G20321_1120W2_C18ospectra}. (c) Integrated CO emission with sensitivity 5.21 $\kkms$ with similar symbols are Figure \ref{fig:G203.21-11.20W2outflow_parabola}a.}
\label{fig:appendix_G215.87-17.62N}
\end{figure*}

\bibliography{sample63}{}

\begin{thebibliography}{}
\expandafter\ifx\csname natexlab\endcsname\relax\def\natexlab#1{#1}\fi
\providecommand{\url}[1]{\href{#1}{#1}}
\providecommand{\dodoi}[1]{doi:~\href{http://doi.org/#1}{\nolinkurl{#1}}}
\providecommand{\doeprint}[1]{\href{http://ascl.net/#1}{\nolinkurl{http://ascl.net/#1}}}
\providecommand{\doarXiv}[1]{\href{https://arxiv.org/abs/#1}{\nolinkurl{https://arxiv.org/abs/#1}}}

\bibitem[{{Arce} {et~al.}(2007){Arce}, {Shepherd}, {Gueth}, {Lee}, {Bachiller},
  {Rosen}, \& {Beuther}}]{2007prpl.conf..245A}
{Arce}, H.~G., {Shepherd}, D., {Gueth}, F., {et~al.} 2007, in Protostars and
  Planets V, ed. B.~{Reipurth}, D.~{Jewitt}, \& K.~{Keil}, 245.
\newblock \doarXiv{astro-ph/0603071}

\bibitem[{{Astropy Collaboration} {et~al.}(2013){Astropy Collaboration},
  {Robitaille}, {Tollerud}, {Greenfield}, {Droettboom}, {Bray}, {Aldcroft},
  {Davis}, {Ginsburg}, {Price-Whelan}, {Kerzendorf}, {Conley}, {Crighton},
  {Barbary}, {Muna}, {Ferguson}, {Grollier}, {Parikh}, {Nair}, {Unther},
  {Deil}, {Woillez}, {Conseil}, {Kramer}, {Turner}, {Singer}, {Fox}, {Weaver},
  {Zabalza}, {Edwards}, {Azalee Bostroem}, {Burke}, {Casey}, {Crawford},
  {Dencheva}, {Ely}, {Jenness}, {Labrie}, {Lim}, {Pierfederici}, {Pontzen},
  {Ptak}, {Refsdal}, {Servillat}, \& {Streicher}}]{2013A&A...558A..33A}
{Astropy Collaboration}, {Robitaille}, T.~P., {Tollerud}, E.~J., {et~al.} 2013,
  \aap, 558, A33, \dodoi{10.1051/0004-6361/201322068}

\bibitem[{{Audard} {et~al.}(2014){Audard}, {{\'A}brah{\'a}m}, {Dunham},
  {Green}, {Grosso}, {Hamaguchi}, {Kastner}, {K{\'o}sp{\'a}l}, {Lodato},
  {Romanova}, {Skinner}, {Vorobyov}, \& {Zhu}}]{2014prpl.conf..387A}
{Audard}, M., {{\'A}brah{\'a}m}, P., {Dunham}, M.~M., {et~al.} 2014, in
  Protostars and Planets VI, ed. H.~{Beuther}, R.~S. {Klessen}, C.~P.
  {Dullemond}, \& T.~{Henning}, 387,
  \dodoi{10.2458/azu\_uapress\_9780816531240-ch017}

\bibitem[{{Bachiller} {et~al.}(2000){Bachiller}, {Gueth}, {Guilloteau},
  {Tafalla}, \& {Dutrey}}]{2000A&A...362L..33B}
{Bachiller}, R., {Gueth}, F., {Guilloteau}, S., {Tafalla}, M., \& {Dutrey}, A.
  2000, \aap, 362, L33.
\newblock \doarXiv{astro-ph/0009439}

\bibitem[{{Baek} {et~al.}(2020){Baek}, {MacFarlane}, {Lee}, {Stamatellos},
  {Herczeg}, {Johnstone}, {Pe{\~n}a}, {Varricatt}, {Hodapp}, {Chen}, \&
  {Kang}}]{baek20}
{Baek}, G., {MacFarlane}, B.~A., {Lee}, J.-E., {et~al.} 2020, \apj, 895, 27,
  \dodoi{10.3847/1538-4357/ab8ad4}

\bibitem[{{Balan{\c{c}}a} {et~al.}(2018){Balan{\c{c}}a}, {Dayou}, {Faure},
  {Wiesenfeld}, \& {Feautrier}}]{2018MNRAS.479.2692B}
{Balan{\c{c}}a}, C., {Dayou}, F., {Faure}, A., {Wiesenfeld}, L., \&
  {Feautrier}, N. 2018, \mnras, 479, 2692, \dodoi{10.1093/mnras/sty1681}

\bibitem[{{Bjerkeli} {et~al.}(2019){Bjerkeli}, {Ramsey}, {Harsono}, {Calcutt},
  {Kristensen}, {van der Wiel}, {J{\o}rgensen}, {Muller}, \&
  {Persson}}]{2019A&A...631A..64B}
{Bjerkeli}, P., {Ramsey}, J.~P., {Harsono}, D., {et~al.} 2019, \aap, 631, A64,
  \dodoi{10.1051/0004-6361/201935948}

\bibitem[{{Blondin} {et~al.}(1990){Blondin}, {Fryxell}, \&
  {Konigl}}]{1990ApJ...360..370B}
{Blondin}, J.~M., {Fryxell}, B.~A., \& {Konigl}, A. 1990, \apj, 360, 370,
  \dodoi{10.1086/169128}

\bibitem[{{Bontemps} {et~al.}(1996){Bontemps}, {Andre}, {Terebey}, \&
  {Cabrit}}]{1996A&A...311..858B}
{Bontemps}, S., {Andre}, P., {Terebey}, S., \& {Cabrit}, S. 1996, \aap, 311,
  858

\bibitem[{{Cabrit} \& {Bertout}(1992)}]{1992A&A...261..274C}
{Cabrit}, S., \& {Bertout}, C. 1992, \aap, 261, 274

\bibitem[{{Charnley} {et~al.}(2001){Charnley}, {Rodgers}, \&
  {Ehrenfreund}}]{2001A&A...378.1024C}
{Charnley}, S.~B., {Rodgers}, S.~D., \& {Ehrenfreund}, P. 2001, \aap, 378,
  1024, \dodoi{10.1051/0004-6361:20011193}

\bibitem[{{Cieza} {et~al.}(2016){Cieza}, {Casassus}, {Tobin}, {Bos},
  {Williams}, {Perez}, {Zhu}, {Caceres}, {Canovas}, {Dunham}, {Hales},
  {Prieto}, {Principe}, {Schreiber}, {Ruiz-Rodriguez}, \&
  {Zurlo}}]{2016Natur.535..258C}
{Cieza}, L.~A., {Casassus}, S., {Tobin}, J., {et~al.} 2016, \nat, 535, 258,
  \dodoi{10.1038/nature18612}

\bibitem[{{Codella} {et~al.}(2014){Codella}, {Maury}, {Gueth}, {Maret},
  {Belloche}, {Cabrit}, \& {Andr{\'e}}}]{2014A&A...563L...3C}
{Codella}, C., {Maury}, A.~J., {Gueth}, F., {et~al.} 2014, \aap, 563, L3,
  \dodoi{10.1051/0004-6361/201323024}

\bibitem[{{Connelley} \& {Reipurth}(2018)}]{connelley18}
{Connelley}, M.~S., \& {Reipurth}, B. 2018, \apj, 861, 145,
  \dodoi{10.3847/1538-4357/aaba7b}

\bibitem[{{Draine}(2006)}]{2006ApJ...636.1114D}
{Draine}, B.~T. 2006, \apj, 636, 1114, \dodoi{10.1086/498130}

\bibitem[{{Draine} \& {Salpeter}(1979)}]{1979ApJ...231...77D}
{Draine}, B.~T., \& {Salpeter}, E.~E. 1979, \apj, 231, 77,
  \dodoi{10.1086/157165}

\bibitem[{{Dunham} {et~al.}(2014){Dunham}, {Arce}, {Mardones}, {Lee},
  {Matthews}, {Stutz}, \& {Williams}}]{2014ApJ...783...29D}
{Dunham}, M.~M., {Arce}, H.~G., {Mardones}, D., {et~al.} 2014, \apj, 783, 29,
  \dodoi{10.1088/0004-637X/783/1/29}

\bibitem[{{Dutta} {et~al.}(2020){Dutta}, {Lee}, {Liu}, {Hirano}, {Liu},
  {Tatematsu}, {Kim}, {Shang}, {Sahu}, {Kim}, {Moraghan}, {Jhan}, {Hsu},
  {Evans}, {Johnstone}, {Ward-Thompson}, {Kuan}, {Lee}, {Lee}, {Traficante},
  {Juvela}, {Vastel}, {Zhang}, {Sanhueza}, {Soam}, {Kwon}, {Bronfman}, {Eden},
  {Goldsmith}, {He}, {Wu}, {Pelkonen}, {Qin}, {Li}, \&
  {Li}}]{2020ApJS..251...20D}
{Dutta}, S., {Lee}, C.-F., {Liu}, T., {et~al.} 2020, \apjs, 251, 20,
  \dodoi{10.3847/1538-4365/abba26}

\bibitem[{{Dutta} {et~al.}(2022{\natexlab{a}}){Dutta}, {Lee}, {Johnstone},
  {Liu}, {Hirano}, {Liu}, {Lee}, {Shang}, {Tatematsu}, {Kim}, {Sahu},
  {Sanhueza}, {di Francesco}, {Jhan}, {Lee}, {Kwon}, {Li}, {Bronfman}, {Liu},
  {Traficante}, {Kuan}, {Hsu}, {Moraghan}, {Liu}, {Eden}, {Soam}, {Luo}, \&
  {(Almasop Team)}}]{2022ApJ...925...11D}
{Dutta}, S., {Lee}, C.-F., {Johnstone}, D., {et~al.} 2022{\natexlab{a}}, \apj,
  925, 11, \dodoi{10.3847/1538-4357/ac3424}

\bibitem[{{Dutta} {et~al.}(2022{\natexlab{b}}){Dutta}, {Lee}, {Hirano}, {Liu},
  {Johnstone}, {Liu}, {Tatematsu}, {Goldsmith}, {Sahu}, {Evans}, {Sanhueza},
  {Kwon}, {Qin}, {Samal}, {Zhang}, {Kim}, {Shang}, {Lee}, {Moraghan}, {Jhan},
  {Li}, {Lee}, {Traficante}, {Juvela}, {Bronfman}, {Eden}, {Soam}, {He}, {Liu},
  {Kuan}, {Pelkonen}, {Luo}, {Yi}, \& {Hsu}}]{2022ApJ...931..130D}
{Dutta}, S., {Lee}, C.-F., {Hirano}, N., {et~al.} 2022{\natexlab{b}}, \apj,
  931, 130, \dodoi{10.3847/1538-4357/ac67a1}

\bibitem[{{Ellerbroek} {et~al.}(2013){Ellerbroek}, {Podio}, {Kaper}, {Sana},
  {Huppenkothen}, {de Koter}, \& {Monaco}}]{2013A&A...551A...5E}
{Ellerbroek}, L.~E., {Podio}, L., {Kaper}, L., {et~al.} 2013, \aap, 551, A5,
  \dodoi{10.1051/0004-6361/201220635}

\bibitem[{{Fischer} {et~al.}(2022){Fischer}, {Hillenbrand}, {Herczeg},
  {Johnstone}, {K{\'o}sp{\'a}l}, \& {Dunham}}]{2022arXiv220311257F}
{Fischer}, W.~J., {Hillenbrand}, L.~A., {Herczeg}, G.~J., {et~al.} 2022, arXiv
  e-prints, arXiv:2203.11257.
\newblock \doarXiv{2203.11257}

\bibitem[{{Frank} {et~al.}(2014){Frank}, {Ray}, {Cabrit}, {Hartigan}, {Arce},
  {Bacciotti}, {Bally}, {Benisty}, {Eisl{\"o}ffel}, {G{\"u}del}, {Lebedev},
  {Nisini}, \& {Raga}}]{2014prpl.conf..451F}
{Frank}, A., {Ray}, T.~P., {Cabrit}, S., {et~al.} 2014, in Protostars and
  Planets VI, ed. H.~{Beuther}, R.~S. {Klessen}, C.~P. {Dullemond}, \&
  T.~{Henning}, 451, \dodoi{10.2458/azu_uapress_9780816531240-ch020}

\bibitem[{{Furlan} {et~al.}(2016){Furlan}, {Fischer}, {Ali}, {Stutz}, {Stanke},
  {Tobin}, {Megeath}, {Osorio}, {Hartmann}, {Calvet}, {Poteet}, {Booker},
  {Manoj}, {Watson}, \& {Allen}}]{2016ApJS..224....5F}
{Furlan}, E., {Fischer}, W.~J., {Ali}, B., {et~al.} 2016, \apjs, 224, 5,
  \dodoi{10.3847/0067-0049/224/1/5}

\bibitem[{{Glassgold} {et~al.}(1991){Glassgold}, {Mamon}, \&
  {Huggins}}]{1991ApJ...373..254G}
{Glassgold}, A.~E., {Mamon}, G.~A., \& {Huggins}, P.~J. 1991, \apj, 373, 254,
  \dodoi{10.1086/170045}

\bibitem[{{Harsono} {et~al.}(2018){Harsono}, {Bjerkeli}, {van der Wiel},
  {Ramsey}, {Maud}, {Kristensen}, \& {J{\o}rgensen}}]{2018NatAs...2..646H}
{Harsono}, D., {Bjerkeli}, P., {van der Wiel}, M. H.~D., {et~al.} 2018, Nature
  Astronomy, 2, 646, \dodoi{10.1038/s41550-018-0497-x}

\bibitem[{{Herczeg} {et~al.}(2017){Herczeg}, {Johnstone}, {Mairs}, {Hatchell},
  {Lee}, {Bower}, {Chen}, {Aikawa}, {Yoo}, {Kang}, {Kang}, {Chen}, {Williams},
  {Bae}, {Dunham}, {Vorobyov}, {Zhu}, {Rao}, {Kirk}, {Takahashi}, {Morata},
  {Lacaille}, {Lane}, {Pon}, {Scholz}, {Samal}, {Bell}, {Graves}, {Lee},
  {Parsons}, {He}, {Zhou}, {Kim}, {Chapman}, {Drabek-Maunder}, {Chung},
  {Eyres}, {Forbrich}, {Hillenbrand}, {Inutsuka}, {Kim}, {Kim}, {Kuan}, {Kwon},
  {Lai}, {Lalchand}, {Lee}, {Lee}, {Long}, {Lyo}, {Qian}, {Scicluna}, {Soam},
  {Stamatellos}, {Takakuwa}, {Tang}, {Wang}, \& {Wang}}]{herczeg17}
{Herczeg}, G.~J., {Johnstone}, D., {Mairs}, S., {et~al.} 2017, \apj, 849, 43,
  \dodoi{10.3847/1538-4357/aa8b62}

\bibitem[{{Hirano} \& {Machida}(2019)}]{2019MNRAS.485.4667H}
{Hirano}, S., \& {Machida}, M.~N. 2019, \mnras, 485, 4667,
  \dodoi{10.1093/mnras/stz740}

\bibitem[{{Hodapp} {et~al.}(2012){Hodapp}, {Chini}, {Watermann}, \&
  {Lemke}}]{hodapp12}
{Hodapp}, K.~W., {Chini}, R., {Watermann}, R., \& {Lemke}, R. 2012, \apj, 744,
  56, \dodoi{10.1088/0004-637X/744/1/56}

\bibitem[{{Hsieh} {et~al.}(2016){Hsieh}, {Lai}, {Belloche}, \&
  {Wyrowski}}]{2016ApJ...826...68H}
{Hsieh}, T.-H., {Lai}, S.-P., {Belloche}, A., \& {Wyrowski}, F. 2016, \apj,
  826, 68, \dodoi{10.3847/0004-637X/826/1/68}

\bibitem[{{Hsieh} {et~al.}(2019){Hsieh}, {Murillo}, {Belloche}, {Hirano},
  {Walsh}, {van Dishoeck}, {J{\o}rgensen}, \& {Lai}}]{2019ApJ...884..149H}
{Hsieh}, T.-H., {Murillo}, N.~M., {Belloche}, A., {et~al.} 2019, \apj, 884,
  149, \dodoi{10.3847/1538-4357/ab425a}

\bibitem[{{Hsu} {et~al.}(2020){Hsu}, {Liu}, {Liu}, {Sahu}, {Hirano}, {Lee},
  {Tatematsu}, {Kim}, {Juvela}, {Sanhueza}, {He}, {Johnstone}, {Qin},
  {Bronfman}, {Chen}, {Dutta}, {Eden}, {Jhan}, {Kim}, {Kuan}, {Kwon}, {Lee},
  {Lee}, {Moraghan}, {Rawlings}, {Shang}, {Soam}, {Thompson}, {Traficante},
  {Wu}, {Yang}, \& {Zhang}}]{2020ApJ...898..107H}
{Hsu}, S.-Y., {Liu}, S.-Y., {Liu}, T., {et~al.} 2020, \apj, 898, 107,
  \dodoi{10.3847/1538-4357/ab9f3a}

\bibitem[{{Hsu} {et~al.}(2022){Hsu}, {Liu}, {Liu}, {Sahu}, {Lee}, {Tatematsu},
  {Kim}, {Hirano}, {Yang}, {Johnstone}, {Liu}, {Juvela}, {Bronfman}, {Chen},
  {Dutta}, {Eden}, {Jhan}, {Kuan}, {Lee}, {Lee}, {Li}, {Liu}, {Qin},
  {Sanhueza}, {Shang}, {Soam}, {Traficante}, \& {Zhou}}]{2022ApJ...927..218H}
---. 2022, \apj, 927, 218, \dodoi{10.3847/1538-4357/ac49e0}

\bibitem[{{Hu} {et~al.}(2019){Hu}, {Zhukovska}, {Somerville}, \&
  {Naab}}]{2019MNRAS.487.3252H}
{Hu}, C.-Y., {Zhukovska}, S., {Somerville}, R.~S., \& {Naab}, T. 2019, \mnras,
  487, 3252, \dodoi{10.1093/mnras/stz1481}

\bibitem[{{Hunter}(2007)}]{2007CSE.....9...90H}
{Hunter}, J.~D. 2007, Computing in Science and Engineering, 9, 90,
  \dodoi{10.1109/MCSE.2007.55}

\bibitem[{{Imai} {et~al.}(2019){Imai}, {Oya}, {Sakai}, {L{\'o}pez-Sepulcre},
  {Watanabe}, \& {Yamamoto}}]{2019ApJ...873L..21I}
{Imai}, M., {Oya}, Y., {Sakai}, N., {et~al.} 2019, \apjl, 873, L21,
  \dodoi{10.3847/2041-8213/ab0c20}

\bibitem[{{Jhan} \& {Lee}(2016)}]{2016ApJ...816...32J}
{Jhan}, K.-S., \& {Lee}, C.-F. 2016, \apj, 816, 32,
  \dodoi{10.3847/0004-637X/816/1/32}

\bibitem[{{Jhan} \& {Lee}(2021)}]{2021ApJ...909...11J}
---. 2021, \apj, 909, 11, \dodoi{10.3847/1538-4357/abd6c5}

\bibitem[{{Jhan} {et~al.}(2022){Jhan}, {Lee}, {Johnstone}, {Liu}, {Liu},
  {Hirano}, {Tatematsu}, {Dutta}, {Moraghan}, {Shang}, {Lee}, {Li}, {Liu},
  {Hsu}, {Kwon}, {Sahu}, {Liu}, {Kim}, {Luo}, {Qin}, {Sanhueza}, {Bronfman},
  {Qizhou}, {Eden}, {Traficante}, {Lee}, \& {Almasop
  Team}}]{2022ApJ...931L...5J}
{Jhan}, K.-S., {Lee}, C.-F., {Johnstone}, D., {et~al.} 2022, \apjl, 931, L5,
  \dodoi{10.3847/2041-8213/ac6a53}

\bibitem[{{Johnstone} {et~al.}(2013){Johnstone}, {Hendricks}, {Herczeg}, \&
  {Bruderer}}]{johnstone13}
{Johnstone}, D., {Hendricks}, B., {Herczeg}, G.~J., \& {Bruderer}, S. 2013,
  \apj, 765, 133, \dodoi{10.1088/0004-637X/765/2/133}

\bibitem[{{Johnstone} {et~al.}(2018){Johnstone}, {Herczeg}, {Mairs},
  {Hatchell}, {Bower}, {Kirk}, {Lane}, {Bell}, {Graves}, \&
  {Aikawa}}]{johnstone18}
{Johnstone}, D., {Herczeg}, G.~J., {Mairs}, S., {et~al.} 2018, \apj, 854, 31,
  \dodoi{10.3847/1538-4357/aaa764}

\bibitem[{{J{\o}rgensen} {et~al.}(2022){J{\o}rgensen}, {Kuruwita}, {Harsono},
  {Haugb{\o}lle}, {Kristensen}, \& {Bergin}}]{2022Natur.606..272J}
{J{\o}rgensen}, J.~K., {Kuruwita}, R.~L., {Harsono}, D., {et~al.} 2022, \nat,
  606, 272, \dodoi{10.1038/s41586-022-04659-4}

\bibitem[{{J{\o}rgensen} {et~al.}(2015){J{\o}rgensen}, {Visser}, {Williams}, \&
  {Bergin}}]{2015A&A...579A..23J}
{J{\o}rgensen}, J.~K., {Visser}, R., {Williams}, J.~P., \& {Bergin}, E.~A.
  2015, \aap, 579, A23, \dodoi{10.1051/0004-6361/201425317}

\bibitem[{{Karnath} {et~al.}(2020){Karnath}, {Megeath}, {Tobin}, {Stutz}, {Li},
  {Sheehan}, {Reynolds}, {Sadavoy}, {Stephens}, {Osorio}, {Anglada},
  {D{\'\i}az-Rodr{\'\i}guez}, \& {Cox}}]{2020ApJ...890..129K}
{Karnath}, N., {Megeath}, S.~T., {Tobin}, J.~J., {et~al.} 2020, \apj, 890, 129,
  \dodoi{10.3847/1538-4357/ab659e}

\bibitem[{{Kim} {et~al.}(2019){Kim}, {Lee}, {Maheswar}, {Kim}, {Soam}, {Saito},
  {Kiyokane}, \& {Kim}}]{2019ApJS..240...18K}
{Kim}, G., {Lee}, C.~W., {Maheswar}, G., {et~al.} 2019, \apjs, 240, 18,
  \dodoi{10.3847/1538-4365/aaf889}

\bibitem[{{Krumholz} {et~al.}(2014){Krumholz}, {Bate}, {Arce}, {Dale},
  {Gutermuth}, {Klein}, {Li}, {Nakamura}, \& {Zhang}}]{2014prpl.conf..243K}
{Krumholz}, M.~R., {Bate}, M.~R., {Arce}, H.~G., {et~al.} 2014, in Protostars
  and Planets VI, ed. H.~{Beuther}, R.~S. {Klessen}, C.~P. {Dullemond}, \&
  T.~{Henning}, 243--266, \dodoi{10.2458/azu_uapress_9780816531240-ch011}

\bibitem[{{Larson}(1969)}]{1969MNRAS.145..271L}
{Larson}, R.~B. 1969, \mnras, 145, 271, \dodoi{10.1093/mnras/145.3.271}

\bibitem[{{Lee}(2020)}]{2020A&ARv..28....1L}
{Lee}, C.-F. 2020, \aapr, 28, 1, \dodoi{10.1007/s00159-020-0123-7}

\bibitem[{{Lee} {et~al.}(2010){Lee}, {Hasegawa}, {Hirano}, {Palau}, {Shang},
  {Ho}, \& {Zhang}}]{2010ApJ...713..731L}
{Lee}, C.-F., {Hasegawa}, T.~I., {Hirano}, N., {et~al.} 2010, \apj, 713, 731,
  \dodoi{10.1088/0004-637X/713/2/731}

\bibitem[{{Lee} {et~al.}(2007{\natexlab{a}}){Lee}, {Ho}, {Hirano}, {Beuther},
  {Bourke}, {Shang}, \& {Zhang}}]{2007ApJ...659..499L}
{Lee}, C.-F., {Ho}, P. T.~P., {Hirano}, N., {et~al.} 2007{\natexlab{a}}, \apj,
  659, 499, \dodoi{10.1086/512540}

\bibitem[{{Lee} {et~al.}(2017{\natexlab{a}}){Lee}, {Ho}, {Li}, {Hirano},
  {Zhang}, \& {Shang}}]{2017NatAs...1E.152L}
{Lee}, C.-F., {Ho}, P. T.~P., {Li}, Z.-Y., {et~al.} 2017{\natexlab{a}}, Nature
  Astronomy, 1, 0152, \dodoi{10.1038/s41550-017-0152}

\bibitem[{{Lee} {et~al.}(2007{\natexlab{b}}){Lee}, {Ho}, {Palau}, {Hirano},
  {Bourke}, {Shang}, \& {Zhang}}]{2007ApJ...670.1188L}
{Lee}, C.-F., {Ho}, P. T.~P., {Palau}, A., {et~al.} 2007{\natexlab{b}}, \apj,
  670, 1188, \dodoi{10.1086/522333}

\bibitem[{{Lee} {et~al.}(2018){Lee}, {Li}, {Hirano}, {Shang}, {Ho}, \&
  {Zhang}}]{2018ApJ...863...94L}
{Lee}, C.-F., {Li}, Z.-Y., {Hirano}, N., {et~al.} 2018, \apj, 863, 94,
  \dodoi{10.3847/1538-4357/aad2da}

\bibitem[{{Lee} {et~al.}(2017{\natexlab{b}}){Lee}, {Li}, {Ho}, {Hirano},
  {Zhang}, \& {Shang}}]{2017SciA....3E2935L}
{Lee}, C.-F., {Li}, Z.-Y., {Ho}, P. T.~P., {et~al.} 2017{\natexlab{b}}, Science
  Advances, 3, e1602935, \dodoi{10.1126/sciadv.1602935}

\bibitem[{{Lee} {et~al.}(2022){Lee}, {Li}, {Shang}, \&
  {Hirano}}]{2022ApJ...927L..27L}
{Lee}, C.-F., {Li}, Z.-Y., {Shang}, H., \& {Hirano}, N. 2022, \apjl, 927, L27,
  \dodoi{10.3847/2041-8213/ac59c0}

\bibitem[{{Lee} {et~al.}(2000){Lee}, {Mundy}, {Reipurth}, {Ostriker}, \&
  {Stone}}]{2000ApJ...542..925L}
{Lee}, C.-F., {Mundy}, L.~G., {Reipurth}, B., {Ostriker}, E.~C., \& {Stone},
  J.~M. 2000, \apj, 542, 925, \dodoi{10.1086/317056}

\bibitem[{{Lee} {et~al.}(2014){Lee}, {Rao}, {Ching}, {Lai}, {Hirano}, {Ho}, \&
  {Hwang}}]{2014ApJ...797L...9L}
{Lee}, C.-F., {Rao}, R., {Ching}, T.-C., {et~al.} 2014, \apjl, 797, L9,
  \dodoi{10.1088/2041-8205/797/1/L9}

\bibitem[{{Lee} {et~al.}(2001){Lee}, {Stone}, {Ostriker}, \&
  {Mundy}}]{2001ApJ...557..429L}
{Lee}, C.-F., {Stone}, J.~M., {Ostriker}, E.~C., \& {Mundy}, L.~G. 2001, \apj,
  557, 429, \dodoi{10.1086/321648}

\bibitem[{{Lee} {et~al.}(2019{\natexlab{a}}){Lee}, {Kwon}, {Jhan}, {Hirano},
  {Hwang}, {Lai}, {Ching}, {Rao}, \& {Ho}}]{2019ApJ...879..101L}
{Lee}, C.-F., {Kwon}, W., {Jhan}, K.-S., {et~al.} 2019{\natexlab{a}}, \apj,
  879, 101, \dodoi{10.3847/1538-4357/ab2458}

\bibitem[{{Lee} {et~al.}(2013){Lee}, {Kim}, {Kim}, {Saito}, {Myers}, \&
  {Kurono}}]{2013ApJ...777...50L}
{Lee}, C.~W., {Kim}, M.-R., {Kim}, G., {et~al.} 2013, \apj, 777, 50,
  \dodoi{10.1088/0004-637X/777/1/50}

\bibitem[{{Lee} {et~al.}(2019{\natexlab{b}}){Lee}, {Lee}, {Baek}, {Aikawa},
  {Cieza}, {Yoon}, {Herczeg}, {Johnstone}, \& {Casassus}}]{2019NatAs...3..314L}
{Lee}, J.-E., {Lee}, S., {Baek}, G., {et~al.} 2019{\natexlab{b}}, Nature
  Astronomy, 3, 314, \dodoi{10.1038/s41550-018-0680-0}

\bibitem[{{Lee} {et~al.}(2020){Lee}, {Johnstone}, {Lee}, {Herczeg}, {Mairs},
  {Varricatt}, {Hodapp}, {Naylor}, {Pe{\~n}a}, {Baek}, {Haas}, {Chini}, \&
  {JCMT Transient Team}}]{lee20}
{Lee}, Y.-H., {Johnstone}, D., {Lee}, J.-E., {et~al.} 2020, \apj, 903, 5,
  \dodoi{10.3847/1538-4357/abb6fe}

\bibitem[{{Lee} {et~al.}(2021){Lee}, {Johnstone}, {Lee}, {Herczeg}, {Mairs},
  {Contreras-Pe{\~n}a}, {Hatchell}, {Naylor}, {Bell}, {Bourke}, {Broughton},
  {Francis}, {Gupta}, {Harsono}, {Liu}, {Park}, {Plovie}, {Moriarty-Schieven},
  {Scholz}, {Sharma}, {Teixeira}, {Wang}, {Aikawa}, {Bower}, {Vivien Chen},
  {Bae}, {Baek}, {Chapman}, {Ping Chen}, {Du}, {Dutta}, {Forbrich}, {Guo},
  {Inutsuka}, {Kang}, {Kirk}, {Kuan}, {Kwon}, {Lai}, {Lalchand}, {Lane}, {Lee},
  {Liu}, {Morata}, {Pearson}, {Pon}, {Sahu}, {Shang}, {Stamatellos}, {Tang},
  {Xu}, {Yoo}, \& {Rawlings}}]{2021ApJ...920..119L}
---. 2021, \apj, 920, 119, \dodoi{10.3847/1538-4357/ac1679}

\bibitem[{{Lef{\`e}vre} {et~al.}(2017){Lef{\`e}vre}, {Cabrit}, {Maury},
  {Gueth}, {Tabone}, {Podio}, {Belloche}, {Codella}, {Maret}, {Anderl},
  {Andr{\'e}}, \& {Hennebelle}}]{2017A&A...604L...1L}
{Lef{\`e}vre}, C., {Cabrit}, S., {Maury}, A.~J., {et~al.} 2017, \aap, 604, L1,
  \dodoi{10.1051/0004-6361/201730766}

\bibitem[{{Luo} {et~al.}(2022){Luo}, {Liu}, {Tatematsu}, {Liu}, {Li}, {di
  Francesco}, {Johnstone}, {Goldsmith}, {Dutta}, {Hirano}, {Lee}, {Li}, {Kim},
  {Won Lee}, {Lee}, {Liu}, {Juvela}, {He}, {Qin}, {Liu}, {Eden}, {Kwon},
  {Sahu}, {Li}, {Xu}, {Zhang}, {Hsu}, {Bronfman}, {Sanhueza}, {Pelkonen},
  {Zhou}, {Liu}, {Gu}, {Wu}, {Mai}, {Falgarone}, \&
  {Shen}}]{2022ApJ...931..158L}
{Luo}, Q.-y., {Liu}, T., {Tatematsu}, K., {et~al.} 2022, \apj, 931, 158,
  \dodoi{10.3847/1538-4357/ac66d9}

\bibitem[{{MacFarlane} {et~al.}(2019{\natexlab{a}}){MacFarlane}, {Stamatellos},
  {Johnstone}, {Herczeg}, {Baek}, {Chen}, {Kang}, \& {Lee}}]{macfarlane19a}
{MacFarlane}, B., {Stamatellos}, D., {Johnstone}, D., {et~al.}
  2019{\natexlab{a}}, arXiv e-prints, arXiv:1906.01960.
\newblock \doarXiv{1906.01960}

\bibitem[{{MacFarlane} {et~al.}(2019{\natexlab{b}}){MacFarlane}, {Stamatellos},
  {Johnstone}, {Herczeg}, {Baek}, {Vivien Chen}, {Kang}, \&
  {Lee}}]{macfarlane19b}
---. 2019{\natexlab{b}}, \mnras, 1508, \dodoi{10.1093/mnras/stz1570}

\bibitem[{{Machida} \& {Basu}(2019)}]{2019ApJ...876..149M}
{Machida}, M.~N., \& {Basu}, S. 2019, \apj, 876, 149,
  \dodoi{10.3847/1538-4357/ab18a7}

\bibitem[{{Machida} \& {Hosokawa}(2013)}]{2013MNRAS.431.1719M}
{Machida}, M.~N., \& {Hosokawa}, T. 2013, \mnras, 431, 1719,
  \dodoi{10.1093/mnras/stt291}

\bibitem[{{McMullin} {et~al.}(2007){McMullin}, {Waters}, {Schiebel}, {Young},
  \& {Golap}}]{2007ASPC..376..127M}
{McMullin}, J.~P., {Waters}, B., {Schiebel}, D., {Young}, W., \& {Golap}, K.
  2007, Astronomical Society of the Pacific Conference Series, Vol. 376, {CASA
  Architecture and Applications}, ed. R.~A. {Shaw}, F.~{Hill}, \& D.~J. {Bell},
  127

\bibitem[{{Mercer} \& {Stamatellos}(2017)}]{2017MNRAS.465....2M}
{Mercer}, A., \& {Stamatellos}, D. 2017, \mnras, 465, 2,
  \dodoi{10.1093/mnras/stw2714}

\bibitem[{{Moraghan} {et~al.}(2016){Moraghan}, {Lee}, {Huang}, \&
  {Vaidya}}]{2016MNRAS.460.1829M}
{Moraghan}, A., {Lee}, C.-F., {Huang}, P.-S., \& {Vaidya}, B. 2016, \mnras,
  460, 1829, \dodoi{10.1093/mnras/stw1089}

\bibitem[{{Ninan} {et~al.}(2013){Ninan}, {Ojha}, {Bhatt}, {Ghosh}, {Mohan},
  {Mallick}, {Tamura}, \& {Henning}}]{ninan13}
{Ninan}, J.~P., {Ojha}, D.~K., {Bhatt}, B.~C., {et~al.} 2013, \apj, 778, 116,
  \dodoi{10.1088/0004-637X/778/2/116}

\bibitem[{{Nony} {et~al.}(2020){Nony}, {Motte}, {Louvet}, {Plunkett},
  {Gusdorf}, {Fechtenbaum}, {Pouteau}, {Lefloch}, {Bontemps}, {Molet}, \&
  {Robitaille}}]{2020A&A...636A..38N}
{Nony}, T., {Motte}, F., {Louvet}, F., {et~al.} 2020, \aap, 636, A38,
  \dodoi{10.1051/0004-6361/201937046}

\bibitem[{{Nozawa} {et~al.}(2006){Nozawa}, {Kozasa}, \&
  {Habe}}]{2006ApJ...648..435N}
{Nozawa}, T., {Kozasa}, T., \& {Habe}, A. 2006, \apj, 648, 435,
  \dodoi{10.1086/505639}

\bibitem[{{Offner} {et~al.}(2009){Offner}, {Klein}, {McKee}, \&
  {Krumholz}}]{2009ApJ...703..131O}
{Offner}, S. S.~R., {Klein}, R.~I., {McKee}, C.~F., \& {Krumholz}, M.~R. 2009,
  \apj, 703, 131, \dodoi{10.1088/0004-637X/703/1/131}

\bibitem[{{Palau} {et~al.}(2014){Palau}, {Zapata}, {Rodr{\'\i}guez}, {Bouy},
  {Barrado}, {Morales-Calder{\'o}n}, {Myers}, {Chapman}, {Ju{\'a}rez}, \&
  {Li}}]{2014MNRAS.444..833P}
{Palau}, A., {Zapata}, L.~A., {Rodr{\'\i}guez}, L.~F., {et~al.} 2014, \mnras,
  444, 833, \dodoi{10.1093/mnras/stu1461}

\bibitem[{{Panoglou} {et~al.}(2012){Panoglou}, {Cabrit}, {Pineau Des
  For{\^e}ts}, {Garcia}, {Ferreira}, \& {Casse}}]{2012A&A...538A...2P}
{Panoglou}, D., {Cabrit}, S., {Pineau Des For{\^e}ts}, G., {et~al.} 2012, \aap,
  538, A2, \dodoi{10.1051/0004-6361/200912861}

\bibitem[{{Park} {et~al.}(2021){Park}, {Lee}, {Contreras Pe{\~n}a},
  {Johnstone}, {Herczeg}, {Lee}, {Lee}, {Bhardwaj}, \&
  {Moriarty-Schieven}}]{2021ApJ...920..132P}
{Park}, W., {Lee}, J.-E., {Contreras Pe{\~n}a}, C., {et~al.} 2021, \apj, 920,
  132, \dodoi{10.3847/1538-4357/ac1745}

\bibitem[{{Plunkett} {et~al.}(2015){Plunkett}, {Arce}, {Mardones}, {van
  Dokkum}, {Dunham}, {Fern{\'a}ndez-L{\'o}pez}, {Gallardo}, \&
  {Corder}}]{2015Plunkett}
{Plunkett}, A.~L., {Arce}, H.~G., {Mardones}, D., {et~al.} 2015, \nat, 527, 70,
  \dodoi{10.1038/nature15702}

\bibitem[{{Podio} {et~al.}(2015){Podio}, {Codella}, {Gueth}, {Cabrit},
  {Bachiller}, {Gusdorf}, {Lee}, {Lefloch}, {Leurini}, {Nisini}, \&
  {Tafalla}}]{2015A&A...581A..85P}
{Podio}, L., {Codella}, C., {Gueth}, F., {et~al.} 2015, \aap, 581, A85,
  \dodoi{10.1051/0004-6361/201525778}

\bibitem[{{Podio} {et~al.}(2016){Podio}, {Codella}, {Gueth}, {Cabrit}, {Maury},
  {Tabone}, {Lef{\`e}vre}, {Anderl}, {Andr{\'e}}, {Belloche}, {Bontemps},
  {Hennebelle}, {Lefloch}, {Maret}, \& {Testi}}]{2016A&A...593L...4P}
---. 2016, \aap, 593, L4, \dodoi{10.1051/0004-6361/201628876}

\bibitem[{{Podio} {et~al.}(2021){Podio}, {Tabone}, {Codella}, {Gueth}, {Maury},
  {Cabrit}, {Lefloch}, {Maret}, {Belloche}, {Andr{\'e}}, {Anderl}, {Gaudel}, \&
  {Testi}}]{2021A&A...648A..45P}
{Podio}, L., {Tabone}, B., {Codella}, C., {et~al.} 2021, \aap, 648, A45,
  \dodoi{10.1051/0004-6361/202038429}

\bibitem[{{Raga} {et~al.}(2002){Raga}, {Vel{\'a}zquez}, {Cant{\'o}}, \&
  {Masciadri}}]{2002A&A...395..647R}
{Raga}, A.~C., {Vel{\'a}zquez}, P.~F., {Cant{\'o}}, J., \& {Masciadri}, E.
  2002, \aap, 395, 647, \dodoi{10.1051/0004-6361:20021180}

\bibitem[{{Riaz} {et~al.}(2018){Riaz}, {Vanaverbeke}, \&
  {Schleicher}}]{2018A&A...614A..53R}
{Riaz}, R., {Vanaverbeke}, S., \& {Schleicher}, D.~R.~G. 2018, \aap, 614, A53,
  \dodoi{10.1051/0004-6361/201732076}

\bibitem[{{Robitaille} \& {Bressert}(2012)}]{2012ascl.soft08017R}
{Robitaille}, T., \& {Bressert}, E. 2012, {APLpy: Astronomical Plotting Library
  in Python}, Astrophysics Source Code Library, record ascl:1208.017.
\newblock \doeprint{1208.017}

\bibitem[{{Rodgers} \& {Charnley}(2003)}]{2003ApJ...585..355R}
{Rodgers}, S.~D., \& {Charnley}, S.~B. 2003, \apj, 585, 355,
  \dodoi{10.1086/345497}

\bibitem[{{Safron} {et~al.}(2015){Safron}, {Fischer}, {Megeath}, {Furlan},
  {Stutz}, {Stanke}, {Billot}, {Rebull}, {Tobin}, {Ali}, {Allen}, {Booker},
  {Watson}, \& {Wilson}}]{safron15}
{Safron}, E.~J., {Fischer}, W.~J., {Megeath}, S.~T., {et~al.} 2015, \apjl, 800,
  L5, \dodoi{10.1088/2041-8205/800/1/L5}

\bibitem[{{Sahu} {et~al.}(2021){Sahu}, {Liu}, {Liu}, {Evans}, {Hirano},
  {Tatematsu}, {Lee}, {Kim}, {Dutta}, {Alina}, {Bronfman}, {Cunningham},
  {Eden}, {Garay}, {Goldsmith}, {He}, {Hsu}, {Jhan}, {Johnstone}, {Juvela},
  {Kim}, {Kuan}, {Kwon}, {Lee}, {Lee}, {Li}, {Li}, {Li}, {Luo}, {Montillaud},
  {Moraghan}, {Pelkonen}, {Qin}, {Ristorcelli}, {Sanhueza}, {Shang}, {Shen},
  {Soam}, {Wu}, {Zhang}, \& {Zhou}}]{2021ApJ...907L..15S}
{Sahu}, D., {Liu}, S.-Y., {Liu}, T., {et~al.} 2021, \apjl, 907, L15,
  \dodoi{10.3847/2041-8213/abd3aa}

\bibitem[{{Santiago-Garc{\'\i}a} {et~al.}(2009){Santiago-Garc{\'\i}a},
  {Tafalla}, {Johnstone}, \& {Bachiller}}]{2009A&A...495..169S}
{Santiago-Garc{\'\i}a}, J., {Tafalla}, M., {Johnstone}, D., \& {Bachiller}, R.
  2009, \aap, 495, 169, \dodoi{10.1051/0004-6361:200810739}

\bibitem[{{Schilke} {et~al.}(1997){Schilke}, {Walmsley}, {Pineau des Forets},
  \& {Flower}}]{1997A&A...321..293S}
{Schilke}, P., {Walmsley}, C.~M., {Pineau des Forets}, G., \& {Flower}, D.~R.
  1997, \aap, 321, 293

\bibitem[{{Shang} {et~al.}(2023){Shang}, {Liu}, {Krasnopolsky}, \&
  {Wang}}]{2023Shang}
{Shang}, H., {Liu}, C.-F., {Krasnopolsky}, R., \& {Wang}, L.-Y. 2023, \apj,
  944, 230, \dodoi{10.3847/1538-4357/aca763}

\bibitem[{{Shu} {et~al.}(1995){Shu}, {Najita}, {Ostriker}, \&
  {Shang}}]{1995ApJ...455L.155S}
{Shu}, F.~H., {Najita}, J., {Ostriker}, E.~C., \& {Shang}, H. 1995, \apjl, 455,
  L155, \dodoi{10.1086/309838}

\bibitem[{{Simon} {et~al.}(2000){Simon}, {Dutrey}, \&
  {Guilloteau}}]{2000ApJ...545.1034S}
{Simon}, M., {Dutrey}, A., \& {Guilloteau}, S. 2000, \apj, 545, 1034,
  \dodoi{10.1086/317838}

\bibitem[{{Skretas} \& {Kristensen}(2022)}]{2022A&A...660A..39S}
{Skretas}, I.~M., \& {Kristensen}, L.~E. 2022, \aap, 660, A39,
  \dodoi{10.1051/0004-6361/202141944}

\bibitem[{{Snell} {et~al.}(1980){Snell}, {Loren}, \&
  {Plambeck}}]{1980ApJ...239L..17S}
{Snell}, R.~L., {Loren}, R.~B., \& {Plambeck}, R.~L. 1980, \apjl, 239, L17,
  \dodoi{10.1086/183283}

\bibitem[{{Stahler}(1988)}]{1988ApJ...332..804S}
{Stahler}, S.~W. 1988, \apj, 332, 804, \dodoi{10.1086/166694}

\bibitem[{{Stamatellos} {et~al.}(2011){Stamatellos}, {Whitworth}, \&
  {Hubber}}]{2011ApJ...730...32S}
{Stamatellos}, D., {Whitworth}, A.~P., \& {Hubber}, D.~A. 2011, \apj, 730, 32,
  \dodoi{10.1088/0004-637X/730/1/32}

\bibitem[{{Stamatellos} {et~al.}(2012){Stamatellos}, {Whitworth}, \&
  {Hubber}}]{2012MNRAS.427.1182S}
---. 2012, \mnras, 427, 1182, \dodoi{10.1111/j.1365-2966.2012.22038.x}

\bibitem[{{Tabone} {et~al.}(2020){Tabone}, {Godard}, {Pineau des For{\^e}ts},
  {Cabrit}, \& {van Dishoeck}}]{2020A&A...636A..60T}
{Tabone}, B., {Godard}, B., {Pineau des For{\^e}ts}, G., {Cabrit}, S., \& {van
  Dishoeck}, E.~F. 2020, \aap, 636, A60, \dodoi{10.1051/0004-6361/201937383}

\bibitem[{{Tafalla} {et~al.}(2015){Tafalla}, {Bachiller}, {Lefloch},
  {Rodr{\'\i}guez-Fern{\'a}ndez}, {Codella}, {L{\'o}pez-Sepulcre}, \&
  {Podio}}]{2015A&A...573L...2T}
{Tafalla}, M., {Bachiller}, R., {Lefloch}, B., {et~al.} 2015, \aap, 573, L2,
  \dodoi{10.1051/0004-6361/201425255}

\bibitem[{{Tafalla} {et~al.}(2017){Tafalla}, {Su}, {Shang}, {Johnstone},
  {Zhang}, {Santiago-Garc{\'\i}a}, {Lee}, {Hirano}, \&
  {Wang}}]{2017A&A...597A.119T}
{Tafalla}, M., {Su}, Y.~N., {Shang}, H., {et~al.} 2017, \aap, 597, A119,
  \dodoi{10.1051/0004-6361/201629493}

\bibitem[{{Takahashi} \& {Ho}(2012)}]{2012ApJ...745L..10T}
{Takahashi}, S., \& {Ho}, P. T.~P. 2012, \apjl, 745, L10,
  \dodoi{10.1088/2041-8205/745/1/L10}

\bibitem[{{Taquet} {et~al.}(2016){Taquet}, {Wirstr{\"o}m}, \&
  {Charnley}}]{2016ApJ...821...46T}
{Taquet}, V., {Wirstr{\"o}m}, E.~S., \& {Charnley}, S.~B. 2016, \apj, 821, 46,
  \dodoi{10.3847/0004-637X/821/1/46}

\bibitem[{{Tobin} {et~al.}(2020){Tobin}, {Sheehan}, {Megeath},
  {D{\'\i}az-Rodr{\'\i}guez}, {Offner}, {Murillo}, {van 't Hoff}, {van
  Dishoeck}, {Osorio}, {Anglada}, {Furlan}, {Stutz}, {Reynolds}, {Karnath},
  {Fischer}, {Persson}, {Looney}, {Li}, {Stephens}, {Chand ler}, {Cox},
  {Dunham}, {Tychoniec}, {Kama}, {Kratter}, {Kounkel}, {Mazur}, {Maud},
  {Patel}, {Perez}, {Sadavoy}, {Segura-Cox}, {Sharma}, {Stephenson}, {Watson},
  \& {Wyrowski}}]{2020ApJ...890..130T}
{Tobin}, J.~J., {Sheehan}, P.~D., {Megeath}, S.~T., {et~al.} 2020, \apj, 890,
  130, \dodoi{10.3847/1538-4357/ab6f64}

\bibitem[{{Wakelam} {et~al.}(2022){Wakelam}, {Coutens}, {Gratier}, {Vidal}, \&
  {Vaytet}}]{2022A&A...666A.191W}
{Wakelam}, V., {Coutens}, A., {Gratier}, P., {Vidal}, T.~H.~G., \& {Vaytet}, N.
  2022, \aap, 666, A191, \dodoi{10.1051/0004-6361/202243459}

\bibitem[{{Wiebe} {et~al.}(2019){Wiebe}, {Molyarova}, {Akimkin}, {Vorobyov}, \&
  {Semenov}}]{2019MNRAS.485.1843W}
{Wiebe}, D.~S., {Molyarova}, T.~S., {Akimkin}, V.~V., {Vorobyov}, E.~I., \&
  {Semenov}, D.~A. 2019, \mnras, 485, 1843, \dodoi{10.1093/mnras/stz512}

\bibitem[{{Yang} {et~al.}(2010){Yang}, {Stancil}, {Balakrishnan}, \&
  {Forrey}}]{2010ApJ...718.1062Y}
{Yang}, B., {Stancil}, P.~C., {Balakrishnan}, N., \& {Forrey}, R.~C. 2010,
  \apj, 718, 1062, \dodoi{10.1088/0004-637X/718/2/1062}

\bibitem[{{Yen} {et~al.}(2017){Yen}, {Koch}, {Takakuwa}, {Krasnopolsky},
  {Ohashi}, \& {Aso}}]{2017ApJ...834..178Y}
{Yen}, H.-W., {Koch}, P.~M., {Takakuwa}, S., {et~al.} 2017, \apj, 834, 178,
  \dodoi{10.3847/1538-4357/834/2/178}

\bibitem[{{Y{\i}ld{\i}z} {et~al.}(2015){Y{\i}ld{\i}z}, {Kristensen}, {van
  Dishoeck}, {Hogerheijde}, {Karska}, {Belloche}, {Endo}, {Frieswijk},
  {G{\"u}sten}, {van Kempen}, {Leurini}, {Nagy}, {P{\'e}rez-Beaupuits},
  {Risacher}, {van der Marel}, {van Weeren}, \&
  {Wyrowski}}]{2015A&A...576A.109Y}
{Y{\i}ld{\i}z}, U.~A., {Kristensen}, L.~E., {van Dishoeck}, E.~F., {et~al.}
  2015, \aap, 576, A109, \dodoi{10.1051/0004-6361/201424538}

\bibitem[{{Yoo} {et~al.}(2017){Yoo}, {Lee}, {Mairs}, {Johnstone}, {Herczeg},
  {Kang}, {Kang}, {Cho}, \& {The JCMT Transient Team}}]{yoo17}
{Yoo}, H., {Lee}, J.-E., {Mairs}, S., {et~al.} 2017, \apj, 849, 69,
  \dodoi{10.3847/1538-4357/aa8c0a}

\bibitem[{{Yoon} {et~al.}(2022){Yoon}, {Herczeg}, {Lee}, {Lee}, {Johnstone},
  {Varricatt}, {Tobin}, {Contreras Pe{\~n}a}, {Mairs}, {Hodapp}, {Manoj},
  {Osorio}, {Megeath}, \& {JCMT Transient Team}}]{2022yoon}
{Yoon}, S.-Y., {Herczeg}, G.~J., {Lee}, J.-E., {et~al.} 2022, \apj, 929, 60,
  \dodoi{10.3847/1538-4357/ac5632}

\bibitem[{{Yvart} {et~al.}(2016){Yvart}, {Cabrit}, {Pineau des For{\^e}ts}, \&
  {Ferreira}}]{2016A&A...585A..74Y}
{Yvart}, W., {Cabrit}, S., {Pineau des For{\^e}ts}, G., \& {Ferreira}, J. 2016,
  \aap, 585, A74, \dodoi{10.1051/0004-6361/201525915}

\bibitem[{{Zapata} {et~al.}(2018){Zapata}, {Fern{\'a}ndez-L{\'o}pez},
  {Rodr{\'\i}guez}, {Garay}, {Takahashi}, {Lee}, \&
  {Hern{\'a}ndez-G{\'o}mez}}]{2018AJ....156..239Z}
{Zapata}, L.~A., {Fern{\'a}ndez-L{\'o}pez}, M., {Rodr{\'\i}guez}, L.~F.,
  {et~al.} 2018, \aj, 156, 239, \dodoi{10.3847/1538-3881/aae51e}

\bibitem[{{Zhang} {et~al.}(2019){Zhang}, {Arce}, {Mardones}, {Cabrit},
  {Dunham}, {Garay}, {Noriega-Crespo}, {Offner}, {Raga}, \&
  {Corder}}]{2019ApJ...883....1Z}
{Zhang}, Y., {Arce}, H.~G., {Mardones}, D., {et~al.} 2019, \apj, 883, 1,
  \dodoi{10.3847/1538-4357/ab3850}

\bibitem[{{Zinnecker} {et~al.}(1998){Zinnecker}, {McCaughrean}, \&
  {Rayner}}]{1998Natur.394..862Z}
{Zinnecker}, H., {McCaughrean}, M.~J., \& {Rayner}, J.~T. 1998, \nat, 394, 862,
  \dodoi{10.1038/29716}

\end{thebibliography}
\bibliographystyle{aasjournal}
\end{document}